\documentclass[11pt]{article}
\pdfoutput=1

\usepackage[usenames,dvipsnames,table]{xcolor}
\usepackage[utf8]{inputenc}

\usepackage{jheppub} 

\usepackage{tikz-cd} 
\usepackage{hyperref}
\usepackage{bookmark}
\usepackage{circuitikz}
\usepgfmodule{shapes}
\usepackage{ytableau}
\usepackage{booktabs}
\usetikzlibrary{decorations.pathmorphing}
\usetikzlibrary{decorations.pathreplacing,decorations.markings}
\usetikzlibrary{backgrounds, arrows,calc,shapes,decorations.pathreplacing, automata,positioning}
\usetikzlibrary{positioning, arrows.meta, calc}
\usepackage{xcolor}
\usetikzlibrary{shapes.misc}
\usepackage[thmmarks,amsmath]{ntheorem}
\newtheorem*{remark}{Remark}

\tikzset{cross/.style={cross out, draw=black, minimum size=5*(#1-\pgflinewidth), inner sep=0pt, outer sep=0pt},
cross/.default={2pt}}

\tikzset{snake it/.style={decorate, decoration=snake}}

\tikzset{mid arrow/.style={postaction={decorate,decoration={
        markings,
        mark = at position .55 with {\arrow[#1]{Straight Barb[width=5pt]}}
      }}}}
      
\tikzset{mid arrowsm/.style={postaction={decorate,decoration={
        markings,
        mark = at position .55 with {\arrow[#1]{Straight Barb[width=3pt]}}
      }}}}
      
\tikzset{middx arrowsm/.style={postaction={decorate,decoration={
        markings,
        mark = at position .7 with {\arrow[#1]{Straight Barb[width=3pt]}}
      }}}}
      
\tikzset{midsx arrowsm/.style={postaction={decorate,decoration={
        markings,
        mark = at position .4 with {\arrow[#1]{Straight Barb[width=3pt]}}
      }}}}

\newtheorem{theorem}{Theorem}

\usepackage{graphicx}
\usepackage{amsmath, amssymb}
\usepackage{youngtab}
\usepackage{changepage}
\usepackage{mathabx} 

\newcommand{\be}{\begin{equation}}
\newcommand{\ee}{\end{equation}}
\newcommand{\ba}{\begin{array}}
\newcommand{\ea}{\end{array}}
\newcommand{\bea}{\begin{eqnarray}}
\newcommand{\eea}{\end{eqnarray}}
\newcommand{\bpic}{\begin{tikzpicture}}
\newcommand{\epic}{\end{tikzpicture}}

\newcommand{\bn}{\begin{enumerate}}
\newcommand{\en}{\end{enumerate}}

\def\CH{{\cal H}}
\def\CI{{\cal I}}

\def\cN{{\cal N}}

\def\CS{{\cal S}}

\def\cW{{\cal W}}

\def\D{\Delta}

\def\M{{\mathfrak M}}

\def\qt{\tilde{q}}

\def\pt{\tilde{p}}

\title{Hilbert Series and Superconformal Indices of the Improved Bifundamentals}

\author[1,2]{Sergio Benvenuti}
\author[1,2]{Gabriel Pedde Ungureanu}

\affiliation[1]{INFN, Sezione di Trieste, Via Valerio 2, 34127 Trieste, Italy}
\affiliation[2]{International School of Advanced Studies (SISSA), Via Bonomea 265, 34136 Trieste, Italy}
\emailAdd{benve79@gmail.com, gpeddeun@sissa.it}

\abstract{We explore the structure of the moduli space of vacua of \emph{Improved Bifundamentals}, a recently introduced class of superconformal field theories (SCFTs). Utilising the Hilbert Series, computed as a specific limit of the Superconformal Index, we establish that the moduli spaces of these theories are simple algebraic varieties, presenting a single connected component for three of the families studied ($FT_N, FC_N, FH_N$) and a main branch plus simple branches generated by singlets for the remaining families ($FM_N, FE_N$).}

\begin{document}

\maketitle

\section{Introduction and summary}
\label{intro}

Recent investigations have introduced novel classes of superconformal field theories (SCFTs) with four supercharges, termed \emph{Improved Bifundamentals} \cite{Aprile:2018oau, Pasquetti:2019tix, Pasquetti:2019hxf, Benvenuti:2024mpn}. Among them, the $FE_N$ series exists in both 4d and 3d, while the $FM_N$, $FH_N$, $FC_N$, and $FT_N$ families are purely three-dimensional. These Improved Bifundamentals exhibit intriguing star-triangle dualities \cite{Benvenuti:2024mpn} and serve as fundamental building blocks for constructing quivers relevant to various generalisations of 3d mirror symmetry \cite{Hwang:2021ulb, Comi:2022aqo, Benvenuti:2023qtv}, the compactification of 6d E-strings on Riemann surfaces \cite{Pasquetti:2019tix}, and Hanany--Witten brane setups with four supercharges \cite{Benvenuti:2023qtv}.

This paper aims to elucidate the structure of the moduli space of vacua for the Improved Bifundamentals. Our analysis reveals that the moduli spaces are irreducible algebraic varieties, exhibiting a single branch rather than the multiple intersecting branches typically found in SCFT moduli spaces. Specifically, for $FT_N$, $FC_N$, and $FH_N$, the moduli space consists of a single branch generated by operators in rank-2 representations of the infrared (IR) non-abelian global symmetry. In contrast, for $FM_N$ and $FE_N$, in addition to this \emph{main branch}, we identify additional simple branches associated with vacuum expectation values of the $B_{ij}$ operators, which are singlets under the non-abelian global symmetry.

To explore the moduli space of vacua, we employ the Hilbert series (HS), a powerful mathematical tool initially introduced in the context of 4d supersymmetric quantum field theories in \cite{Benvenuti:2006qr, Feng:2007ur}.\footnote{The plethystic program has notably been applied to study the moduli space of supersymmetric QCD \cite{Gray_2008} and instanton moduli spaces \cite{Benvenuti:2010pq, Hanany_2013}. Its versatility has also led to various applications in both 4d \cite{Hanany:2010qu} and 3d \cite{Hanany:2015via, Hanany:2011db, Cremonesi:2015dja} supersymmetric gauge theories; for a recent overview of further applications, see \cite{Bao_2022}.} In this context, the HS enumerates holomorphic functions on the moduli space, which are in one-to-one correspondence with chiral ring operators. Consequently, the HS also encodes the relations among these operators, which can be either classical or quantum in nature. In three dimensions, the presence of local monopole operators \cite{Borokhov_2002, Borokhov_2002sac} complicates the computation of the Hilbert series, generally requiring a summation over monopole sectors. While this problem has been solved for 3d $\mathcal{N}=4$ gauge theories \cite{Cremonesi_2014}, it remains largely open for 3d $\mathcal{N}=2$ theories (see \cite{Hanany:2015via, Cremonesi:2015dja}).

Our aim is to compute the Hilbert series (HS) of the moduli space, including contributions from monopole operators and quantum relations. The strategy we pursue is to extract the HS from a suitable limit of the superconformal index (SCI), i.e.\ the partition function of the theory localised on $\mathbf{S}^2 \times \mathbf{S}^1$~\cite{Bhattacharya:2008,Imamura:2011,Kapustin:2011}. The rationale is standard: the coordinate ring of the moduli space is finitely generated by $\tfrac12$-BPS gauge-invariant chiral operators — bare and dressed monopoles included — subject to $F$-term and quantum relations. These operators sit in short multiplets of the superconformal algebra and are therefore counted by the index, graded by their $R$-charge and global-symmetry quantum numbers~\cite{Cordova:2016,Gadde:2020yah}.

Whether the HS can in fact be recovered from the SCI depends on the amount of supersymmetry and on the branch one wishes to probe, and several obstructions are known. For $3d$ $\mathcal{N}=4$ theories, \cite{Razamat:2014} identified two distinct limits of the SCI reproducing the Coulomb and Higgs branches separately. Already at eight supercharges, however, isolating operators supported on mixed branches is subtle~\cite{Bachas:2019jaa}. A related difficulty arises in class $\mathcal{S}$: for the untwisted theories considered in~\cite{Cremonesi:2014vla}, agreement between the Higgs-branch HS and the Hall--Littlewood limit of the index was found at genus zero but not at higher genus. Genus zero is not sufficient in full generality, as counterexamples with twisted punctures are known~\cite{Kang:2022}. For $3d$ $\mathcal{N}=2$ theories with global $U(1)$ symmetries, an analogous branch-by-branch analysis was developed in~\cite{Hanany:2015via}, but here an additional obstruction appears: cancellations between bosonic and fermionic contributions to the index can obscure chiral-ring generators, so that the index undercounts the moduli-space HS.

For the class of theories studied in this paper, we argue — and verify in the explicit examples below — that no such accidental cancellations occur. As a consequence, the appropriate limit of the SCI reproduces the HS of either the full moduli space or, in the cases where mixed branches are present, of its main branch. We first summarise our results; the limit is defined precisely in Section~\ref{sec:HS_limit}, and its practical evaluation is described in Section~\ref{sec:practical}.

\subsection*{Summary of the results}
For the $FT_N$ theory of \cite{Aprile:2018oau}, which is the only series whose moduli spaces are complete intersections, we produce a simple formula for the unrefined HS for any $N$, analogous to the formula for the Higgs branch of $T[SU(N)]$ \cite{Hanany:2011db}:
$$\mathcal{HS}_{FT_N}(t)= PE\left[\chi_{[1,0,\ldots,0,1]_{SU(N)_L}}t+\chi_{[1,0,\ldots,0,1]_{SU(N)_R}} t-\sum_{j=2}^Nt^j \right] \overset{unr}{=}
\frac{\prod_{j=1}^{N-1}\sum_{h=0}^j t^h}{(1-t)^{2N^2-N-1}}\,.$$
Note that the full moduli space of $T[SU(N)]$ is a union of multiple branches, whereas the moduli space of $FT_N$, obtained by flipping the adjoint of $T[SU(N)]$ generating the Higgs or Coulomb branch, possesses a single branch.

For the other improved bifundamentals (if $N \geq 3$), we are not able to write the fully unrefined HS, but we compute the HS expressed in the so-called \emph{Euler form} (in a single fugacity):
    \be
    \mathcal{HS}(t; \mathcal{M})= \frac{h_0 +h_1 t+\dots +h_d t^d}{(1-t)^{\dim_{\mathcal{M}}}},
    \ee
and write the form explicitly for $N \leq 3$.

The improved bifundamentals at $N=1$ are simple free theories\footnote{In this paper, we consider versions of the $FM_N$ and $FE_N$ theories without the $B_{11}$ singlet, as appropriate for studying the isolated $FM_N$ and $FE_N$ theories. This implies that $FM_1$ and $FE_1$ are equivalent to $2$ and $4$ free chiral superfields, respectively, rather than cubic Wess--Zumino models with $3$ and $5$ chiral superfields.}, rendering the computation of the HS trivial. For $N=2$, thanks to non-generic symmetry enhancements, we can write the fully unrefined HS as sums over characters.

We observe that since the number of chiral ring generators grows quadratically with $N$, the dimension of the moduli space and the degree of the numerator also exhibit quadratic growth with $N$. This fact allows us to extrapolate these quantities for generic $N$ based on the $N=1, 2, 3$ results. Let us write the unrefined HS for $N=1, 2, 3$ in the following table:

\definecolor{headbg}{RGB}{220, 230, 242}
\begin{center}
\resizebox{\textwidth}{!}{
\begin{tabular}{>{\bfseries}c c c c}
\toprule
\rowcolor{headbg}
 & $N=1$ & $N=2$ & $N=3$ \\
\midrule
$FT_N$ 
    & $1$ 
    & $\dfrac{1+t}{(1-t)^5}$ 
    & $\dfrac{(1+t)(1+t+t^2)}{(1-t)^{14}}$ \\[0.6em]

$FC_N$ 
    & $\dfrac{1}{1-t}$ 
    & $\dfrac{1+3t+t^2}{(1-t)^7}$ 
    & $\dfrac{1+8t+26t^2+39t^3+26t^4+8t^5+t^6}{(1-t)^{17}}$ \\[0.6em]

$FH_N$ 
    & $\dfrac{1}{(1-t)^2}$ 
    & $\dfrac{(1+t)(1+4t+t^2)}{(1-t)^{11}}$ 
    & $\dfrac{(1+t)^3(1+3t+t^2)(1+8t+22t^2+8t^3+t^4)}{(1-t)^{26}}$ \\[0.6em]

$FM_N$ 
    & $\dfrac{1}{(1-t)^2}$ 
    & $\dfrac{(1+t)(1+5t+t^2)}{(1-t)^{8}}$ 
    & $\dfrac{(1+t)^2(1+14t+72t^2+133t^3+72t^4+14t^5+t^6)}{(1-t)^{18}}$ \\[0.6em]

$FE_N$ 
    & $\dfrac{1}{(1-t)^4}$ 
    & $\dfrac{(1+t)(1+9t+19t^2+9t^3+t^4)}{(1-t)^{16}}$ 
    & $\dfrac{(1+t)^2(1+26t+288t^2+1716t^3+5970t^4+12545t^5+16071t^6+\cdots+t^{12})}{(1-t)^{36}}$ \\
\bottomrule
\end{tabular}
}
\end{center}
\vspace{0.2cm}

The extrapolation to generic $N$ yields the following results for the complex dimension of the moduli space, the number of generators, and the degree of the numerator of the HS:

\begin{center}
\renewcommand{\arraystretch}{1.5}
\resizebox{\textwidth}{!}{%
\begin{tabular}{>{\bfseries}c c c c c c}
\toprule
\rowcolor{headbg}
  \text{Ref.}
  & \textbf{Theory}
  & $\boldsymbol{\dim_{\mathcal{M}}}$
  & \textbf{Generators}
  & \textbf{Numerator degree} $\boldsymbol{d}$
  & \textbf{Global symmetry} ($N \geq 3$) \\
\midrule
  \eqref{ftsun}
  & $FT_N$
  & $2N^2-N-1$
  & $2N^2-2$
  & $N(N-1)/2$
  & $S({\color{blue!70!black}U(N)}\times{\color{red!70!black}U(N)})$ \\[0.3em]

  \eqref{fcsun}
  & $FC_N$
  & $2N^2-1$
  & $3N^2-2$
  & $N(N-1)$
  & $S({\color{blue!70!black}U(N)}\times{\color{red!70!black}U(N)})\times U(1)$ \\[0.3em]

  \eqref{fhsun}
  & $FH_N$
  & $3N^2-1$
  & $5N^2-N-2$
  & $3N(N-1)/2$
  & ${\color{blue!70!black}SU(N)}\times{\color{red!70!black}USp(2N)}\times U(1)^2$ \\[0.3em]

  \eqref{fmsun}
  & $FM_N$
  & $2N^2$
  & $4N^2-2$
  & $N^2-1$
  & $S({\color{blue!70!black}U(N)}\times{\color{red!70!black}U(N)})\times U(1)^2$ \\[0.3em]

  \eqref{feuspn}
  & $FE_N$
  & $4N^2$
  & $8N^2-2N-2$
  & $(2N+1)(N-1)$
  & ${\color{blue!70!black}USp(2N)}\times{\color{red!70!black}USp(2N)}\times U(1)^2$ \\

\bottomrule
\end{tabular}}
\end{center}

The five families are related by RG flows triggered by vacuum expectation values together with real mass deformations; the resulting pattern is summarised in Figure~\ref{fig:theory-relations}.

\begin{figure}[h]
\centering\begin{tikzpicture}[
    scale=0.75, transform shape,
    block/.style={
        draw=#1!60!black,
        fill=#1!12,
        rounded corners=6pt,
        align=center,
        inner sep=10pt,
        font=\small,
        anchor=north,
        minimum width=6.5cm,
        line width=0.8pt
    },
    block/.default=gray,
    arrowLine/.style={
        draw=gray!60!black,
        thick,
        ->,
        >=Stealth,
        shorten >=2pt,
        shorten <=2pt
    }
]

\node[block=purple] (FEN) at (0,0) {
    \textbf{\large $FE_N$} \enspace
    \textcolor{blue!70!black}{$USp(2N)$} $\times$
    \textcolor{red!70!black}{$USp(2N)$} $\times U(1)^2$
    \\[0.4em]
    \textcolor{gray!60!black}{\rule{5cm}{0.3pt}}\\[0.3em]
    $\dim_{\mathcal{M}} = 4N^2$
    \quad Gens: $8N^2 - 2N - 2$
    \quad Deg: $(2N+1)(N-1)$
};

\node[block=orange, below right=1.8cm and -4.6cm of FEN] (FHN) {
    \textbf{\large $FH_N$} \enspace
    \textcolor{blue!70!black}{$SU(N)$} $\times$
    \textcolor{red!70!black}{$USp(2N)$} $\times U(1)^2$
    \\[0.4em]
    \textcolor{gray!60!black}{\rule{5cm}{0.3pt}}\\[0.3em]
    $\dim_{\mathcal{M}} = 3N^2 - 1$
    \quad Gens: $5N^2 - N - 2$
    \quad Deg: $3N(N-1)/2$
};

\node[block=teal, below left=1.8cm and -4.6cm of FEN] (FMN) {
    \textbf{\large $FM_N$} \enspace
    $S(\textcolor{blue!70!black}{U(N)} \times \textcolor{red!70!black}{U(N)}) \times U(1)^2$
    \\[0.4em]
    \textcolor{gray!60!black}{\rule{5cm}{0.3pt}}\\[0.3em]
    $\dim_{\mathcal{M}} = 2N^2$
    \quad Gens: $4N^2 - 2$
    \quad Deg: $N^2 - 1$
};

\node[block=cyan, below=1.8cm of $(FMN.south)!0.5!(FHN.south)$] (FCN) {
    \textbf{\large $FC_N$} \enspace
    $S(\textcolor{blue!70!black}{U(N)} \times \textcolor{red!70!black}{U(N)}) \times U(1)$
    \\[0.4em]
    \textcolor{gray!60!black}{\rule{5cm}{0.3pt}}\\[0.3em]
    $\dim_{\mathcal{M}} = 2N^2 - 1$
    \quad Gens: $3N^2 - 2$
    \quad Deg: $N(N-1)$
};

\node[block=blue, below=1.5cm of FCN] (FTN) {
    \textbf{\large $FT_N$} \enspace
    $S(\textcolor{blue!70!black}{U(N)} \times \textcolor{red!70!black}{U(N)})$
    \\[0.4em]
    \textcolor{gray!60!black}{\rule{5cm}{0.3pt}}\\[0.3em]
    $\dim_{\mathcal{M}} = 2N^2 - N - 1$
    \quad Gens: $2N^2 - 2$
    \quad Deg: $N(N-1)/2$
};

\draw[arrowLine] (FEN.south) -- (FMN.north);
\draw[arrowLine] (FEN.south) -- (FHN.north);
\draw[arrowLine] (FMN.south) -- (FCN.north);
\draw[arrowLine] (FHN.south) -- (FCN.north);
\draw[arrowLine] (FCN.south) -- (FTN.north);

\end{tikzpicture}
\caption{Diagram with the relations among the different theories. The arrows correspond to RG flows triggered by VEVs \textit{together} with real mass deformations.
Each block shows the global symmetry (for $N \ge 3$), the complex dimension of the moduli space, the number of generators, and the degree of the numerator of the HS in Euler form.}
\label{fig:theory-relations}
\end{figure}

\subsection*{Further symmetry enhancements at $N=2$}
The improved bifundamentals at $N=2$ exhibit an IR symmetry larger than the generic-$N$ IR global symmetry, as discussed for $FE_2$ in \cite{Hwang:2020ddr}. One way to argue for this enhanced symmetry is through the \emph{exceptional} 3d dualities of \cite{Amariti:2018wht, Benvenuti:2018bav}. In Section \ref{n=2impbif}, we derive simple formulae for the fully unrefined Hilbert series at $N=2$ as sums over characters of representations of the enhanced global symmetry group, generalising the Hilbert series for the $FT_2$ theory found in \cite{Benini:2018bhk}.

In addition to $FE_2$ and $FM_2$, we consider the $\widetilde{FE}_{2}$ and $\widetilde{FM}_{2}$ theories, defined by flipping the singlet $B_{2,1}$ in $FE_2$ and $FM_2$, respectively. This increases the dimension of the moduli space by one. The moduli spaces of the five theories $\widetilde{FE}_{2}$, $\widetilde{FM}_{2}$, $FH_2$, $FC_2$, and $FT_2$ are complex cones over \emph{irreducible Hermitian symmetric spaces}\footnote{We are grateful to Richard Eager for useful correspondence about this point.} \cite{Eager:2018czl, Eager:2019zrc}.

A \emph{Hermitian symmetric space} (HSS) of compact type is a compact complex 
manifold $B = H/K$, where $H$ is a compact simple Lie group and $K$ a 
closed subgroup of maximal rank whose centre contains a circle --- equivalently, 
the centraliser of a $U(1)\subset H$ --- such that $B$ 
carries an $H$-invariant Hermitian metric and each point is a fixed point 
of a global involutive isometry. The central $U(1) \subset K$ endows the 
tangent space with a complex structure, making $B$ a K\"{a}hler manifold. 
These spaces were classified by Cartan \cite{Cartan1926, Cartan1935} into four infinite families and 
two exceptional cases:
\begin{equation}
    \frac{SU(p+q)}{S(U(p)\times U(q))}, \quad 
    \frac{SO(2n)}{U(n)}, \quad 
    \frac{USp(2n)}{U(n)}, \quad 
    \frac{SO(p+2)}{SO(p)\times U(1)}, \quad 
    \frac{E_6}{SO(10)\times U(1)}, \quad 
    \frac{E_7}{E_6\times U(1)}.
\end{equation}
What makes these spaces especially natural as moduli spaces is their 
representation-theoretic rigidity: writing $\mathcal{M}$ for the affine 
cone over $H/K$ in its minimal homogeneous embedding, the coordinate ring 
$\mathbb{C}[\mathcal{M}]$ decomposes as a \emph{multiplicity-free} 
sum of irreducible $H$-representations~\cite{Landsberg:2001,Weyman:2003},
\begin{equation}
    \mathbb{C}[\mathcal{M}] = \bigoplus_{k=0}^{\infty} V_{k\lambda}^{*},
\end{equation}
where $R=V_{\lambda}^{*}$ is the irreducible representation defining the embedding and $V_{k\lambda}^{*}$ is its $k$-th Cartan power (the highest-weight summand of $\mathrm{Sym}^{k}R$). 
This is precisely the structure captured by a Hilbert series of the form 
$\mathrm{HS}(t) = \sum_{k=0}^{\infty} \chi_{V_{k\lambda}^{*}}\, t^k$, 
which is what we observe for the five theories listed above. The 
generators of the chiral ring are identified with the elements of $R$, 
and the relations they satisfy cut out $\mathcal{M}$ inside the free 
ring $\mathrm{Sym}^\bullet(R)$. For the five theories at hand, the 
relevant spaces and their generator representations are given below.\footnote{Three of the five spaces carry a well-established
magic-square signature: curved $\beta\gamma$ systems on the cones 
over $\mathrm{Gr}(2,4)$, $SO(10)/U(5)$, and 
$E_6/(SO(10)\times U(1))$ exhibit hidden affine 
$\hat{\mathfrak{so}}(8)$, $\hat{\mathfrak{e}}_6$, and 
$\hat{\mathfrak{e}}_7$ symmetry respectively, where the last case 
is precisely the pure spinor sector of the superstring 
\cite{Eager:2019zrc}. This fits the general pattern of 
Landsberg--Manivel \cite{Landsberg:2001} projective geometries associated to the 
Freudenthal--Tits magic square \cite{Freudenthal1954, Tits1966} arising as moduli spaces of 
supersymmetric gauge theories \cite{Eager:2018czl}. The remaining 
spaces $\mathrm{Gr}(2,5)$ and $\mathrm{Gr}(2,6)$ belong to the 
same family of rank-two cominuscule Hermitian symmetric spaces and 
complete the Grassmannian chain 
$\mathrm{Gr}(2,4)\subset\mathrm{Gr}(2,5)\subset\mathrm{Gr}(2,6)$ 
and the branching $SU(5)\subset SO(10)$.}
\begin{equation}
\begin{array}{cccc}
    \toprule
    \text{Ref.} & \text{Theory} & B = H/K & R \\
    \midrule
    \eqref{ftsu2} & FT_2   & SU(4)/S(U(2){\times}U(2))   & \mathbf{6}  \\
    \eqref{fcsu2} & FC_2   & SU(5)/S(U(2){\times}U(3))   & \mathbf{10} \\
    \eqref{fh2} &  FH_2   & SO(10)/U(5)                  & \mathbf{16} \\
    \eqref{fm2} &  \widetilde{FM}_2 & SU(6)/S(U(2){\times}U(4)) & \mathbf{15} \\
    \eqref{fe2} &  \widetilde{FE}_2 & E_6/(SO(10){\times}U(1))  & \mathbf{27} \\
    \bottomrule
\end{array}
\end{equation}

Let us summarise the results in Figure \ref{fig:hasse-N2}, where the symmetric space is shown in orange, the HS in blue, and $\dim_{\mathcal{M}}$ denotes the complex dimension of the conical moduli space.

\begin{figure}[h]
\centering
\begin{tikzpicture}[
    scale=0.82, transform shape,
    block/.style={
        draw=#1!60!black,
        fill=#1!10,
        rounded corners=6pt,
        align=center,
        inner sep=10pt,
        font=\small,
        anchor=north,
        minimum width=5.4cm,
        text width=5.1cm,
        line width=0.8pt
    },
    block/.default=gray,
    arrowLine/.style={
        draw=gray!50!black,
        thick,
        ->,
        >=Stealth,
        shorten >=2pt,
        shorten <=2pt
    }
]

\newcommand{\hilbert}[1]{{\color{blue!70!black}#1}}
\newcommand{\coset}[1]{{\color{orange!80!black}#1}}
\newcommand{\sep}{\textcolor{gray!50!black}{\rule{4.7cm}{0.3pt}}\\[0.2em]}

\node[block=purple] (FE2_tilde) at (-4.2, 0) {
    \textbf{$\widetilde{FE}_{2}$} \enspace
    \coset{$\mathcal{M}=\mathcal{C}\!\left[\tfrac{E_6}{SO(10)\times U(1)}\right]$}
    \\[0.35em]\sep
    \hilbert{$\displaystyle\sum_{k=0}^{\infty}
        \chi_{[k,0,0,0,0,0]_{E_6}}\, t^k$}
    \\[0.35em]
    $27$ gen,\; $27$ eqs,\; $\dim_{\mathcal{M}}=17$
};

\node[block=purple] (FE2) at (4.2, 0) {
    \textbf{$FE_{2}$}
    \\[0.35em]\sep
    \hilbert{$\displaystyle\sum_{k,l=0}^{\infty}
        \chi_{[k,0,0,0,l]_{so(10)}}\,t_a^k\, t_d^l$}
    \\[0.35em]
    $26$ gen,\; $27$ eqs,\; $\dim_{\mathcal{M}}=16$
};

\node[block=teal] (FM2_tilde) at (-5.8, -5) {
    \textbf{$\widetilde{FM}_{2}$} \enspace
    \coset{$\mathcal{M}=\mathcal{C}\!\left[\tfrac{SU(6)}{S(U(2)\times U(4))}\right]$}
    \\[0.35em]\sep
    \hilbert{$\displaystyle\sum_{k=0}^{\infty}
        \chi_{[0,k,0,0,0]_{su(6)}}\, t^k$}
    \\[0.35em]
    $15$ gen,\; $15$ eqs,\; $\dim_{\mathcal{M}}=9$
};

\node[block=orange] (FH2) at (0, -5) {
    \textbf{$FH_{2}$} \enspace
    \coset{$\mathcal{M}=\mathcal{C}\!\left[\tfrac{SO(10)}{U(5)}\right]$}
    \\[0.35em]\sep
    \hilbert{$\displaystyle\sum_{k=0}^{\infty}
        \chi_{[0,0,0,0,k]_{so(10)}}\, t^k$}
    \\[0.35em]
    $16$ gen,\; $10$ eqs,\; $\dim_{\mathcal{M}}=11$
};

\node[block=teal] (FM2) at (5.8, -5) {
    \textbf{$FM_{2}$}
    \\[0.35em]\sep
    \hilbert{$\displaystyle\sum_{k,l=0}^{\infty}
        \chi_{[l,k,0;\,l]_{su(4)\times su(2)}}\,t_a^k\, t_d^l$}
    \\[0.35em]
    $14$ gen,\; $15$ eqs,\; $\dim_{\mathcal{M}}=8$
};

\node[block=cyan] (FC2) at (0, -10.2) {
    \textbf{$FC_{2}$} \enspace
    \coset{$\mathcal{M}=\mathcal{C}\!\left[\tfrac{SU(5)}{S(U(2)\times U(3))}\right]$}
    \\[0.35em]\sep
    \hilbert{$\displaystyle\sum_{k=0}^{\infty}
        \chi_{[0,k,0,0]_{su(5)}}\, t^k$}
    \\[0.35em]
    $10$ gen,\; $5$ eqs,\; $\dim_{\mathcal{M}}=7$
};

\node[block=blue] (FT2) at (0, -15) {
    \textbf{$FT_{2}$} \enspace
    \coset{$\mathcal{M}=\mathcal{C}\!\left[\tfrac{SU(4)}{S(U(2)\times U(2))}\right]$}
    \\[0.35em]\sep
    \hilbert{$\displaystyle\sum_{k=0}^{\infty}
        \chi_{[0,k,0]_{su(4)}}\, t^k$}
    \\[0.35em]
    $6$ gen,\; $1$ eq,\; $\dim_{\mathcal{M}}=5$
};

\draw[arrowLine] (FE2_tilde.south) -- (FM2_tilde.north);
\draw[arrowLine] (FE2_tilde.south) -- (FH2.north);
\draw[arrowLine] (FE2.south)       -- (FM2.north);
\draw[arrowLine] (FE2.south)       -- (FH2.north);
\draw[arrowLine] (FM2_tilde.south) -- (FC2.west);
\draw[arrowLine] (FH2.south)       -- (FC2.north);
\draw[arrowLine] (FM2.south)       -- (FC2.east);
\draw[arrowLine] (FC2.south)       -- (FT2.north);

\end{tikzpicture}
\caption{Hasse diagram of the $N=2$ theories. {\color{orange!80!black}Orange} labels
indicate the associated moduli space as a coset, and {\color{blue!70!black}blue} expressions
give the corresponding Hilbert series. Arrows denote RG flows and specialisations
between theories.}
\label{fig:hasse-N2}
\end{figure}

\subsection*{Organisation of the paper}
In Section \ref{impbif}, we review the class of theories under consideration and highlight their most relevant features.

Section \ref{sec:computation} defines the limit of the superconformal index that yields the Hilbert series, describes how the indices are computed and how the limit is evaluated in practice, and how the closed forms of the Hilbert series are obtained from the resulting data.

Section \ref{genNimpbif} provides the Hilbert series as a function of two variables, effectively unrefining the non-abelian fugacities. It also presents the Euler form of the HS, explores its key properties, and provides a detailed list of chiral ring generators (already listed in \cite{Aprile:2018oau, Pasquetti:2019tix, Pasquetti:2019hxf, Benvenuti:2024mpn}) along with the relations they satisfy. For $FT_N$, we also derive the closed form of the HS for all $N$ by evaluating the limit of the index directly on the localisation formula of the UV quiver (Section~\ref{sec:FT_limit}).

Section \ref{n=2impbif} gives special attention to the lowest-rank ($N=2$) examples, which exhibit unique symmetry enhancement patterns. In this section, we leverage the enlarged IR global symmetries to express the fully refined Hilbert series as a sum over representations of the enhanced global symmetry group.

Appendix \ref{appHS} reviews various algebro-geometric concepts useful for working with Hilbert series, as well as the plethystic exponential and logarithm.

Appendices \ref{AppN=3} and \ref{AppN=4} report our computations of the SCI and HS for the improved bifundamentals at $N=3$ and $N=4$.

\section{Short review of the \emph{Improved Bifundamental} theories}
\label{impbif}

The \emph{Improved Bifundamentals} are a class of superconformal field theories (SCFTs) with four supercharges, introduced in \cite{Aprile:2018oau, Pasquetti:2019tix, Pasquetti:2019hxf, Benvenuti:2024mpn}, that admit quiver UV Lagrangian descriptions. They share interesting properties such as symmetry enhancement in the IR, (self-)duality of their UV completions under 3d $\mathcal{N}=2$ mirror symmetry \cite{Intriligator:1996ex, Aharony_1997}, and a rich spectrum of protected operators.
In the following sections, we present the theories in the chronological order they were discovered.

\subsection{The \texorpdfstring{$FT_N$}{ftn} theory}

We begin by considering the $FT_N \equiv FT[SU(N)]$ theory \cite{Aprile:2018oau}, i.e. the flipped version of the $T[SU(N)]$ theory, which arises from the study of S-duality and Dirichlet boundary conditions in 4d $\mathcal{N}=4$ SYM \cite{Gaiotto_2009}. We remind the reader that the theory has the following Lagrangian UV completion:

\begin{equation}
\label{ftsun}
\begin{tikzpicture}[
    thick,
    node distance=2cm,
    gauge/.style={
        circle, draw=black, fill=white,
        minimum size=8mm, inner sep=1pt,
        font=\small
    },
    flavour/.style={
        rectangle, draw=blue!70!black, fill=blue!6,
        minimum size=8mm, inner sep=2pt,
        font=\small, text=blue!70!black
    },
    arr/.style={
        -{Stealth[scale=0.8]}, shorten >=1pt, shorten <=1pt
    },
    biArr/.style={
        draw=black!70
    }
]

\node[gauge] (g1) at (0,0)   {$1$};
\node[gauge] (g2) at (2,0)   {$2$};
\node        (dt) at (3.6,0) {$\cdots$};
\node[gauge] (gn) at (5.2,0) {$\scriptstyle N{-}1$};
\node[flavour](f1) at (7.2,0) {$N$};

\draw[arr, biArr, transform canvas={yshift= 3.5pt}] (g1) -- (g2);
\draw[arr, biArr, transform canvas={yshift=-3.5pt}] (g2) -- (g1);

\draw[arr, biArr, transform canvas={yshift= 3.5pt}] (g2) -- (dt);
\draw[arr, biArr, transform canvas={yshift=-3.5pt}] (dt) -- (g2);

\draw[arr, biArr, transform canvas={yshift= 3.5pt}] (dt) -- (gn);
\draw[arr, biArr, transform canvas={yshift=-3.5pt}] (gn) -- (dt);

\draw[arr, biArr, transform canvas={yshift= 3.5pt}] (gn) -- (f1);
\draw[arr, biArr, transform canvas={yshift=-3.5pt}] (f1) -- (gn);

\node at (1,   0.45) {$b_1$};
\node at (2.8, 0.45) {$b_2$};
\node at (4.4, 0.45) {$b_{N-2}$};
\node at (6.2, 0.45) {$b_{N-1}$};

\node at (1,   0) {\LARGE$\times$};
\node at (2.8, 0) {\LARGE$\times$};
\node at (4.4, 0) {\LARGE$\times$};
\node at (6.2, 0) {\LARGE$\times$};

\draw (g2) to[out=65,in=115, looseness=3] (g2);
\node at (2, 1) {$a_2$};

\draw (gn) to[out=65,in=115, looseness=3] (gn);
\node at (5.2, 1) {$a_{N-1}$};

\draw[blue!70!black] (f1) to[out=60,in=120, looseness=3] (f1);
\node[blue!70!black] at (7.2, 1) {$a_N$};

\node at (3.6,-1.2) {$\mathcal{W} = a_2 b_1\tilde{b}_1 
    + \displaystyle\sum_{i=2}^{N-1} b_i(a_i+a_{i+1})\tilde{b}_i 
    + \sum_{i=1}^{N-1}\mathrm{Flip}[b_i\tilde{b}_i]$};

\node[right] at (8.0, 0) {
    \renewcommand{\arraystretch}{1.3}
    \begin{tabular}{>{\small}c >{\small}c}
        \toprule
        & $R = R_0 + \tau\, Q_\tau$ \\
        \midrule
        $b_i,\,\tilde{b}_i$ & $\tau/2$ \\
        $a_i$               & $2-\tau$  \\
        \bottomrule
    \end{tabular}
};

\end{tikzpicture}
\end{equation}

In this diagram, the circular node labelled $i$ represents a $U(i)$ gauge group, for $i=1,\dots,N-1$, the square blue node represents a global $SU(N)$ symmetry, an outgoing (ingoing) arrow represents an (anti-)fundamental field, and the crosses on the links between the nodes represent flipping fields, i.e., gauge singlets that couple linearly to the gauge-invariant products $b_i \tilde{b}_i$.
The field $b_i \;(\tilde{b}_i)$ transforms in the bifundamental representation $\Box \otimes \overline{\Box}$ ($\overline{\Box}\otimes \Box$ resp.) under $U(i)\times U(i+1)$ (where $\Box$ ($\overline{\Box} $) denotes the (anti-)fundamental representation), while $a_i$ transforms in the adjoint representation of the corresponding $U(i)$ gauge node.
The global symmetries of the theory include an $SU(N)$ flavour symmetry, $N-1$ $U(1)$ topological symmetries (one for each gauge node), as well as a $U(1)_\tau$ global symmetry and a $U(1)_R$ R-symmetry.
The table on the right-hand side of \eqref{ftsun} lists the R-charges $R$ of the UV fields. They are given by the trial R-charge $R_0$ mixed with the $U(1)_{\tau}$ symmetry, where $Q_{\tau}$ denotes the charge of each field under this symmetry and $\tau$ is the corresponding mixing parameter.

A notable feature of the theory is that at the IR fixed point, the global symmetry is enhanced to: 
\[
{\color{blue}SU(N)} \times {\color{red}SU(N)} \times U(1)_\tau \times U(1)_R.
\]
This enhancement is related to the self-mirror \cite{Intriligator:1996ex} property of the $\mathcal{N}=4$ theory, first studied in \cite{Gaiotto_2009}, which implies that both the Coulomb and Higgs branches exhibit an $SU(N)$ global symmetry.

Moreover, at the strongly coupled IR fixed point, the operator spectrum contains two fields, $A_L$ and $A_R$, in the adjoint representations of the two $SU(N)$ global symmetries, respectively. The first operator is the UV traceless matrix $a_N = A_L$, in the adjoint representation of the manifest $\color{blue} SU(N)$, while the second is constructed by collecting the $N(N-1)$ gauge-invariant monopoles with Goddard--Nuyts--Olive (GNO) \cite{Goddard:1976qe} charges $\mathfrak{M}^{+0\dots}$, $\mathfrak{M}^{++0\dots}$, \dots, their conjugates with negative GNO charges, and the $N-1$ traces $\text{Tr}(a_i)$ for $i = 1, \dots, N-1$.

For example, for $N=4$, we have:
\[
A_R = 
\begin{bmatrix}
  0 & \mathfrak{M}^{+00}  & \mathfrak{M}^{++0} & \mathfrak{M}^{+++}  \\
  \mathfrak{M}^{-00} &  0 & \mathfrak{M}^{0+0} & \mathfrak{M}^{0++} \\
  \mathfrak{M}^{--0} & \mathfrak{M}^{0-0} &  0 & \mathfrak{M}^{00+} \\
  \mathfrak{M}^{---}  & \mathfrak{M}^{0--}  &\mathfrak{M}^{00-}  & 0 \\
\end{bmatrix}
+ \sum_{i=1}^3 \text{Tr}(a_{i}) \mathcal{D}_i
\]
where $\mathcal{D}_i$ are the traceless diagonal generators of $SU(N)$.

\subsection{The \texorpdfstring{$FM_N$}{fmn} theory}

Next we turn to the first ``improved'' bifundamental discovered, the $FM_N \equiv FM[SU(N)]$ theory \cite{Pasquetti:2019tix}. The theory has the following Lagrangian UV completion:

\be
\label{fmsun}
\resizebox{\hsize}{!}{
\begin{tikzpicture}[
    thick,
    gauge/.style={
        circle, draw=black, fill=white,
        minimum size=8mm, inner sep=1pt, font=\small
    },
    manifest/.style={
        rectangle, draw=blue!70!black, fill=blue!6,
        minimum size=8mm, inner sep=2pt,
        font=\small, text=blue!70!black
    },
    flavour/.style={
        rectangle, draw=red!70!black, fill=red!6,
        minimum size=7mm, inner sep=2pt,
        font=\small, text=red!70!black
    },
    arr/.style={-{Stealth[scale=0.8]}, shorten >=1pt, shorten <=1pt},
    barr/.style={draw=black!70}
]

\node[gauge]    (g1) at (0,0)     {$1$};
\node[gauge]    (g2) at (2.5,0)   {$2$};
\node           (dt) at (4,0)     {$\cdots$};
\node[gauge]    (g3) at (5.5,0)   {$\scriptstyle N{-}1$};
\node[manifest] (m3) at (7.5,0)   {$N$};

\node[flavour] (x1) at (-0.75,-1.7) {$1$};
\node[flavour] (x2) at (1.25,-1.7)  {$1$};
\node[flavour] (x3) at (3.25,-1.7)     {$1$};
\node[flavour] (x4) at (6.5,-1.7)   {$1$};

\draw[arr,barr,transform canvas={yshift= 3.5pt}] (g1) -- (g2);
\draw[arr,barr,transform canvas={yshift=-3.5pt}] (g2) -- (g1);
\node at (1.25, 0.45) {$b_1$};
\node at (1.25, 0) {\huge$\times$};

\draw[arr,barr,transform canvas={yshift= 3.5pt}] (g2) -- (dt);
\draw[arr,barr,transform canvas={yshift=-3.5pt}] (dt) -- (g2);
\node at (3.2, 0.45) {$b_2$};
\node at (3.2, 0) {\huge$\times$};

\draw[arr,barr,transform canvas={yshift= 3.5pt}] (dt) -- (g3);
\draw[arr,barr,transform canvas={yshift=-3.5pt}] (g3) -- (dt);
\node at (4.75, 0.45) {$b_{N-2}$};
\node at (4.75, 0) {\huge$\times$};

\draw[arr,barr,transform canvas={yshift= 3.5pt}] (g3) -- (m3);
\draw[arr,barr,transform canvas={yshift=-3.5pt}] (m3) -- (g3);
\node at (6.5, 0.45) {$b_{N-1}$};
\node at (6.5, 0) {\huge$\times$};

\draw (g2) to[out=65,in=115,looseness=3] (g2);
\node at (2.5, 0.85) {$a_2$};

\draw (g3) to[out=65,in=115,looseness=3] (g3);
\node at (5.5, 0.95) {$a_{N-1}$};

\draw[blue!70!black] (m3) to[out=65,in=115,looseness=3] (m3);
\node[blue!70!black] at (7.5, 0.95) {$a_N$};

\draw[arr,barr] (x1) to[bend right=5] (g1);
\draw[arr,barr] (g1) to[bend right=5] (x1);
\node at (-1,-0.8) {$d_1$};
\node[rotate=-60] at (-0.35,-0.8) {\LARGE$\times$};

\draw[arr,barr] (g1) to[bend left=5] (x2);
\draw[arr,barr] (x2) to[bend left=5] (g1);
\node at (1.1,-0.9) {$v_1$};

\draw[arr,barr] (x2) to[bend right=5] (g2);
\draw[arr,barr] (g2) to[bend right=5] (x2);
\node at (2.4,-0.9) {$d_2$};
\node[rotate=-55] at (1.85,-0.9) {\LARGE$\times$};

\draw[arr,barr] (g2) to[bend left=5] (x3);
\draw[arr,barr] (x3) to[bend left=5] (g2);
\node at (3.25,-0.85) {$v_2$};

\draw[arr,barr] (4.6, -1.5) to[bend right=5] (g3);
\draw[arr,barr] (g3) to[bend right=5] (4.5, -1.5);
\node at (4.35,-0.85) {$d_{N-1}$};
\node[rotate=-50] at (5,-0.8) {\LARGE$\times$};

\draw[arr,barr] (g3) to[bend left=5] (x4);
\draw[arr,barr] (x4) to[bend left=5] (g3);
\node at (6.5,-0.6) {$v_{N-1}$};

\draw[arr,barr] (m3) to[bend left=5] (x4);
\draw[arr,barr] (x4) to[bend left=5] (m3);
\node at (7.5,-0.85) {$d_N$};
\node[rotate=-50] at (7,-0.8) {\LARGE$\times$};

\node at (3.5,-4) {$\begin{aligned}
    \mathcal{W} =\;& \sum_{j=1}^{N-1}\!\left[v_j b_j d_{j+1}
        +\tilde{v}_j\tilde{b}_j\tilde{d}_{j+1}\right]
        + a_2 b_1\tilde{b}_1
        + \sum_{j=2}^{N-1} b_j(a_j+a_{j+1})\tilde{b}_j
        + \sum_{i=1}^{N-1}\mathrm{Flip}[b_i\tilde{b}_i]\\
    &+\left(\mathfrak{M}^{\pm00\cdots}
        +\mathfrak{M}^{0\pm0\cdots}+\cdots\right)
        +\sum_{i=1}^{N}\mathrm{Flip}[d_i\tilde{d}_i]
\end{aligned}$};

\node[right] at (8.5, -0.6) {%
    \renewcommand{\arraystretch}{1.4}
    \begin{tabular}{>{\small}c >{\small}c}
        \toprule
        & $R=R_0+\tau Q_\tau+\Delta Q_{\Delta}$ \\
        \midrule
        $b_i,\,\tilde{b}_i$ & $\tau/2$ \\
        $a_i$               & $2-\tau$ \\
        $v_i,\,\tilde{v}_i$ & $2-\tfrac{i-N+2}{2}\tau-\Delta$ \\
        $d_i,\,\tilde{d}_i$ & $\Delta+\tfrac{i-N}{2}\tau$ \\
        \bottomrule
    \end{tabular}%
};

\end{tikzpicture}

}
\ee
where the blue node denotes an $SU(N)$ flavour symmetry, while the $N$ red nodes denote $U(1)$ flavour symmetries.
This theory, similarly to the $FT_N$ theory, possesses a self-mirror property, which exchanges the manifest $SU(N)$ flavour symmetry with the emergent $SU(N)$ one. The full global symmetry of the theory at the IR fixed point is:
\[
S({\color{blue}U(N)} \times {\color{red}U(N)}) \times U(1)_{\tau} \times U(1)_{\Delta} \times U(1)_R.
\]

The operator spectrum (chiral ring generators) at the IR fixed point consists of two fields, $A_L$ and $A_R$, which transform under the representations $\mathbf{Adj} \otimes \mathbf{1}$ and $\mathbf{1} \otimes \mathbf{Adj}$ of the ${\color{blue}SU(N)} \times {\color{red}SU(N)} $ global symmetries, respectively. Additionally, there are two fields, $\Pi$ and $\tilde{\Pi}$, in the representations $\Box \otimes \overline{\Box}$ and $\overline{\Box} \otimes \Box$, and a collection of singlets, denoted by $B_{n,m}$, with R-charges given by:
\[
R(B_{n,m}) = 2n - 2\Delta + (m-n)\tau.
\]
In particular, the singlets $B_{1,m}$ correspond to the \textit{flippers} of $d_{N-m+1}\tilde{d}_{N-m+1}$ for $m=1,2,\dots,N$, while the singlets $B_{n,m}$ correspond to the gauge singlets $v_{N-m}\,a_{N-m}^{\,n-2}\,\tilde{v}_{N-m}$, for $n=2,3,\dots,N$ (reducing to $v_{N-m}\tilde{v}_{N-m}$ for $n=2$), whose R-charges match the formula above.

This theory is often referred to as an ``improved'' or ``generalised'' bifundamental because, compared to the usual bifundamentals of $SU(N) \times SU(N)$, it includes two bifundamental fields in the spectrum and an additional $U(1)$ global symmetry. The usual bifundamentals, on the other hand, have only the global symmetry $S(U(N)\times U(N)) \times U(1)_A \times U(1)_R$. The improved bifundamental is particularly relevant to studies of mirror symmetry for $\mathcal{N}=2$ theories, such as $U(N)$ adjoint SQCD with flavours and vanishing superpotential \cite{Benvenuti:2023qtv}.
The singlet $B_{1,1}$, which flips $d_N \tilde{d}_N$, should be removed from the interacting theory because its scaling dimension lies below the unitarity bound; it therefore becomes a free scalar decoupled from the IR theory. We have checked this explicitly for $N=2$ and expect it to hold for $N>2$ as well.
Thus, in the theories studied below, we always remove this singlet and write the corrected superpotential by dropping the term $\mathrm{Flip}[d_N \tilde{d}_N]=B_{1,1}d_N \tilde{d}_N$. This statement applies to the isolated theory, but may not hold when the theory is part of a larger quiver, where the scaling dimension of $B_{1,1}$ should be determined by $Z$-extremisation.\footnote{Equivalently, $F$-maximisation: extremising the $\mathbf{S}^{3}$ partition function $Z$ and maximising the free energy $F=-\log|Z|$ are one and the same 3d procedure~\cite{Jafferis:2012}. We use the name $Z$-extremisation throughout; its 4d counterpart is $a$-maximisation.}

\subsection{The \texorpdfstring{$FE_N$}{feusp} theory}

The next theory we consider is the $FE_N$ theory, also called $FE[USp(2N)]$ in the literature. This theory was originally constructed as a 4d $\mathcal{N}=1$ class $\mathcal{S}$ theory, resulting from the compactification of the 6d $(1,0)$ rank-$N$ E-string SCFT on a torus with flux \cite{Pasquetti:2019hxf}. Further compactification of the resulting $\mathcal{N}=1$ theory on $\mathbf{S}^1$ yields the $FE_N$ theory in 3d, with the following Lagrangian UV completion:

\begin{equation}
\label{feuspn}
\begin{tikzpicture}[
    thick,
    gauge/.style={
        circle, draw=black, fill=white,
        minimum size=8mm, inner sep=1pt, font=\small
    },
    manifest/.style={
        rectangle, draw=blue!70!black, fill=blue!6,
        minimum size=8mm, inner sep=2pt,
        font=\small, text=blue!70!black
    },
    flavour/.style={
        rectangle, draw=red!70!black, fill=red!6,
        minimum size=7mm, inner sep=2pt,
        font=\small, text=red!70!black
    },
    arr/.style={-{Stealth[scale=0.8]}, shorten >=1pt, shorten <=1pt},
    barr/.style={draw=black!70}
]

\node[gauge]    (g1) at (0,0)   {$C_1$};
\node[gauge]    (g2) at (2.5,0) {$C_2$};
\node           (dt) at (4,0)   {$\cdots$};
\node[gauge]    (g3) at (5.5,0) {$C_{N-1}$};
\node[manifest] (m1) at (7.5,0) {$C_N$};

\node[flavour] (f1) at (-0.75,-1.5) {$C_1$};
\node[flavour] (f2) at (1.25,-1.5)  {$C_1$};
\node[flavour] (f3) at (3.25,-1.5)  {$C_1$};
\node[flavour] (f4) at (6.5,-1.5)   {$C_1$};

\draw[barr] (g1) -- (g2);
\node at (1.25, 0.5) {$b_1$};
\node at (1.25, 0) {\LARGE$\times$};

\draw[barr] (g2) -- (dt);
\node at (3.2, 0.5) {$b_2$};
\node at (3.2, 0) {\LARGE$\times$};

\draw[barr] (dt) -- (g3);
\node at (4.55, 0.5) {$b_{N-2}$};
\node at (4.75, 0) {\LARGE$\times$};

\draw[barr] (g3) -- (m1);
\node at (6.5, 0.5) {$b_{N-1}$};
\node at (6.5, 0) {\LARGE$\times$};

\draw (g2) to[out=65,in=115,looseness=3] (g2);
\node at (2.5, 0.85) {$a_2$};

\draw (g3) to[out=65,in=115,looseness=3] (g3);
\node at (5.5, 1.15) {$a_{N-1}$};

\draw[blue!70!black] (m1) to[out=65,in=115,looseness=3] (m1);
\node[blue!70!black] at (7.5, 0.95) {$a_N$};

\draw[barr] (f1) -- (g1);
\node at (-1,-0.8) {$d_1$};
\node[rotate=-40] at (-0.4,-0.8) {\LARGE$\times$};

\draw[barr] (g1) -- (f2);
\node at (1,-0.8) {$v_1$};

\draw[barr] (f2) -- (g2);
\node at (2.2,-1) {$d_2$};
\node[rotate=-55] at (1.85,-0.75) {\LARGE$\times$};

\draw[barr] (g2) -- (f3);
\node at (3.2,-0.85) {$v_2$};
\draw[barr] (4.7, -1.3) -- (g3);
\node at (5.3,-1.3) {$d_{N-1}$};
\node[rotate=-50] at (5,-0.8) {\LARGE$\times$};

\draw[barr] (g3) -- (f4);
\node at (6.4,-0.65) {$v_{N-1}$};

\draw[barr] (m1) -- (f4);
\node at (7.45,-0.85) {$d_N$};
\node[rotate=50] at (6.95,-0.8) {\LARGE$\times$};

\node at (3.5,-3.5) {$\begin{aligned}
    \mathcal{W} =\;&
        b_1^2 a_2
        + \sum_{i=2}^{N-1} b_i^2(a_i + a_{i+1})
        + \sum_{i=1}^{N-1}\!\left[v_i b_i d_{i+1}
        + \mathrm{Flip}[b_i^2]\right] \\
    &+ \sum_{j=1}^{N} \mathrm{Flip}[d_j^2]
        + \left(\mathfrak{M}^{100\cdots0}
        + \mathfrak{M}^{010\cdots0}
        + \cdots
        + \mathfrak{M}^{00\cdots01}\right)
\end{aligned}$};

\node[right] at (8.5,-0.7) {%
    \renewcommand{\arraystretch}{1.4}
    \begin{tabular}{>{\small}c >{\small}c}
        \toprule
        & $R = R_0 + \tau Q_\tau + \Delta Q_\Delta$ \\
        \midrule
        $b_i$       & $\tau/2$ \\
        $a_i$       & $2-\tau$ \\
        $v_i$       & $2-\Delta-\tfrac{i-N+2}{2}\tau$ \\
        $d_i$       & $\Delta+\tfrac{i-N}{2}\tau$ \\
        \bottomrule
    \end{tabular}%
};

\end{tikzpicture}
\end{equation}

where $C_k$ denotes a $USp(2k)$ symmetry.

Compared to its 4d version, the 3d theory has additional linear monopole terms in the superpotential, which necessarily arise upon compactification \cite{Amariti:2018wht, Benini:2017dud}. The UV global symmetry is
\[
{\color{blue}USp(2N)} \times {\color{red}USp(2)^N} \times U(1)_{\Delta} \times U(1)_{\tau} \times U(1)_R,
\]
which is enhanced in the IR to:
\[
{\color{blue}USp(2N)}  \times {\color{red}USp(2N)} \times U(1)^2\times U(1)_R.
\]
This enhancement can be understood through a self-duality that exchanges the manifest $USp(2N)$ symmetry with the emergent one. The $FM_N$ theory can be reached from $FE_N$ by turning on Coulomb-branch vacuum expectation values that break $USp(2N)$ to $U(N)$, together with real-mass deformations, as described in \cite{Bottini:2021vms}.

In the IR spectrum of the theory, we observe two fields, $A_L$ and $A_R$, which transform under the antisymmetric traceless representation of each $USp(2N)$ global symmetry. Additionally, there is a field $\Pi$ in the representation $\mathbf{2N}\otimes \mathbf{2N}$ under $USp(2N) \times USp(2N)$ (where bold numbers label the representation of the corresponding dimension), along with a collection of singlets $B_{n,m}$ with R-charges given by:
\[
R(B_{n,m}) = 2n - 2\Delta + (m-n)\tau.
\]
In particular, the singlets $B_{1,m}$ correspond to the flippers of $d_{N-m+1}^2$ for $m = 1, 2, \dots, N$, while the singlets $B_{n,m}$ correspond to the dressed gauge singlets $v_{N-m}\,a_{N-m}^{\,n-2}\,v_{N-m}$, for $n=2,3,\dots,N$ (reducing to $v^2_{N-m}$ for $n=2$), whose R-charges match the formula above.
As in the $FM_N$ theory, the flipper $B_{1,1}$ should be removed because its scaling dimension lies below the 3d unitarity bound $R \ge 1/2$ for a chiral operator, as we checked through $Z$-extremisation for $N=2$. Because the $FE_N$ theory has a 4d $\mathcal{N}=1$ origin, we can also perform 4d $a$-maximisation on the corresponding theory $FE_N^{4d}$, where the unitarity bound instead reads $\Delta = \frac{3}{2}R \ge 1$, i.e.\ $R \ge 2/3$. We find that $R(B_{1,1}) \le 2/3$, with equality only for $N=1$, when the theory is a Wess--Zumino model. Moreover, the R-charges of the other singlets $B_{n,m}$ decrease with $N$, so they eventually drop below the bound as well and should likewise be flipped to obtain a consistent theory. This carries a caveat that we wish to make explicit: for $N$ large enough that some $B_{n,m}$ with $(n,m)\neq(1,1)$ violates the bound, the consistent theory is no longer the one obtained by flipping $B_{1,1}$ alone, which is the definition adopted throughout Section~\ref{genNimpbif}. Our generic-$N$ results should therefore be read as statements about the theory with $B_{1,1}$ removed, which is the physical SCFT only as long as the remaining singlets stay above the bound.
In the rest of the paper, whenever we consider the $FE_N$ and $FM_N$ theories, we use versions in which the singlet $B_{1,1}$ is removed and the corresponding term $\mathrm{Flip}[d_N^2]=B_{1,1}d_N^2$ (or $\mathrm{Flip}[d_N \tilde{d}_N]=B_{1,1}d_N \tilde{d}_N$ for the $FM_N$ theory) is dropped from the superpotential.
 
\subsection{The \texorpdfstring{$FC_N$}{fcn} theory}

Another theory of interest is $FC_N$, which arises from a specific real-mass deformation of the $FM_N$ theory \cite{Benvenuti:2024mpn}. This deformation gives mass to half of the bifundamentals present in the saw of the $FM_N$ theory. A possible UV completion is given below:

\be
\label{fcsun}
\resizebox{\hsize}{!}{
\begin{tikzpicture}[
    thick,
    gauge/.style={
        circle, draw=black, fill=white,
        minimum size=8mm, inner sep=1pt, font=\small
    },
    manifest/.style={
        rectangle, draw=blue!70!black, fill=blue!6,
        minimum size=8mm, inner sep=2pt,
        font=\small, text=blue!70!black
    },
    flavour/.style={
        rectangle, draw=red!70!black, fill=red!6,
        minimum size=7mm, inner sep=2pt,
        font=\small, text=red!70!black
    },
    arr/.style={-{Stealth[scale=0.8]}, shorten >=1pt, shorten <=1pt},
    barr/.style={draw=black!70}
]

\node[gauge]    (g1) at (0,0)     {$1$};
\node[gauge]    (g2) at (2.5,0)   {$2$};
\node           (dt) at (4,0)     {$\cdots$};
\node[gauge]    (g3) at (5.5,0)   {$\scriptstyle N{-}1$};
\node[manifest] (m3) at (7.5,0)   {$N$};

\node[flavour] (x1) at (-0.75,-1.7) {$1$};
\node[flavour] (x2) at (1.25,-1.7)  {$1$};
\node[flavour] (x3) at (3.25,-1.7)  {$1$};
\node[flavour] (x4) at (6.5,-1.7)   {$1$};

\draw[arr,barr,transform canvas={yshift= 3.5pt}] (g1) -- (g2);
\draw[arr,barr,transform canvas={yshift=-3.5pt}] (g2) -- (g1);
\node at (1.25, 0.45) {$b_1$};
\node at (1.25, 0) {\huge$\times$};

\draw[arr,barr,transform canvas={yshift= 3.5pt}] (g2) -- (dt);
\draw[arr,barr,transform canvas={yshift=-3.5pt}] (dt) -- (g2);
\node at (3.2, 0.45) {$b_2$};
\node at (3.2, 0) {\huge$\times$};

\draw[arr,barr,transform canvas={yshift= 3.5pt}] (dt) -- (g3);
\draw[arr,barr,transform canvas={yshift=-3.5pt}] (g3) -- (dt);
\node at (4.75, 0.45) {$b_{N-2}$};
\node at (4.75, 0) {\huge$\times$};

\draw[arr,barr,transform canvas={yshift= 3.5pt}] (g3) -- (m3);
\draw[arr,barr,transform canvas={yshift=-3.5pt}] (m3) -- (g3);
\node at (6.5, 0.45) {$b_{N-1}$};
\node at (6.5, 0) {\huge$\times$};

\draw (g2) to[out=65,in=115,looseness=3] (g2);
\node at (2.5, 0.85) {$a_2$};

\draw (g3) to[out=65,in=115,looseness=3] (g3);
\node at (5.5, 0.95) {$a_{N-1}$};

\draw[blue!70!black] (m3) to[out=65,in=115,looseness=3] (m3);
\node[blue!70!black] at (7.5, 0.95) {$a_N$};

\draw[arr,barr] (x1) to[bend right=5] (g1);
\node at (-1.05,-0.75) {$d_1$};

\draw[arr,barr] (g1) to[bend left=5] (x2);
\node at (1.0,-0.85) {$v_1$};

\draw[arr,barr] (x2) to[bend right=5] (g2);
\node at (2.35,-0.85) {$d_2$};

\draw[arr,barr] (g2) to[bend left=5] (x3);
\node at (3.25,-0.8) {$v_2$};

\draw[arr,barr] (4.6,-1.5) to[bend right=5] (g3);
\node at (4.3,-0.8) {$d_{N-1}$};

\draw[arr,barr] (g3) to[bend left=5] (x4);
\node at (6.45,-0.55) {$v_{N-1}$};

\draw[arr,barr] (x4) to[bend left=5] (m3);
\node at (7.55,-0.8) {$d_N$};

\node at (3.5,-4) {$\begin{aligned}
    \mathcal{W} =\;& \sum_{j=1}^{N-1}v_j b_j d_{j+1}
        + a_2 b_1\tilde{b}_1
        + \sum_{j=2}^{N-1} b_j(a_j+a_{j+1})\tilde{b}_j
        + \sum_{i=1}^{N-1}\mathrm{Flip}[b_i\tilde{b}_i]\\
    &+\left(\mathfrak{M}^{+00\cdots}
        +\mathfrak{M}^{0+0\cdots}+\cdots\right)
\end{aligned}$};

\node[right] at (8.5,-0.6) {%
    \renewcommand{\arraystretch}{1.4}
    \begin{tabular}{>{\small}c >{\small}c}
        \toprule
        & $R=R_0+\tau Q_\tau+\Delta Q_{\Delta}$ \\
        \midrule
        $b_i,\,\tilde{b}_i$ & $\tau/2$ \\
        $a_i$               & $2-\tau$ \\
        $v_i$ & $2-\tfrac{i-N+2}{2}\tau-\Delta$ \\
        $d_i$ & $\Delta+\tfrac{i-N}{2}\tau$ \\
        \bottomrule
    \end{tabular}%
};

\end{tikzpicture}
}
\ee

Note that, as a result of the deformation, the superpotential contains only $(N-1)$ cubic terms from the closed triangles and $(N-1)$ monopole terms. In total, we have integrated out $2N-1$ chiral fields but have $2N-2$ fewer superpotential terms, resulting in one less global $U(1)$ symmetry compared to the $FM_N$ theory.

The $FC_N$ theory also exhibits interesting duality properties derived from the self-duality of $FM_N$. A consequence of these dualities is the claim that the global symmetry is enhanced in the IR to ${\color{blue}SU(N)} \times {\color{red}SU(N)} \times U(1)_\Delta \times U(1)_\tau \times U(1)_R$. In the IR spectrum, we still have two fields, $A_L$ and $A_R$, in the representations $\mathbf{Adj} \otimes \mathbf{1}$ and $\mathbf{1} \otimes \mathbf{Adj}$ of the $SU(N) \times SU(N)$ global symmetries, but a single bifundamental field, $\Pi$, in the representation $\Box\otimes \overline{\Box}$, and \textit{no} additional non-abelian singlets.

\subsection{The \texorpdfstring{$FH_N$}{fhn} theory}
\label{FHN}

The final theory we study is $FH_N$, which can be derived from the $FE_N$ theory by turning on a Coulomb-branch vacuum expectation value that breaks the gauge group from $USp$ to $U$, together with real-mass deformations. These deformations break one of the two global $USp(2N)$ symmetries to $U(N)$ while preserving the other \cite{Benvenuti:2024mpn}. A possible Lagrangian UV completion, with global symmetry ${\color{red}SU(2)}^N\times {\color{blue} U(N)} \times U(1)_{\tau}\times U(1)_R$, is as follows:

\be
\label{fhsun}
\resizebox{\hsize}{!}{
\begin{tikzpicture}[
    thick,
    gauge/.style={
        circle, draw=black, fill=white,
        minimum size=8mm, inner sep=1pt, font=\small
    },
    manifest/.style={
        rectangle, draw=blue!70!black, fill=blue!6,
        minimum size=8mm, inner sep=2pt,
        font=\small, text=blue!70!black
    },
    flavour/.style={
        rectangle, draw=red!70!black, fill=red!6,
        minimum size=7mm, inner sep=2pt,
        font=\small, text=red!70!black
    },
    arr/.style={-{Stealth[scale=0.8]}, shorten >=1pt, shorten <=1pt},
    barr/.style={draw=black!70}
]

\node[gauge]    (g1) at (0,0)   {$1$};
\node[gauge]    (g2) at (2.5,0) {$2$};
\node           (dt) at (4,0)   {$\cdots$};
\node[gauge]    (g3) at (5.5,0) {$\scriptstyle N{-}1$};
\node[manifest] (m3) at (7.5,0) {$N$};

\node[flavour] (x0) at (-0.75,-1.7) {$C_1$};
\node[flavour] (x1) at (1.25,-1.7)  {$C_1$};
\node[flavour] (x2) at (3.1,-1.7)   {$C_1$};
\node[flavour] (x3) at (6.5,-1.7)   {$C_1$};

\draw[arr,barr,transform canvas={yshift= 3.5pt}] (g1) -- (g2);
\draw[arr,barr,transform canvas={yshift=-3.5pt}] (g2) -- (g1);
\node at (1.25, 0.45) {$b_1$};
\node at (1.25, 0) {\huge$\times$};

\draw[arr,barr,transform canvas={yshift= 3.5pt}] (g2) -- (dt);
\draw[arr,barr,transform canvas={yshift=-3.5pt}] (dt) -- (g2);
\node at (3.2, 0.45) {$b_2$};
\node at (3.2, 0) {\huge$\times$};

\draw[arr,barr,transform canvas={yshift= 3.5pt}] (dt) -- (g3);
\draw[arr,barr,transform canvas={yshift=-3.5pt}] (g3) -- (dt);
\node at (4.75, 0.45) {$b_{N-2}$};
\node at (4.75, 0) {\huge$\times$};

\draw[arr,barr,transform canvas={yshift= 3.5pt}] (g3) -- (m3);
\draw[arr,barr,transform canvas={yshift=-3.5pt}] (m3) -- (g3);
\node at (6.5, 0.45) {$b_{N-1}$};
\node at (6.5, 0) {\huge$\times$};

\draw (g2) to[out=65,in=115,looseness=3] (g2);
\node at (2.5, 0.85) {$a_2$};

\draw (g3) to[out=65,in=115,looseness=3] (g3);
\node at (5.5, 0.95) {$a_{N-1}$};

\draw[blue!70!black] (m3) to[out=65,in=115,looseness=3] (m3);
\node[blue!70!black] at (7.5, 0.95) {$a_N$};


\draw[arr,barr] (x0) to[bend left=5] (g1);
\node at (-0.65,-0.8) {$d_1$};

\draw[arr,barr] (g1) to[bend right=5] (x1);
\node at (0.9,-0.75) {$v_1$};

\draw[arr,barr] (x1) to[bend left=5] (g2);
\node at (2.05,-0.82) {$d_2$};

\draw[arr,barr] (g2) to[bend right=5] (x2);
\node at (2.95,-0.72) {$v_2$};

\draw[arr,barr] (4.65,-1.55) to[bend left=5] (g3);
\node at (4.45,-0.82) {$d_{N-1}$};

\draw[arr,barr] (g3) to[bend right=5] (x3);
\node at (6.35,-0.65) {$v_{N-1}$};

\draw[arr,barr] (x3) to[bend left=5] (m3);
\node at (7.6,-0.82) {$d_N$};

\node at (3.5,-4) {$\begin{aligned}
    \mathcal{W} =\;& \sum_{j=1}^{N-1} v_j b_j d_{j+1}
        + a_2 b_1\tilde{b}_1
        + \sum_{j=2}^{N-1} b_j(a_j+a_{j+1})\tilde{b}_j
        + \sum_{i=1}^{N-1}\mathrm{Flip}[b_i\tilde{b}_i]\\
    &+\left(\mathfrak{M}^{\pm00\cdots}
        +\mathfrak{M}^{0\pm0\cdots}+\cdots\right)
\end{aligned}$};

\node[right] at (8.5,-0.6) {%
    \renewcommand{\arraystretch}{1.4}
    \begin{tabular}{>{\small}c >{\small}c}
        \toprule
        & $R=R_0+\tau Q_\tau+\Delta Q_{\Delta}$ \\
        \midrule
        $b_i,\,\tilde{b}_i$ & $\tau/2$ \\
        $a_i$               & $2-\tau$ \\
        $v_i$ & $2-\tfrac{i-N+2}{2}\tau-\Delta$ \\
        $d_i$ & $\Delta+\tfrac{i-N}{2}\tau$ \\
        \bottomrule
    \end{tabular}%
};

\end{tikzpicture}
}
\ee

The self-duality of the $FE_N$ theory induces a duality for $FH_N$ that exchanges the manifest $\color{blue}U(N)$ global symmetry with the emergent $\color{red}USp(2N)$. Consequently, we can write a dual frame with UV symmetry ${\color{blue}U(1)}^N\times {\color{red} USp(2N)} \times U(1)_{\tau}\times U(1)_R$ as follows:

\be
\resizebox{\hsize}{!}{
\begin{tikzpicture}[
    thick,
    gauge/.style={
        circle, draw=black, fill=white,
        minimum size=8mm, inner sep=1pt, font=\small
    },
    manifest/.style={
        rectangle, draw=red!70!black, fill=red!6,
        minimum size=8mm, inner sep=2pt,
        font=\small, text=red!70!black
    },
    flavour/.style={
        rectangle, draw=blue!70!black, fill=blue!6,
        minimum size=7mm, inner sep=2pt,
        font=\small, text=blue!70!black
    },
    arr/.style={-{Stealth[scale=0.8]}, shorten >=1pt, shorten <=1pt},
    barr/.style={draw=black!70}
]

\node[gauge]    (g1) at (0,0)   {$C_1$};
\node[gauge]    (g2) at (2.5,0) {$C_2$};
\node           (dt) at (4,0)   {$\cdots$};
\node[gauge]    (g3) at (5.5,0) {$C_{N-1}$};
\node[manifest] (m3) at (7.5,0) {$C_N$};

\node[flavour] (x0) at (-0.75,-1.7) {$1$};
\node[flavour] (x1) at (1.25,-1.7)  {$1$};
\node[flavour] (x2) at (3.1,-1.7)   {$1$};
\node[flavour] (x3) at (6.5,-1.7)   {$1$};

\draw[barr] (g1) -- (g2);
\node at (1.25, 0.45) {$b_1$};
\node at (1.25, 0) {\huge$\times$};

\draw[barr] (g2) -- (dt);
\node at (3.2, 0.45) {$b_2$};
\node at (3.2, 0) {\huge$\times$};

\draw[barr] (dt) -- (g3);
\node at (4.55, 0.45) {$b_{N-2}$};
\node at (4.55, 0) {\huge$\times$};

\draw[barr] (g3) -- (m3);
\node at (6.5, 0.45) {$b_{N-1}$};
\node at (6.5, 0) {\huge$\times$};

\draw (g2) to[out=65,in=115,looseness=3] (g2);
\node at (2.5, 0.85) {$a_2$};

\draw (g3) to[out=65,in=115,looseness=3] (g3);
\node at (5.5, 1.15) {$a_{N-1}$};

\draw[red!70!black] (m3) to[out=65,in=115,looseness=3] (m3);
\node[red!70!black] at (7.5, 0.95) {$a_N$};


\draw[arr,barr] (x0) to[bend left=5] (g1);
\node at (-0.65,-0.8) {$d_1$};

\draw[arr,barr] (g1) to[bend right=5] (x1);
\node at (0.9,-0.75) {$v_1$};

\draw[arr,barr] (x1) to[bend left=5] (g2);
\node at (2.05,-0.82) {$d_2$};

\draw[arr,barr] (g2) to[bend right=5] (x2);
\node at (2.95,-0.72) {$v_2$};

\draw[arr,barr] (4.65,-1.55) to[bend left=5] (g3);
\node at (4.45,-0.82) {$d_{N-1}$};

\draw[arr,barr] (g3) to[bend right=5] (x3);
\node at (6.35,-0.55) {$v_{N-1}$};

\draw[arr,barr] (x3) to[bend left=5] (m3);
\node at (7.6,-0.82) {$d_N$};

\node at (3.5,-4) {$\begin{aligned}
    \mathcal{W} =\;& \sum_{j=1}^{N-1} v_j b_j d_{j+1}
        + a_2 b_1^2
        + \sum_{j=2}^{N-1} b_j^2(a_j+a_{j+1})
        + \sum_{i=1}^{N-1}\mathrm{Flip}[b_i^2]\\
    &+\left(\mathfrak{M}^{100\cdots}
        +\mathfrak{M}^{010\cdots}+\cdots\right)
\end{aligned}$};

\node[right] at (8.5,-0.6) {%
    \renewcommand{\arraystretch}{1.4}
    \begin{tabular}{>{\small}c >{\small}c}
        \toprule
        & $R=R_0+\tau Q_\tau+\Delta Q_{\Delta}$ \\
        \midrule
        $b_i$  & $\tau/2$ \\
        $a_i$  & $2-\tau$ \\
        $v_i$  & $2-\tfrac{i-N+2}{2}\tau-\Delta$ \\
        $d_i$  & $\Delta+\tfrac{i-N}{2}\tau$ \\
        \bottomrule
    \end{tabular}%
};

\end{tikzpicture}
}
\ee

As a consequence of this duality, the global symmetry of the theory must enhance in the IR to ${\color{blue}U(N)}\times {\color{red}USp(2N)}\times U(1)_{\tau}\times U(1)_R$, where $U(1)_{\Delta}\subset U(N)$. In the IR spectrum we now have a field $A_L$ in the adjoint representation of $SU(N)$, a field $A_R$ in the antisymmetric traceless representation of $USp(2N)$ and a bifundamental $\Pi$ in the representation $\mathbf{2N}\otimes \overline{\Box}$ of $USp(2N)\times U(N)$, and \textit{no} additional non-abelian singlets.

\section{Computation of the index and of its Hilbert-series limit}
\label{sec:computation}

In this section we describe how the Hilbert series presented in the rest of the paper are obtained from the superconformal index. In Section~\ref{sec:HS_limit} we define the limit of the index that we use, and in Section~\ref{sec:practical} we explain how it is evaluated on the UV quivers of Section~\ref{impbif}, and how the closed forms of the HS are obtained from the resulting data.

\subsection{The Hilbert series as a limit of the index}
\label{sec:HS_limit}

\paragraph{The index and the charge decomposition.}
We use the superconformal index on $\mathbf{S}^2\times\mathbf{S}^1$ \cite{Bhattacharya:2008,Imamura:2011,Kapustin:2011,Kim:2009wb} (see \cite{Gadde:2020yah} for a review), defined as a Witten index with respect to a supercharge $\mathcal{Q}$ and its conjugate,
\be
\mathcal{I}\bigl(x,\psi_\tau,\chi_\Delta,\{\mu\}\bigr)
=\mathrm{Tr}\,(-1)^F\,x^{E+j_3}\,\psi_\tau^{\,-Q_\tau}\,\chi_\Delta^{\,Q_\Delta}\,\prod\mu^{f},
\ee
where $\psi_\tau$ and $\chi_\Delta$ are the fugacities for the abelian symmetries $U(1)_{-\tau}$ and $U(1)_\Delta$, $\{\mu\}$ collectively denotes the fugacities
of the non-abelian flavour symmetries, and $E$ in the exponent of $x$ is the scaling dimension. Only states annihilated by $\mathcal{Q}$ and $\mathcal{Q}^\dagger$, i.e.\ with $\{\mathcal{Q},\mathcal{Q}^\dagger\}=E-R-j_3=0$, contribute, so that $x^{E+j_3}=x^{R+2j_3}$.
We normalise the abelian charges so that the adjoint-like generators ($A_L$, $A_R$; the traceless antisymmetrics for $FH_N$ and $FE_N$)
have $(Q_\tau,Q_\Delta)=(-1,0)$ and the bifundamentals ($\Pi$, and $\tilde\Pi$ for $FM_N$) have $(Q_\tau,Q_\Delta)=(0,1)$.
For $FT_N$ there is no $U(1)_\Delta$ and $\chi_\Delta$ is absent.

The superconformal R-symmetry is the UV R-symmetry mixed with $U(1)_\tau$ (and $U(1)_\Delta$), so the R-charge of any operator
depends linearly on the mixing parameters. We split it as
\be
R=(\tau-2)\,Q_\tau+\Delta\,Q_\Delta+w ,
\label{eq:w_def}
\ee
which defines $w$ as a number independent of $\tau$ and $\Delta$. By construction, $w=0$ for $A_L$, $A_R$, $\Pi$ and $\tilde\Pi$.
In terms of
\be
t_a\equiv\psi_\tau\,x^{2-\tau},\qquad t_d\equiv\chi_\Delta\,x^{\Delta},
\ee
every state contributes to the index as
\be
x^{R+2j_3}\,\psi_\tau^{\,-Q_\tau}\chi_\Delta^{\,Q_\Delta}=t_a^{\,-Q_\tau}\,t_d^{\,Q_\Delta}\,x^{\,w+2j_3},
\ee
so $\mathcal{I}$ is a function of $t_a$, $t_d$, $x$ and $\{\mu\}$.

\paragraph{The performed limit.}
For the class of theories we consider in the present paper, we perform the rescaling
\be
\boxed{x\to x\,y,\qquad \psi_\tau\to\psi_\tau\,y^{\tau-2},\qquad \chi_\Delta\to\chi_\Delta\,y^{-\Delta}},
\label{eq:rescaling}
\ee
(only the first two for $FT_N$), under which $t_a\to t_a$, $t_d\to t_d$ and $x\to x\,y$. Each state is therefore weighted by $y^{\,w+2j_3}$, and
\be
\CH\CS(t_a,t_d,\{\mu\})=\lim_{y\to0}\mathcal{I}\bigl(x\,y,\ \psi_\tau\,y^{\tau-2},\ \chi_\Delta\,y^{-\Delta},\ \{\mu\}\bigr)
=\mathcal{I}\big|_{x^0,\ t_a,t_d\ \text{fixed}} .
\label{eq:HS_limit}
\ee
The limit exists provided $w+2j_3\ge0$ for every state contributing to the index, and it retains exactly the states with $w+2j_3=0$.
In all the indices computed in this paper, every term satisfies $w+2j_3\ge0$ to the order computed, and all the terms with $w+2j_3=0$ have positive coefficients,
consistent with their being bosonic with $j_3=0$: they are the chiral-ring operators whose R-charge is saturated by their $U(1)_\tau\times U(1)_\Delta$ charges.
Because $w$ is additive and non-negative on the chiral-ring generators, these are precisely the polynomials in the generators with $w=0$.
\begin{itemize}
	\item For $FT_N$, $FC_N$ and $FH_N$, all the chiral-ring generators (the adjoint-like generators and the bifundamental $\Pi$) have $w=0$, and they generate the whole chiral ring. The limit \eqref{eq:HS_limit} therefore gives the Hilbert series of the full moduli space.
	\item For $FM_N$ and $FE_N$, the singlets $B_{n,m}$ have $R=2n-2\Delta+(m-n)\tau$, i.e.\ $(Q_\tau,Q_\Delta)=(m-n,-2)$ and $w=2m>0$.
	They, and all their products with other generators, are projected out. The limit \eqref{eq:HS_limit} gives the Hilbert series of the branch generated by the adjoints (antisymmetrics) and the bifundamentals, i.e. what we call the \textit{main branch}.
\end{itemize}
As stressed in the introduction, this rests on the absence of cancellations between bosonic and fermionic contributions at $w+2j_3=0$. For $FT_N$ this is guaranteed by the derivation of Section~\ref{sec:FT_limit}, in which all fermionic letters drop out of the integrand. For the other families it cannot be established from the index alone; we have checked that the resulting series are consistent with the chiral-ring generators and relations identified in Sections~\ref{genNimpbif} and~\ref{n=2impbif}.

\subsection{Practical evaluation}
\label{sec:practical}

\paragraph{The localisation formula.}
For each improved bifundamental we start from one of its UV Lagrangian completions reviewed in Section~\ref{impbif}. Its superconformal index is computed by localisation \cite{Imamura:2011,Kapustin:2011,Kim:2009wb} as
\be
\mathcal{I}=\sum_{m}\frac{1}{|W_m|}\oint\prod_{a=1}^{\mathrm{rk}\,G}\frac{dz_a}{2\pi i z_a}\;
Z_{\mathrm{cl}}(z,m)\,Z_{\mathrm{vec}}(z,m)\prod_{\Phi}Z_{\Phi}(z,m),
\label{eq:SCI_loc}
\ee
where $m$ runs over the GNO fluxes of the gauge group $G$ (one integer $m_a$ for each Cartan generator of each gauge node), $|W_m|$ is the order of the Weyl group of the subgroup of $G$ left unbroken by $m$, $Z_{\mathrm{cl}}$ contains the topological and Chern--Simons couplings, and the one-loop determinants $Z_{\mathrm{vec}}$, $Z_\Phi$ of the vector and chiral multiplets are given explicitly in Section~\ref{sec:FT_limit}. The R-charge of each UV chiral is the UV trial R-charge $R_0$ mixed with $U(1)_\tau$ and $U(1)_\Delta$, as specified in the quiver diagrams of Section~\ref{impbif}. In the numerical evaluation we fix the mixing parameters $\tau$ and $\Delta$ to rational values close to the result of $Z$-extremisation (when the latter converges), and refine the index with the abelian fugacities $\psi_\tau$, $\chi_\Delta$; the values used are reported case by case in Section~\ref{n=2impbif} and Appendices~\ref{AppN=3}--\ref{AppN=4}.

\paragraph{Expansion and truncation of the flux sum.}
We compute the index as a power series in $x$ up to a given order $x^n$. The contribution of a monopole flux sector $m$ starts at order $x^{\epsilon(m)}$, with $\epsilon(m)$ the R-charge of the corresponding bare monopole, which grows with $|m|$, so that only \textit{finitely many} flux sectors contribute up to order $x^n$. We therefore truncate the sum over $m$ to a window
\be
|m_a|\le M\qquad\text{for each component } m_a\in\mathbb{Z} \text{ of the flux of each gauge node},
\ee
with $M$ chosen large enough that the sectors left out do not contribute up to the order computed. The only exception is $FH_4$, for which we retain only the zero-flux sector; as discussed in Appendix~\ref{AppN=4}, this is justified a posteriori by the fact that all the chiral-ring generators already appear at zero flux.
The truncated index is then a polynomial in $x$ and in $\psi_\tau$, $\chi_\Delta$, and in the non-abelian flavour fugacities whenever these are turned on.

\paragraph{Taking the limit on the truncated index.}
The limit~\eqref{eq:HS_limit} is applied to this truncated expansion: we perform the substitution~\eqref{eq:rescaling} and keep the coefficient of $y^0$. In terms of $t_a=\psi_\tau x^{2-\tau}$ and $t_d=\chi_\Delta x^{\Delta}$, the index at order $x^n$ contains the full contribution of the monomials $t_a^{\,i}t_d^{\,j}$ with $i(2-\tau)+j\Delta\le n$. Only these coefficients of the HS are reliably determined by an index computed to order $x^n$, and only these are used below. 

\paragraph{Why the limit is taken on the index and not on the integrand.}
It would be tempting to take the limit~\eqref{eq:HS_limit} directly on the integrand of~\eqref{eq:SCI_loc}, before performing the gauge integral. Except for $FT_N$, this is \textit{not} allowed. After the rescaling~\eqref{eq:rescaling}, a bosonic letter of a UV chiral $\Phi$ with gauge weight $\rho$ enters the integrand as $1/(1-X\,y^{\,w_\Phi+|\rho(m)|}z^{\rho})$, with $X$ independent of $y$, and its pole sits at $|z^\rho| \; \propto \; y^{-(w_\Phi+|\rho(m)|)}$. The index is defined by a contour that separates the poles of the chirals of weight $+\rho$ from those of weight $-\rho$~\cite{Rains:2005lkc,Dolan:2008qi,Spiridonov:2009za,Krattenthaler:2011da}.
\begin{itemize}
	\item If $w_\Phi\ge0$ for all UV chirals, as for $FT_N$ ($w_\Phi\in\{0,1\}$), the poles move away from the unit circle as $y\to0$, staying on their own side of it. The integral is the same analytic function of $y$, and the limit commutes with the gauge integral (Section~\ref{sec:FT_limit}).
	\item In $FC_N$, $FH_N$, $FM_N$ and $FE_N$ the fields $d_i$ with $i<N$ have $w_{d_i}=i-N<0$. In the flux sectors with $|\rho(m)|<|w_{d_i}|$ their poles, which lie outside the unit circle at $y=1$ (since $R_{d_i}>0$), move \emph{inside} it as $y\to0$. The integral over $|z|=1$ then jumps by the residues of the poles that have crossed the contour, and the limit taken inside the integral gives a wrong result. Extracting the coefficient of $y^0$ from the Laurent expansion of the integrand at small $y$ fails for the same reason.\footnote{A minimal example is a $U(1)$ gauge theory with two chirals of charge $\pm1$, keeping only the bosonic letters at zero flux: $\oint\frac{dz}{2\pi i z}\frac{1}{(1-az)(1-b/z)}=\frac{1}{1-ab}$ for $|a|,|b|<1$, the PE of the meson. With $a=A\,y^{-1}$ and $b=B\,y$, mimicking $w_d=-1$ and $w_v=+1$, the meson has $w=0$ and should survive the limit. Yet for $y<|A|$ the pole at $z=y/A$ enters the unit circle and the integral over $|z|=1$ vanishes: the limit taken inside the integral loses the meson.}
\end{itemize}
For these theories the limit must therefore be taken on the index itself, i.e.\ after the gauge integral has been performed with the correct contour. This is what we do: the index is expanded order by order in $x$, and only then is the limit~\eqref{eq:HS_limit} applied to its coefficients. This is the only order of operations that is allowed for all the theories considered in this paper.

\paragraph{From the coefficients to closed forms.}
The data obtained in this way are a finite number of coefficients of the HS. The closed forms presented in the following sections are obtained as follows.
\begin{enumerate}
	\item The two slices $t_d=0$ and $t_a=0$ are generated by the adjoint-like generators and by the bifundamentals respectively, and their HS are known in closed form (Section~\ref{genNimpbif}). They fix the exponents of the denominator in the ansatz
	\be
	\CH\CS(t_a,t_d)=\frac{\sum_{i,j}a_{i,j}\,t_a^{\,i}\,t_d^{\,j}}{(1-t_a)^{d_a}\,(1-t_d)^{d_d}},
	\ee
	together with the numerator coefficients $a_{i,0}$ and $a_{0,j}$.
	\item The remaining coefficients $a_{i,j}$ are fixed by matching the computed coefficients of the HS. When the available data do not fix all of them, we additionally impose that the numerator be palindromic (the Gorenstein property), and we state explicitly where this is done.
	\item The resulting HS is checked against the chiral-ring generators and relations read off from its plethystic logarithm, which must agree with the operators and relations identified in~\cite{Aprile:2018oau, Pasquetti:2019tix, Pasquetti:2019hxf, Benvenuti:2024mpn}.
\end{enumerate}
For $N=2$ the enlarged IR symmetry allows us to resum the index directly into characters of the enhanced symmetry (Section~\ref{n=2impbif}). For $FT_N$ the HS is derived for all $N$ in Section~\ref{sec:FT_limit}. For the other four families, the generic-$N$ formulae of Section~\ref{genNimpbif} are conjectures obtained by extrapolating the structure of the $N\le3$ results, and checked against the partial results at $N=4$ (the denominators, the bidegree of the numerator and the sum rules of its coefficients); they are not derived from the index for generic $N$.

\section{Moduli spaces of improved bifundamentals at generic N}
\label{genNimpbif}

In this section, we study the Hilbert series of the theories of interest at generic $N$. For $FC_N$, $FH_N$, $FM_N$ and $FE_N$ the formulae below are extrapolated from the HS at $N\le3$ and checked against the partial results at $N=4$, which are extracted from the superconformal indices as described in Section~\ref{sec:computation}; we focus on the HS and leave a detailed study of the indices for future work. The exception is $FT_N$, for which we evaluate the limit directly on the integrand of the index and derive the closed form of the HS for all $N$.

We examine the case $N = 2$ separately in Section \ref{n=2impbif}, highlighting its enhanced IR symmetry.

In addition to the rank-$2$ generators (adjoints or antisymmetrics and bifundamentals), the final two theories, $FM_N$ and $FE_N$, have generators that are singlets under the non-abelian global symmetries. These singlets generate simple branches in the moduli space of vacua. In this section, we disregard these branches and focus on the rank-$2$ generators, which generate an algebraic variety that we call the \emph{main branch}.

\subsection{\texorpdfstring{$FT_N$}{ftsun}}
We begin with the simplest family, $FT_N$.

\begin{equation}
\begin{tikzpicture}[
    thick,
    node distance=2cm,
    gauge/.style={
        circle, draw=black, fill=white,
        minimum size=8mm, inner sep=1pt,
        font=\small
    },
    flavour/.style={
        rectangle, draw=blue!70!black, fill=blue!6,
        minimum size=8mm, inner sep=2pt,
        font=\small, text=blue!70!black
    },
    arr/.style={
        -{Stealth[scale=0.8]}, shorten >=1pt, shorten <=1pt
    },
    biArr/.style={
        draw=black!70
    }
]

\node[gauge] (g1) at (0,0)   {$1$};
\node[gauge] (g2) at (2,0)   {$2$};
\node        (dt) at (3.6,0) {$\cdots$};
\node[gauge] (gn) at (5.2,0) {$\scriptstyle N{-}1$};
\node[flavour](f1) at (7.2,0) {$N$};

\draw[arr, biArr, transform canvas={yshift= 3.5pt}] (g1) -- (g2);
\draw[arr, biArr, transform canvas={yshift=-3.5pt}] (g2) -- (g1);

\draw[arr, biArr, transform canvas={yshift= 3.5pt}] (g2) -- (dt);
\draw[arr, biArr, transform canvas={yshift=-3.5pt}] (dt) -- (g2);

\draw[arr, biArr, transform canvas={yshift= 3.5pt}] (dt) -- (gn);
\draw[arr, biArr, transform canvas={yshift=-3.5pt}] (gn) -- (dt);

\draw[arr, biArr, transform canvas={yshift= 3.5pt}] (gn) -- (f1);
\draw[arr, biArr, transform canvas={yshift=-3.5pt}] (f1) -- (gn);

\node at (1,   0.45) {$b_1$};
\node at (2.8, 0.45) {$b_2$};
\node at (4.4, 0.45) {$b_{N-2}$};
\node at (6.2, 0.45) {$b_{N-1}$};

\node at (1,   0) {\LARGE$\times$};
\node at (2.8, 0) {\LARGE$\times$};
\node at (4.4, 0) {\LARGE$\times$};
\node at (6.2, 0) {\LARGE$\times$};

\draw (g2) to[out=65,in=115, looseness=3] (g2);
\node at (2, 1) {$a_2$};

\draw (gn) to[out=65,in=115, looseness=3] (gn);
\node at (5.2, 1) {$a_{N-1}$};

\draw[blue!70!black] (f1) to[out=60,in=120, looseness=3] (f1);
\node[blue!70!black] at (7.2, 1) {$a_N$};

\node at (3.6,-1.2) {$\mathcal{W} = a_2 b_1\tilde{b}_1 
    + \displaystyle\sum_{i=2}^{N-1} b_i(a_i+a_{i+1})\tilde{b}_i 
    + \sum_{i=1}^{N-1}\mathrm{Flip}[b_i\tilde{b}_i]$};

\node[right] at (8.0, 0) {
    \renewcommand{\arraystretch}{1.3}
    \begin{tabular}{>{\small}c >{\small}c}
        \toprule
        $b_i,\,\tilde{b}_i$ & $\tau/2$ \\
        $a_i$               & $2-\tau$  \\
        \bottomrule
    \end{tabular}
};

\end{tikzpicture}
\label{ftsun_genN}
\end{equation}

For $FT_N$, the refined Hilbert series is given by
\be
\CH\CS_{FT_N} = PE[\chi_{(adj_{L},\mathbf{1})} t +\chi_{(\mathbf{1},adj_{R})}t - \sum_{j=2}^N t^j]
\label{ftHS}
\ee
where $\chi_{(adj_{L},\mathbf{1})}$ and $\chi_{(\mathbf{1},adj_{R})}$ represent the characters of the adjoint representations of $SU(N)_L$ and $SU(N)_R$, respectively. The function $PE$ denotes the Plethystic Exponential. We derive \eqref{ftHS} from the superconformal index at the end of this subsection.
 
\paragraph{The unrefined Hilbert series.}
Setting all flavour fugacities to unity, the refined Hilbert series \eqref{ftHS} collapses to
\begin{equation}\label{HSftuNUR}
    \CH\CS_{FT_N}(t) \;=\; \frac{\displaystyle\prod_{j=1}^{N-1}\sum_{h=0}^{j} t^{h}}{(1-t)^{\,2N^{2}-N-1}}\,.
\end{equation}
The numerator is palindromic, which confirms that the moduli space is a Calabi--Yau cone of complex dimension $2N^{2}-N-1$.
 
\paragraph{Generators and relations from the plethystic logarithm.}
The main structural result of this subsection is obtained by taking the Plethystic Logarithm (PL) of the refined Hilbert series \eqref{ftHS}:
\begin{equation}\label{eq:PL-FTN}
    \mathrm{PL}\!\left[\CH\CS_{FT_N}(t)\right]
    \;=\; \chi_{(\mathrm{adj}_{L},\mathbf{1})}\,t \;+\; \chi_{(\mathbf{1},\mathrm{adj}_{R})}\,t
    \;-\; \sum_{j=2}^{N} t^{\,j}\,.
\end{equation}
This expression admits a transparent reading: the moduli space is generated by the $2(N^{2}-1)$ components of two adjoint-valued matrices $A_{L}$ and $A_{R}$, transforming in the adjoint of $SU(N)_{L}$ and $SU(N)_{R}$ respectively, subject to the $N-1$ relations
\begin{equation}\label{eq:trace-relations}
    \mathrm{tr}_{L}\, A_{L}^{k} \;=\; \mathrm{tr}_{R}\, A_{R}^{k}\,,
    \qquad k = 2, \dots, N\,,
\end{equation}
where $\mathrm{tr}_{L}$ ($\mathrm{tr}_{R}$) denotes the trace over $SU(N)_{L}$ ($SU(N)_{R}$) indices. The termination of the plethystic logarithm in \eqref{eq:PL-FTN} confirms that no higher syzygies appear: the moduli space is therefore a \emph{complete intersection} of degree $N!$. Analogous results were obtained in \cite{Hanany:2011db} for the Higgs branch of the $\mathcal{N}=4$ $T[SU(N)]$ theory.
 
\paragraph{The $h$-sequence and a recursive pattern.}
Expanding the numerator of \eqref{HSftuNUR} as
\begin{equation}
    \prod_{j=1}^{N-1}\sum_{h=0}^{j} t^{h}
    \;=\; \sum_{i=0}^{N(N-1)/2} h_{i}\, t^{\,i}\,,
\end{equation}
one reads off the $h$-sequence $h=(h_{0},h_{1},\dots,h_{d})$ of the moduli space, with $d = N(N-1)/2$. The first few entries are
\begin{equation}
    h \;=\; \left(1,\; N-1,\; \tfrac{1}{2}(N+1)(N-2),\; \tfrac{1}{6}\,N(N^{2}-7),\; \dots\right)\,.
\end{equation}
 
\begin{remark}[First-difference recursion]\label{rem:h-recursion}
	We observe that the first-difference sequence of the first half of $h$,
	\begin{equation}
		\begin{aligned}
			\hat{h} \;=\; \bigl(h_{0},\; h_{1}-h_{0},\; h_{2}-h_{1},\; &\dots,\; h_{\lfloor d/2\rfloor}-h_{\lfloor d/2\rfloor-1}\bigr)
			\\
			&\;=\;\left(1,\; N-2,\; \tfrac{1}{2}N(N-3),\; \tfrac{1}{6}(N-1)(N^{2}-2N-6),\; \dots\right)\,,
		\end{aligned}
	\end{equation}
	coincides, for $k < N$, with the $h$-sequence of $FT[SU(N-1)]$. This is in fact an exact consequence of the factorisation $[N]!_{t} = [N-1]!_{t}\,\bigl(1+t+\dots+t^{N-1}\bigr)$, which implies $h_{k}(N)-h_{k-1}(N) = h_{k}(N-1)-h_{k-N}(N-1)$: the identity holds exactly for $k<N$ and fails for $k \geq N$ (the first violation inside the first half of the sequence occurs at $k=N=5$).
\end{remark}

\paragraph{Mahonian numbers.}
The coefficients $h_{i}$ of the numerator admit a classical combinatorial interpretation. Since $\prod_{j=1}^{N-1}\sum_{h=0}^{j} t^{h} = [N]!_{t}$, where the $q$-factorial is defined by
\begin{equation}
	[n]!_{q} \;=\; \prod_{k=1}^{n-1}\bigl(1 + q + q^{2} + \cdots + q^{k}\bigr)\,,
\end{equation}
one has
\begin{equation}
	h_{i} \;=\; \frac{1}{i!}\,\frac{d^{\,i}}{dt^{\,i}}\,[N]!_{t}\,\bigg|_{t=0}\,.
\end{equation}
The $h_{i}$ are the \emph{Mahonian numbers}, first studied by MacMahon \cite{MacMahon:1913}; they count permutations of $N$ elements by major index (equivalently, by number of inversions).
 
\paragraph{Counting BPS operators.}
The coefficient of $t^{n}$ in the unrefined Hilbert series \eqref{HSftuNUR} equals the number of independent gauge-invariant operators of $U(1)_{\tau}$ charge $n$. For low $N$ one finds the closed-form expressions
\begin{align}
    N=1: \quad & 1\,, \\[2pt]
    N=2: \quad & \tfrac{1}{12}\,(n+1)(n+2)^{2}(n+3)\,, \\[2pt]
    N=3: \quad & \tfrac{1}{2075673600}\,(n+1)(n+2)(n+3)(n+4)(n+5)(n+6)(n+7) \nonumber \\
               & \phantom{\tfrac{1}{2075673600}}\times(n+8)(n+9)(n+10)(2n+11)\bigl(n(n+11)+52\bigr)\,, \\
    \cdots \nonumber
\end{align}
The $N=1$ case reflects the fact that $FT_{1}$ is trivial, with a zero-dimensional moduli space consisting of a single point. The jump in complexity from $N=2$ to $N=3$ is notable: while $N=2$ gives a fully factorised quartic, $N=3$ already involves an irreducible quadratic factor $n(n+11)+52$, signalling that the Hilbert polynomial ceases to split into linear factors for $N\geq 3$.

\paragraph{The full Hilbert series as the limit of the rescaled index}\label{sec:FT_limit}
We now show that, for $FT_N$, the limit \eqref{eq:HS_limit} can be evaluated directly on the localisation formula of the UV quiver \eqref{ftsun_genN}, and that this derives \eqref{ftHS} for all $N$.

\paragraph{Evaluation on the integrand.}
The rescaling \eqref{eq:rescaling} acts on fugacities only. A UV chiral $\Phi$ in the
representation with weights $\rho$ has charges $(Q_\tau^\Phi,Q_\Delta^\Phi,w_\Phi)$ defined as in \eqref{eq:w_def}, and in the flux sector $m$ contributes \cite{Imamura:2011,Kapustin:2011,Kim:2009wb}
\be
\prod_{\rho\in\Phi}(-1)^{\rho(m)}\bigl(x^{1-w_\Phi}\,t_a^{Q^\Phi_\tau}t_d^{-Q^\Phi_\Delta}\,z^{-\rho}\bigr)^{\frac{|\rho(m)|}{2}}
\frac{\bigl(t_a^{Q^\Phi_\tau}t_d^{-Q^\Phi_\Delta}\,x^{2-w_\Phi+|\rho(m)|}\,z^{-\rho};x^2\bigr)_\infty}
{\bigl(t_a^{-Q^\Phi_\tau}t_d^{Q^\Phi_\Delta}\,x^{w_\Phi+|\rho(m)|}\,z^{\rho};x^2\bigr)_\infty},
\ee
while the vector multiplet gives $\prod_\alpha x^{-|\alpha(m)|/2}\bigl(1-x^{|\alpha(m)|}z^\alpha\bigr)$.
Here $z=(z_1,\dots,z_{\mathrm{rk}\,G})$ are the gauge fugacities, $z_a=e^{i\sigma_a}$ with $\sigma_a$ the holonomy of the gauge field along $\mathbf{S}^1$ in the Cartan of the gauge group $G$, $z^\rho\equiv\prod_a z_a^{\rho_a}$ and $\rho(m)\equiv\sum_a\rho_a m_a$ (similarly $\alpha(m)$); they are integrated over with the Haar measure $\oint\prod_a\frac{dz_a}{2\pi i z_a}$, divided by the order of the Weyl group of the subgroup of $G$ left unbroken by $m$. The non-abelian flavour fugacities $\{\mu\}$ enter in the same way as $z$ and are suppressed. In all the quivers considered here the matter is vector-like at each unitary node ($b_i$ and $\tilde b_i$, $d_i$ and $v_i$) and in real representations at the symplectic nodes, so the weights of the chirals come in pairs $\pm\rho$ with equal $|\rho(m)|$: the sign $(-1)^{\rho(m)}$ and the gauge charge $z^{-\rho|\rho(m)|/2}$ of the zero-point factor cancel identically in every flux sector.
Provided that at each fixed order in $t_a,t_d$ only finitely many flux sectors contribute (for $FT_N$ this is shown below), the limit may be taken inside the flux sum and the gauge integral.
When all UV fields have $0\le w_\Phi<2$ the result is particularly simple, because
\be
\bigl(X\,x^{a};x^2\bigr)_\infty\ \xrightarrow{\ x\to0\ }\ \begin{cases}1, & a>0,\\ 1-X, & a=0.\end{cases}
\ee
All fermionic letters (which carry $x^{2-w_\Phi+|\rho(m)|}$) drop out, and so do all bosons with $w_\Phi+|\rho(m)|>0$; a flux sector survives only if its zero-point power of $x$, $\epsilon(m)=\sum_\Phi\sum_{\rho\in\Phi}\frac{1-w_\Phi}{2}|\rho(m)|-\sum_\alpha\frac{|\alpha(m)|}{2}$, vanishes.
When some UV fields lie outside this range, as for $d_i$ with $i<N$ ($w_{d_i}=i-N<0$) and $v_i$ with $i\le N-2$ ($w_{v_i}=N-i\ge2$) in $FC_N$, $FH_N$, $FM_N$ and $FE_N$, higher terms of the $q$-Pochhammers and fermionic letters can also contribute at $x^0$, so the limit cannot be taken term-by-term. Moreover, for $w_\Phi<0$ poles of the integrand cross the integration contour as $y\to0$, so the limit must be taken on the index itself, after the gauge integral (see Section~\ref{sec:computation}).

For the quiver \eqref{ftsun_genN} of $FT_N$ (gauge group $\prod_{i=1}^{N-1}U(i)$, flavour $SU(N)_L$, fluxes $m_i=(m_{i,1},\dots,m_{i,i})$),
the UV fields have
\be
\begin{array}{c|c|c|c}
	\Phi & Q_\tau & w & \text{representation} \\ \hline
	b_i,\ \tilde b_i & \tfrac12 & 1 & U(i)\times U(i+1)\ \text{bifundamental}\ (U(N)\to SU(N)_L) \\
	a_i,\ i<N & -1 & 0 & U(i)\ \text{traceless adjoint} \\
	a_N & -1 & 0 & SU(N)_L\ \text{adjoint} \\
	\beta_i=\mathrm{Flip}[b_i\tilde b_i] & -1 & 0 & \text{singlet}
\end{array}
\ee
All $w_\Phi\in\{0,1\}$, so the simplifications above apply.
\begin{itemize}
	\item \emph{Matter.} Every fermion and every $b_i,\tilde b_i$ (which have $w=1$) drop out, together with all $F$-term relations. Among the remaining letters, at node $i$ the components of $a_i$ with $\rho(m)=0$, together with $\beta_i$, form exactly the adjoint of the residual gauge group $H_{m_i}=\prod_{\text{blocks}}U(k)\subset U(i)$ left unbroken by $m_i$; $a_N$ survives entirely.
	\item \emph{Zero-point powers of $x$.} The roots of the traceless $a_i$ cancel those of the $U(i)$ vector multiplet, so $\epsilon(m)=0$ identically and every flux sector contributes.
	\item \emph{Zero-point powers of $t$.} Setting $t\equiv t_a$, the zero-point factors $t^{\,Q^\Phi_\tau|\rho(m)|/2}$ of the chirals give the monopole $t^{\Delta_{\mathcal{C}}(m)}$ with
	\be
	\begin{split}
		\Delta_{\mathcal{C}}(m)={}&\frac12\sum_{i=1}^{N-2}\sum_{a=1}^{i}\sum_{b=1}^{i+1}|m_{i,a}-m_{i+1,b}|+\frac N2\sum_{a=1}^{N-1}|m_{N-1,a}|\\
		&-\sum_{i=1}^{N-1}\sum_{a<b}|m_{i,a}-m_{i,b}| ,
	\end{split}
	\ee
	which is the conformal dimension of the bare monopole of the $\mathcal{N}=4$ theory $T[SU(N)]$ \cite{Borokhov_2002sac,Cremonesi_2014,Gaiotto_2009}. Since $T[SU(N)]$ is a good theory, $\Delta_{\mathcal{C}}(m)\ge1$ for all $m\neq0$ and grows with $|m|$, so at each order in $t$ only finitely many flux sectors contribute.
\end{itemize}
The gauge integral of the surviving letters is the Molien integral of the adjoint of $H_m$, i.e.\ the dressing factor of the monopole formula \cite{Cremonesi_2014}:
\be
P(t;m)=\prod_{i=1}^{N-1}\ \prod_{\text{blocks of }m_i}\ \prod_{j=1}^{k}\frac{1}{1-t^{j}} ,
\ee
with $k$ the size of each block of equal fluxes. Including the topological fugacities $\omega_i$ of the gauge nodes and summing over fluxes modulo Weyl reflections, $m_{i,1}\ge\dots\ge m_{i,i}$, we obtain
\be
\begin{split}
	\lim_{y\to0}\mathcal{I}_{FT_N}\bigl(x\,y,\ \psi_\tau\,y^{\tau-2}\bigr)
	&=\mathrm{PE}\bigl[\chi_{\mathrm{adj}_L}\,t\bigr]\ \sum_{m}\ t^{\Delta_{\mathcal{C}}(m)}\,P(t;m)\prod_{i=1}^{N-1}\omega_i^{\sum_a m_{i,a}}\\
	&=\mathrm{PE}\bigl[\chi_{\mathrm{adj}_L}\,t\bigr]\times\CH\CS\bigl[\mathcal{C}\bigl(T[SU(N)]\bigr)\bigr](t,\omega),
\end{split}
\ee
where the second factor is the monopole formula of \cite{Cremonesi_2014} for the Coulomb branch of $T[SU(N)]$, in the grading where $t$ counts the $\mathcal{N}=4$ conformal dimension $\Delta_{\mathcal{C}}$.
The $\omega_i$ combine into the Cartan of $SU(N)_R$, and this Coulomb branch is the nilpotent cone $\mathcal{N}_{\mathfrak{sl}(N)}$. Its coordinate ring is the polynomial ring
on $\mathfrak{sl}(N)$ modulo the $N-1$ Casimirs, a complete intersection, so
$\CH\CS[\mathcal{N}_{\mathfrak{sl}(N)}]=\mathrm{PE}\bigl[\chi_{\mathrm{adj}_R}\,t-\sum_{j=2}^{N}t^{j}\bigr]$~\cite{Hanany:2011db}. Altogether,
\be
\begin{split}
	\lim_{y\to0}\mathcal{I}_{FT_N}\bigl(x\,y,\ \psi_\tau\,y^{\tau-2}\bigr)
	&=\mathrm{PE}\Bigl[\chi_{\mathrm{adj}_L}\,t+\chi_{\mathrm{adj}_R}\,t-\sum_{j=2}^{N}t^{j}\Bigr]\\
	&\xrightarrow{\ \text{unrefined}\ }\ \frac{\prod_{j=2}^{N}(1-t^{j})}{(1-t)^{2(N^2-1)}}
	=\frac{\prod_{j=1}^{N-1}\sum_{h=0}^{j}t^{h}}{(1-t)^{2N^2-N-1}} ,
\end{split}
\ee
which reproduces \eqref{ftHS} and \eqref{HSftuNUR}. For $FT_N$ the limit thus derives exactly the closed form of the Hilbert series.
We have checked the flux sum explicitly against the closed form for $N=2,3,4$, through orders $t^{12}$, $t^{10}$ and $t^{6}$ respectively. 

\subsection{\texorpdfstring{$FC_N$}{fcsun}}
We now move to a theory which actually contains a bifundamental in the spectrum.

\be
\resizebox{\hsize}{!}{
\begin{tikzpicture}[
    thick,
    gauge/.style={
        circle, draw=black, fill=white,
        minimum size=8mm, inner sep=1pt, font=\small
    },
    manifest/.style={
        rectangle, draw=blue!70!black, fill=blue!6,
        minimum size=8mm, inner sep=2pt,
        font=\small, text=blue!70!black
    },
    flavour/.style={
        rectangle, draw=red!70!black, fill=red!6,
        minimum size=7mm, inner sep=2pt,
        font=\small, text=red!70!black
    },
    arr/.style={-{Stealth[scale=0.8]}, shorten >=1pt, shorten <=1pt},
    barr/.style={draw=black!70}
]

\node[gauge]    (g1) at (0,0)     {$1$};
\node[gauge]    (g2) at (2.5,0)   {$2$};
\node           (dt) at (4,0)     {$\cdots$};
\node[gauge]    (g3) at (5.5,0)   {$\scriptstyle N{-}1$};
\node[manifest] (m3) at (7.5,0)   {$N$};

\node[flavour] (x1) at (-0.75,-1.7) {$1$};
\node[flavour] (x2) at (1.25,-1.7)  {$1$};
\node[flavour] (x3) at (3.25,-1.7)  {$1$};
\node[flavour] (x4) at (6.5,-1.7)   {$1$};

\draw[arr,barr,transform canvas={yshift= 3.5pt}] (g1) -- (g2);
\draw[arr,barr,transform canvas={yshift=-3.5pt}] (g2) -- (g1);
\node at (1.25, 0.45) {$b_1$};
\node at (1.25, 0) {\huge$\times$};

\draw[arr,barr,transform canvas={yshift= 3.5pt}] (g2) -- (dt);
\draw[arr,barr,transform canvas={yshift=-3.5pt}] (dt) -- (g2);
\node at (3.2, 0.45) {$b_2$};
\node at (3.2, 0) {\huge$\times$};

\draw[arr,barr,transform canvas={yshift= 3.5pt}] (dt) -- (g3);
\draw[arr,barr,transform canvas={yshift=-3.5pt}] (g3) -- (dt);
\node at (4.75, 0.45) {$b_{N-2}$};
\node at (4.75, 0) {\huge$\times$};

\draw[arr,barr,transform canvas={yshift= 3.5pt}] (g3) -- (m3);
\draw[arr,barr,transform canvas={yshift=-3.5pt}] (m3) -- (g3);
\node at (6.5, 0.45) {$b_{N-1}$};
\node at (6.5, 0) {\huge$\times$};

\draw (g2) to[out=65,in=115,looseness=3] (g2);
\node at (2.5, 0.85) {$a_2$};

\draw (g3) to[out=65,in=115,looseness=3] (g3);
\node at (5.5, 0.95) {$a_{N-1}$};

\draw[blue!70!black] (m3) to[out=65,in=115,looseness=3] (m3);
\node[blue!70!black] at (7.5, 0.95) {$a_N$};

\draw[arr,barr] (x1) to[bend right=5] (g1);
\node at (-1.05,-0.75) {$d_1$};

\draw[arr,barr] (g1) to[bend left=5] (x2);
\node at (1.0,-0.85) {$v_1$};

\draw[arr,barr] (x2) to[bend right=5] (g2);
\node at (2.35,-0.85) {$d_2$};

\draw[arr,barr] (g2) to[bend left=5] (x3);
\node at (3.25,-0.8) {$v_2$};

\draw[arr,barr] (4.6,-1.5) to[bend right=5] (g3);
\node at (4.3,-0.8) {$d_{N-1}$};

\draw[arr,barr] (g3) to[bend left=5] (x4);
\node at (6.45,-0.55) {$v_{N-1}$};

\draw[arr,barr] (x4) to[bend left=5] (m3);
\node at (7.55,-0.8) {$d_N$};

\node at (3.5,-4) {$\begin{aligned}
    \mathcal{W} =\;& \sum_{j=1}^{N-1}v_j b_j d_{j+1}
        + a_2 b_1\tilde{b}_1
        + \sum_{j=2}^{N-1} b_j(a_j+a_{j+1})\tilde{b}_j
        + \sum_{i=1}^{N-1}\mathrm{Flip}[b_i\tilde{b}_i]\\
    &+\left(\mathfrak{M}^{+00\cdots}
        +\mathfrak{M}^{0+0\cdots}+\cdots\right)
\end{aligned}$};

\node[right] at (8.5,-0.6) {%
    \renewcommand{\arraystretch}{1.4}
    \begin{tabular}{>{\small}c >{\small}c}
        \toprule
        $b_i,\,\tilde{b}_i$ & $\tau/2$ \\
        $a_i$               & $2-\tau$ \\
        $v_i$ & $2-\tfrac{i-N+2}{2}\tau-\Delta$ \\
        $d_i$ & $\Delta+\tfrac{i-N}{2}\tau$ \\
        \bottomrule
    \end{tabular}%
};

\end{tikzpicture}
}
\ee

\paragraph{Bigraded structure of the unrefined HS.}
Unrefining the non-abelian fugacities but retaining the two abelian gradings $t_a$ (associated to the adjoints) and $t_d$ (associated to the bifundamental), the Hilbert series of $FC_N$ takes the form
\begin{equation}\label{fcnHS}
    \CH\CS_{FC_N}(t_a,t_d) \;=\; \frac{\mathrm{Num}(t_a,t_d)}{(1-t_d)^{N^{2}}\,(1-t_a)^{\,2N^{2}-N-1}}\,,
    \qquad
    \mathrm{Num}(t_a,t_d) \;\equiv\; \sum_{i,j=0}^{N(N-1)} a_{i,j}\, t_a^{\,i}\, t_d^{\,j}\,.
\end{equation}
The two exponents in the denominator reflect the structure of the moduli space on the two ``pure'' branches:
\begin{itemize}
    \item \emph{Bifundamental branch} ($t_a=0$).\ The subring generated by $\Pi$ is freely generated, so
    \begin{equation}
        \CH\CS_{FC_N}(0,t_d) \;=\; \frac{1}{(1-t_d)^{N^{2}}}\,,
        \qquad a_{0,j} = \delta_{j,0}\,.
    \end{equation}
    \item \emph{Adjoint branch} ($t_d=0$).\ The subring generated by $(A_L, A_R)$ reproduces the full chiral ring of $FT_N$,
    \begin{equation}
        \CH\CS_{FC_N}(t_a,0) \;=\; \CH\CS_{FT_N}(t_a) \;=\; \frac{\prod_{j=1}^{N-1}\sum_{h=0}^{j} t_a^{h}}{(1-t_a)^{\,2N^{2}-N-1}}\,,
    \end{equation}
    from which one reads off $a_{i,0} = \frac{1}{i!}\,\partial_{t}^{i}[N]!_{t}\big|_{t=0}$.
\end{itemize}
 
\paragraph{Degree of the numerator.}
Explicit SCI computations yield $d[2]=2$ and $d[3]=6$ for the highest degree of $\mathrm{Num}(t_a,t_d)$ in each of $t_a$ and $t_d$ (the two being equal, so that the total degree is $2d[N]$). Together with $d[1]=0$ and the expectation that $d[N]$ is quadratic in $N$, these values fix
\begin{equation}
    d[N] \;=\; N(N-1)
\end{equation}
for generic $N$.
 
\paragraph{Constraints from diagonal and boundary specialisations.}
The numerator satisfies three useful identities, which we have verified explicitly for $N=2,3$ and expect to hold for all $N$. On the diagonal $t_a = t_d = t$, one finds
\begin{equation}\label{eq:diag-factorisation}
    \mathrm{Num}(t,t) \;=\; \sum_{i,j} a_{i,j}\, t^{\,i+j}
    \;=\; (1-t)^{N(N-1)}\,\sum_{k=0}^{N^{2}-N} c_{k}\, t^{\,k}\,,
    \qquad c_{k} = c_{N^{2}-N-k}\,.
\end{equation}
On the two boundary specialisations,
\begin{equation}\label{eq:boundary-specialisations}
    \mathrm{Num}(1,t_d) \;=\; N!\,(1-t_d)^{N(N-1)}\,,
    \qquad
    \mathrm{Num}(t_a,1) \;=\; (1-t_a)^{N(N-1)}\,,
\end{equation}
which translate into the following row and column sum rules for the coefficient matrix $\{a_{i,j}\}$:
\begin{equation}\label{eq:sum-rules}
    \sum_{i} a_{i,j} \;=\; (-1)^{j}\, N!\, \binom{N(N-1)}{j}\,,
    \qquad
    \sum_{j} a_{i,j} \;=\; (-1)^{i}\, \binom{N(N-1)}{i}\,.
\end{equation}
 
\paragraph{Palindromic symmetry of columns.}
The coefficient matrix exhibits a column-wise palindromic symmetry,
\begin{equation}\label{eq:palindrome}
    a_{j,k} \;=\; a_{k-j+N(N-1)/2,\, k}
    \quad \text{for } j \leq k + N(N-1)/2\,,
    \qquad
    a_{j,k} \;=\; 0 \quad \text{for } j > k + N(N-1)/2\,,
\end{equation}
which lifts to the functional identity
\begin{equation}
    \mathrm{Num}(t_a,t_d) \;=\; t_a^{\,N(N-1)/2}\, \mathrm{Num}\!\left(1/t_a,\, t_a t_d\right)\,.
\end{equation}
 
\paragraph{Explicit coefficient matrices.}
For low $N$ the coefficient matrices are
\begin{equation}
    N=1:\ \begin{pmatrix} 1 \end{pmatrix}\,,
    \qquad
    N=2:\ \begin{pmatrix}
        1 & 0 & 0 \\
        1 & -4 & 1 \\
        0 & 0 & 1
    \end{pmatrix}\,,
\end{equation}
\begin{equation}
    N=3:\ \begin{pmatrix}
         1 &  0 &  0 &   0 &  0 &  0 & 0 \\
         2 & -9 &  0 &   1 &  0 &  0 & 0 \\
         2 &-18 & 45 & -14 &  0 &  0 & 0 \\
         1 & -9 & 45 & -94 & 45 & -9 & 1 \\
         0 &  0 &  0 & -14 & 45 &-18 & 2 \\
         0 &  0 &  0 &   1 &  0 & -9 & 2 \\
         0 &  0 &  0 &   0 &  0 &  0 & 1
    \end{pmatrix}\,.
\end{equation}
Inspection of the upper-left corner for general $N$ reveals the pattern
\begin{equation}
    \{a_{j,k}\} \;=\;
    \begin{pmatrix}
        1 & 0 & 0 & 0 & 0 & \cdots \\[2pt]
        N-1 & -N^{2} & 0 & 0 & \cdots & \\[2pt]
        \tfrac{1}{2}(N-2)(N+1) & -N^{2}(N-1) & & \cdots & & \\[2pt]
        \tfrac{1}{6}N(N^{2}-7) & -\tfrac{N^{2}}{2}(N-2)(N+1) & \cdots & & & \\[2pt]
        \vdots & & \ddots & & &
    \end{pmatrix}\,,
\end{equation}
exhibiting two notable regularities:
\begin{itemize}
    \item The second column equals $-N^{2}$ times the first column, which is the truncated Mahonian sequence, with the central entry of the palindrome omitted.
    \item The second row reads $(N-1,\,-N^{2},\,0,\dots,0,\,1,\,0,\dots,0)$, where the isolated $1$ sits in position $N$. This entry corresponds precisely to the syzygy $t_a t_d^{\,N}$ that appears in the plethystic logarithm below.
\end{itemize}
 
\paragraph{The fully unrefined Hilbert series.}
Setting $t_a = t_d = t$ in \eqref{fcnHS} and using \eqref{eq:diag-factorisation}, one obtains
\begin{equation}
    \CH\CS_{FC_N}(t,t) \;=\; \frac{\sum_{k=0}^{N^{2}-N} c_{k}\, t^{\,k}}{(1-t)^{\,2N^{2}-1}}\,,
\end{equation}
with numerators for low $N$ given by
\begin{align}
    N=1:\quad & 1\,, \nonumber\\
    N=2:\quad & 1 + 3t + t^{2}\,, \\
    N=3:\quad & 1 + 8t + 26t^{2} + 39t^{3} + 26t^{4} + 8t^{5} + t^{6}\,. \nonumber
\end{align}
The coefficients of the numerator form a Stanley-Iarrobino (SI) sequence (see App.~\ref{appHS} for a review of $h$-sequences and SI sequences): it is the Hilbert function of a Gorenstein Artinian
ring~\cite{tHarima}. Following ~\cite{Stanley:1978}, this Gorenstein property is the algebraic counterpart of the Calabi--Yau property of the moduli space. Expanding to the first few orders,
\begin{equation}
    \CH\CS_{FC_N}(t,t) \;=\; 1 + (3N^{2}-2)\,t + \left[\binom{3N^{2}-1}{2} - (N^{2}+1)\right] t^{2} + \cdots\,.
\end{equation}
 
\paragraph{Generators and relations.}
The leading terms in the plethystic logarithm of the refined Hilbert series read
\begin{equation}\label{eq:PL-FCN}
    \begin{aligned}
    \mathrm{PL}\!\left[\CH\CS_{FC_N}\right](t_a,t_d) \;=\;{}& \chi_{(\mathbf{N},\overline{\mathbf{N}})}\,t_d
      \;+\; \bigl(\chi_{(\mathrm{adj}_L,\mathbf{1})} + \chi_{(\mathbf{1},\mathrm{adj}_R)}\bigr)\,t_a
      \;-\; \sum_{k=2}^{N} t_a^{\,k} \\
    &{}-\; \chi_{(\mathbf{N},\overline{\mathbf{N}})}\,t_a t_d
      \;+\; t_a\, t_d^{\,N}
      \;+\; \chi_{(\mathbf{N},\overline{\mathbf{N}})}\,t_a^{\,N}\, t_d
      \;+\; \cdots\,,
    \end{aligned}
\end{equation}
where $\chi_R$ denotes the $SU(N)_L\times SU(N)_R$ character of the representation $R$. From \eqref{eq:PL-FCN} we read off the following chiral-ring data.
 
\emph{Generators} ($3N^{2}-2$ in total):
\begin{itemize}
    \item the bifundamental $\Pi$ at order $t_d$,
    \item the two adjoints $A_L, A_R$ at order $t_a$.
\end{itemize}
 
\emph{Relations:}
\begin{equation}
    \begin{aligned}
        A_L\, \Pi &\;=\; \Pi\, A_R\,,
        && \bigl(N^{2} \text{ relations at order } t_a t_d\bigr)\,, \\[2pt]
        \mathrm{tr}_{L} A_L^{k} &\;=\; \mathrm{tr}_{R} A_R^{k}\,,
        && \bigl(\text{at order } t_a^{k},\ k=2,\dots,N\bigr)\,.
    \end{aligned}
\end{equation}
 
Unlike the case of $FT_N$, the plethystic logarithm of $\CH\CS_{FC_N}$ is \emph{not} a finite sum: higher-order syzygies (such as those at orders $t_a t_d^{N}$ and $t_a^{N} t_d$, appearing with positive coefficients in~\eqref{eq:PL-FCN}) persist indefinitely. The moduli space of $FC_N$ is therefore not a complete intersection. The set of generators and the leading relations obtained here agree with the results of~\cite{Benvenuti:2024mpn}.

\subsection{\texorpdfstring{$FH_N$}{fhn}}

\be
\resizebox{\hsize}{!}{
\begin{tikzpicture}[
    thick,
    gauge/.style={
        circle, draw=black, fill=white,
        minimum size=8mm, inner sep=1pt, font=\small
    },
    manifest/.style={
        rectangle, draw=blue!70!black, fill=blue!6,
        minimum size=8mm, inner sep=2pt,
        font=\small, text=blue!70!black
    },
    flavour/.style={
        rectangle, draw=red!70!black, fill=red!6,
        minimum size=7mm, inner sep=2pt,
        font=\small, text=red!70!black
    },
    arr/.style={-{Stealth[scale=0.8]}, shorten >=1pt, shorten <=1pt},
    barr/.style={draw=black!70}
]

\node[gauge]    (g1) at (0,0)   {$1$};
\node[gauge]    (g2) at (2.5,0) {$2$};
\node           (dt) at (4,0)   {$\cdots$};
\node[gauge]    (g3) at (5.5,0) {$\scriptstyle N{-}1$};
\node[manifest] (m3) at (7.5,0) {$N$};

\node[flavour] (x0) at (-0.75,-1.7) {$C_1$};
\node[flavour] (x1) at (1.25,-1.7)  {$C_1$};
\node[flavour] (x2) at (3.1,-1.7)   {$C_1$};
\node[flavour] (x3) at (6.5,-1.7)   {$C_1$};

\draw[arr,barr,transform canvas={yshift= 3.5pt}] (g1) -- (g2);
\draw[arr,barr,transform canvas={yshift=-3.5pt}] (g2) -- (g1);
\node at (1.25, 0.45) {$b_1$};
\node at (1.25, 0) {\huge$\times$};

\draw[arr,barr,transform canvas={yshift= 3.5pt}] (g2) -- (dt);
\draw[arr,barr,transform canvas={yshift=-3.5pt}] (dt) -- (g2);
\node at (3.2, 0.45) {$b_2$};
\node at (3.2, 0) {\huge$\times$};

\draw[arr,barr,transform canvas={yshift= 3.5pt}] (dt) -- (g3);
\draw[arr,barr,transform canvas={yshift=-3.5pt}] (g3) -- (dt);
\node at (4.75, 0.45) {$b_{N-2}$};
\node at (4.75, 0) {\huge$\times$};

\draw[arr,barr,transform canvas={yshift= 3.5pt}] (g3) -- (m3);
\draw[arr,barr,transform canvas={yshift=-3.5pt}] (m3) -- (g3);
\node at (6.5, 0.45) {$b_{N-1}$};
\node at (6.5, 0) {\huge$\times$};

\draw (g2) to[out=65,in=115,looseness=3] (g2);
\node at (2.5, 0.85) {$a_2$};

\draw (g3) to[out=65,in=115,looseness=3] (g3);
\node at (5.5, 0.95) {$a_{N-1}$};

\draw[blue!70!black] (m3) to[out=65,in=115,looseness=3] (m3);
\node[blue!70!black] at (7.5, 0.95) {$a_N$};


\draw[arr,barr] (x0) to[bend left=5] (g1);
\node at (-0.65,-0.8) {$d_1$};

\draw[arr,barr] (g1) to[bend right=5] (x1);
\node at (0.9,-0.75) {$v_1$};

\draw[arr,barr] (x1) to[bend left=5] (g2);
\node at (2.05,-0.82) {$d_2$};

\draw[arr,barr] (g2) to[bend right=5] (x2);
\node at (2.95,-0.72) {$v_2$};

\draw[arr,barr] (4.65,-1.55) to[bend left=5] (g3);
\node at (4.45,-0.82) {$d_{N-1}$};

\draw[arr,barr] (g3) to[bend right=5] (x3);
\node at (6.35,-0.65) {$v_{N-1}$};

\draw[arr,barr] (x3) to[bend left=5] (m3);
\node at (7.6,-0.82) {$d_N$};

\node at (3.5,-4) {$\begin{aligned}
    \mathcal{W} =\;& \sum_{j=1}^{N-1} v_j b_j d_{j+1}
        + a_2 b_1\tilde{b}_1
        + \sum_{j=2}^{N-1} b_j(a_j+a_{j+1})\tilde{b}_j
        + \sum_{i=1}^{N-1}\mathrm{Flip}[b_i\tilde{b}_i]\\
    &+\left(\mathfrak{M}^{\pm00\cdots}
        +\mathfrak{M}^{0\pm0\cdots}+\cdots\right)
\end{aligned}$};

\node[right] at (8.5,-0.6) {%
    \renewcommand{\arraystretch}{1.4}
    \begin{tabular}{>{\small}c >{\small}c}
        \toprule
        $b_i,\,\tilde{b}_i$ & $\tau/2$ \\
        $a_i$               & $2-\tau$ \\
        $v_i$ & $2-\tfrac{i-N+2}{2}\tau-\Delta$ \\
        $d_i$ & $\Delta+\tfrac{i-N}{2}\tau$ \\
        \bottomrule
    \end{tabular}%
};

\end{tikzpicture}
}
\ee

Unrefining the $SU(N)_L \times USp(2N)_R$ fugacities and just keeping the pair
$(t_a,t_d)$ of $U(1)_\tau \times U(1)_\D$ fugacities that keep track, respectively, of the adjoint-like and
bifundamental-like generators, the Hilbert series of $FH_N$ takes the
form
\begin{equation}
\label{eq:fhnHS}
\mathcal{HS}_{FH_N}(t_a,t_d)
=\frac{\displaystyle\sum_{i,j=0}^{\frac{3}{2}N(N-1)} a_{i,j}\,t_a^{\,i}\,t_d^{\,j}}
      {(1-t_d)^{2N^2-\frac{1}{2}N(N-1)}\,
       (1-t_a)^{\,\dim[ SU(N)]\,+\,\dim\{\mathrm{asym}[USp(2N)]\}\,-\,(N-1)}}\, .
\end{equation}
The numerator is a polynomial of bidegree
$\bigl(\tfrac{3}{2}N(N-1),\tfrac{3}{2}N(N-1)\bigr)$ whose
coefficients $a_{i,j}$ we now analyse.

\subsubsection*{The two complete-intersection slices}

Setting $t_a=0$ isolates the bifundamental sector.  The resulting
series is a complete intersection generated by $2N^2$ 
operators $\Pi^{a,A}$ (with $a,b$ a $USp(2N)$ index and $A,B$ an
$SU(N)$ index), subject to the $\tfrac{1}{2}N(N-1)$ quadratic
relations $\Omega_{ab}\,\Pi^{a,A}\Pi^{b,B}=0$:
\begin{equation}
\label{eq:HSta0}
\mathcal{HS}_{FH_N}(0,t_d)
=\frac{\sum_{j=0}^{\frac{3}{2}N(N-1)} a_{0,j}\,t_d^{\,j}}
      {(1-t_d)^{2N^2-\frac{1}{2}N(N-1)}}
=\frac{(1-t_d^{\,2})^{\frac{1}{2}N(N-1)}}{(1-t_d)^{2N^2}}
=\frac{(1+t_d)^{\frac{1}{2}N(N-1)}}{(1-t_d)^{2N^2-\frac{1}{2}N(N-1)}}\, .
\end{equation}
In particular
$a_{0,j}=\binom{N(N-1)/2}{\,j\,}$.

Dually, setting $t_d=0$ isolates the ``adjoint'' sector.  It is again
a complete intersection, with $3N^2-N-2$ generators -- the $SU(N)$
adjoint $A_L$ and the traceless antisymmetric of the emergent
$USp(2N)$, $A_R$ -- subject to the $N-1$ trace relations
$\mathrm{tr}_L A_L^{\,j}=\mathrm{tr}_R A_R^{\,j}$, $j=2,\dots,N$:
\begin{equation}
\label{eq:HStd0}
\mathcal{HS}_{FH_N}(t_a,0)
=\frac{\prod_{j=2}^{N}(1-t_a^{\,j})}{(1-t_a)^{3N^2-N-2}}
=\frac{\prod_{j=1}^{N-1}\sum_{h=0}^{j} t_a^{\,h}}
      {(1-t_a)^{3N^2-2N-1}}\, .
\end{equation}

\subsubsection*{Explicit results for small $N$}

For $N=1,2,3$, the matrices of numerator coefficients
$\bigl(a_{i,j}\bigr)$ in~\eqref{eq:fhnHS} are
\begin{equation}
\label{eq:aij-matrices}
\begin{aligned}
N=1:\quad &(1)\, ,\\[4pt]
N=2:\quad &
\begin{pmatrix}
 1 &  1 &  0 &  0 \\
 1 & -7 &  0 &  0 \\
 0 &  0 &  7 & -1 \\
 0 &  0 & -1 & -1
\end{pmatrix},\\[4pt]
N=3:\quad &
\begin{pmatrix}
 1 &   3 &    3 &    1 &    0 &     0 &    0 &    0 &   0 &   0 \\
 2 & -12 &  -42 &  -20 &    0 &     0 &    0 &    0 &   0 &   0 \\
 2 & -30 &  105 &  205 &    9 &    -3 &    0 &    0 &   0 &   0 \\
 1 & -15 &  153 & -547 & -318 &    54 &    0 &    0 &   0 &   0 \\
 0 &   0 &    0 & -202 & 1191 &    75 &  -59 &    3 &   0 &   0 \\
 0 &   0 &   -3 &   59 &  -75 & -1191 &  202 &    0 &   0 &   0 \\
 0 &   0 &    0 &    0 &  -54 &   318 &  547 & -153 &  15 &  -1 \\
 0 &   0 &    0 &    0 &    3 &    -9 & -205 & -105 &  30 &  -2 \\
 0 &   0 &    0 &    0 &    0 &     0 &   20 &   42 &  12 &  -2 \\
 0 &   0 &    0 &    0 &    0 &     0 &   -1 &   -3 &  -3 &  -1
\end{pmatrix}.
\end{aligned}
\end{equation}
All numerators are anti-palindromic.  In factored form,
\begin{equation}
\label{eq:num-factored}
\begin{aligned}
N=1:\quad &\mathrm{Num}(t,t)=1,\\
N=2:\quad &\mathrm{Num}(t,t)=(1+t)(1+4t+t^{2})(1-t)^3,\\
N=3:\quad &\mathrm{Num}(t,t)=(1+t)^{3}(1+3t+t^{2})(1+8t+22t^{2}+8t^{3}+t^{4})(1-t)^9,
\end{aligned}
\end{equation}
where we have set $t_a=t_d\equiv t$.

\subsubsection*{Structure of the coefficient matrix}

The $a_{i,j}$ fill a $\bigl(\tfrac{3}{2}N(N-1)+1\bigr)\times
\bigl(\tfrac{3}{2}N(N-1)+1\bigr)$ matrix.  Its upper-left corner has
the closed form
\begin{equation}
\label{eq:aij-corner}
\{a_{i,j}\}=
\begin{pmatrix}
 1 & \tfrac{1}{2}N(N-1) & \binom{N(N-1)/2}{2} & \binom{N(N-1)/2}{3} & \cdots \\[4pt]
 N-1 & \tfrac{1}{2}N(N^{2}-6N+1) & \cdots & & \\[4pt]
 \tfrac{1}{2}(N-2)(N+1) & \cdots & & & \\[4pt]
 \tfrac{1}{6}N(N^{2}-7) & \cdots & & & \\[4pt]
 \vdots & & & &
\end{pmatrix}.
\end{equation}

\subsubsection*{Sum rules}

We have identified four sum rules obeyed by the numerator.  The
specialisations $t_a=1$, $t_d=1$, and $t_a=t_d=t$ give
\begin{align}
\label{eq:sumrule1}
\mathrm{Num}(1,t) &= N!\,(1-t)^{\frac{3}{2}N(N-1)},\\[2pt]
\label{eq:sumrule2}
\mathrm{Num}(t,1) &= 2^{N(N-1)/2}\,(1-t)^{\frac{3}{2}N(N-1)},\\[2pt]
\label{eq:sumrule3}
\mathrm{Num}(t,t) &= \Biggl(\sum_{k=0}^{\frac{3}{2}N(N-1)} c_{k}\,t^{k}\Biggr)
                     (1-t)^{\frac{3}{2}N(N-1)},
\qquad c_{k}=c_{\frac{3}{2}N(N-1)-k}.
\end{align}
Finally, along every $k$-diagonal of $\{a_{i,j}\}$ (with $k=0$ the
main diagonal and $k>0$, $k<0$ denoting the diagonals above, resp.
below), the coefficients sum to zero.

\subsubsection*{Generators and relations}

The moduli space has $5N^{2}-N-2$ generators and
$\tfrac{1}{2}N(N-1)+2N^{2}+1$ quadratic relations, coming from
\[
\Pi\,\Pi=0,\qquad \Pi\,A_R = A_L\,\Pi,\qquad A_L^{\,2}=A_R^{\,2}.
\]
Accordingly, the low-order expansion of the Hilbert series reads
\begin{equation}
\label{eq:HS-expansion}
\mathcal{HS}_{FH_N}(t)
= 1 + (5N^{2}-N-2)\,t
   + \Bigl[\tbinom{5N^{2}-N-1}{2}-\bigl(\tfrac{1}{2}N(N-1)+2N^{2}+1\bigr)\Bigr]t^{2}
   + \cdots .
\end{equation}
The diagonal-specialised numerator coefficients $c_k$ defined
in~\eqref{eq:sumrule3} (i.e.\ those appearing in
$\mathrm{Num}(t,t)=(\sum_k c_k t^{k})(1-t)^{\frac{3}{2}N(N-1)}$, not
the coefficients of the HS expansion above) take the closed forms,
for $N>1$,
\[
c_{0}=1,\qquad
c_{1}=2N^{2}-N-1,\qquad
c_{2}=2N^{4}-2N^{3}-3N^{2}+N-1.
\]

The fully refined plethystic logarithm is
\begin{equation}
\label{eq:PL}
\begin{aligned}
\mathrm{PL}\!\left[\mathcal{HS}_{FH_N}\right](t_a,t_d)
&= \chi_{(\Box,\,\mathbf{2N})}\,t_d
 + \bigl(\chi_{(\mathrm{adj}_L,\mathbf{1})}
        +\chi_{(\mathbf{1},\mathrm{asym}_R)}\bigr)\,t_a
 - \sum_{k=2}^{N} t_a^{\,k} \\[2pt]
&\quad - \chi_{(\mathrm{asym}_L,\mathbf{1})}\,t_d^{\,2}
       - \chi_{(\Box,\,\mathbf{2N})}\,t_a t_d
       + \chi_{(\mathrm{sym}_L,\mathbf{1})}\,t_a\,t_d^{\,2}
       + \cdots ,
\end{aligned}
\end{equation}
where $\chi_R$ is the character of the representation $R$ under
$SU(N)_L\times USp(2N)_R$.  Reading~\eqref{eq:PL} off order by
order, one recognises: at order $t_d$ the bifundamental
$\Pi^{a,A}$; at order $t_a$ the $SU(N)$ adjoint $(A_L)^{A}_{\;B}$
together with the traceless antisymmetric $A_R^{\,ab}$ of
$USp(2N)$.  These fields satisfy
\begin{equation}
\label{eq:relations}
\begin{aligned}
\Omega_{ab}\,\Pi^{a,A}\Pi^{b,B}&=0
&& \bigl(\tfrac{1}{2}N(N-1)\ \text{relations at order } t_d^{\,2}\bigr),\\
\mathrm{tr}_L A_L^{\,k}&=\mathrm{tr}_R A_R^{\,k}
&& \bigl(\text{at order } t_a^{\,k},\ k=2,\dots,N\bigr),\\
\Pi^{a,B}(A_L)^{A}_{\;B}&=\Pi^{b,A}A_R^{\,ba}
&& \bigl(2N^{2}\ \text{relations at order } t_a t_d\bigr).
\end{aligned}
\end{equation}

The chiral-ring generators and relations identified above agree with
those of~\cite{Benvenuti:2024mpn}.  As in the previous cases, the
moduli space is \emph{not} a complete intersection: the relations
in~\eqref{eq:relations} are not independent, but obey higher-order
syzygies.

\subsection{\texorpdfstring{$FM_N$}{fmsun}}
\label{fmsun_hs}

\be
\resizebox{\hsize}{!}{
\begin{tikzpicture}[
    thick,
    gauge/.style={
        circle, draw=black, fill=white,
        minimum size=8mm, inner sep=1pt, font=\small
    },
    manifest/.style={
        rectangle, draw=blue!70!black, fill=blue!6,
        minimum size=8mm, inner sep=2pt,
        font=\small, text=blue!70!black
    },
    flavour/.style={
        rectangle, draw=red!70!black, fill=red!6,
        minimum size=7mm, inner sep=2pt,
        font=\small, text=red!70!black
    },
    arr/.style={-{Stealth[scale=0.8]}, shorten >=1pt, shorten <=1pt},
    barr/.style={draw=black!70}
]

\node[gauge]    (g1) at (0,0)     {$1$};
\node[gauge]    (g2) at (2.5,0)   {$2$};
\node           (dt) at (4,0)     {$\cdots$};
\node[gauge]    (g3) at (5.5,0)   {$\scriptstyle N{-}1$};
\node[manifest] (m3) at (7.5,0)   {$N$};

\node[flavour] (x1) at (-0.75,-1.7) {$1$};
\node[flavour] (x2) at (1.25,-1.7)  {$1$};
\node[flavour] (x3) at (3.25,-1.7)     {$1$};
\node[flavour] (x4) at (6.5,-1.7)   {$1$};

\draw[arr,barr,transform canvas={yshift= 3.5pt}] (g1) -- (g2);
\draw[arr,barr,transform canvas={yshift=-3.5pt}] (g2) -- (g1);
\node at (1.25, 0.45) {$b_1$};
\node at (1.25, 0) {\huge$\times$};

\draw[arr,barr,transform canvas={yshift= 3.5pt}] (g2) -- (dt);
\draw[arr,barr,transform canvas={yshift=-3.5pt}] (dt) -- (g2);
\node at (3.2, 0.45) {$b_2$};
\node at (3.2, 0) {\huge$\times$};

\draw[arr,barr,transform canvas={yshift= 3.5pt}] (dt) -- (g3);
\draw[arr,barr,transform canvas={yshift=-3.5pt}] (g3) -- (dt);
\node at (4.75, 0.45) {$b_{N-2}$};
\node at (4.75, 0) {\huge$\times$};

\draw[arr,barr,transform canvas={yshift= 3.5pt}] (g3) -- (m3);
\draw[arr,barr,transform canvas={yshift=-3.5pt}] (m3) -- (g3);
\node at (6.5, 0.45) {$b_{N-1}$};
\node at (6.5, 0) {\huge$\times$};

\draw (g2) to[out=65,in=115,looseness=3] (g2);
\node at (2.5, 0.85) {$a_2$};

\draw (g3) to[out=65,in=115,looseness=3] (g3);
\node at (5.5, 0.95) {$a_{N-1}$};

\draw[blue!70!black] (m3) to[out=65,in=115,looseness=3] (m3);
\node[blue!70!black] at (7.5, 0.95) {$a_N$};

\draw[arr,barr] (x1) to[bend right=5] (g1);
\draw[arr,barr] (g1) to[bend right=5] (x1);
\node at (-1,-0.8) {$d_1$};
\node[rotate=-60] at (-0.35,-0.8) {\LARGE$\times$};

\draw[arr,barr] (g1) to[bend left=5] (x2);
\draw[arr,barr] (x2) to[bend left=5] (g1);
\node at (1.1,-0.9) {$v_1$};

\draw[arr,barr] (x2) to[bend right=5] (g2);
\draw[arr,barr] (g2) to[bend right=5] (x2);
\node at (2.4,-0.9) {$d_2$};
\node[rotate=-55] at (1.85,-0.9) {\LARGE$\times$};

\draw[arr,barr] (g2) to[bend left=5] (x3);
\draw[arr,barr] (x3) to[bend left=5] (g2);
\node at (3.25,-0.85) {$v_2$};

\draw[arr,barr] (4.6, -1.5) to[bend right=5] (g3);
\draw[arr,barr] (g3) to[bend right=5] (4.5, -1.5);
\node at (4.35,-0.85) {$d_{N-1}$};
\node[rotate=-50] at (5,-0.8) {\LARGE$\times$};

\draw[arr,barr] (g3) to[bend left=5] (x4);
\draw[arr,barr] (x4) to[bend left=5] (g3);
\node at (6.5,-0.6) {$v_{N-1}$};

\draw[arr,barr] (m3) to[bend left=5] (x4);
\draw[arr,barr] (x4) to[bend left=5] (m3);
\node at (7.5,-0.85) {$d_N$};
\node[rotate=-50] at (7,-0.8) {\LARGE$\times$};

\node at (3.5,-4) {$\begin{aligned}
    \mathcal{W} =\;& \sum_{j=1}^{N-1}\!\left[v_j b_j d_{j+1}
        +\tilde{v}_j\tilde{b}_j\tilde{d}_{j+1}\right]
        + a_2 b_1\tilde{b}_1
        + \sum_{j=2}^{N-1} b_j(a_j+a_{j+1})\tilde{b}_j
        + \sum_{i=1}^{N-1}\mathrm{Flip}[b_i\tilde{b}_i]\\
    &+\left(\mathfrak{M}^{\pm00\cdots}
        +\mathfrak{M}^{0\pm0\cdots}+\cdots\right)
        +\sum_{i=1}^{N}\mathrm{Flip}[d_i\tilde{d}_i]
\end{aligned}$};

\node[right] at (8.5, -0.6) {%
    \renewcommand{\arraystretch}{1.4}
    \begin{tabular}{>{\small}c >{\small}c}
        \toprule
        $b_i,\,\tilde{b}_i$ & $\tau/2$ \\
        $a_i$               & $2-\tau$ \\
        $v_i,\,\tilde{v}_i$ & $2-\tfrac{i-N+2}{2}\tau-\Delta$ \\
        $d_i,\,\tilde{d}_i$ & $\Delta+\tfrac{i-N}{2}\tau$ \\
        \bottomrule
    \end{tabular}%
};

\end{tikzpicture}

}
\ee

The Hilbert series of the branch of $FM_N$ generated by the
bifundamentals $\Pi,\tilde{\Pi}$ and the adjoints $A_L,A_R$,
after unrefining the $SU(N)_L \times SU(N)_R \times U(1)_{\text{baryonic}}$
fugacities and keeping the pair $(t_a,t_d)$ of $U(1)_\tau \times U(1)_\D$ fugacities, takes the form
\begin{equation}
\label{eq:hsfmn}
\mathcal{HS}_{FM_N}(t_a,t_d)
=\frac{\displaystyle\sum_{i=0}^{N(N-1)}\sum_{j=0}^{N^{2}-1}
       a_{i,j}\,t_a^{\,i}\,t_d^{\,j}}
      {(1-t_d)^{N^{2}+1}\,(1-t_a)^{2N^{2}-N-1}}\, .
\end{equation}

\subsubsection*{The two limiting sectors}

The sector at $t_a=0$ is generated by the $2N^{2}$ bifundamental
components of $\Pi$ and $\tilde{\Pi}$, subject to the
$2(N^{2}-1)$ quadratic relations $\Pi\tilde{\Pi}=0$:
\begin{equation}
\label{eq:FMN-ta0}
\mathcal{HS}_{FM_N}(0,t_d)
=\frac{\sum_{j=0}^{(N-1)^{2}} a_{0,j}\,t_d^{\,j}}
      {(1-t_d)^{N^{2}+1}}\, .
\end{equation}
This sector is \emph{not} Calabi--Yau: already for $N=2,3$ the
numerator fails to be palindromic.

Dually, the sector at $t_d=0$ is generated by the $2N^{2}-2$ components
of the adjoints $A_L$ and $A_R$ and satisfies
\begin{equation}
\label{eq:FMN-td0}
\mathcal{HS}_{FM_N}(t_a,0)
=\frac{\sum_{j=0}^{N(N-1)/2} a_{j,0}\,t_a^{\,j}}
      {(1-t_a)^{2N^{2}-N-1}}\, ,
\end{equation}
where the $a_{j,0}$ are Mahonian numbers (as for $FT_N$).  This
sector \emph{is} of Calabi--Yau type.

\subsubsection*{Explicit results for small $N$}

For $N=1,2,3$, the numerator coefficients
$a_{i,j}$ of~\eqref{eq:hsfmn} are
\begin{equation}
\label{eq:FMN-aij}
\begin{aligned}
N=1:\quad &(1),\\[4pt]
N=2:\quad &
\begin{pmatrix}
 1 &  3 &  0 & 0\\
 1 & -5 & -5 & 1\\
 0 &  0 &  3 & 1
\end{pmatrix},\\[4pt]
N=3:\quad &
\begin{pmatrix}
 1 &   8 &  20 &   10 &    1 &    0 &   0 &   0 & 0\\
 2 &  -2 & -86 & -140 &  -16 &    2 &   0 &   0 & 0\\
 2 & -20 &  23 &  360 &  272 &  -38 &   1 &   0 & 0\\
 1 & -10 &  66 & -170 & -574 & -170 &  66 & -10 & 1\\
 0 &   0 &   1 &  -38 &  272 &  360 &  23 & -20 & 2\\
 0 &   0 &   0 &    2 &  -16 & -140 & -86 &  -2 & 2\\
 0 &   0 &   0 &    0 &    1 &   10 &  20 &   8 & 1
\end{pmatrix}.
\end{aligned}
\end{equation}
In every case the numerator is palindromic,
$a_{i,j}=a_{N(N-1)-i,\,N^{2}-1-j}$, and exhibits triangular
regions of zeros in the bottom-left and top-right corners.

\subsubsection*{Sum rules}

Setting the variables equal or to unity, the numerator factorises:
\begin{align}
\label{eq:FMN-sr1}
\mathrm{Num}(t,t) &= (1-t)^{N(N-1)}\,(1+t)^{N-1}\,
                     \sum_{j=0}^{N(N-1)} b_j\,t^{\,j},\\[2pt]
\label{eq:FMN-sr2}
\mathrm{Num}(1,t) &= N!\,(1-t)^{N(N-1)}\,(1+t)^{N-1},\\[2pt]
\label{eq:FMN-sr3}
\mathrm{Num}(t,1) &= C_N\,(t-1)^{N(N-1)},
\end{align}
where $C_N=\sum_j a_{0,j}$ and the auxiliary coefficients $b_j$ are
themselves palindromic.

\subsubsection*{Fully unrefined Hilbert series}

Setting $t_a=t_d=t$, the fully unrefined Hilbert series reads
\begin{equation}
\label{eq:FMN-unrefined}
\mathcal{HS}_{FM_N}(t,t)
=\frac{\sum_{j=0}^{N^{2}-1} c_j\,t^{\,j}}{(1-t)^{2N^{2}}}\, ,
\end{equation}
so that the denominator has degree $2N^{2}$ and the numerator
degree $N^{2}-1$.  For $N=1,2,3$,
\begin{equation}
\label{eq:FMN-unrefined-small}
\begin{aligned}
N=1:\quad & 1,\\
N=2:\quad & (1+t)(1+5t+t^{2}),\\
N=3:\quad & (1+t)^{2}\bigl(1+14t+72t^{2}+133t^{3}+72t^{4}+14t^{5}+t^{6}\bigr).
\end{aligned}
\end{equation}

\subsubsection*{Generators and relations}

The structure of the chiral ring is most transparent from the
plethystic logarithm of~\eqref{eq:hsfmn}:
\begin{equation}
\label{eq:FMN-PL}
\begin{aligned}
\mathrm{PL}\!\left[\mathcal{HS}_{FM_N}\right](t_a,t_d)
&= \bigl(\chi_{(\Box,\overline{\Box})}
        +\chi_{(\overline{\Box},\Box)}\bigr)\,t_d
 + \bigl(\chi_{(\mathrm{adj}_L,\mathbf{1})}
        +\chi_{(\mathbf{1},\mathrm{adj}_R)}\bigr)\,t_a
 - \sum_{k=2}^{N} t_a^{\,k} \\[2pt]
&\quad - \bigl(\chi_{(\mathrm{adj}_L,\mathbf{1})}
              +\chi_{(\mathbf{1},\mathrm{adj}_R)}\bigr)\,t_d^{\,2}
       - \bigl(\chi_{(\Box,\overline{\Box})}
              +\chi_{(\overline{\Box},\Box)}\bigr)\,t_a t_d \\[2pt]
&\quad + \mathcal{O}\!\bigl(t_a t_d\,(t_a+t_d)\bigr),
\end{aligned}
\end{equation}
where $\chi_R$ is the character of the representation $R$ under
$SU(N)_L\times SU(N)_R$.  Reading~\eqref{eq:FMN-PL} off order by
order, the generators of the main branch of the moduli space are:
\begin{itemize}
\item the two bifundamentals $\Pi$ and $\tilde{\Pi}$
      ($2N^{2}$ components at order $t_d$);
\item the two adjoints $A_L$ and $A_R$, transforming in the adjoint
      of $SU(N)_L$ and $SU(N)_R$ respectively
      ($2N^{2}-2$ components at order $t_a$).
\end{itemize}
These satisfy the relations
\begin{equation}
\label{eq:FMN-relations}
\begin{aligned}
\bigl(\Pi\tilde{\Pi}\bigr)\big|_{\mathrm{traceless}}
  =0=\bigl(\tilde{\Pi}\Pi\bigr)\big|_{\mathrm{traceless}},\qquad &
  \bigl(2N^{2}-2\ \text{relations at order } t_d^{\,2}\bigr),\\[2pt]
\mathrm{tr}_L A_L^{\,k}=\mathrm{tr}_R A_R^{\,k},\qquad &
  \bigl(\text{at order } t_a^{\,k},\ k=2,\dots,N\bigr),\\[2pt]
A_L\,\Pi=\Pi\,A_R,\qquad &
  \bigl(N^{2}\ \text{relations at order } t_a t_d\bigr),\\[2pt]
\tilde{\Pi}\,A_L=A_R\,\tilde{\Pi},\qquad &
  \bigl(N^{2}\ \text{relations at order } t_a t_d\bigr),
\end{aligned}
\end{equation}
where $\mathrm{tr}_L$ ($\mathrm{tr}_R$) denotes the trace over the
$SU(N)_L$ ($SU(N)_R$) indices.  The chiral-ring generators and
relations identified above agree with the proposal
of~\cite{Pasquetti:2019tix}.

\subsection{\texorpdfstring{$FE_N$}{feusp}}

\begin{equation}
\begin{tikzpicture}[
    thick,
    gauge/.style={
        circle, draw=black, fill=white,
        minimum size=8mm, inner sep=1pt, font=\small
    },
    manifest/.style={
        rectangle, draw=blue!70!black, fill=blue!6,
        minimum size=8mm, inner sep=2pt,
        font=\small, text=blue!70!black
    },
    flavour/.style={
        rectangle, draw=red!70!black, fill=red!6,
        minimum size=7mm, inner sep=2pt,
        font=\small, text=red!70!black
    },
    arr/.style={-{Stealth[scale=0.8]}, shorten >=1pt, shorten <=1pt},
    barr/.style={draw=black!70}
]

\node[gauge]    (g1) at (0,0)   {$C_1$};
\node[gauge]    (g2) at (2.5,0) {$C_2$};
\node           (dt) at (4,0)   {$\cdots$};
\node[gauge]    (g3) at (5.5,0) {$C_{N-1}$};
\node[manifest] (m1) at (7.5,0) {$C_N$};

\node[flavour] (f1) at (-0.75,-1.5) {$C_1$};
\node[flavour] (f2) at (1.25,-1.5)  {$C_1$};
\node[flavour] (f3) at (3.25,-1.5)  {$C_1$};
\node[flavour] (f4) at (6.5,-1.5)   {$C_1$};

\draw[barr] (g1) -- (g2);
\node at (1.25, 0.5) {$b_1$};
\node at (1.25, 0) {\LARGE$\times$};

\draw[barr] (g2) -- (dt);
\node at (3.2, 0.5) {$b_2$};
\node at (3.2, 0) {\LARGE$\times$};

\draw[barr] (dt) -- (g3);
\node at (4.55, 0.5) {$b_{N-2}$};
\node at (4.75, 0) {\LARGE$\times$};

\draw[barr] (g3) -- (m1);
\node at (6.5, 0.5) {$b_{N-1}$};
\node at (6.5, 0) {\LARGE$\times$};

\draw (g2) to[out=65,in=115,looseness=3] (g2);
\node at (2.5, 0.85) {$a_2$};

\draw (g3) to[out=65,in=115,looseness=3] (g3);
\node at (5.5, 1.15) {$a_{N-1}$};

\draw[blue!70!black] (m1) to[out=65,in=115,looseness=3] (m1);
\node[blue!70!black] at (7.5, 0.95) {$a_N$};

\draw[barr] (f1) -- (g1);
\node at (-1,-0.8) {$d_1$};
\node[rotate=-40] at (-0.4,-0.8) {\LARGE$\times$};

\draw[barr] (g1) -- (f2);
\node at (1,-0.8) {$v_1$};

\draw[barr] (f2) -- (g2);
\node at (2.2,-1) {$d_2$};
\node[rotate=-55] at (1.85,-0.75) {\LARGE$\times$};

\draw[barr] (g2) -- (f3);
\node at (3.2,-0.85) {$v_2$};
\draw[barr] (4.7, -1.3) -- (g3);
\node at (5.3,-1.3) {$d_{N-1}$};
\node[rotate=-50] at (5,-0.8) {\LARGE$\times$};

\draw[barr] (g3) -- (f4);
\node at (6.4,-0.65) {$v_{N-1}$};

\draw[barr] (m1) -- (f4);
\node at (7.45,-0.85) {$d_N$};
\node[rotate=50] at (6.95,-0.8) {\LARGE$\times$};

\node at (3.5,-3.5) {$\begin{aligned}
    \mathcal{W} =\;&
        b_1^2 a_2
        + \sum_{i=2}^{N-1} b_i^2(a_i + a_{i+1})
        + \sum_{i=1}^{N-1}\!\left[v_i b_i d_{i+1}
        + \mathrm{Flip}[b_i^2]\right] \\
    &+ \sum_{j=1}^{N} \mathrm{Flip}[d_j^2]
        + \left(\mathfrak{M}^{100\cdots0}
        + \mathfrak{M}^{010\cdots0}
        + \cdots
        + \mathfrak{M}^{00\cdots01}\right)
\end{aligned}$};

\node[right] at (8.5,-0.7) {%
    \renewcommand{\arraystretch}{1.4}
    \begin{tabular}{>{\small}c >{\small}c}
        \toprule
        $b_i$       & $\tau/2$ \\
        $a_i$       & $2-\tau$ \\
        $v_i$       & $2-\Delta-\tfrac{i-N+2}{2}\tau$ \\
        $d_i$       & $\Delta+\tfrac{i-N}{2}\tau$ \\
        \bottomrule
    \end{tabular}%
};

\end{tikzpicture}
\end{equation}

For the branch of the moduli space of $FE_N$ generated by the
bifundamental $\Pi$ and by the two traceless antisymmetric fields
$A_L,A_R$, unrefining the $USp(2N)_L\times USp(2N)_R$ fugacities to
the pair $(t_a,t_d)$, the Hilbert series takes the form
\begin{equation}
\label{eq:feusphs}
\mathcal{HS}_{FE_N}(t_a,t_d)
=\frac{\sum_{i=0}^{2N(N-1)}\sum_{j=0}^{(2N+1)(N-1)}
       a_{i,j}\,t_a^{\,i}\,t_d^{\,j}}
      {(1-t_d)^{2N^{2}+N+1}\,(1-t_a)^{4N^{2}-3N-1}}\, .
\end{equation}

\subsubsection*{The two limiting sectors}

The sector at $t_a=0$ is generated by the $4N^{2}$ components of the
bifundamental $\Pi^{a,A}$, subject to the $2N(2N-1)-2$ quadratic
relations
\[
\Omega_{ab}\,\Pi^{a,A}\Pi^{b,B}=0=\Omega_{AB}\,\Pi^{a,A}\Pi^{b,B},
\]
where $a,b$ and $A,B$ are $USp(2N)_L$ and $USp(2N)_R$ indices,
respectively.  Its Hilbert series is
\begin{equation}
\label{eq:FEN-ta0}
\mathcal{HS}_{FE_N}(0,t_d)
=\frac{\sum_{j=0}^{N^{2}-1} a_{0,j}\,t_d^{\,j}}
      {(1-t_d)^{2N^{2}+N+1}}
=\frac{1+(2N^{2}-N-1)\,t_d+\cdots}{(1-t_d)^{2N^{2}+N+1}}\, .
\end{equation}
At least for $N<4$ the numerator is palindromic, so this sector
is of Calabi--Yau type.

Dually, the sector at $t_d=0$ is generated by the two traceless
antisymmetrics $A_L$ and $A_R$, subject to the $N-1$ trace
relations $\mathrm{tr}_L A_L^{\,j}=\mathrm{tr}_R A_R^{\,j}$ for
$j=2,\dots,N$.  The Hilbert series reads
\begin{equation}
\label{eq:FEN-td0}
\begin{aligned}
\mathcal{HS}_{FE_N}(t_a,0)
&=\frac{\prod_{j=2}^{N}(1-t_a^{\,j})}{(1-t_a)^{4N^{2}-2N-2}}\\
&=1+(4N^{2}-2N-2)\,t_a+N\bigl(8N^{3}-8N^{2}-4N+3\bigr)\,t_a^{\,2}+\cdots .
\end{aligned}
\end{equation}
The order-$t_a$ term counts the generators $A_L,A_R$; the
order-$t_a^{\,2}$ term counts their symmetric product minus the single
quadratic relation $\mathrm{tr}_L A_L^{\,2}=\mathrm{tr}_R A_R^{\,2}$.
\subsubsection*{Explicit results for small $N$}

The numerator coefficients $a_{i,j}$ of~\eqref{eq:feusphs} fill a
$\bigl(2N(N-1)+1\bigr)\times\bigl((2N+1)(N-1)+1\bigr)$ matrix,
which we find to be palindromic under
$a_{i,j}=a_{\,2N(N-1)-i,\,(2N+1)(N-1)-j}$.
For $N=3$ the matrix is completely fixed by expanding the Hilbert
series up to total degree $14$ in $(t_a,t_d)$, which determines all
entries with $i+j\le 14$, the remaining ones following from the
palindromic symmetry $a_{i,j}=a_{12-i,\,14-j}$.  For $N=1,2,3$ the
coefficients are
\setcounter{MaxMatrixCols}{20}
\begin{equation}
	\label{eq:FEN-aij-small}
	\begin{aligned}
		N=1:\quad &(1),\\[4pt]
		N=2:\quad &
		\begin{pmatrix}
			1 &   5 &   5 &   1 &   0 & 0\\
			1 & -11 & -29 &  -9 &   0 & 0\\
			0 &   0 &  36 &  36 &   0 & 0\\
			0 &   0 &  -9 & -29 & -11 & 1\\
			0 &   0 &   1 &   5 &   5 & 1
		\end{pmatrix},\\[6pt]
		N=3:\quad &
		\resizebox{0.85\textwidth}{!}{$
		\begin{pmatrix}
			1 &  14 &   77 &   204 &    280 &    204 &      77 &      14 &       1 &      0 &      0 &     0 &    0 &   0 & 0 \\
			2 &  -8 & -307 & -1602 &  -3360 &  -3294 &   -1541 &    -328 &     -26 &      0 &      0 &     0 &    0 &   0 & 0 \\
			2 & -44 &   59 &  3654 &  14589 &  21648 &   13705 &    3614 &     325 &      0 &      0 &     0 &    0 &   0 & 0 \\
			1 & -22 &  459 & -1520 & -26473 & -70030 &  -67060 &  -24490 &   -2720 &     14 &      1 &     0 &    0 &   0 & 0 \\
			0 &   0 &  -10 & -2080 &  16723 & 112070 &  183050 &  104326 &   17747 &   -160 &    -26 &     0 &    0 &   0 & 0 \\
			0 &   0 &  -24 &   796 &    967 & -82390 & -267988 & -261506 &  -78952 &  -2062 &    556 &   -22 &    1 &   0 & 0 \\
			0 &   0 &    3 &   -30 &  -2531 &  23868 &  202392 &  358324 &  202392 &  23868 &  -2531 &   -30 &    3 &   0 & 0 \\
			0 &   0 &    1 &   -22 &    556 &  -2062 &  -78952 & -261506 & -267988 & -82390 &    967 &   796 &  -24 &   0 & 0 \\
			0 &   0 &    0 &     0 &    -26 &   -160 &   17747 &  104326 &  183050 & 112070 &  16723 & -2080 &  -10 &   0 & 0 \\
			0 &   0 &    0 &     0 &      1 &     14 &   -2720 &  -24490 &  -67060 & -70030 & -26473 & -1520 &  459 & -22 & 1 \\
			0 &   0 &    0 &     0 &      0 &      0 &     325 &    3614 &   13705 &  21648 &  14589 &  3654 &   59 & -44 & 2 \\
			0 &   0 &    0 &     0 &      0 &      0 &     -26 &    -328 &   -1541 &  -3294 &  -3360 & -1602 & -307 &  -8 & 2 \\
			0 &   0 &    0 &     0 &      0 &      0 &       1 &      14 &      77 &    204 &    280 &   204 &   77 &  14 & 1
		\end{pmatrix}$}.
	\end{aligned}
\end{equation}

\subsubsection*{Sum rules}

As for $FM_N$, the numerator factorises on the diagonal and on the two
single-variable specialisations.  Writing
$\mathrm{Num}(t_a,t_d)=\sum_{i,j}a_{i,j}t_a^{\,i}t_d^{\,j}$ for the
numerator of~\eqref{eq:feusphs}, we find
\begin{align}
	\label{eq:FEN-sr1}
	\mathrm{Num}(t,t) &= (1-t)^{2N(N-1)}\,(1+t)^{N-1}\,
	\sum_{j=0}^{2N(N-1)} b_j\,t^{\,j},\\[2pt]
	\label{eq:FEN-sr2}
	\mathrm{Num}(1,t) &= N!\,(1-t)^{2N(N-1)}\,(1+t)^{N-1},\\[2pt]
	\label{eq:FEN-sr3}
	\mathrm{Num}(t,1) &= C_N\,(1-t)^{2N(N-1)},
\end{align}
where $C_N=\sum_j a_{0,j}$ and the auxiliary coefficients $b_j$ are
themselves palindromic; $C_1=1$, $C_2=12$, $C_3=872$.  Equivalently, in
terms of column and row sums of the coefficient matrix,
\begin{equation}
	\label{eq:FEN-row-col-sums}
	\sum_i a_{i,j} \;=\; N!\sum_{m=0}^{N-1}(-1)^{\,j-m}\binom{2N(N-1)}{\,j-m\,}\binom{N-1}{m},
	\qquad
	\sum_{j} a_{i,j}
	= C_N\,(-1)^{i}\binom{2N(N-1)}{i}\, .
\end{equation}
We have verified~\eqref{eq:FEN-sr1}--\eqref{eq:FEN-row-col-sums}
for $N=2,3$; the $N=2$ case reproduces~\eqref{eq:FE2-sum-rules}
and~\eqref{eq:FE2-row-col-sums}.  Note that $2N(N-1)>0$ for $N\ge2$
implies $\sum_{i,j}a_{i,j}=\mathrm{Num}(1,1)=0$.  The $b_j$
in~\eqref{eq:FEN-sr1} are precisely the coefficients $c_j$
of~\eqref{eq:FEN-unrefined} stripped of their $(1+t)^{N-1}$ factor.
As linear conditions on the $a_{i,j}$, \eqref{eq:FEN-sr2}
fixes each of the $(2N+1)(N-1)+1$ column sums and~\eqref{eq:FEN-sr3}
each of the $2N(N-1)+1$ row sums---one relation being shared, since both
total to $\mathrm{Num}(1,1)$---while the divisibility in~\eqref{eq:FEN-sr1}
imposes $(N-1)(2N+1)$ conditions on the anti-diagonal sums, of which
$2N(N-1)-2+(N-1)$ are independent of the previous ones (the vanishing of
$\mathrm{Num}(t,t)$ and of its first derivative at $t=1$ already follow from
the row and column sums).  For $N=3$ this gives $15+13-1+12=39$ constraints,
all of which~\eqref{eq:FEN-aij-small} satisfies.

\subsubsection*{Fully unrefined Hilbert series}

Setting $t_a=t_d=t$, the fully unrefined Hilbert series reads
\begin{equation}
	\label{eq:FEN-unrefined}
	\mathcal{HS}_{FE_N}(t,t)
	=\frac{\sum_{j=0}^{(N-1)(2N+1)} c_j\,t^{\,j}}{(1-t)^{4N^{2}}}\, ,
\end{equation}
describing a complex cone of dimension $4N^{2}$ whose numerator has
degree $(2N+1)(N-1)$.  For $N=1,2,3$,
\begin{equation}
	\label{eq:FEN-unrefined-small}
	\begin{aligned}
		N=1:\quad & 1,\\
		N=2:\quad & (1+t)\,(1+9t+19t^{2}+9t^{3}+t^{4}),\\
		N=3:\quad & (1+t)^{2}\bigl(1+26t+288t^{2}+1716t^{3}+5970t^{4}+12545t^{5}
		+16071t^{6}\\
		&\hspace{4.8em}+12545t^{7}+5970t^{8}+1716t^{9}+288t^{10}+26t^{11}+t^{12}\bigr).
	\end{aligned}
\end{equation}

Equation~\eqref{eq:FEN-unrefined-small} provides a stringent consistency
check on the matrix~\eqref{eq:FEN-aij-small}.  Setting $t_a=t_d=t$ collapses
the $13\times15$ array onto the $27$ anti-diagonal sums
$\sum_{i+j=n}a_{i,j}$, and the result must be divisible by
$(1-t)^{2N(N-1)}=(1-t)^{12}$, because the denominator
of~\eqref{eq:feusphs} carries $(2N^{2}+N+1)+(4N^{2}-3N-1)=48$ powers
while the cone has dimension $4N^{2}=36$.  We have checked that the
division is exact, with vanishing remainder, and that the quotient is
the palindromic degree-$14$ polynomial displayed above.  Neither the
vanishing of the $12$ remainder coefficients nor the $7$ independent
palindromicity constraints on the quotient involves any residual freedom
in~\eqref{eq:FEN-aij-small}, so together they furnish $19$ nontrivial tests
of the entries.

\subsubsection*{Generators and relations}

The structure of the chiral ring is most transparent from the
plethystic logarithm of~\eqref{eq:feusphs}:
\begin{equation}
\label{eq:FEN-PL}
\begin{aligned}
\mathrm{PL}\!\left[\mathcal{HS}_{FE_N}\right](t_a,t_d)
&= \bigl(\chi_{(\mathrm{asym}_L,\mathbf{1})}
        +\chi_{(\mathbf{1},\mathrm{asym}_R)}\bigr)\,t_a
 + \chi_{(\mathbf{2N},\mathbf{2N})}\,t_d
 - \sum_{k=2}^{N} t_a^{\,k} \\[2pt]
&\quad - \bigl(\chi_{(\mathrm{asym}_L,\mathbf{1})}
              +\chi_{(\mathbf{1},\mathrm{asym}_R)}\bigr)\,t_d^{\,2}
       - \chi_{(\mathbf{2N},\mathbf{2N})}\,t_a t_d \\[2pt]
&\quad + \mathcal{O}\!\bigl(t_a t_d\,(t_a+t_d)\bigr),
\end{aligned}
\end{equation}
where $\chi_R$ is the character of the representation $R$ under
$USp(2N)_L\times USp(2N)_R$.  Reading~\eqref{eq:FEN-PL} off order by
order, the generators of the (main branch of the) moduli space are:
\begin{itemize}
\item the bifundamental $\Pi^{a,A}$, in the $(\mathbf{2N},\mathbf{2N})$
      of $USp(2N)_L\times USp(2N)_R$ (at order $t_d$);
\item the two traceless antisymmetrics $A_L$ and $A_R$, in the
      antisymmetric of $USp(2N)_L$ and $USp(2N)_R$ respectively
      (at order $t_a$).
\end{itemize}
These generators satisfy the relations
\begin{equation}
\label{eq:FEN-relations}
\begin{aligned}
\Omega_{ab}\,\Pi^{a,A}\Pi^{b,B}
  =0=\Omega_{AB}\,\Pi^{a,A}\Pi^{b,B},\qquad &
  \bigl(4N^{2}-2N-2\ \text{relations at order } t_d^{\,2}\bigr),\\[2pt]
\mathrm{tr}_L A_L^{\,k}=\mathrm{tr}_R A_R^{\,k},\qquad &
  \bigl(\text{at order } t_a^{\,k},\ k=2,\dots,N\bigr),\\[2pt]
\Omega_{ac}\,A_L^{\,ab}\,\Pi^{c,A}=\Omega_{BC}\,A_R^{\,AB}\,\Pi^{b,C},\qquad &
  \bigl(4N^{2}\ \text{relations at order } t_a t_d\bigr).
\end{aligned}
\end{equation}
The chiral-ring generators agree with those
of~\cite{Pasquetti:2019hxf}, and their quantum relations agree
with the proposal of~\cite{Benvenuti:2024mpn}.  Since the number of
relations greatly exceeds the codimension, the moduli space is
manifestly not a complete intersection.



\section{Moduli spaces of \texorpdfstring{$N=2$}{n=2} improved bifundamentals: symmetry enhancements}
\label{n=2impbif}

As already mentioned, the case $N=2$ is special and deserves a separate section. As we will show, the IR fixed points of these theories exhibit further global-symmetry enhancement. We find relatively simple formulae for their Hilbert series in terms of characters of the larger global symmetry.

\subsection{Index resummation for $N=2$}
For the five theories discussed in this section, the index may be rewritten as an infinite series in even powers of the fugacity $x$:
\begin{equation}
    \mathcal{I}(x,t) = \sum_{n=0}^{\infty} \mathcal{I}^{(n)}(t)\, x^{2n} = \mathcal{I}^{(0)}(t) + \mathcal{I}^{(1)}(t)x^2 + \mathcal{O}(x^4)
    \label{eq:resum}
\end{equation}
where $t$ is a second fugacity defined as $t = \psi\, x^{2-\tau}$, with $\psi$ being the fugacity
for the single abelian $U(1)_\tau$ global symmetry present in the IR and $\tau$ the corresponding mixing with the $R$--charge, to be determined through $Z$-extremisation.

Some comments are in order.

\paragraph{The leading coefficient: Hilbert series of the moduli space.}
At fixed $t$, the coefficient of $x^0$ is the limit~\eqref{eq:HS_limit} specialised to $N=2$, with the single abelian fugacity $t=t_a$. It receives contributions only from the chiral-ring operators (not only the generators) whose R-charge is saturated by their $U(1)_\tau$ charge, i.e.\ with $w=0$ and $j_3=0$. Their generating function
is the Hilbert series of the moduli space $\mathcal{M}$ of vacua of the theory for $FT_2$, $FC_2$ and $FH_2$, and of its main branch for $FM_2$ and $FE_2$, whose singlets $B_{n,m}$ have $w>0$ and are projected out:
\begin{equation}
    \mathcal{I}^{(0)}(t) \;=\; \mathrm{HS}_{\mathcal{M}}(t)
    \;=\; \sum_{k\geq 0}\, \dim\!\big[\mathbb{C}[\mathcal{M}]\big]_{k}\, t^{\,k}\,,
\end{equation}
where $\mathbb{C}[\mathcal{M}]_{k}$ denotes the space of chiral-ring elements of $U(1)_\tau$ charge $k$. 

\paragraph{The first subleading coefficient: currents dressed by the 
chiral ring.}
The net content of the subleading coefficient $\mathcal{I}^{(1)}(t)$, present at order $x^{2}$, may be written schematically as

\begin{equation}
    \mathcal{I}^{(1)}(t) \;=\; -\,\chi_{\rm adj}(t)\cdot\mathcal{I}^{(0)}(t)
    \;+\;\Delta_{\rm Koszul}(t)\,,
    \label{eq:I1}
\end{equation}

where $\chi_{\rm adj}(t)$ is the character of the fermionic 
component of the flavour-current multiplet in the adjoint of the 
enhanced global symmetry. The two terms admit the following 
interpretation:
\begin{itemize}
    \item The first term is the naive ``current $\times$ chiral ring'' 
          tower, built by multiplying the flavour-current fermion with 
          arbitrary chiral-ring generators as if the latter were free 
          polynomial variables.
    \item The Koszul correction $\Delta_{\rm Koszul}(t)$ removes the 
          over-counting produced by the chiral-ring relations \cite{Benvenuti:2006qr, Weyman:2003}: if the 
          generators satisfy relations (e.g.\ the Plücker relations for 
          a Grassmannian), then products of the current fermion with 
          those relations must be subtracted, and so on through the 
          syzygies of the chiral ring. In the five theories we study, 
          at order $x^2$ only the leading Koszul correction contributes; higher syzygies of the cominuscule (i.e. HSS) BGG complex are pushed to $O(x^4)$ and beyond.\footnote{Explicit BGG resolutions and closed-form Hilbert series for the coordinate rings of Hermitian symmetric spaces were constructed in \cite{Enright:2004}.}
\end{itemize}

In practice we see that for our theories we may write this coefficient as
\begin{equation}
    \mathcal{I}^{(1)}(t) = - \sum_{k=0}^{\infty} C^{(1)}_k t^k\,,
\end{equation}
where the coefficients $C^{(1)}_k$ are specified by the following table:
\begin{equation}
\begin{array}{c | @{\hspace{1cm}} c}
    \toprule
    \text{Theory} &  C_k^{(1)} =  \,[\chi_{\rm adj}(t)\cdot\mathcal{I}^{(0)}(t)]_k
    \;-[\Delta_{\rm Koszul}(t)]_k\, \\
    \midrule
    FT_2    &  \chi_{[1, k, 1]_{su(4)}} - \chi_{[1, k-2, 1]_{su(4)}}  \\
    FC_2   & \chi_{[1, k, 0, 1]_{su(5)}} - \chi_{[1, k-2, 1, 0]_{su(5)}}  \\
    FH_2                 &\chi_{[0, 1, 0, 0,k]_{so(10)}} - \chi_{[0, 0, 1, 0, k-2]_{so(10)}}\\
    \widetilde{FM}_2  & \chi_{[1, k, 0,0, 1]_{su(6)}} - \chi_{[1, k-2, 1, 0,0]_{su(6)}} \\
    \widetilde{FE}_2  & \chi_{[k,0,0,0,0,1]_{e_6}}-\chi_{[k-2,1,0,0,0,0]_{e_6}} \\
    \bottomrule
\end{array}
\end{equation}

\paragraph{Higher-order structure and plethystic organisation.}
Iterating the analysis, $\mathcal{I}^{(n)}(t)$ collects all BPS operators with $w+2j_3=2n$, i.e.\ with $E+j_3=(2-\tau)k+2n$ for an operator contributing to $t^k$, organised into symmetric products of ``single-letter'' contributions. A more 
compact way to phrase this is that \eqref{eq:resum} admits a plethystic exponential representation
\begin{equation}
    \mathcal{I}(x,t) \;=\; \mathrm{PE}\!\left[f(x,t)\right]
    \;\equiv\;\exp\!\left(\sum_{k\geq 1}\frac{1}{k}\,f(x^{k},t^{k})\right)\,,
    \qquad
    f(x,t)\;=\;f_{0}(t) + f_{1}(t)\,x^{2} + \mathcal{O}(x^{4})\,,
    \label{eq:PE}
\end{equation}
with $f_{0}(t)$ the generating function of single-letter chiral-ring generators (so that 
$\mathrm{PE}[f_{0}]=\mathcal{I}^{(0)}$ reproduces the moduli-space Hilbert series) and 
$f_{1}(t)$ encoding currents and recombination data. 

\subsection{ \texorpdfstring{$FT_2$}{ftsu2} }
This theory has a UV description as $U(1)$ with $(2_q, 2_{\qt})$ flavours and a superpotential $\cW=Flip[q_i \qt_j]$:

\be \label{ftsu2}
\resizebox{0.6\hsize}{!}{
\begin{tikzpicture}[
    thick,
    gauge/.style={
        circle, draw=black, fill=white,
        minimum size=8mm, inner sep=1pt, font=\small
    },
    manifest/.style={
        rectangle, draw=blue!70!black, fill=blue!6,
        minimum size=8mm, inner sep=2pt,
        font=\small, text=blue!70!black
    },
    arr/.style={-{Stealth[scale=0.8]}, shorten >=1pt, shorten <=1pt},
    barr/.style={draw=black!70}
]

\node[gauge]    (g1) at (0,0)   {$1$};
\node[manifest] (g2) at (2.5,0) {$2$};

\draw[arr,barr,transform canvas={yshift= 3.5pt}] (g1) -- (g2);
\draw[arr,barr,transform canvas={yshift=-3.5pt}] (g2) -- (g1);
\node at (1.25, 0.45) {$q,\,\tilde{q}$};
\node at (1.25, 0) {\huge$\times$};

\draw[blue!70!black] (g2) to[out=65,in=115,looseness=3] (g2);
\node[blue!70!black] at (2.5, 0.9) {$a_2$};

\node at (1.25,-1.5) {$\mathcal{W} = a_2\, q\tilde{q} + \mathrm{Flip}[q\tilde{q}]$};

\node[right] at (3.5,-0.4) {%
    \renewcommand{\arraystretch}{1.4}
    \begin{tabular}{>{\small}c >{\small}c}
        \toprule
        $q,\,\tilde{q}$ & $\tau/2$ \\
        $a_i$           & $2-\tau$ \\
        \bottomrule
    \end{tabular}%
};

\end{tikzpicture}
}
\ee

\subsubsection*{Duality chain and IR symmetry enhancement}

Applying Aharony duality~\cite{Aharony:1997gp} to~\eqref{ftsu2}
maps it onto a $U(1)$ gauge theory with $(2_Q,2_{\tilde{Q}})$
flavours and superpotential
$\mathcal{W}=\mathrm{Flip}[\mathfrak{M}^{\pm}]$.  By eq.~(5.7)
of~\cite{Benvenuti:2018bav}, the latter is in turn dual to $SU(2)$
with four fundamental flavours and $\mathcal{W}=0$.  It follows that
the IR global symmetry of the abelian theory in~\eqref{ftsu2} is
$SU(4)\times U(1)$, as was already established by a different route
in~\cite{Benini:2018bhk}.

\subsubsection*{Refined and unrefined Hilbert series}

The refined Hilbert series was proposed
in~\cite{Benini:2018bhk} in the compact form
\begin{equation}
\label{eq:HSftu2}
\mathcal{HS}_{FT_2}(t)
= \mathrm{PE}\!\left[\chi_{[0,1,0]_{su(4)}}\,t - t^{2}\right]
= \sum_{k=0}^{\infty}\chi_{[0,k,0]_{su(4)}}\,t^{k},
\qquad t\equiv x^{2-\tau},
\end{equation}
with $\chi_R$ the character of the $SU(4)$ representation $R$
specified by its Dynkin labels.  The representation $[0,k,0]$ has
dimension
$\binom{k+2}{2}\binom{k+3}{2}/3=(k+1)(k+2)^{2}(k+3)/12$.

Setting all $SU(4)$ fugacities to unity, the unrefined series resums
to
\begin{equation}
\label{eq:HSftu2UR}
\mathcal{HS}_{FT_2}(t)
= \mathrm{PE}\!\left[6t-t^{2}\right]
= \frac{1+t}{(1-t)^{5}}\, ,
\end{equation}
and the plethystic logarithm follows immediately:
\begin{equation}
\label{eq:PLftu2}
\mathrm{PL}\!\left[\mathcal{HS}_{FT_2}\right](t)
= \chi_{[0,1,0]_{su(4)}}\,t - t^{2}.
\end{equation}

\subsubsection*{Generators, relations and geometry}

From~\eqref{eq:PLftu2} the moduli space is seen to be a complete
intersection.  Its six generators, appearing at order $t$, are the
four flippers of the mesons together with the two monopoles; they
transform in the $\mathbf{6}=[0,1,0]$ of $SU(4)$.  The single
quadratic relation at order $t^{2}$ projects out the singlet in the
symmetric square of the $\mathbf{6}$, and can be written as
\[
\mathrm{tr}\,A_L^{\,2}=\mathrm{tr}\,A_R^{\,2},
\]
where $A_{L/R}$ are the two adjoints of $FT_2$.  Because only
$SU(4)$ representations of even quadrality appear, the faithful IR
symmetry is in fact $SO(6)\times U(1)$~\cite{Benini:2018bhk}.

Geometrically, the moduli space is the complex cone over an
irreducible Hermitian symmetric space of compact type
$\mathcal{B}=H/K$ with $H=SU(4)$ and $K=S(U(2)\times U(2))$,
i.e.\ the complex Grassmannian $\mathrm{Gr}(2,4)$~\cite{Braun:2017}.

\subsubsection*{Decomposition under the generic-$N$ symmetry}

Under the generic-$N$ global symmetry
$SU(2)\times SU(2)\times U(1)_{\tau}$, the fundamental and the
antisymmetric of $SU(4)$ branch as
\begin{align}
\label{eq:branch-fund}
\mathbf{4} &= [1,0,0]_{SU(4)} = [1;1]_{SU(2)\times SU(2)},\\[2pt]
\label{eq:branch-antisym}
\mathbf{6} &= [0,1,0]_{SU(4)}
           = \bigl([2;0]+[0;2]\bigr)_{SU(2)\times SU(2)}.
\end{align}
In particular the Hilbert series~\eqref{eq:HSftu2} can be rewritten
as
\begin{equation}
\label{eq:HSftu2B}
\mathcal{HS}_{FT_2}(t)
= \mathrm{PE}\!\left[
   \bigl(\chi_{[2;0]_{su(2)\times su(2)}}
        +\chi_{[0;2]_{su(2)\times su(2)}}\bigr)\,t - t^{2}\right].
\end{equation}
The general branching rule for the chiral-ring representations is
\begin{equation}
\label{eq:branch-general}
[0,k,0]_{su(4)}
= \sum_{k_1,k_2=0}^{\infty} [2k_1;2k_2]_{su(2)\times su(2)}
   \sum_{s=0}^{\lfloor k/2\rfloor}\delta_{k_1+k_2,\,k-2s}\, ,
\end{equation}
which, substituted into~\eqref{eq:HSftu2}, yields\footnote{%
Using~\eqref{eq:branch-general} in~\eqref{eq:HSftu2} and noting
that, for any $k_1,k_2\in\mathbb{N}$,
\[
\sum_{k=0}^{\infty}\sum_{s=0}^{\lfloor k/2\rfloor}
  \delta_{k_1+k_2,\,k-2s}\,t^{k}
= \sum_{k,s=0}^{\infty}\delta_{k_1+k_2,\,k-2s}\,t^{k}
= \sum_{s=0}^{\infty}t^{\,k_1+k_2+2s}.
\]}
\begin{equation}
\label{eq:HSftu2r}
\mathcal{HS}_{FT_2}(t)
= \sum_{k_1,k_2,m=0}^{\infty}
  \chi_{[2k_1;2k_2]_{su(2)\times su(2)}}\,t^{\,k_1+k_2+2m}.
\end{equation}
We have checked this expression against the superconformal index
of~\eqref{eq:indftu2} up to order $x^{10}$.

\subsubsection*{Superconformal R-charge and index}

$Z$-extremisation~\cite{Jafferis:2012,Willett:2011gp} fixes the
superconformal R-charge of the two adjoints to high precision,
\begin{equation}
\label{eq:tau-FT2}
\tau = 1.339192\ldots
\end{equation}
For the index computation we rounded this value to $\tau=134/100$,
obtaining
\begin{equation}
\label{eq:indftu2}
\begin{aligned}
\mathcal{I}_{FT_2} = \;&
    1 + 6\psi\,x^{33/50} + 20\psi^{2}\,x^{33/25}
      + 50\psi^{3}\,x^{99/50} - 16\,x^{2}
      + 105\psi^{4}\,x^{66/25}\\
  & - 64\psi\,x^{133/50} - \psi^{-2}\,x^{67/25}
      + 196\psi^{5}\,x^{33/10} - 160\psi^{2}\,x^{83/25}
      + 20\psi^{-1}\,x^{167/50}\\
  & + 336\psi^{6}\,x^{99/25} - 320\psi^{3}\,x^{199/50}
      + 88\,x^{4} + 540\psi^{7}\,x^{231/50}
      - 560\psi^{4}\,x^{116/25}\\
  & + 156\psi\,x^{233/50} - 16\psi^{-2}\,x^{117/25}
      + 825\psi^{8}\,x^{132/25} - 896\psi^{5}\,x^{53/10}
      + 192\psi^{2}\,x^{133/25}\\
  & - 64\psi^{-1}\,x^{267/50}
      + 1210\psi^{9}\,x^{297/50} - 1344\psi^{6}\,x^{149/25}
      + 164\psi^{3}\,x^{299/50} + 64\,x^{6}\\
  & + 6\psi^{-3}\,x^{301/50} + 1716\psi^{10}\,x^{33/5}
      - 1920\psi^{7}\,x^{331/50} + 40\psi^{4}\,x^{166/25}\\
  & + 448\psi\,x^{333/50} + 20\psi^{-2}\,x^{167/25}
      + 2366\psi^{11}\,x^{363/50} - 2640\psi^{8}\,x^{182/25}\\
  & - 212\psi^{5}\,x^{73/10} + 992\psi^{2}\,x^{183/25}
      - 276\psi^{-1}\,x^{367/50}
      + 3185\psi^{12}\,x^{198/25}\\
  & - 3520\psi^{9}\,x^{397/50} - 624\psi^{6}\,x^{199/25}
      + 1664\psi^{3}\,x^{399/50} - 785\,x^{8}\\
  & + 4200\psi^{13}\,x^{429/50} - 4576\psi^{10}\,x^{43/5}
      - 1228\psi^{7}\,x^{431/50} + 2432\psi^{4}\,x^{216/25}\\
  & - 824\psi\,x^{433/50} + 256\psi^{-2}\,x^{217/25}
      + 5440\psi^{14}\,x^{231/25} - 5824\psi^{11}\,x^{463/50}\\
  & - 2056\psi^{8}\,x^{232/25} + 3264\psi^{5}\,x^{93/10}
      - 320\psi^{2}\,x^{233/25} + 288\psi^{-1}\,x^{467/50}\\
  & + 6936\psi^{15}\,x^{99/10} - 7280\psi^{12}\,x^{248/25}
      - 3140\psi^{9}\,x^{497/50} + 4128\psi^{6}\,x^{249/25}\\
  & + 504\psi^{3}\,x^{499/50} - 1072\,x^{10}
      + \mathcal{O}\!\bigl(x^{501/50}\bigr).
\end{aligned}
\end{equation}
Here $\psi$ is the fugacity for the global $U(1)_{\tau}$, introduced
to track the $\tau$-dependence.

\subsubsection*{Reading the index}

The index~\eqref{eq:indftu2} admits a clean physical decomposition.
The chiral-ring operators contribute as
$\chi_{[0,k,0]_{SU(4)}}\,x^{k(2-\tau)}$, while marginal operators
and the fermionic components of conserved currents sit at even
powers of $x$ and are independent of $\psi$.

The $-16\,x^{2}$ term is the hallmark of the symmetry enhancement to
$\mathfrak{su}(4)\oplus\mathfrak{u}(1)$.  Its origin is the
following.  The spin-$\tfrac{1}{2}$ components of the conserved
currents supply
\[
(q_i)_S(\bar q^{\,j})_F ,\quad
(\tilde q^{\,i})_S(\bar{\tilde q}_j)_F
\quad (\text{8 components}),
\qquad
(a_i)_S(\bar a^{\,j})_F \quad (\text{16 components}),
\]
with $i,j=1,2$ and $a_1$ the gauge singlet flipping $q\tilde q$ in
the superpotential.  The monopole sectors with GNO flux $\pm1$ add
$\mathfrak{M}^{\pm}(\bar a_i)_F$ (8 further gauge-invariant fermionic
components) at order $x^{2}$.  Sixteen of these current components
recombine with the sixteen scalars of the superpotential terms,
leaving the sixteen fermionic components that assemble into the
adjoint of $SO(6)$ together with the $U(1)$ current -- precisely the
spin-$\tfrac{1}{2}$, $\Delta=\tfrac{3}{2}$, $R=1$ components of the
flavour current~\cite{Cordova:2016}, which are the only ones that
contribute to the index.

One also reads off $\tfrac{1}{4}$-BPS operators of schematic form
$[\mathrm{current}]\times[\mathrm{chiral}]$ at orders
$x^{2n+k(2-\tau)}$ with $n,k\in\mathbb{N}_{+}$, as well as further
$\tfrac{1}{4}$-BPS operators (outside the chiral ring) at orders
$x^{2n-k(2-\tau)}$ with $n>1$.  As an example, the fermionic term
$-\psi^{-2}x^{67/25}$ arises as follows.  Recombination at order
$x^{67/50}$ between $(q_i\tilde q^{\,j})_S$ and $(a_i)_F$, driven by
the cubic superpotential, annihilates those contributions in the
index; squaring this relation (with due care for statistics and
product-representation decomposition) tracks the residual pieces at
order $x^{67/25}$.

Concretely, the 16 scalar terms

$$
(a_1)_F(a_2)_F\ [(\mathbf{1}, \mathbf{3})],\quad
(a_2)_F^{\,2}\ [(\mathbf{1}, \mathbf{3})],\quad
(q_i\tilde q^{\,j})_S^{\,2}\
[(\mathbf{1}, \mathbf{1})+(\mathbf{3}, \mathbf{3})]
$$

are subject to the F-term relation

$$
\frac{\partial\mathcal W}{\partial a_1}
=\operatorname{Tr}(q\tilde q)=0.
$$

Since \(a_1\) flips only the trace of the meson, this relation removes
precisely the \((\mathbf{1},\mathbf{1})\) component of
\(q_i\tilde q^{\,j}\).  Hence the scalar contributions reduce from

$$
3+3+(1+9)=16
$$

to

$$
3+3+9=15
$$

independent terms.  These 15 scalar contributions recombine with the
16 fermionic terms

$$
(a_1)_F(q_i\tilde q^{\,j})_S\
[(\mathbf{2},\mathbf{2})],\quad
(a_2)_F(q_i\tilde q^{\,j})_S\
[(\mathbf{2},\mathbf{2})+(\mathbf{2},\mathbf{4})],
$$

where the representations refer to \(SU(2)_L\times SU(2)_R\).
The net imbalance is therefore one surviving fermionic mode, giving
the \(-\psi^{-2}x^{67/25}\) contribution to the index.

\subsubsection{Index resummation}
First, defining $t = \psi x^{33/50}$, we can rewrite the index as follows:

\begin{equation}
	\mathcal{I}_{FT_2}(x, t) = \mathcal{I}^{(0)}(t) + x^2 \mathcal{I}^{(1)}(t) + x^4 \mathcal{I}^{(2)}(t) + x^6 \mathcal{I}^{(3)}(t) + \mathcal{O}(x^8)
\end{equation}
where the functions $\mathcal{I}^{(k)}(t)$ can be resummed as
\begin{align}
	\mathcal{I}^{(0)}(t) &= \frac{1+t}{(1-t)^5} = \mathcal{HS}_{FT_2}(t)\\[1ex]
	\mathcal{I}^{(1)}(t) &= -\frac{16}{(1-t)^4} \\[1ex]
	\mathcal{I}^{(2)}(t) &= -\frac{1 - 24t - 2t^2 + 72t^3 - 15t^4}{t^2(1-t)^4} \\[1ex]
	\mathcal{I}^{(3)}(t) &= \frac{-16 + 224t^2 - 128t^3 - 176t^4 + 64t^5}{t^2(1-t)^4}
\end{align}
In terms of the $SU(4)$ representations, the functions $\mathcal{I}^{(k)}(t)$ can be written as an exact sum over irreducible representations $\chi_{[q_1, p, q_2]}$ as follows:
\begin{align}
	\mathcal{I}^{(0)}(t) &= \sum_{k=0}^{\infty} \chi_{[0, k, 0]} t^k \\[1ex]
	\mathcal{I}^{(1)}(t) &= - \sum_{k=0}^{\infty} \left( \chi_{[1, k, 1]} - \chi_{[1, k-2, 1]} \right) t^k \\[1ex]
	\mathcal{I}^{(2)}(t) &= - \frac{\chi_{[0, 0, 0]}}{t^2} + \sum_{k=-1}^{\infty} C^{(2)}_k t^k \\[1ex]
	\mathcal{I}^{(3)}(t) &= - \frac{\chi_{[1, 0, 1]} + \chi_{[0, 0, 0]}}{t^2} - \frac{\chi_{[1, 1, 1]}}{t} + \sum_{k=0}^{\infty} C^{(3)}_k t^k
\end{align}
When evaluating the sum for $\mathcal{I}^{(1)}(t)$ at $k=0$ and $k=1$, we formally encounter characters with negative Dynkin labels, namely $\chi_{[1, -2, 1]}$ and $\chi_{[1, -1, 1]}$. According to the Weyl character formula, these are resolved via Weyl reflections outside the fundamental chamber. If a weight lies on the boundary of a Weyl chamber, its character vanishes ($\chi_{[1, -1, 1]} = 0$). If it can be reflected back into the fundamental chamber by a single Weyl reflection, it acquires a minus sign, yielding $\chi_{[1, -2, 1]} = -\chi_{[0, 0, 0]} = -1$.

The coefficients $C^{(2)}_k$ and $C^{(3)}_k$ are finite combinations of characters encoding the higher-order syzygies between the chiral ring and the conserved-current multiplets. Define the discrete difference operators
\begin{align}
	\Delta_1(k) &= \chi_{[1, k, 1]} - \chi_{[1, k-2, 1]} \\[1ex]
	\Delta_3(k) &= \chi_{[0, k+2, 0]} - 3\chi_{[0, k+1, 0]} + 3\chi_{[0, k, 0]} - \chi_{[0, k-1, 0]}
\end{align}
the exact character combinations are explicitly given by:
\begin{align}
	C^{(2)}_k &= 4 \Delta_1(k+1) - 6 \Delta_1(k) - 14 \Delta_3(k) - 2 \chi_{[0, 0, 0]} \\[1ex]
	C^{(3)}_k &= 4 \Delta_1(k) + 12 \Delta_1(k-1) - 26 \Delta_1(k-2) + 8 \Delta_1(k-3)
\end{align}

\subsubsection{The \texorpdfstring{$T_2$}{tsu2} theory}
Flipping one of the two adjoints in the IR spectrum yields the $\cN=4$ theory $T_2 \equiv T[SU(2)]$ \cite{Gaiotto_2009}. The first few orders in the expansion of the resulting index are
\begin{equation}
\begin{split}     \mathcal{I}_{T_2}= & 1+3 \psi x^{33/50}+5 \psi ^2 x^{33/25}+3 \psi^{-1} x^{67/50}+7 \psi ^3 x^{99/50}-7 x^2+9 \psi ^4 x^{66/25}-4 \psi 
   x^{133/50}\\
   &+5 \psi^{-2} x^{67/25}+11 \psi ^5 x^{33/10}-4 \psi ^2 x^{83/25}-4 \psi^{-1} x^{167/50}+13 \psi ^6
   x^{99/25}-4 \psi ^3 x^{199/50}+7 x^4\\
   &+7 \psi^{-3} x^{201/50}+15 \psi ^7 x^{231/50}-4 \psi ^4 x^{116/25}-20 \psi x^{233/50}-4\psi^{-2} x^{117/25}+O\left(x^{5}\right)
   \end{split}
   \label{indtu2}
\end{equation}
where again we turned on the fugacity $\psi$ for the single $U(1)$ global symmetry, in order to keep track of the charges of short multiplets appearing in the index.
As the index suggests, the theory has two identical branches, a Higgs branch and a Coulomb branch, which are orthogonal and exchanged by mirror symmetry.
The Hilbert series for each of the two branches reads \cite{Hanany:2011db}:
\be\CH\CS^{CB}_{T_2}= \CH\CS^{HB}_{T_2}= PE[(\chi_{[2]_{su(2)}}) x^{a}- x^{2a}], \ee
with $a=\tau$ for one branch and $a=2-\tau$ for the other. The unrefined series gives
\be\CH\CS^{unr}_{T_2}= PE[3x^{a}- x^{2a}]=1+3x^a+5x^{2a}+7x^{3a}+\dots, \ee

Both the Coulomb- and Higgs-branch Hilbert series can be obtained from the index \eqref{indtu2} by taking the following limits:
\be
\begin{aligned}
    \CH\CS^{CB}_{T_2}=&\lim_{y\to0} \mathcal{I}_{T_2}(x\to x y, \psi \to \psi y^{\tau-2})=1+3 \psi  x^{33/50}+5 \psi ^2 x^{33/25}+7 \psi ^3
   x^{99/50}+\dots \\
    \CH\CS^{HB}_{T_2}=&\lim_{y\to0} \mathcal{I}_{T_2}(x\to x y, \psi \to \psi y^{\tau})=1+\frac{3 x^{67/50}}{\psi }+\frac{5 x^{67/25}}{\psi
   ^2}+\frac{7 x^{201/50}}{\psi
   ^3}+\dots \\
\end{aligned}
\ee
In this case, the procedure reconstructs only a single branch of the full moduli space. As we will see below, however, in some special cases it yields the HS of the full moduli space.

\subsubsection{Exact Hilbert series as a limit of the index}
The Hilbert series can be obtained by taking an appropriate limit of the index, as noted for 3d $\mathcal{N}=4$ theories in \cite{Razamat:2014} and for $\mathcal{N}=2$ theories in \cite{Hanany:2015via}. In general, this limit does not give the full Hilbert series, but only its restriction to either the Higgs or the Coulomb branch. For theories with a single branch, such as $FT_N$, however, it yields the HS of the full moduli space. Consider the index \eqref{eq:indftu2} for $N=2$ and make the shifts $x \to x\, y$ and $\psi \to \psi \, y^{\tau-2}$. Taking the limit $y \to 0$ then yields the HS of the full moduli space:
\begin{equation}
    \begin{split}
    \lim_{y \to 0} \mathcal{I}_{FT_2}(x \,y, \psi  \,y^{\tau-2})&=1+6\psi   x^{33/50}+20\psi ^2 x^{33/25}+50 \psi  ^3 x^{99/50}+105\psi  ^4 x^{66/25}+O\left(x^{151/50}\right)\\
    &= \mathcal{HS}_{FT_2}(\psi \, x^{2-\tau})
    \end{split}
\end{equation}

\subsection{\texorpdfstring{$FC_2$}{fcsu2}}
We now consider $FC_2$, which admits the following abelian UV completion:

\be
\label{fcsu2}
\resizebox{0.75\hsize}{!}{
\begin{tikzpicture}[
    thick,
    gauge/.style={
        circle, draw=black, fill=white,
        minimum size=8mm, inner sep=1pt, font=\small
    },
    manifest/.style={
        rectangle, draw=blue!70!black, fill=blue!6,
        minimum size=8mm, inner sep=2pt,
        font=\small, text=blue!70!black
    },
    flavour/.style={
        rectangle, draw=red!70!black, fill=red!6,
        minimum size=7mm, inner sep=2pt,
        font=\small, text=red!70!black
    },
    arr/.style={-{Stealth[scale=0.8]}, shorten >=1pt, shorten <=1pt},
    barr/.style={draw=black!70}
]

\node[gauge]    (g1) at (0,0)    {$1$};
\node[manifest] (m1) at (2.5,0)  {$2$};
\node[flavour]   (x1) at (-1,-1.8) {$1$};
\node[flavour]   (x2) at (1.5,-1.8) {$1$};

\draw[arr,barr,transform canvas={yshift= 3.5pt}] (g1) -- (m1);
\draw[arr,barr,transform canvas={yshift=-3.5pt}] (m1) -- (g1);
\node at (1.25, 0.45) {$b_1,\,\tilde{b}_1$};
\node at (1.25, 0) {\huge$\times$};

\draw[blue!70!black] (m1) to[out=65,in=115,looseness=3] (m1);
\node[blue!70!black] at (2.5, 0.95) {$a_2$};

\draw[arr,barr] (x1) to[bend left=5] (g1);
\node at (-0.9,-0.75) {$d_1$};

\draw[arr,barr] (g1) to[bend right=5] (x2);
\node at (0.9,-0.75) {$v_1$};

\draw[arr,barr] (x2) to[bend right=5] (m1);
\node at (2.4,-0.85) {$d_2$};

\node at (0.75,-3.2) {$\mathcal{W} =
    v_1 b_1 d_2
    + a_2 b_1\tilde{b}_1
    + \mathfrak{M}^{+}
    + \mathrm{Flip}[b_1\tilde{b}_1]$};

\node[right] at (3.5,-0.5) {%
    \renewcommand{\arraystretch}{1.4}
    \begin{tabular}{>{\small}c >{\small}c}
        \toprule
        $b_1,\,\tilde{b}_1$ & $\tau/2$ \\
        $a_i$               & $2-\tau$ \\
        $v_1$               & $2-\tfrac{\tau}{2}-\Delta$ \\
        $d_i$               & $\Delta+\tfrac{i-2}{2}\tau$ \\
        \bottomrule
    \end{tabular}%
};

\end{tikzpicture}
}
\ee

We can rewrite the $FC_2$ theory as $U(1)$ with $(3_q,2_{\qt}+1_{\pt})$ flavours and 
 $\cW=\M^+ + Flip[q_i \qt_j]$ as follows:
 \begin{center}
\be
\tikzstyle{flavour}=[rectangle,draw=red!50,thick,inner sep = 0pt, minimum size = 6mm]
\tikzstyle{manifest}=[rectangle,draw=blue!50,thick,inner sep = 0pt, minimum size = 6mm]
\tikzstyle{gauge}=[circle,draw=black!50,thick,inner sep = 0pt, minimum size = 6mm]
\tikzset{->-/.style={decoration={
  markings,
  mark=at position .5 with {\arrow{>}}},postaction={decorate}}}
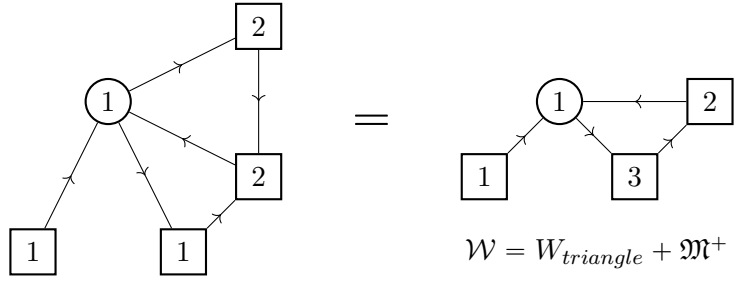
\begin{tikzpicture}

    \begin{scope}
    \node at (0,1) (g1) [gauge,black] {$1$};
    \node at (2,0) (f1) [flavour,black] {$2$};
    \node at (2,2) (f5) [flavour,black] {$2$};
    \node at (-1,-1)(f2) [flavour,black] {$1$};
    \node at (1,-1) (f3) [flavour,black] {$1$};
    \draw (1,-2) node{$\mathcal{W} = W_{triangles}+\mathfrak{M}^{+}$};
    
    \draw[->-] (f1)--(g1);
    \draw[->-] (g1)--(f3); 
    \draw[->-] (f3)--(f1);
    \draw[->-] (f2)--(g1);
    \draw[->-] (f5)--(f1);
    \draw[->-] (g1)--(f5);

    \draw (3.5,.7) node {\huge$=$};
    \end{scope}

    \begin{scope}
    \node at (6,1) (g1) [gauge,black] {$1$};
    \node at (8,1) (f1) [flavour,black] {$2$};
    \node at (5,0)(f2) [flavour,black] {$1$};
    \node at (7,0) (f3) [flavour,black] {$3$};

    \draw[->-] (f1)--(g1);
    \draw[->-] (g1)--(f3); 
    \draw[->-] (f3)--(f1);
    \draw[->-] (f2)--(g1);

    \node at (6.5,-1) {$\mathcal{W} = W_{triangle}+\mathfrak{M}^{+}$};
    \end{scope}
    
\end{tikzpicture}
\ee
\end{center}

where the first step can be reached by observing that the flipper of $b_1 \tilde{b}_1$, together with the adjoint $a_2$ of $SU(2)$, can be seen as a bifundamental of $SU(2)\times SU(2)$, from which the second equality follows.
Thus, the UV global symmetry is actually $\mathfrak{su}(3) \times \mathfrak{su}(2) \times \mathfrak{u}(1)^2$.

\subsubsection*{Duality chain and IR symmetry enhancement}

Combining eqs.~(5.2) and~(5.5) of~\cite{Benvenuti:2018bav}, one finds
that $U(1)$ with $(3_q,2_{\tilde q}+1_{\tilde p})$ flavours and
$\mathcal{W}=\mathfrak{M}^{+}$ is dual to $SU(2)_{1/2}$ with
$3_Q+2_P$ fundamentals and $\mathcal{W}=\mathrm{Flip}[Q_iP_j]$,
under the mapping
\[
\{q\tilde q,\ q\tilde p,\ \mathfrak{M}^{-}\}\;\longleftrightarrow\;
\{\mathcal{F}[QP],\ QQ,\ PP\}.
\]
Moving the six flippers to the left-hand side, the $FC_2$ theory is
dual to $SU(2)_{1/2}$ with $3_Q+2_P$ fundamentals and
$\mathcal{W}=0$.  Consequently, the IR global symmetry of the
abelian gauge theory is $\mathfrak{su}(5)\oplus\mathfrak{u}(1)$.

\subsubsection*{Generators, relations and geometry}

The chiral-ring generators
\[
\bigl\{\mathfrak{M}^{-},\,d_2,\,d_1\tilde b_1,\,d_1 v_1,\,a_1,\,a_2\bigr\},
\]
with $a_1$ the flipper of $b_1\tilde b_1$, transform in the
$\mathbf{10}$ of $SU(5)$.  In $FC_2$ language, they decompose into
two $\mathfrak{su}(2)$ adjoints and one bifundamental:
\begin{align}
A_L &= \{\mathfrak{M}^{-},\,d_1 v_1,\,a_1\},\qquad
A_R  = \{a_2\},\\
\Pi  &= \{d_2,\,d_1\tilde b_1\}.
\end{align}
The fact that these operators lie in a single irreducible
representation forces $\Delta+\tau=2$, as can also be checked through
$Z$-extremisation.

The ten generators satisfy five quadratic relations of two types,
\begin{equation}
\label{eq:FC2-relations}
\mathrm{tr}_L A_L^{\,2}=\mathrm{tr}_R A_R^{\,2}\quad
\text{(one relation)},
\qquad
A_L\,\Pi=\Pi\,A_R\quad
\text{(four relations)}.
\end{equation}

Geometrically, the moduli space is the complex cone over the
Grassmannian $\mathrm{Gr}(2,5)$~\cite{Braun:2017}, realised as the Hermitian symmetric
space of compact type $SU(5)/S\bigl(U(3)\times U(2)\bigr)$, of
rank~$2$.

\subsubsection*{Refined and unrefined Hilbert series}

From the superconformal index we extrapolate the compact form
\begin{equation}
\label{eq:HSfCu2}
\mathcal{HS}_{FC_2}(t)
= \sum_{k=0}^{\infty}\chi_{[0,k,0,0]_{su(5)}}\,t^{k}
\quad\overset{\mathrm{unr}}{=}\quad
\frac{1+3t+t^{2}}{(1-t)^{7}}\, ,
\end{equation}
which we have verified against the SCI expansion (with
$t\equiv x^{\Delta}$) up to the first 12 terms.  The representation
$[0,k,0,0]_{su(5)}$ has dimension
$\binom{k+4}{3}\binom{k+3}{3}/4$, which grows as $k^{6}$, so the
moduli space has dimension $7$.  It cannot be a complete
intersection: accordingly, the plethystic logarithm
of~\eqref{eq:HSfCu2} is an infinite series,
\begin{equation}
\label{eq:FC2-PL}
\begin{aligned}
\mathrm{PL}\!\left[\mathcal{HS}_{FC_2}\right](t)
&= \chi_{[0,1,0,0]}\,t - \chi_{[0,0,0,1]}\,t^{2}
 + \chi_{[1,0,0,0]}\,t^{3} - \chi_{[0,0,1,0]}\,t^{4}\\
&\quad + \chi_{[1,0,0,1]}\,t^{5}
       - \bigl(\chi_{[0,1,0,1]}+\chi_{[0,1,0,0]}\bigr)\,t^{6}\\
&\quad + \bigl(\chi_{[1,0,1,0]}+\chi_{[1,0,0,2]}
              +\chi_{[0,0,0,1]}\bigr)\,t^{7}\\
&\quad - 270\,t^{8} + 640\,t^{9} - 1524\,t^{10}
       + \mathcal{O}(t^{11}),
\end{aligned}
\end{equation}
where from order $t^{8}$ onwards we display only the total dimension
of the contributing representations.

A clean pattern emerges: representations enter with a plus sign at
odd powers of $t$ and with a minus sign at even powers, the latter
being conjugate representations of the form $\overline{X}$.  From
the leading terms one recognises the ten baryons $B_{ij}$, sitting
in the two-index antisymmetric of $SU(5)$ at order $t$, subject to
the five quadratic relations
\[
\epsilon^{i_1 i_2 i_3 i_4 i_5}\,B_{i_1 i_2}B_{i_3 i_4}=0
\]
at order $t^{2}$.

\subsubsection*{Branching to the generic-$N$ symmetry}

Under $\mathfrak{su}(5)\to\mathfrak{su}(4)$ one has
\begin{equation}
\label{eq:FC2-branch-su4}
[0,k,0,0]_{su(5)} = \sum_{h=0}^{k}[k-h,h,0]_{su(4)}.
\end{equation}
Using $\Delta+\tau=2$ and $t\equiv x^{\Delta}$,
eq.~\eqref{eq:HSfCu2} may be rewritten as
\begin{equation}
\label{eq:HSfCu2B}
\mathcal{HS}_{FC_2}
= \sum_{m,k=0}^{\infty}
  \chi_{[m,k,0]_{su(4)}}\,x^{m\Delta+k(2-\tau)}.
\end{equation}
Unrefining the $\mathfrak{su}(4)$ fugacities and setting
$t_d=x^{\Delta}$, $t_a=x^{2-\tau}$, the series resums to the closed
form\footnote{If the unrefined Hilbert series is written as
$\mathcal{HS}(t_a,t_d)$, switching off one of the two fugacities
yields
\[
\mathcal{HS}(0,t_d)=\frac{1}{(1-t_d)^{N^{2}}},\qquad
\mathcal{HS}(t_a,0)=\frac{\prod_{k=2}^{N}(1-t_a^{\,k})}
                         {(1-t_a)^{2(N^{2}-1)}}.
\]}
\begin{equation}
\label{eq:HSFC22F}
\mathcal{HS}_{FC_2}(t_a,t_d)
= \sum_{m,k=0}^{\infty}\dim[m,k,0]_{su(4)}\,t_a^{\,k}t_d^{\,m}
= \frac{1+t_a-4t_at_d+t_at_d^{\,2}+t_a^{\,2}t_d^{\,2}}
       {(1-t_a)^{5}\,(1-t_d)^{4}}\, .
\end{equation}
Writing the numerator of~\eqref{eq:HSFC22F} as
$N(t_a,t_d)=\sum_{j,k}a_{j,k}\,t_a^{\,j}t_d^{\,k}$, the matrix of
coefficients is
\begin{equation}
\label{eq:FC2-aij}
\{a_{j,k}\}=
\begin{pmatrix}
 1 &  0 & 0\\
 1 & -4 & 1\\
 0 &  0 & 1
\end{pmatrix},
\end{equation}
and satisfies the sum rules
\begin{equation}
\label{eq:FC2-sumrules}
\begin{aligned}
N(t,t) &= 1+t-4t^{2}+t^{3}+t^{4}=(1+3t+t^{2})(1-t)^{2},\\
N(1,t) &= 2\,(1-t)^{2},\\
N(t,1) &= (1-t)^{2}.
\end{aligned}
\end{equation}

The general branching rule for $\mathfrak{su}(4)\to\mathfrak{su}(2)^{2}$
reads
\begin{equation}
\label{eq:FC2-branch-general}
\begin{aligned}
[m,k,0]_{su(4)}
&= \sum_{k_1,k_2=0}^{\infty}\sum_{h=0}^{\lfloor m/2\rfloor}
    [2k_1+m-2h;\,2k_2+m-2h]_{su(2)\times su(2)}\\
&\quad \times
    \Bigl(\theta(m-2h)\!\sum_{s=0}^{k}\delta_{k_1+k_2,\,k-s}
         +\delta_{m,2h}\!\sum_{s=0}^{\lfloor k/2\rfloor}
          \delta_{k_1+k_2,\,k-2s}\Bigr),
\end{aligned}
\end{equation}
with $\theta(x)=1$ for $x>0$ and $\theta(x)=0$ otherwise.  Using
this, the Hilbert series can finally be expressed in
$SU(2)\times SU(2)\times U(1)_{\Delta}\times U(1)_{\tau}$
representations as
\begin{equation}
\label{eq:HSfcu2B}
\mathcal{HS}_{FC_2}(t_a,t_d)
= \mathrm{PE}\!\left[
    \bigl(\chi_{[2;0]}+\chi_{[0;2]}\bigr)\,t_a
  + \chi_{[1;1]}\,t_d
  - t_a^{\,2}
  - \chi_{[1;1]}\,t_a t_d + \cdots\right].
\end{equation}

\subsubsection*{Superconformal R-charge and index}

$Z$-extremisation gives
\begin{equation}
\label{eq:FC2-tau}
\tau = 2-\Delta = 1.312\ldots
\end{equation}
hinting that a combination of the two $U(1)$ symmetries is spurious.
For the index computation we rounded this value to $\tau=131/100$.
The first few orders of the refined index read
\begin{equation}
\label{eq:FC2-refined-index}
\mathcal{I}_{FC_2}
= 1 + x^{69/100}\chi_{[0,1,0,0]_{su(5)}}
    + x^{69/50}\chi_{[0,2,0,0]_{su(5)}}
    - x^{2}\bigl(\chi_{[1,0,0,1]_{su(5)}}+1\bigr)
    + \mathcal{O}\!\bigl(x^{101/50}\bigr),
\end{equation}
where the baryon contributions appear at orders $x^{k\Delta}$,
$k=1,2,3,\dots$, and the conserved currents of
$\mathfrak{su}(5)\oplus\mathfrak{u}(1)$ enter at order $x^{2}$.

Turning on the fugacity $\psi$ for $U(1)_{\tau}$ to track the
$\Delta=2-\tau$ dependence, the unrefined index expands as
\begin{equation}
\label{eq:FC2-unref-index}
\begin{aligned}
\mathcal{I}^{\mathrm{unr}}_{FC_2}
= \;&1 + 10\psi\,x^{69/100} + 50\psi^{2}\,x^{69/50}
     - 25\,x^{2} + 175\psi^{3}\,x^{207/100}
     - 5\psi^{-2}\,x^{131/50}\\
   &- 175\psi\,x^{269/100} + 490\psi^{4}\,x^{69/25}
     + 15\psi^{-1}\,x^{331/100} - 675\psi^{2}\,x^{169/50}\\
   &+ 1176\psi^{5}\,x^{69/20} + 5\psi^{-3}\,x^{393/100}
     + 250\,x^{4} - 1925\psi^{3}\,x^{407/100}\\
   &+ 2520\psi^{6}\,x^{207/50} + 1010\psi\,x^{469/100}
     - 4550\psi^{4}\,x^{119/25} + 4950\psi^{7}\,x^{483/100}\\
   &- 250\psi^{-1}\,x^{531/100} + 2600\psi^{2}\,x^{269/50}
     - 9450\psi^{5}\,x^{109/20} + 9075\psi^{8}\,x^{138/25}\\
   &- 775\,x^{6} + 5225\psi^{3}\,x^{607/100}
     - 17850\psi^{6}\,x^{307/50} + 15730\psi^{9}\,x^{621/100}\\
   &- \psi^{-5}\,x^{131/20} + 245\psi^{-2}\,x^{331/50}
     - 640\psi\,x^{669/100} + 8890\psi^{4}\,x^{169/25}\\
   &- 31350\psi^{7}\,x^{683/100} + 26026\psi^{10}\,x^{69/10}
     + 15\psi^{-4}\,x^{181/25}\\
   &+ 300\psi^{-1}\,x^{731/100} + 1925\psi^{2}\,x^{369/50}
     + 13300\psi^{5}\,x^{149/20}\\
   &- 51975\psi^{8}\,x^{188/25} + 41405\psi^{11}\,x^{759/100}
     - 170\psi^{-3}\,x^{793/100}\\
   &- 2450\,x^{8} + \mathcal{O}\!\bigl(x^{801/100}\bigr).
\end{aligned}
\end{equation}

\subsubsection*{Chiral-ring residue}

As a consistency check that~\eqref{eq:HSfCu2} fully captures the
chiral ring, we subtract it from~\eqref{eq:FC2-unref-index}; what
remains is
\begin{equation}
\label{eq:FC2-red-index}
\begin{aligned}
\mathcal{I}^{\mathrm{red}}_{FC_2}
= \;&1 - 25\,x^{2} - 5\psi^{-2}\,x^{131/50}
     - 175\psi\,x^{269/100} + 15\psi^{-1}\,x^{331/100}
     - 675\psi^{2}\,x^{169/50}\\
   &+ 5\psi^{-3}\,x^{393/100} + 250\,x^{4}
     - 1925\psi^{3}\,x^{407/100} + 1010\psi\,x^{469/100}
     - 4550\psi^{4}\,x^{119/25}\\
   &- 250\psi^{-1}\,x^{531/100} + 2600\psi^{2}\,x^{269/50}
     - 9450\psi^{5}\,x^{109/20} - 775\,x^{6}\\
   &+ 5225\psi^{3}\,x^{607/100} - 17850\psi^{6}\,x^{307/50}
     - \psi^{-5}\,x^{131/20} + 245\psi^{-2}\,x^{331/50}\\
   &- 640\psi\,x^{669/100} + 8890\psi^{4}\,x^{169/25}
     - 31350\psi^{7}\,x^{683/100} + 15\psi^{-4}\,x^{181/25}\\
   &+ 300\psi^{-1}\,x^{731/100} + 1925\psi^{2}\,x^{369/50}
     + 13300\psi^{5}\,x^{149/20} - 51975\psi^{8}\,x^{188/25}\\
   &- 170\psi^{-3}\,x^{793/100} - 2450\,x^{8}
     + \mathcal{O}\!\bigl(x^{801/100}\bigr).
\end{aligned}
\end{equation}
The surviving terms fall into three physical classes: $\psi$-independent
contributions from conserved currents and marginal operators;
$\tfrac{1}{4}$-BPS operators of schematic form
$[\mathrm{current}]\times[\mathrm{chiral}]$ at orders
$x^{2n+k(2-\tau)}$; and further $\tfrac{1}{4}$-BPS operators outside
the chiral ring, signalled by negative powers of $\psi$.

\subsubsection{Index resummation}
First, defining $t = \psi x^{69/100}$ absorbs all fractional powers. The series then organises exactly into even powers of $x$ as follows:

\begin{equation}
	\mathcal{I}_{FC_2}^{unr}(x, t) = \mathcal{I}^{(0)}(t) + x^2 \mathcal{I}^{(1)}(t) + x^4 \mathcal{I}^{(2)}(t) + \mathcal{O}(x^6)
\end{equation}
where the functions $\mathcal{I}^{(k)}(t)$ can be resummed into closed rational forms:
\begin{align}
	\mathcal{I}^{(0)}(t) &= \frac{1+3t+t^2}{(1-t)^7} = \mathcal{HS}_{FC_2}(t,t) \\[1ex]
	\mathcal{I}^{(1)}(t) &= - \frac{25(1+t)}{(1-t)^6} \\[1ex]
	\mathcal{I}^{(2)}(t) &= \frac{-5 + 45t + 85t^2 - 165t^3 - 85t^4 + 30t^5 - 5t^6}{t^2(1-t)^6}
\end{align}

In terms of the global $SU(5)$ symmetry, the functions $\mathcal{I}^{(k)}(t)$ can be written as an exact sum over irreducible representations (denoted by their Dynkin labels $\chi_{[q_1, q_2, q_3, q_4]}$). The chiral ring is generated by symmetric insertions $\chi_{[0, k, 0, 0]}$, while the current multiplet transforms in the adjoint representation $\mathbf{24}$ ($\chi_{[1, 0, 0, 1]}$). The exact character combinations encoding the syzygies can be written as:
\begin{align}
	\mathcal{I}^{(0)}(t) &= \sum_{k=0}^{\infty} \chi_{[0, k, 0, 0]} t^k \\[1ex]
	\mathcal{I}^{(1)}(t) &= - \sum_{k=0}^{\infty} \left( \chi_{[1, k, 0, 1]} - \chi_{[1, k-2, 1, 0]} \right) t^k \equiv - \sum_{k=0}^{\infty} C_k^{(1)}t^k \\[1ex]
	\mathcal{I}^{(2)}(t) &= - \frac{\chi_{[1, 0, 0, 0]}}{t^2} + \frac{\chi_{[2, 0, 0, 0]}}{t} + \sum_{k=0}^{\infty} C^{(2)}_k t^k
\end{align}

The formula for $\mathcal{I}^{(1)}(t)$ incorporates the $U(1)_\tau$ singlet constraint through the Weyl character formula. When $k=1$, the subtracted representation is $\chi_{[1, -1, 1, 0]}$, which lies on the boundary of a Weyl chamber and vanishes identically, leaving $175$. When $k=0$, we encounter $\chi_{[1, -2, 1, 0]}$. Reflecting this weight across the second simple root $s_2$ maps it exactly to the trivial representation $[0,0,0,0]$ with a minus sign. Therefore:
\begin{equation}
	C^{(1)}_0 = \chi_{[1, 0, 0, 1]} - (-\chi_{[0, 0, 0, 0]}) = \mathbf{24} + \mathbf{1} = 25
\end{equation}
which reproduces the exact dimension of the conserved current multiplet.

In the unrefined limit, the dimensions of the combinations $C^{(k)}_k$ follow exact polynomials in $k$ derived from the rational generating functions:
\begin{align}
	\dim C^{(1)}_k &= \frac{5}{24} (k+1)(k+2)(k+3)(k+4)(2k+5) \\[2ex]
	\dim C^{(2)}_k &= \frac{5}{6} (k+2)(k+3) ( 50 + 47k + 5k^2 - k^3 )
\end{align}

The negative leading coefficient in $\dim C^{(2)}_k$ ($- \frac{5}{6} k^5$) confirms that this tower consists entirely of higher-order syzygies removing redundant degrees of freedom. In terms of characters, this tower is dominated by the highest-weight representation extracted from the tensor product of two current multiplets, $\chi_{[2, k, 0, 2]}$, accompanied by a cascade of subtracted syzygies. For instance, at $k=0$, the dimension $250$ decomposes exactly into the sum of this tensor and the chiral ring excitation: $\chi_{[2, 0, 0, 2]} + \chi_{[0, 2, 0, 0]} = 200 + 50 = 250$.

\subsection{\texorpdfstring{$FH_2$}{fh2}}

Consider the following UV completion of $FH_2$:

\be
\label{fh2}
\resizebox{0.75\hsize}{!}{
\begin{tikzpicture}[
    thick,
    gauge/.style={
        circle, draw=black, fill=white,
        minimum size=8mm, inner sep=1pt, font=\small
    },
    manifest/.style={
        rectangle, draw=blue!70!black, fill=blue!6,
        minimum size=8mm, inner sep=2pt,
        font=\small, text=blue!70!black
    },
    flavour/.style={
        rectangle, draw=red!70!black, fill=red!6,
        minimum size=7mm, inner sep=2pt,
        font=\small, text=red!70!black
    },
    arr/.style={-{Stealth[scale=0.8]}, shorten >=1pt, shorten <=1pt},
    barr/.style={draw=black!70}
]

\node[gauge]    (g1) at (0,0)     {$1$};
\node[manifest] (m1) at (2.5,0)   {$2$};
\node[flavour]   (x1) at (-1,-1.8) {$C_1$};
\node[flavour]   (x2) at (1.5,-1.8) {$C_1$};

\draw[arr,barr,transform canvas={yshift= 3.5pt}] (g1) -- (m1);
\draw[arr,barr,transform canvas={yshift=-3.5pt}] (m1) -- (g1);
\node at (1.25, 0.45) {$b_1,\,\tilde{b}_1$};
\node at (1.25, 0) {\huge$\times$};

\draw[blue!70!black] (m1) to[out=65,in=115,looseness=3] (m1);
\node[blue!70!black] at (2.5, 0.95) {$a_2$};

\draw[arr,barr] (g1) to[bend right=5] (x1);
\node at (-0.85,-0.7) {$d_1$};

\draw[arr,barr] (x2) to[bend right=5] (g1);
\node at (0.95,-0.7) {$v_1$};

\draw[arr,barr] (m1) to[bend left=5] (x2);
\node at (2.45,-0.85) {$d_2$};

\node at (0.75,-3.2) {$\mathcal{W} =
    v_1 b_1 d_2
    + a_2 b_1\tilde{b}_1
    + \mathfrak{M}^{+} + \mathfrak{M}^{-}
    + \mathrm{Flip}[b_1\tilde{b}_1]$};

\node[right] at (3.5,-0.5) {%
    \renewcommand{\arraystretch}{1.4}
    \begin{tabular}{>{\small}c >{\small}c}
        \toprule
        $b_1,\,\tilde{b}_1$ & $\tau/2$ \\
        $a_i$               & $2-\tau$ \\
        $v_1$               & $2-\tfrac{\tau}{2}-\Delta$ \\
        $d_i$               & $\Delta+\tfrac{i-2}{2}\tau$ \\
        \bottomrule
    \end{tabular}%
};

\end{tikzpicture}
}
\ee

\subsubsection*{UV symmetry and self-dual frames}

The manifest UV global symmetry of $FH_2$ is
$\mathfrak{su}(4)\oplus\mathfrak{su}(2)\oplus\mathfrak{u}(1)$,
which enhances in the IR to $\mathfrak{so}(10)\oplus\mathfrak{u}(1)$.
This can be seen once one realises
that the $SU(2)$ adjoint $a_2$, together with the flipper of
$b_1\tilde b_1$, can be reorganised into an
$SU(2)\times SU(2)$ bifundamental, as shown below:
\begin{equation}
\label{eq:FH2-quiver-UV}
\tikzstyle{flavour}=[rectangle,draw=red!50,thick,inner sep=0pt,minimum size=6mm]
\tikzstyle{manifest}=[rectangle,draw=blue!50,thick,inner sep=0pt,minimum size=6mm]
\tikzstyle{gauge}=[circle,draw=black!50,thick,inner sep=0pt,minimum size=6mm]
\tikzset{->-/.style={decoration={markings,
  mark=at position .5 with {\arrow{>}}},postaction={decorate}}}
\begin{tikzpicture}
  \begin{scope}
    \node at (0,1)   (g1) [gauge,black]  {$1$};
    \node at (2,0)   (f1) [flavour,black] {$2$};
    \node at (2,2)   (f5) [flavour,black] {$2$};
    \node at (-1,-1) (f2) [flavour,black] {$2$};
    \node at (1,-1)  (f3) [flavour,black] {$2$};
    \draw (1,-2) node{$\mathcal{W}=W_{\text{triangles}}+\mathfrak{M}^{\pm}$};
    \draw[->-] (f1)--(g1);
    \draw[->-] (g1)--(f3);
    \draw[->-] (f3)--(f1);
    \draw[->-] (f2)--(g1);
    \draw[->-] (f5)--(f1);
    \draw[->-] (g1)--(f5);
    \draw (3.5,.7) node {\huge$=$};
  \end{scope}
  \begin{scope}
    \node at (6,1) (g1) [gauge,black]  {$1$};
    \node at (8,1) (f1) [flavour,black] {$2$};
    \node at (5,0) (f2) [flavour,black] {$2$};
    \node at (7,0) (f3) [flavour,black] {$4$};
    \draw[->-] (f1)--(g1);
    \draw[->-] (g1)--(f3);
    \draw[->-] (f3)--(f1);
    \draw[->-] (f2)--(g1);
    \node at (6.5,-1){$\mathcal{W}=W_{\text{triangle}}+\mathfrak{M}^{\pm}$};
  \end{scope}
\end{tikzpicture}
\end{equation}
It follows that the true UV symmetry is
$\mathfrak{su}(4)\oplus\mathfrak{su}(2)^{2}\oplus\mathfrak{u}(1)$.
In analogy with the observation of~\cite{Hwang:2020ddr} for
$FE_2\equiv FE[USp(4)]$, we expect $FH_2$ to admit $19$ further
self-dual frames, since
\[
20 = \frac{|W(\mathfrak{so}(10))|}
          {|W(\mathfrak{su}(4)\oplus\mathfrak{su}(2)^{2})|},
\]
where the numerator and denominator are the orders of the Weyl
groups of the IR and UV global symmetries, respectively.

\subsubsection*{Mirror UV completion}

As anticipated in Section~\ref{FHN}, $FH_2$ admits another UV
completion with gauge group $SU(2)$, descending from the mirror
self-duality of $FE_2$.  For completeness we record the mirror
quiver:
\begin{equation}
\label{eq:FH2-quiver-mirror}
\resizebox{0.75\hsize}{!}{
\begin{tikzpicture}[
    thick,
    gauge/.style={circle,draw=black,fill=white,minimum size=8mm,inner sep=1pt,font=\small},
    manifest/.style={rectangle,draw=red!70!black,fill=red!6,minimum size=8mm,inner sep=2pt,font=\small,text=red!70!black},
    flavour/.style={rectangle,draw=blue!70!black,fill=blue!6,minimum size=7mm,inner sep=2pt,font=\small,text=blue!70!black},
    arr/.style={-{Stealth[scale=0.8]},shorten >=1pt,shorten <=1pt},
    barr/.style={draw=black!70}
]
  \node[gauge]    (g1) at (0,0)      {$C_1$};
  \node[manifest] (m1) at (2.5,0)    {$C_2$};
  \node[flavour]   (x1) at (-1,-1.8)  {$1$};
  \node[flavour]   (x2) at (1.5,-1.8) {$1$};
  \draw[barr] (g1) -- (m1);
  \node at (1.25, 0.4) {$b_1$};
  \node at (1.25, 0)   {\huge$\times$};
  \draw[red!70!black] (m1) to[out=65,in=115,looseness=3] (m1);
  \node[red!70!black] at (2.5,0.95) {$a_2$};
  \draw[arr,barr] (g1) to[bend right=5] (x1); \node at (-0.85,-0.7){$d_1$};
  \draw[arr,barr] (x2) to[bend right=5] (g1); \node at ( 0.95,-0.7){$v_1$};
  \draw[arr,barr] (m1) to[bend left=5]  (x2); \node at ( 2.45,-0.85){$d_2$};
  \node at (0.75,-3.2) {$\mathcal{W}
     = v_1 b_1 d_2 + a_2 b_1^{\,2}
     + \mathfrak{M} + \mathrm{Flip}[b_1^{\,2}]$};
  \node[right] at (3.5,-0.5) {%
    \renewcommand{\arraystretch}{1.4}
    \begin{tabular}{>{\small}c >{\small}c}
      \toprule
      $b_1$ & $\tau/2$ \\
      $a_i$ & $2-\tau$ \\
      $v_1$ & $2-\tfrac{\tau}{2}-\Delta$ \\
      $d_i$ & $\Delta+\tfrac{i-2}{2}\tau$ \\
      \bottomrule
    \end{tabular}};
\end{tikzpicture}}
\end{equation}
This mirror duality was discussed, up to flippers,
in~\cite{Benvenuti:2018bav}.

\subsubsection*{Superconformal R-charge and index}

$Z$-extremisation fixes
\begin{equation}
\label{eq:FH2-tau}
\tau = 1.326,\qquad \Delta = 0.674,
\end{equation}
so that $\Delta+\tau=2$, confirming that one combination of the two
$U(1)$ fugacities is spurious and the theory carries a single
effective $U(1)$ in the IR.  Rounding to $\tau=133/100$ for the
numerical computation, the refined index reads
\begin{equation}
\label{eq:FH2-refined-index}
\begin{aligned}
\mathcal{I}_{FH_2} = \;&
    1 + \chi_{[0,0,0,0,1]_{so(10)}}\,\psi\,x^{67/100}
      + \chi_{[0,0,0,0,2]_{so(10)}}\,\psi^{2}\,x^{67/50}\\
  & - \bigl(\chi_{[0,1,0,0,0]_{so(10)}}+1\bigr)\,x^{2}
    + \chi_{[0,0,0,0,3]_{so(10)}}\,\psi^{3}\,x^{201/100}
    + \mathcal{O}\!\bigl(x^{133/50}\bigr).
\end{aligned}
\end{equation}
Unrefining the $\mathfrak{so}(10)$ fugacities, the index expands up
to order $x^{6}$ as
\begin{equation}
\label{eq:FH2-unref-index}
\begin{aligned}
\mathcal{I}^{\mathrm{unr}}_{FH_2} = \;&
    1 + 16\psi\,x^{67/100} + 126\psi^{2}\,x^{67/50}
      - 46\,x^{2} + 672\psi^{3}\,x^{201/100}\\
  & - 10\psi^{-2}\,x^{133/50} - 560\psi\,x^{267/100}
      + 2772\psi^{4}\,x^{67/25} - 3576\psi^{2}\,x^{167/50}\\
  & + 9504\psi^{5}\,x^{67/20} + 16\psi^{-3}\,x^{399/100}
      + 943\,x^{4} - 16080\psi^{3}\,x^{401/100}\\
  & + 28314\psi^{6}\,x^{201/50} + 120\psi^{-2}\,x^{233/50}
      + 8064\psi\,x^{467/100} - 57420\psi^{4}\,x^{117/25}\\
  & + 75504\psi^{7}\,x^{469/100} - 864\psi^{-1}\,x^{533/100}
      + 39604\psi^{2}\,x^{267/50} - 173712\psi^{5}\,x^{107/20}\\
  & + 184041\psi^{8}\,x^{134/25} - 144\psi^{-3}\,x^{599/100}
      - 11086\,x^{6} + \mathcal{O}\!\bigl(x^{601/100}\bigr).
\end{aligned}
\end{equation}
The contributions proportional to $\psi^{k}\,x^{k\Delta}$ come from
the fields in the $[0,0,0,0,k]$ of $\mathfrak{so}(10)$,
$k=1,2,3,\dots$.  The $-46\,x^{2}$ piece is the conserved-current
signature of the symmetry enhancement
\[
USp(2)^{2}\times U(2)\times U(1)^{2}\ \longrightarrow\ Spin(10)\times U(1),
\qquad \dim = 45+1 = 46.
\]

\subsubsection*{Refined and unrefined Hilbert series}

From the index one extrapolates the compact form of the full
refined Hilbert series,
\begin{equation}
\label{eq:HSmb2}
\mathcal{HS}_{FH_2}(t)
= \sum_{k=0}^{\infty}\chi_{[0,0,0,0,k]_{so(10)}}\,t^{k},
\qquad t\equiv x^{\Delta}.
\end{equation}
The unrefined series resums to a rational function,
\begin{equation}
\label{eq:unrHSmb2}
\begin{aligned}
\mathcal{HS}^{\mathrm{unr}}_{FH_2}(t)
&= \frac{1+5t+5t^{2}+t^{3}}{(1-t)^{11}}\\[2pt]
&= 1+16t+126t^{2}+672t^{3}+2772t^{4}+9504t^{5}+28314t^{6}\\
&\quad +75504t^{7}+184041t^{8}+\mathcal{O}(t^{9})\\[2pt]
&= \sum_{n=4}^{\infty}\frac{1}{1!\,2!\,3!\,4!}
    \sum_{1\le i,j,k,l\le n}\bigl|\det V(i,j,k,l)\bigr|\,t^{n-4},
\end{aligned}
\end{equation}
where $V(i,j,k,l)$ is the $4\times 4$ Vandermonde matrix.

\subsubsection*{Generators, relations and geometry}

The Euler form of~\eqref{eq:unrHSmb2} identifies the moduli space as
a cone of complex dimension $11$ (the exponent of the denominator).
It is the affine cone over the ten-dimensional orthogonal Grassmannian
$OG(5,10)_{+}$ (see e.g.~\cite{Mukai:1995}), whose base is the
Hermitian symmetric space of compact type $SO(10)/U(5)$, of rank~$2$.

The same conclusion follows from the large-$k$ asymptotics
of~\eqref{eq:HSmb2}: the representations $[0,0,0,0,k]_{so(10)}$ have
dimension $\binom{k+5}{5}\binom{k+7}{5}/21\sim k^{10}$, giving a
moduli-space dimension $10+1=11$.

The structure of the chiral ring is encoded in the plethystic
logarithm,
\begin{equation}
\label{eq:FH2-PL}
\begin{aligned}
\mathrm{PL}\!\left[\mathcal{HS}_{FH_2}\right](t)
&= \chi_{[0,0,0,0,1]_{so(10)}}\,t
 - \chi_{[1,0,0,0,0]_{so(10)}}\,t^{2}
 + \chi_{[0,0,0,0,1]_{so(10)}}\,t^{3}\\
&\quad - \chi_{[0,1,0,0,0]_{so(10)}}\,t^{4}
       + \chi_{[1,0,0,0,1]_{so(10)}}\,t^{5}
       + \mathcal{O}(t^{6}).
\end{aligned}
\end{equation}
Reading off~\eqref{eq:FH2-PL}: there are $16$ chiral generators, in
the spinorial of $SO(10)$ (at order $t$), subject to $10$ quadratic
relations in the fundamental of $SO(10)$ (at order $t^{2}$).  Since
the series does not terminate, the relations are not independent and
the moduli space is not a complete intersection.  This is also
consistent with the simple consideration
\[
\#\text{relations}=10 \neq 5=\mathrm{codim},
\]
forbidding a complete-intersection structure.

\subsubsection*{Branching to the generic-$N$ symmetry}

To connect with the generic-$N$ analysis, we break $\mathfrak{so}(10)$
along the chain
\begin{equation}
\label{eq:FH2-chain}
\mathfrak{so}(10)\oplus\mathfrak{u}(1)_{\Delta}
\ \longrightarrow\ \mathfrak{so}(8)\oplus\mathfrak{u}(1)_{\tau}
                  \oplus\mathfrak{u}(1)_{\Delta}
\ \longrightarrow\ \mathfrak{usp}(4)\oplus\mathfrak{su}(2)
                  \oplus\mathfrak{u}(1)_{\tau}
                  \oplus\mathfrak{u}(1)_{\Delta}.
\end{equation}
Under the chain~\eqref{eq:FH2-chain}, the spinor of $\mathfrak{so}(10)$ decomposes as
\begin{equation}
\label{eq:FH2-v-branch}
\begin{aligned}
\relax[0,0,0,0,1]_{so(10)}\,x^{\Delta}
&= [1,0;1]_{usp(4)\times su(2)}\,x^{\Delta}\\
&\quad +\bigl([0,1;0]_{usp(4)\times su(2)}
            +[0,0;2]_{usp(4)\times su(2)}\bigr)\,x^{2-\tau},
\end{aligned}
\end{equation}
matching the chiral-ring content of $FH_2$ at generic $N$: the bifundamental $\Pi\in[1,0;1]$ at order $x^{\Delta}$, together with the traceless antisymmetric $A_R\in[0,1;0]$ and the $\mathfrak{su}(2)$ adjoint $A_L\in[0,0;2]$ at order $x^{2-\tau}$.
More generally,
\begin{equation}
\label{eq:FH2-general-branch}
[0,0,0,0,k]_{so(10)}\,x^{k\Delta}
= \sum_{k_1,k_2=0}^{\infty}[0,0,k_1,k_2]_{so(8)}
  \,x^{k_1\Delta+k_2(2-\tau)}\,\delta_{k_1+k_2,\,k}.
\end{equation}
Substituting~\eqref{eq:FH2-general-branch} into~\eqref{eq:HSmb2} and
unrefining the $\mathfrak{so}(8)$ fugacities yields a resummed
two-variable Hilbert series in the abelian fugacities
$t_a\equiv x^{2-\tau}$ and $t_d\equiv x^{\Delta}$:
\begin{equation}
\label{eq:HSMB22F}
\sum_{k_1,k_2=0}^{\infty}\dim[0,0,k_1,k_2]_{so(8)}\,t_d^{\,k_1}t_a^{\,k_2}
= \frac{1+t_a+t_d-7t_a t_d+7t_a^{\,2}t_d^{\,2}
        -t_a^{\,3}t_d^{\,2}-t_a^{\,2}t_d^{\,3}-t_a^{\,3}t_d^{\,3}}
       {(1-t_a)^{7}\,(1-t_d)^{7}}.
\end{equation}

\subsubsection*{Structure of the two-variable numerator}

The numerator $N(t_a,t_d)=\sum_{j,k}a_{j,k}\,t_a^{\,j}t_d^{\,k}$
of~\eqref{eq:HSMB22F} is \emph{anti-palindromic},
$a_{j,k}=-a_{3-j,\,3-k}$, giving the symmetric, block-diagonal
coefficient matrix
\begin{equation}
\label{eq:FH2-aij}
\{a_{j,k}\}=
\begin{pmatrix}
 1 &  1 &  0 &  0\\
 1 & -7 &  0 &  0\\
 0 &  0 &  7 & -1\\
 0 &  0 & -1 & -1
\end{pmatrix}.
\end{equation}
Two immediate consequences are $\sum_{j,k}a_{j,k}=0$ and the
symmetry $t_a\leftrightarrow t_d$ of the Hilbert series -- the latter
reflecting the triality of $\mathfrak{so}(8)$.  Setting
$t_a=t_d\equiv t$, the numerator factorises as
\begin{equation}
\label{eq:FH2-num-diagonal}
-t^{6}-2t^{5}+7t^{4}-7t^{2}+2t+1
= (1-t)^{3}\bigl(1+5t+5t^{2}+t^{3}\bigr),
\end{equation}
which recovers the unrefined Hilbert series~\eqref{eq:unrHSmb2}.
Both single-variable specialisations yield the same result,
\begin{equation}
\label{eq:FH2-sum-rules}
N(1,t) = N(t,1) = 2\,(1-t)^{3}.
\end{equation}

\subsubsection{Index resummation}

Setting $t=\psi\,x^{67/100}$, the unrefined index may be written as
\begin{equation}
\mathcal{I}_{FH_2}^{unr}(x,t) = \mathcal{I}^{(0)}(t) + x^2 \mathcal{I}^{(1)}(t) + x^4 \mathcal{I}^{(2)}(t) + \mathcal{O}(x^6)
\end{equation}
where the functions $I^{(k)}(t)$ can be resummed into closed rational forms:
\begin{align}
\mathcal{I}^{(0)}(t) &= \frac{1 + 5t + 5t^2 + t^3}{(1-t)^{11}}  = \mathcal{HS}_{FH_2} \\[1ex]
\mathcal{I}^{(1)}(t) &= - \frac{46 +100t+46t^2}{(1-t)^{10}} \\[1ex]
\mathcal{I}^{(2)}(t) &= \frac{t^8-10 t^7+45 t^6-136 t^5-701 t^4-166 t^3+493 t^2+100 t-10}{(1-t)^{10}
   t^2}
\end{align}

In terms of the global $SO(10)$ symmetry, the functions $I^{(k)}(t)$ can be written as an exact sum over irreducible representations (denoted by their Dynkin labels $\chi_{[a1,a2,a3,a4,a5]}$). In this theory, the chiral ring $I^{(0)}(t)$ is generated by the spinor representation $\mathbf{16}$ ($\chi_{[0,0,0,1,0]}$), while the current multiplet transforms in the adjoint $\mathbf{45}$ ($\chi_{ [0,1,0,0,0]}$). The exact character combinations are:

\begin{align}
    \mathcal{I}^{(0)}(t) &= \sum_{k=0}^{\infty} \chi_{[0, 0, 0, 0, k]}\, t^k \\[1ex]
    \mathcal{I}^{(1)}(t) &= - \sum_{k=0}^{\infty} \left( \chi_{[0, 1, 0, 0, k]} - \chi_{[0, 0, 1, 0, k-2]} \right) t^k
    \equiv - \sum_{k=0}^{\infty} C_k^{(1)}\, t^k \\[1ex]
    \mathcal{I}^{(2)}(t) &= - \frac{\chi_{[1,0,0,0,0]}}{t^2} + \sum_{k=0}^{\infty} C_k^{(2)}\, t^k
\end{align}
The formula for $\mathcal{I}^{(1)}(t)$ incorporates the $U(1)_\tau$ singlet constraint through the
Weyl character formula. When $k = 1$, the subtracted representation is $\chi_{[0,0,1,0,-1]}$,
whose $\rho$-shifted weight $[0,0,1,0, -1] + \rho = [1,1,2,1,0]$ has a vanishing fifth
Dynkin label. Since this places it on the wall of the fundamental Weyl chamber associated
with the simple root $\alpha_4$, the character vanishes identically, leaving
$C_1^{(1)} = \chi_{[0,1,0,0,1]}$ of dimension $560$.
When $k = 0$, we encounter $\chi_{[0,0,1,0,-2]}$.
The shifted weight $[0,0,1,0,-2] + \rho = [1,1,2,1,-1]$ has a negative fifth label.
A single reflection $s_5$ across the fifth simple root maps it to
$[1,1,1,1,1] = \rho$, the weight of the trivial representation, with a minus sign.
Therefore:
\begin{equation}
    C_0^{(1)} = \chi_{[0,1,0,0,0]} - \bigl(-\chi_{[0,0,0,0,0]}\bigr) = \mathbf{45} + \mathbf{1} = 46
\end{equation}
which reproduces the exact dimension of the conserved current multiplet of the $SO(10)$ theory.

In the unrefined limit, the dimensions of the combination $C_k^{(1)}$ follow an exact
polynomial in $k$ derived from the rational generating function:
\begin{equation}
    \dim C_k^{(1)} \;=\;
    \frac{1}{7560}\,(4k^2 + 32k + 69)\,(k+1)(k+2)(k+3)(k+4)(k+5)(k+6)(k+7)
\end{equation}
The quadratic factor $4k^2+32k+69 = 4(k+4)^2+5$ is strictly positive for all $k \geq 0$,
consistent with $C_k^{(1)}$ being a genuine (non-negative) character combination at every
level. Notably, $\mathcal{I}^{(2)}(t)$ carries only a single polar term at order $t^{-2}$
corresponding to the vector representation $\mathbf{10}$: the potential $t^{-1}$ singularity
is absent because the first two coefficients of the numerator $N(t) = -10 + 100\,t + \cdots$
satisfy the exact cancellation $N_1 + 10\,N_0 = 100 - 100 = 0$.
The regular part $\sum_k C_k^{(2)}\,t^k$ begins at $k = 0$ with dimension
\begin{equation}
    \dim C_0^{(2)} \;=\;
    \chi_{[0,2,0,0,0]} + \chi_{[2,0,0,0,0]} + \chi_{[0,0,1,0,0]} - \chi_{[0,0,0,0,0]}
    \;=\; \mathbf{770} + \mathbf{54} + \mathbf{120} - \mathbf{1} \;=\; 943
\end{equation}
The dominant contribution $\chi_{[0,2,0,0,0]} = \mathbf{770}$ is the highest-weight
representation appearing in the symmetric square of the current multiplet
$\mathbf{45} \otimes_s \mathbf{45}$, accompanied by the symmetric traceless tensor
$\chi_{[2,0,0,0,0]} = \mathbf{54}$, the three-form $\chi_{[0,0,1,0,0]} = \mathbf{120}$,
and a subtracted singlet accounting for trace redundancies.

\subsection{\texorpdfstring{$FM_2$}{fmsu2}}
Consider the $FM_2$ theory, which admits the following Lagrangian UV completion:

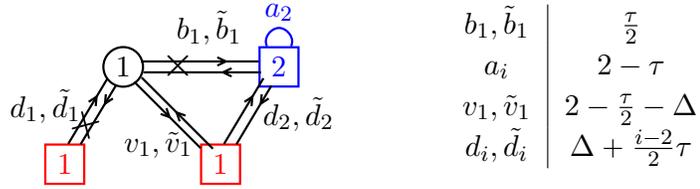
\be
\label{fm2}
\resizebox{0.85\hsize}{!}{
\begin{tikzpicture}[
    thick,
    gauge/.style={
        circle, draw=black, fill=white,
        minimum size=8mm, inner sep=1pt, font=\small
    },
    manifest/.style={
        rectangle, draw=blue!70!black, fill=blue!6,
        minimum size=8mm, inner sep=2pt,
        font=\small, text=blue!70!black
    },
    flavour/.style={
        rectangle, draw=red!70!black, fill=red!6,
        minimum size=7mm, inner sep=2pt,
        font=\small, text=red!70!black
    },
    arr/.style={-{Stealth[scale=0.8]}, shorten >=1pt, shorten <=1pt},
    barr/.style={draw=black!70}
]

\node[gauge]    (g1) at (0,0)      {$1$};
\node[manifest] (m1) at (2.5,0)    {$2$};
\node[flavour]   (x1) at (-1,-1.8)  {$1$};
\node[flavour]   (x2) at (1.5,-1.8) {$1$};

\draw[arr,barr,transform canvas={yshift= 3.5pt}] (g1) -- (m1);
\draw[arr,barr,transform canvas={yshift=-3.5pt}] (m1) -- (g1);
\node at (1.25, 0.48) {$b_1,\,\tilde{b}_1$};
\node at (1.25, 0) {\huge$\times$};

\draw[blue!70!black] (m1) to[out=65,in=115,looseness=3] (m1);
\node[blue!70!black] at (2.5, 0.95) {$a_2$};

\draw[arr,barr] (x1) to[bend left=8]  (g1);
\draw[arr,barr] (g1) to[bend left=8]  (x1);
\node at (-1.3,-0.75) {$d_1,\,\tilde{d}_1$};
\node[rotate=55] at (-0.52,-0.95) {\huge$\times$};

\draw[arr,barr] (g1) to[bend right=8] (x2);
\draw[arr,barr] (x2) to[bend right=8] (g1);
\node at (0.35,-1.35) {$v_1,\,\tilde{v}_1$};

\draw[arr,barr] (x2) to[bend left=8]  (m1);
\draw[arr,barr] (m1) to[bend left=8]  (x2);
\node at (2.9,-1) {$d_2,\,\tilde{d}_2$};

\node at (0.75,-3.3) {$\mathcal{W} =
    v_1 b_1 d_2
    + \tilde{v}_1\tilde{b}_1\tilde{d}_2
    + a_2 b_1\tilde{b}_1
    + \mathfrak{M}^{+} + \mathfrak{M}^{-}
    + \mathrm{Flip}[b_1\tilde{b}_1]
    + \mathrm{Flip}[d_1\tilde{d}_1]$};

\node[right] at (3.5,-0.5) {%
    \renewcommand{\arraystretch}{1.4}
    \begin{tabular}{>{\small}c >{\small}c}
        \toprule
        $b_1,\,\tilde{b}_1$     & $\tau/2$ \\
        $a_i$                   & $2-\tau$ \\
        $v_1,\,\tilde{v}_1$     & $2-\tfrac{\tau}{2}-\Delta$ \\
        $d_i,\,\tilde{d}_i$     & $\Delta+\tfrac{i-2}{2}\tau$ \\
        \bottomrule
    \end{tabular}%
};

\end{tikzpicture}
}
\ee

\subsubsection*{Duality chain and IR symmetry enhancement}

Naively, the generic-$N$ analysis would predict an IR symmetry
$S(U(2)_L\times U(2)_R)\times U(1)_{\Delta}\times U(1)_{\tau}$ for
this theory.  However, by eq.~(5.3) of~\cite{Benvenuti:2018bav}, a
suitably flipped version of $FM_2$ is dual to $SU(2)$ with
$4_q+2_p$ fundamentals and $\mathcal{W}=\mathrm{Flip}[pp]$, with
the R-charges of $q$ and $p$ fixed to $1-\tau/2$ and
$\Delta-1+\tau/2$, respectively.  The resulting R-charge assignments
for the gauge-invariant bilinears are
\begin{equation}
\label{eq:FM2-Rcharges}
R[qq]=2-\tau,\qquad
R[\mathcal{F}[pp]]=4-2\Delta-\tau,\qquad
R[pq]=\Delta,\qquad
R[\mathfrak{M}]=2-2\Delta+\tau.
\end{equation}
Consequently, the IR global symmetry of the abelian theory is
$\mathfrak{su}(4)\oplus\mathfrak{su}(2)\oplus\mathfrak{u}(1)^{2}$,
rather than the generic-$N$ expectation above.

\subsubsection*{Chiral-ring generators}

The $16$ chiral-ring generators (the $8$ mesons together with the
$8$ flippers of mesons) decompose under
$SU(4)\times SU(2)$ as
\begin{equation}
\label{eq:FM2-generators-rep}
(\mathbf{6},\mathbf{1})_{2-\tau}\;\oplus\;
(\mathbf{4},\mathbf{2})_{\Delta}\;\oplus\;
(\mathbf{1},\mathbf{1})_{4-2\Delta-\tau}\;\oplus\;
(\mathbf{1},\mathbf{1})_{2-2\Delta+\tau},
\end{equation}
with the subscripts denoting the $U(1)_{\Delta}\times U(1)_{\tau}$
charges.  In $FM_2$ language, the IR spectrum contains two
$SU(2)_L\times SU(2)_R$ bifundamentals $\Pi$ and $\tilde{\Pi}$, two
adjoints $A_L$ and $A_R$ of $SU(2)_L$ and $SU(2)_R$ respectively,
and the two singlets $B_{1,2}$ (the flipper of
$d_1\tilde d_1$) and $B_{2,1}\equiv v_1\tilde v_1$.

\subsubsection*{Superconformal R-charges and index}

$Z$-extremisation locates the minimum of the $S^{3}$ partition
function at
\begin{equation}
\label{eq:FM2-Rcharges-num}
\Delta = 0.969,\qquad \tau = 1.277.
\end{equation}
For the numerical computation we fixed $\Delta=82/100$ and
$\tau=131/100$, and turn on the fugacities $\psi_{\tau},\chi_{\Delta}$
for the two $U(1)$'s.  The unrefined index then expands as
\begin{equation}
\label{eq:FM2-unref-index}
\begin{aligned}
\mathcal{I}_{FM_2}(x)
= \;&1 + 6\psi_{\tau}\,x^{69/100}
      + 8\chi_{\Delta}\,x^{41/50}
      + \psi_{\tau}\chi_{\Delta}^{-2}\,x^{21/20}\\
   &+ 20\psi_{\tau}^{2}\,x^{69/50}
      + 40\chi_{\Delta}\psi_{\tau}\,x^{151/100}
      + 30\chi_{\Delta}^{2}\,x^{41/25}
      + \chi_{\Delta}^{-2}\psi_{\tau}^{-1}\,x^{167/100}\\
   &+ 6\psi_{\tau}^{2}\chi_{\Delta}^{-2}\,x^{87/50}
      - 20\,x^{2}
      + 50\psi_{\tau}^{3}\,x^{207/100}
      + \psi_{\tau}^{2}\chi_{\Delta}^{-4}\,x^{21/10}\\
   &- 8\chi_{\Delta}\psi_{\tau}^{-1}\,x^{213/100}
      + 120\chi_{\Delta}\psi_{\tau}^{2}\,x^{11/5}
      + 135\chi_{\Delta}^{2}\psi_{\tau}\,x^{233/100}
      + 20\psi_{\tau}^{3}\chi_{\Delta}^{-2}\,x^{243/100}\\
   &+ 80\chi_{\Delta}^{3}\,x^{123/50}
      - \psi_{\tau}^{-2}\,x^{131/50}
      - 98\psi_{\tau}\,x^{269/100}
      + \chi_{\Delta}^{-4}\,x^{68/25}
      + 105\psi_{\tau}^{4}\,x^{69/25}\\
   &+ 6\psi_{\tau}^{3}\chi_{\Delta}^{-4}\,x^{279/100}
      - 136\chi_{\Delta}\,x^{141/50}
      + 280\chi_{\Delta}\psi_{\tau}^{3}\,x^{289/100}\\
   &- 45\chi_{\Delta}^{2}\psi_{\tau}^{-1}\,x^{59/20}
      + 378\chi_{\Delta}^{2}\psi_{\tau}^{2}\,x^{151/50}
      - 16\psi_{\tau}\chi_{\Delta}^{-2}\,x^{61/20}\\
   &+ 50\psi_{\tau}^{4}\chi_{\Delta}^{-2}\,x^{78/25}
      + x^{63/20}\bigl(336\chi_{\Delta}^{3}\psi_{\tau}
                     +\psi_{\tau}^{3}\chi_{\Delta}^{-6}\bigr)\\
   &+ 175\chi_{\Delta}^{4}\,x^{82/25}
      + 10\psi_{\tau}^{-1}\,x^{331/100}
      + \chi_{\Delta}^{-4}\psi_{\tau}^{-2}\,x^{167/50}
      - 285\psi_{\tau}^{2}\,x^{169/50}\\
   &+ 196\psi_{\tau}^{5}\,x^{69/20}
      + 20\psi_{\tau}^{4}\chi_{\Delta}^{-4}\,x^{87/25}
      - 512\chi_{\Delta}\psi_{\tau}\,x^{351/100}
      + 560\chi_{\Delta}\psi_{\tau}^{4}\,x^{179/50}\\
   &- 454\chi_{\Delta}^{2}\,x^{91/25}
      - \chi_{\Delta}^{-2}\psi_{\tau}^{-1}\,x^{367/100}
      + 840\chi_{\Delta}^{2}\psi_{\tau}^{3}\,x^{371/100}
      - 64\psi_{\tau}^{2}\chi_{\Delta}^{-2}\,x^{187/50}\\
   &+ x^{377/100}\bigl(\psi_{\tau}\chi_{\Delta}^{-6}
                      -144\chi_{\Delta}^{3}\psi_{\tau}^{-1}\bigr)
      + 105\psi_{\tau}^{5}\chi_{\Delta}^{-2}\,x^{381/100}\\
   &+ x^{96/25}\bigl(6\psi_{\tau}^{4}\chi_{\Delta}^{-6}
                    +896\chi_{\Delta}^{3}\psi_{\tau}^{2}\bigr)
      + 16\psi_{\tau}\chi_{\Delta}^{-1}\,x^{387/100}\\
   &+ 700\chi_{\Delta}^{4}\psi_{\tau}\,x^{397/100}
      + 166\,x^{4}
      + \mathcal{O}\!\bigl(x^{401/100}\bigr).
\end{aligned}
\end{equation}
The $-20\,x^{2}$ conserved-current contribution witnesses the
symmetry enhancement
\[
SU(2)\times U(1)^{4}\ \longrightarrow\ SU(4)\times SU(2)\times U(1)^{2},
\qquad \dim = 15+3+1+1 = 20.
\]
Reorganising the refined index into $SU(4)\times SU(2)$ characters,
\begin{equation}
\label{eq:FM2-refined-index}
\begin{aligned}
\mathcal{I}_{FM_2}(x)
= \;& -1
    + \sum_{n,k=0}^{\infty}
       \chi_{[n,k,0;n]_{su(4)\times su(2)}}\,x^{n\Delta+k(2-\tau)}
    + \sum_{l,m=0}^{\infty}
       x^{l(4-2\Delta-\tau)+m(2-2\Delta+\tau)}\\
  & + \sum_{k,l=1}^{\infty}
       \chi_{[0,k,0;0]_{su(4)\times su(2)}}\,x^{l(4-2\Delta-\tau)+k(2-\tau)}\\
  & - \bigl(\chi_{[1,0,1;0]_{su(4)\times su(2)}}
           - \chi_{[0,0,0;2]_{su(4)\times su(2)}} - 2\bigr)\,x^{2}
    + \cdots,
\end{aligned}
\end{equation}
with $\chi_R$ the character of the representation $R$ under
$SU(4)\times SU(2)$.

\subsubsection*{Hilbert series and branch structure}

From~\eqref{eq:FM2-refined-index} one extrapolates the all-order
Hilbert series of the full moduli space:
\begin{equation}
\label{eq:fm2hs}
\begin{aligned}
\mathcal{HS}_{FM_2}
= \;& \sum_{n,k=0}^{\infty}
       \chi_{[n,k,0;n]_{su(4)\times su(2)}}\,x^{n\Delta+k(2-\tau)}
    + \sum_{l,m=0}^{\infty}
       x^{l(4-2\Delta-\tau)+m(2-2\Delta+\tau)}\\
  & + \sum_{k,l=0}^{\infty}
       \chi_{[0,k,0;0]_{su(4)\times su(2)}}\,x^{l(4-2\Delta-\tau)+k(2-\tau)}\\
  & - \sum_{k=0}^{\infty}
       \chi_{[0,k,0;0]_{su(4)\times su(2)}}\,x^{k(2-\tau)}
    - \sum_{l=0}^{\infty} x^{l(4-2\Delta-\tau)}.
\end{aligned}
\end{equation}
The moduli space is therefore the union of three branches:
\begin{itemize}
\item a \emph{main branch}, generated by $\{A_L,A_R,\Pi,\tilde{\Pi}\}$;
\item a \emph{second branch}, generated by $\{B_{1,2},B_{2,1}\}$;
\item a \emph{third branch}, generated by $\{A_L,A_R,B_{2,1}\}$.
\end{itemize}
The first two branches intersect only at the origin; the third has
nontrivial intersection with both.  Specifically, the third branch is
isomorphic to $\mathbb{C}$ times the moduli space of $FT_2$.  The
unrefined Hilbert series of the second and third branches read,
respectively,
\begin{equation}
\label{eq:FM2-HS-23}
\frac{1}{(1-x^{4-2\Delta-\tau})(1-x^{2-2\Delta+\tau})},
\qquad
\frac{1+x^{2-\tau}}{(1-x^{4-2\Delta-\tau})\,(1-x^{2-\tau})^{5}},
\end{equation}
while the resummed intersection contributions -- the last two terms
of~\eqref{eq:fm2hs} -- give
\begin{equation}
\label{eq:FM2-HS-intersections}
-\frac{1+x^{2-\tau}}{(1-x^{2-\tau})^{5}}
\qquad\text{and}\qquad
-\frac{1}{1-x^{4-2\Delta-\tau}}.
\end{equation}

\subsubsection*{The main branch}

Setting $t_a\equiv x^{2-\tau}$, $t_d\equiv x^{\Delta}$, the unrefined
Hilbert series of the main branch is
\begin{equation}
\label{eq:hsfm2main}
\mathcal{HS}^{\mathrm{main}}_{FM_2}(t_a,t_d)
= \frac{1+3t_d+t_a\bigl(1-5t_d-5t_d^{\,2}+t_d^{\,3}\bigr)
        +t_a^{\,2}t_d^{\,2}(3+t_d)}
       {(1-t_a)^{5}\,(1-t_d)^{5}}\, .
\end{equation}
Writing the numerator as
$N(t_a,t_d)=\sum_{j,k}a_{j,k}\,t_a^{\,j}t_d^{\,k}$, the coefficient
matrix is
\begin{equation}
\label{eq:FM2-aij}
\{a_{j,k}\}=
\begin{pmatrix}
 1 &  3 &  0 & 0\\
 1 & -5 & -5 & 1\\
 0 &  0 &  3 & 1
\end{pmatrix},
\end{equation}
and obeys the sum rules
\begin{equation}
\label{eq:FM2-sumrules}
\begin{aligned}
N(t,t) &= (t-1)^{2}\bigl(1+6t+6t^{2}+t^{3}\bigr),\\
N(1,t) &= 2(t-1)^{2}(t+1),\\
N(t,1) &= 4(t-1)^{2}.
\end{aligned}
\end{equation}
The last two specialisations translate into the row/column sum rules
\begin{equation}
\label{eq:FM2-row-col-sums}
\sum_{j}a_{j,k}
 = 2(-1)^{k}\biggl[\binom{2}{k}-\binom{2}{k-1}\biggr],
\qquad
\sum_{k}a_{j,k}
 = 4(-1)^{j}\binom{2}{j},
\end{equation}
with the convention $\binom{2}{-1}=\binom{2}{3}=0$.

\subsubsection*{Generators and relations of the main branch}

The plethystic logarithm of the main-branch Hilbert series
in~\eqref{eq:fm2hs} reads
\begin{equation}
\label{eq:FM2-PL-main}
\begin{aligned}
\mathrm{PL}\!\left[\sum_{n,k=0}^{\infty}
  \chi_{[n,k,0;n]_{su(4)\times su(2)}}\,x^{n\Delta+k(2-\tau)}\right]
=\;& \chi_{[0,1,0;0]}\,x^{2-\tau}
   + \chi_{[1,0,0;1]}\,x^{\Delta}
   - x^{4-2\tau}\\
  & - \chi_{[0,0,1;1]}\,x^{2-\tau+\Delta}
    - \chi_{[0,1,0;0]}\,x^{2\Delta}\\
  & + \chi_{[1,0,0;1]}\,x^{\Delta+4-2\tau}
    + \mathcal{O}\!\bigl(x^{2\Delta+2-\tau}\bigr).
\end{aligned}
\end{equation}
Packaging the adjoints $A_L,A_R$ into a two-index antisymmetric
$\Phi_{ab}$ of $SU(4)$ (singlet under $SU(2)$), and the bifundamentals
$\Pi,\tilde{\Pi}$ into $\Psi_{a;A}$ in the
$(\Box,\mathbf{2})$ of $SU(4)\times SU(2)$ ($a,b$ $SU(4)$ indices,
$A$ an $SU(2)$ index), one reads off $\Phi_{ab}$ at order $x^{2-\tau}$
and $\Psi_{a;A}$ at order $x^{\Delta}$.  They satisfy the relations
\begin{equation}
\label{eq:FM2-main-relations}
\begin{aligned}
\epsilon^{abcd}\Phi_{ab}\Phi_{cd}=0,\qquad &
  \bigl(1\ \text{relation at order } x^{4-2\tau}\bigr),\\[2pt]
\epsilon^{AB}\Psi_{a;A}\Psi_{b;B}=0,\qquad &
  \bigl(6\ \text{relations at order } x^{2\Delta}\bigr),\\[2pt]
\epsilon^{abcd}\Phi_{ab}\Psi_{c;A}=0,\qquad &
  \bigl(8\ \text{relations at order } x^{2-\tau+\Delta}\bigr).
\end{aligned}
\end{equation}
Since~\eqref{eq:FM2-PL-main} does not terminate, these relations are
not independent: the main branch is not a complete intersection.

\subsubsection{A flipped theory: $\widetilde{FM}_{2}$}
\label{sec:FM2-flipped}

Consider the theory obtained from $FM_2$ by flipping the singlet
$B_{2,1}\equiv v_1\tilde v_1$; we denote it $\widetilde{FM}_{2}$.
$Z$-extremisation returns
\begin{equation}
\label{eq:FM2tilde-Rcharges}
\Delta = 2-\tau \approx 0.7374,
\end{equation}
and we claim that the IR global symmetry enhances to $U(6)$.  As a
check, fixing $\Delta=37/50$, the first orders of the refined index
read
\begin{equation}
\label{eq:FM2tilde-refined-index}
\begin{aligned}
\mathcal{I}_{\widetilde{FM}_{2}}
= \;&1 + \mathbf{15}\,\chi\,x^{37/50}
      + \overline{\mathbf{105}}{}'\,\chi^{2}\,x^{37/25}
      + \chi^{-3}\,x^{89/50}
      - (\mathbf{35}+1)\,x^{2}\\
  &\;+ \mathbf{490}\,\chi^{3}\,x^{111/50}
      - \mathbf{384}\,\chi\,x^{137/50}
      + \mathbf{1764}'\,\chi^{4}\,x^{74/25}
      + \mathcal{O}\!\bigl(x^{151/50}\bigr),
\end{aligned}
\end{equation}
where $\chi$ is the fugacity for the single $U(1)$ of the IR, and the
bold numbers denote $SU(6)$ characters.  The conserved currents
contributing with a minus sign at order $x^{2}$ fill the adjoint of
$SU(6)\times U(1)$.

The moduli space of $\widetilde{FM}_{2}$ consists of two branches
with trivial intersection:
\begin{itemize}
\item a copy of $\mathbb{C}$, freely generated by the remaining
      singlet $B_{1,2}$ (contributing at order $x^{4-3\Delta}$);
\item a main branch generated by the $\mathbf{15}$ of $SU(6)$
      (order $x^{\Delta}$), subject to $15$ quadratic relations.
\end{itemize}
The Hilbert series of the main branch takes the compact form
\begin{equation}
\label{eq:FM2tilde-HS}
\mathcal{HS}_{\widetilde{FM}_{2}}
= \sum_{k=0}^{\infty}\chi_{[0,k,0,0,0]_{SU(6)}}\,t^{k}
\quad\overset{\mathrm{unr}}{=}\quad
\frac{1+6t+6t^{2}+t^{3}}{(1-t)^{9}}\, ,
\end{equation}
which is exactly the Hilbert series of the complex cone over the
Grassmannian $\mathrm{Gr}(2,6)$~\cite{Braun:2017}, realised
geometrically as the Hermitian symmetric space of compact type
$SU(6)/S\bigl(U(4)\times U(2)\bigr)$.

Unrefining with respect to the $SU(6)$ fugacities, the index expands
up to $x^{8}$ as
\begin{equation}
\label{eq:FM2tilde-unref-index}
\begin{aligned}
\mathcal{I}^{\mathrm{unr}}_{\widetilde{FM}_{2}}(x;\chi)
= \;&1 + 15\chi\,x^{37/50} + 105\chi^{2}\,x^{37/25}
      + \chi^{-3}\,x^{89/50} - 36\,x^{2}\\
   &+ 490\chi^{3}\,x^{111/50} - 384\chi\,x^{137/50}
      + 1764\chi^{4}\,x^{74/25}\\
   &+ 21\chi^{-1}\,x^{163/50} - 2100\chi^{2}\,x^{87/25}
      + \chi^{-6}\,x^{89/25}
      + 5292\chi^{5}\,x^{37/10}\\
   &+ 558\,x^{4} - 8064\chi^{3}\,x^{211/50}
      + 13860\chi^{6}\,x^{111/25}
      + 36\chi^{-2}\,x^{113/25}\\
   &+ 3690\chi\,x^{237/50} - 24696\chi^{4}\,x^{124/25}
      + 32670\chi^{7}\,x^{259/50}\\
   &- 384\chi^{-1}\,x^{263/50} + \chi^{-9}\,x^{267/50}
      + 14526\chi^{2}\,x^{137/25}\\
   &- 64512\chi^{5}\,x^{57/10} - 71\chi^{-3}\,x^{289/50}
      + 70785\chi^{8}\,x^{148/25}\\
   &- 3536\,x^{6} + 42651\chi^{3}\,x^{311/50}
      - 149688\chi^{6}\,x^{161/25}\\
   &+ 135\chi^{-2}\,x^{163/25} + 143143\chi^{9}\,x^{333/50}
      - 12216\chi\,x^{337/50}\\
   &+ 103236\chi^{4}\,x^{174/25} + 36\chi^{-4}\,x^{176/25}
      + \chi^{-12}\,x^{178/25}\\
   &- 316800\chi^{7}\,x^{359/50} + 2562\chi^{-1}\,x^{363/50}
      + 273273\chi^{10}\,x^{37/5}\\
   &- 24576\chi^{2}\,x^{187/25} + 217476\chi^{5}\,x^{77/10}
      - 72\chi^{-3}\,x^{389/50}\\
   &- 622908\chi^{8}\,x^{198/25} + 3465\,x^{8}
      + \mathcal{O}\!\bigl(x^{401/50}\bigr).
\end{aligned}
\end{equation}

\subsubsection*{Index resummation for $\widetilde{FM}_2$}

Setting $t=\chi\,x^{37/50}$, the unrefined index may be written as
\begin{equation}
\mathcal{I}_{\widetilde{FM}_2}^{unr}(x,t) = \mathcal{I}^{(0)}(t) + x^2 \mathcal{I}^{(1)}(t) + x^4 \mathcal{I}^{(2)}(t) + \mathcal{O}(x^6)
\end{equation}
where the functions $I^{(k)}(t)$ can be resummed into closed rational forms:
\begin{align}
\mathcal{I}^{(0)}(t) &= \frac{1+6t+6t^2+t^3}{(1-t)^9} = \mathcal{HS}_{\widetilde{FM}_2} \\[1ex]
\mathcal{I}^{(1)}(t) &= -12 \frac{3+8t+3t^2}{(1-t)^8}\\[1ex]
\mathcal{I}^{(2)}(t) &= \frac{1-8t+49t^2+334t^3-116t^4-602t^5+13t^6-8t^7+t^8}{(1-t)^8 t^3}
\end{align}

In terms of the global $SU(6)$ symmetry, the functions $\mathcal{I}^{(k)}(t)$ can be written as exact sums over irreducible representations, denoted by their Dynkin labels $\chi_{[a_1,a_2,a_3,a_4,a_5]}$. In this theory, the chiral ring $\mathcal{I}^{(0)}(t)$ is generated by the representation $\mathbf{15}$ ($\chi_{[0,1,0,0,0]}$), while the current multiplet transforms in the adjoint $\mathbf{35}$ ($\chi_{[1,0,0,0,1]}$). The exact character combinations are

\begin{align}
\mathcal{I}^{(0)}(t) &= \sum_{k=0}^{\infty} \chi_{[0, k,  0, 0,0]} t^k \\[1ex]
\mathcal{I}^{(1)}(t) &=  - \sum_{k=0}^{\infty} \left( \chi_{[1, k, 0,0, 1]} - \chi_{[1, k-2, 1, 0,0]} \right) t^k \equiv - \sum_{k=0}^{\infty} C_k^{(1)}t^k \\[1ex] 
\mathcal{I}^{(2)}(t) &=  \frac{\chi_{[0,0,0,0,0]}}{t^3} +\frac{\chi_{[2,0,0,0,0]}}{t} + \sum_{k=0}^{\infty} C_k^{(2)}\, t^k
\end{align}

The formula for $\mathcal{I}^{(1)}(t)$ incorporates the $U(1)_\tau$ singlet constraint through the Weyl character formula. 

For $SU(6)$, $\rho = [1,1,1,1,1]$. The Weyl character formula gives $\chi_\lambda = 0$ whenever $\lambda + \rho$ lies on a wall of the Weyl chamber, i.e., when any Dynkin label of $\lambda$ is $-1$ (since then $\lambda + \rho$ has a zero label, placing it on a wall and the character vanishes after antisymmetrisation).

When $k=1$: the subtracted term is $\chi_{[1,-1,1,0,0]}$. The second Dynkin label is $-1$, so $(\lambda+\rho)_2 = 0$, placing it on the wall of the Weyl chamber associated with simple root $\alpha_2$. The character vanishes identically, leaving
$$C_1^{(1)} = \chi_{[1,1,0,0,1]}$$
of dimension $\mathbf{384}$.

When $k=0$: the subtracted term is $\chi_{[1,-2,1,0,0]}$. The shifted weight is $[1,-2,1,0,0]+\rho = [2,-1,2,1,1]$, which has a negative second label. A single reflection $s_2$ across the second simple root maps
$$s_2([2,-1,2,1,1]) = [2,-1,2,1,1] + (-1)\cdot(-\alpha_2)  = [1,1,1,1,1] = \rho,$$
the Weyl vector, which is the shifted weight of the trivial representation $\chi_{[0,0,0,0,0]}$. Since $s_2$ introduces a sign of $-1$, the Weyl formula gives $\chi_{[1,-2,1,0,0]} = -\chi_{[0,0,0,0,0]} = -\mathbf{1}$. Therefore:
$$C_0^{(1)} = \chi_{[1,0,0,0,1]} - \bigl(-\chi_{[0,0,0,0,0]}\bigr) = \mathbf{35} + \mathbf{1} = 36,$$ consistent with the dimension of the conserved current multiplet of the $SU(6)$ theory together with a singlet.

In the unrefined limit, the dimensions of the combination $C_k^{(1)}$ follow an exact polynomial in $k$ derived from the rational generating function. Expanding $-\mathcal{I}^{(1)}(t) = 12\,\frac{3+8t+3t^2}{(1-t)^8}$, the coefficients are:
$$\dim C_k^{(1)} = \frac{1}{30}(k+1) (k+2) (k+3)^3 (k+4) (k+5)$$
This dimension is strictly positive for all $k \geq 0$, consistent with $C_k^{(1)}$ being a genuine (non-negative) character combination at every level.

Regarding $\mathcal{I}^{(2)}(t)$, the regular part $\sum_k C_k^{(2)}\,t^k$ begins at $k=0$ with dimension

$$\dim C_0^{(2)} = 558$$

This admits the decomposition
$$C_0^{(2)} = \chi_{[2,0,0,0,2]} + \chi_{[0,1,0,1,0]} - \chi_{[1,0,0,0,1]} - \chi_{[0,0,0,0,0]} = \mathbf{405} + \mathbf{189} - \mathbf{35} - \mathbf{1} = 558$$

The dominant contribution $\chi_{[2,0,0,0,2]} = \mathbf{405}$ is the highest-weight representation appearing in the symmetric square of the current multiplet: indeed $\mathrm{Sym}^2(\mathbf{35}) = \mathbf{405} + \mathbf{189} + \mathbf{35} + \mathbf{1}$, so $C_0^{(2)}$ is precisely this symmetric square with both the adjoint $\mathbf{35}$ and the singlet $\mathbf{1}$ subtracted, accounting for gauge-invariance and trace redundancies in the operator spectrum.

\subsubsection{Single-branch moduli space: $FFM_{2}$}
\label{sec:FFM2single}

We observed earlier that the moduli space of the $FM_2$ theory consists of several
branches with non-trivial intersections. A natural question is whether, by flipping
some or all of the singlets in the IR spectrum, we can obtain a single-branch moduli
space. This strategy parallels the one employed for the $T[SU(N)]$ theory, where
flipping the adjoints yields the $FT_N$ theory, whose moduli space is a single
branch and, in fact, a complete intersection.

Since the chiral ring contains no operators of the form $\Pi\tilde{\Pi}B_{1,2}$ or
$\Pi\tilde{\Pi}B_{2,1}$ (with $\Pi$ and $\tilde{\Pi}$ denoting the two bifundamentals),
we begin by flipping the two singlets $B_{1,2}$ and $B_{2,1}$. One can verify via
$Z$-extremisation that the flipped singlets lie above the unitarity bound in the
resulting theory. With the same R-charge assignments as before, the unrefined index
up to order $x^3$ reads
\begin{equation*}
    \begin{split}
        \mathcal{I}_{FM_2}^{flip}=& 1 +x^{33/100} +x^{33/50} +6 x^{69/100} \\
        &+8 x^{41/50} +x^{19/20} +x^{99/100} +6 x^{51/50} +8 x^{23/20} +x^{32/25} +x^{33/25} \\
        &+6 x^{27/20} +20 x^{69/50} +8 x^{37/25} +40 x^{151/100} +x^{161/100} +36 x^{41/25} \\
        &+x^{33/20} +6 x^{42/25} +20 x^{171/100} +8x^{177/100} +8 x^{181/100} +40 x^{46/25} \\
        &-8 x^{187/100} +x^{19/10} +x^{97/50} +36 x^{197/100} +x^{99/50} -20 x^2 +6 x^{201/100} \\
        &+20 x^{51/25} +50 x^{207/100} +8 x^{21/10} -8 x^{213/100} +8 x^{107/50} +40 x^{217/100} \\
        &+112 x^{11/5} +x^{223/100} +x^{227/100} +36 x^{23/10} +x^{231/100} +136 x^{233/100} \\
        &+6 x^{117/50} -6 x^{59/25} +20 x^{237/100} +50 x^{12/5} +8 x^{243/100} +112 x^{123/50} \\
        &+8 x^{247/100} -8 x^{249/100} +40 x^{5/2} +112 x^{253/100} -39 x^{64/25} +36 x^{259/100} \\
        &+x^{13/5} -x^{131/50} +36 x^{263/100} +x^{66/25} +136 x^{133/50} +6 x^{267/100} \\
        &-134 x^{269/100} +20 x^{27/10} +8 x^{68/25} +50 x^{273/100} +113 x^{69/25} \\
        &+112 x^{279/100} +8 x^{14/5} -152 x^{141/50} +40 x^{283/100} +x^{57/20} +112 x^{143/50} \\
        &+241 x^{289/100} +36 x^{73/25} +x^{293/100} -65 x^{59/20} +36 x^{74/25} +x^{297/100} \\
        &+136 x^{299/100} +6 x^3 +O\left(x^{301/100}\right).
    \end{split}
\end{equation*}

We claim that the resulting moduli space consists of a single branch. As a check,
consider the unrefined Hilbert series of the putative single branch:
\begin{equation*}
    \begin{split}
        \mathcal{HS}(t_a,t_d,t_e,t_f)=& \sum_{n,k,l,m=0}^{\infty} [n,k,0;n]\,t_d^n t_a^k t_e^l t_f^m \\
        &\overset{\text{unr}}{=} 1 +x^{33/100} +x^{33/50} +6 x^{69/100} \\
        &+8 x^{41/50} +x^{19/20} +x^{99/100} +6 x^{51/50} +8x^{23/20} +x^{32/25} +x^{33/25} \\
        &+6 x^{27/20} +20 x^{69/50} +8 x^{37/25} +40x^{151/100} +x^{161/100} +36 x^{41/25} \\
        &+x^{33/20} +6 x^{42/25} +20 x^{171/100} +8 x^{177/100} +8 x^{181/100} +40 x^{46/25} \\
        &+x^{19/10} +x^{97/50} +36x^{197/100} +x^{99/50} +6 x^{201/100} +20 x^{51/25} \\
        &+50 x^{207/100} +8 x^{21/10} +8x^{107/50} +40 x^{217/100} +120 x^{11/5} +x^{223/100} \\
        &+x^{227/100} +36 x^{23/10} +x^{231/100} +155 x^{233/100} +6 x^{117/50} +20 x^{237/100} \\
        &+50 x^{12/5} +8 x^{243/100} +120 x^{123/50} +8 x^{247/100} +40 x^{5/2} +120 x^{253/100} \\
        &+x^{64/25} +36x^{259/100} +x^{13/5} +36 x^{263/100} +x^{66/25} +155 x^{133/50} \\
        &+6 x^{267/100} +20 x^{27/10} +8 x^{68/25} +50 x^{273/100} +113 x^{69/25} \\
        &+120 x^{279/100} +8 x^{14/5} +40 x^{283/100} +x^{57/20} +120 x^{143/50} +281 x^{289/100} \\
        &+36 x^{73/25} +x^{293/100} +36 x^{74/25} +x^{297/100} +155 x^{299/100} +6 x^3 \\
        &+O\left(x^{301/100}\right),
    \end{split}
\end{equation*}
with $t_a=x^{2-\tau}$, $t_d=x^{\Delta}$, $t_e=x^{2\Delta-2+\tau}$ and
$t_f=x^{2\Delta-\tau}$.

Comparing this with the index, we find complete agreement up to order $x^2$. The
first apparent mismatch occurs at order $x^{11/5}$, where the HS contains
$120\,x^{11/5}$ while the index contains $112\,x^{11/5}$. The index, however, also
includes a term $-8\,x^{187/100}$, coming from a $1/4$-BPS operator that
is not part of the chiral ring. Combining this with $x^{33/100}=x^{2\Delta-\tau}$
produces the $1/4$-BPS contribution $-8\,x^{11/5}$, which precisely accounts for
the discrepancy.

Overall, the index and the HS are in very good agreement, and we expect this
agreement to persist to all orders in $x$ once contributions from operators outside
the chiral ring are properly included, in line with the previous example. We are
thus led to claim that the flipped theory ($FFM_2$) has a single-branch moduli
space, with chiral-ring generators
$\{\Pi\tilde{\Pi},\,A_L,\,A_R,\,\mathcal{F}[B_{1,2}],\,\mathcal{F}[B_{2,1}]\}$.

The closed-form expression for the unrefined HS is
\be
     \mathcal{HS}(t_a,t_d,t_e,t_f)=\frac{t_a^2 \left(t_d+3\right) t_d^2+t_a \left(t_d^3-5 t_d^2-5 t_d+1\right)+3
   t_d+1}{\left(t_a-1\right){}^5 \left(t_d-1\right){}^5 \left(t_e-1\right)
   \left(t_f-1\right)}\,.
\ee
Switching off the fugacities associated with the flippers ($t_e=t_f=0$) recovers
the unrefined HS of the main branch of the unflipped theory in
\eqref{eq:hsfm2main}. This shows that the moduli space of $FFM_2$ is the product
of $\mathbb{C}^2$ with the branch of the unflipped theory generated by
$\{A_L,\,A_R,\,\Pi,\,\tilde{\Pi}\}$.

\subsection{\texorpdfstring{$FE_2$}{feusp4} }
Consider now the $FE_2$ theory, having the following UV completion:

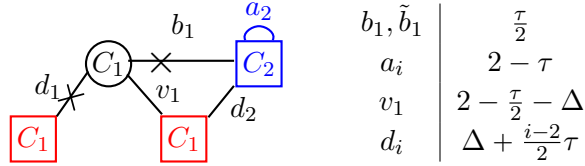
\be
\label{fe2}
\resizebox{0.75\hsize}{!}{
\begin{tikzpicture}[
    thick,
    gauge/.style={
        circle, draw=black, fill=white,
        minimum size=8mm, inner sep=1pt, font=\small
    },
    manifest/.style={
        rectangle, draw=blue!70!black, fill=blue!6,
        minimum size=8mm, inner sep=2pt,
        font=\small, text=blue!70!black
    },
    flavour/.style={
        rectangle, draw=red!70!black, fill=red!6,
        minimum size=7mm, inner sep=2pt,
        font=\small, text=red!70!black
    },
    arr/.style={-{Stealth[scale=0.8]}, shorten >=1pt, shorten <=1pt},
    barr/.style={draw=black!70}
]

\node[gauge]    (g1) at (0,0)      {$C_1$};
\node[manifest] (m1) at (2.5,0)    {$C_2$};
\node[flavour]   (x1) at (-1,-1.8)  {$C_1$};
\node[flavour]   (x2) at (1.5,-1.8) {$C_1$};

\draw[barr] (g1) -- (m1);
\node at (1.25, 0.42) {$b_1$};
\node at (1.25, 0) {\huge$\times$};

\draw[blue!70!black] (m1) to[out=65,in=115,looseness=3] (m1);
\node[blue!70!black] at (2.5, 0.95) {$a_2$};

\draw[barr] (x1) to (g1);
\node at (-1,-0.7) {$d_1$};
\node[rotate=55] at (-0.62,-0.95) {\huge$\times$};

\draw[barr] (g1) to (x2);
\node at (0.95,-0.75) {$v_1$};

\draw[barr] (x2) to(m1);
\node at (2.45,-0.85) {$d_2$};

\node at (0.75,-3.2) {$\mathcal{W} =
    v_1 b_1 d_2
    + b_1^2 a_2
    + \mathrm{Flip}[b_1^2]
    + \mathrm{Flip}[d_1^2]
    + \mathfrak{M}$};

\node[right] at (3.5,-0.5) {%
    \renewcommand{\arraystretch}{1.4}
    \begin{tabular}{>{\small}c >{\small}c}
        \toprule
        $b_1$  & $\tau/2$ \\
        $a_i$  & $2-\tau$ \\
        $v_1$  & $2-\tfrac{\tau}{2}-\Delta$ \\
        $d_i$  & $\Delta+\tfrac{i-2}{2}\tau$ \\
        \bottomrule
    \end{tabular}%
};

\end{tikzpicture}
}
\ee

\subsubsection*{Mirror self-duality and IR symmetry enhancement}

The $FE_2$ theory enjoys a self-duality under mirror symmetry, which
exchanges the manifest $C_2$ global symmetry with an emergent $C_2$
one.  The naive IR global symmetry is therefore
$USp(4)^{2}\times U(1)_{\tau}\times U(1)_{\Delta}$.  As we shall see,
at the strongly-coupled fixed point the symmetry actually enhances to
\[
Spin(10)\times U(1)_{\tau}\times U(1)_{\Delta},
\]
i.e.\ an extra $U(1)$ Cartan emerges along the RG flow -- a
phenomenon already observed for the corresponding $4d$
$\mathcal{N}=1$ theory in~\cite{Hwang:2020ddr}, and a general feature
of the $N=2$ theories considered in this paper.

\subsubsection*{Chiral-ring generators}

The IR spectrum contains a bifundamental $\Pi$ in the
$(\mathbf{4},\mathbf{4})$ of $USp(4)_L\times USp(4)_R$, two
traceless antisymmetric fields $A_L,A_R$ of $USp(4)_L$ and
$USp(4)_R$, and two singlets:
\[
B_{1,2}\ \text{(flipping $d_1^{\,2}$)},\qquad
B_{2,1}\equiv v_1^{\,2}.
\]

\subsubsection*{Superconformal R-charges and index}

$Z$-extremisation gives
\begin{equation}
\label{eq:FE2-Rcharges}
\tau = 1.215\ldots,\qquad \Delta = 0.874\ldots
\end{equation}
For the numerical index we fix $\Delta=85/100$ and
$\tau=12/10$, and turn on fugacities $\psi_{\tau},\chi_{\Delta}$ for
the two $U(1)$ symmetries.  Unrefining the non-abelian $USp(4)$
fugacities yields
\begin{equation}
\label{eq:FE2-unref-index}
\begin{aligned}
\mathcal{I}_{FE_2}
= \;&1 + 10\psi_{\tau}\,x^{4/5}
      + 16\chi_{\Delta}\,x^{17/20}
      + \psi_{\tau}\chi_{\Delta}^{-2}\,x^{11/10}
      + \chi_{\Delta}^{-2}\psi_{\tau}^{-1}\,x^{3/2}\\
   &+ 54\psi_{\tau}^{2}\,x^{8/5}
      + 144\chi_{\Delta}\psi_{\tau}\,x^{33/20}
      + 126\chi_{\Delta}^{2}\,x^{17/10}
      + 10\psi_{\tau}^{2}\chi_{\Delta}^{-2}\,x^{19/10}\\
   &- 47\,x^{2}
      - 16\chi_{\Delta}\psi_{\tau}^{-1}\,x^{41/20}
      + \psi_{\tau}^{2}\chi_{\Delta}^{-4}\,x^{11/5}\\
   &+ x^{12/5}\bigl(210\psi_{\tau}^{3}-\psi_{\tau}^{-2}\bigr)
      + 720\chi_{\Delta}\psi_{\tau}^{2}\,x^{49/20}
      + 1050\chi_{\Delta}^{2}\psi_{\tau}\,x^{5/2}\\
   &+ 672\chi_{\Delta}^{3}\,x^{51/20}
      + \chi_{\Delta}^{-4}\,x^{13/5}
      + 54\psi_{\tau}^{3}\chi_{\Delta}^{-2}\,x^{27/10}
      - 450\psi_{\tau}\,x^{14/5}\\
   &- 720\chi_{\Delta}\,x^{57/20}
      - 210\chi_{\Delta}^{2}\psi_{\tau}^{-1}\,x^{29/10}
      + x^{3}\bigl(10\psi_{\tau}^{3}\chi_{\Delta}^{-4}
                  +\chi_{\Delta}^{-4}\psi_{\tau}^{-2}\bigr)\\
   &- 46\psi_{\tau}\chi_{\Delta}^{-2}\,x^{31/10}
      - 16\chi_{\Delta}^{-1}\,x^{63/20}
      + 660\psi_{\tau}^{4}\,x^{16/5}
      + 2640\chi_{\Delta}\psi_{\tau}^{3}\,x^{13/4}\\
   &+ x^{33/10}\bigl(\psi_{\tau}^{3}\chi_{\Delta}^{-6}
                    +4950\chi_{\Delta}^{2}\psi_{\tau}^{2}\bigr)
      + 5280\chi_{\Delta}^{3}\psi_{\tau}\,x^{67/20}
      + 2772\chi_{\Delta}^{4}\,x^{17/5}\\
   &+ x^{7/2}\bigl(210\psi_{\tau}^{4}\chi_{\Delta}^{-2}
                  -\chi_{\Delta}^{-2}\psi_{\tau}^{-1}\bigr)
      - 2340\psi_{\tau}^{2}\,x^{18/5}
      - 5600\chi_{\Delta}\psi_{\tau}\,x^{73/20}\\
   &+ x^{37/10}\bigl(\psi_{\tau}\chi_{\Delta}^{-6}
                    -5430\chi_{\Delta}^{2}\bigr)
      - 1440\chi_{\Delta}^{3}\psi_{\tau}^{-1}\,x^{15/4}
      + 54\psi_{\tau}^{4}\chi_{\Delta}^{-4}\,x^{19/5}\\
   &- x^{39/10}\bigl(320\psi_{\tau}^{2}\chi_{\Delta}^{-2}
                    +\chi_{\Delta}^{-2}\psi_{\tau}^{-3}\bigr)
      + 16\psi_{\tau}\chi_{\Delta}^{-1}\,x^{79/20}\\
   &+ x^{4}\bigl(1782\psi_{\tau}^{5}+1033\bigr)
      + \mathcal{O}\!\bigl(x^{81/20}\bigr).
\end{aligned}
\end{equation}
The $-47\,x^{2}$ conserved-current contribution signals the
enhancement
\[
USp(4)\times USp(2)^{2}\times U(1)^{2}
\ \longrightarrow\ Spin(10)\times U(1)^{2},
\qquad \dim = 45+1+1 = 47.
\]
Reading the index directly, one identifies $A_L,A_R$ as
$\chi_{[1,0,0,0,0]_{so(10)}}\,x^{2-\tau}$; $\Pi$ as
$\chi_{[0,0,0,0,1]_{so(10)}}\,x^{\Delta}$; and the two singlets
$B_{2,1},B_{1,2}$ as $x^{4-2\Delta-\tau}$ and
$x^{2-2\Delta+\tau}$, respectively.

\subsubsection*{Hilbert series and branch structure}

From the index one extrapolates the all-order Hilbert series
\begin{equation}
\label{eq:fe4hs}
\begin{aligned}
\mathcal{HS}_{FE_2}
= \;& \sum_{k,l=0}^{\infty}\chi_{[k,0,0,0,l]_{so(10)}}
        \,x^{k(2-\tau)+l\Delta}
    + \sum_{k,m=0}^{\infty}\chi_{[k,0,0,0,0]_{so(10)}}
        \,x^{k(2-\tau)+m(4-2\Delta-\tau)}\\
  & + \sum_{n,m=0}^{\infty} x^{m(4-2\Delta-\tau)+n(2-2\Delta+\tau)}
    - \sum_{k=0}^{\infty}\chi_{[k,0,0,0,0]_{so(10)}}\,x^{k(2-\tau)}\\
  & - \sum_{m=0}^{\infty} x^{m(4-2\Delta-\tau)}.
\end{aligned}
\end{equation}
The moduli space therefore consists of three branches:
\begin{itemize}
\item a \emph{main branch}, generated by $\{A_L,A_R,\Pi\}$;
\item a \emph{second branch}, generated by $\{A_L,A_R,B_{2,1}\}$;
\item a \emph{third branch}, generated by $\{B_{1,2},B_{2,1}\}$.
\end{itemize}
The first two branches meet along the subvariety cut out by
$\{A_L,A_R\}$; the second and third branches meet along the locus
cut out by $\{B_{2,1}\}$.  Defining
$t_a=x^{2-\tau},\ t_d=x^{\Delta},\ t_c=x^{4-2\Delta-\tau},\
t_e=x^{2-2\Delta+\tau}$, the unrefined Hilbert series of the second
and third branches are
\begin{equation}
\label{eq:FE2-HS-23}
\mathcal{HS}^{(2)} = \frac{1+t_a}{(1-t_a)^{9}\,(1-t_c)},
\qquad
\mathcal{HS}^{(3)} = \frac{1}{(1-t_c)(1-t_e)}\, ,
\end{equation}
so the third branch is simply $\mathbb{C}^{2}$, while the second is
$\mathbb{C}$ times the polynomial ring generated by $\{A_L,A_R\}$
modulo the single relation $\mathrm{tr}_L A_L^{\,2}=\mathrm{tr}_R A_R^{\,2}$.

\subsubsection*{The main branch}

For the main branch, generated by $\{A_L,A_R,\Pi\}$, the unrefined
Hilbert series takes the closed form
\begin{equation}
\label{eq:unrhsfe4}
\mathcal{HS}^{\mathrm{main}}_{FE_2}(t_a,t_d)
= \frac{\sum_{j,k}a_{j,k}\,t_a^{\,j}t_d^{\,k}}
       {(1-t_a)^{9}\,(1-t_d)^{11}}\, ,
\end{equation}
with numerator coefficient matrix
\begin{equation}
\label{eq:FE2-aij}
\{a_{j,k}\}=
\begin{pmatrix}
 1 &   5 &   5 &   1 &   0 & 0\\
 1 & -11 & -29 &  -9 &   0 & 0\\
 0 &   0 &  36 &  36 &   0 & 0\\
 0 &   0 &  -9 & -29 & -11 & 1\\
 0 &   0 &   1 &   5 &   5 & 1
\end{pmatrix}.
\end{equation}
The matrix is palindromic, $a_{j,k}=a_{4-j,\,5-k}$, and satisfies
$\sum_{j,k}a_{j,k}=0$.  The diagonal specialisation $t_a=t_d=t$
factorises as
\begin{equation}
\label{eq:FE2-num-diag}
N(t,t) = (1-t)^{4}\,(1+t)\bigl(1+9t+19t^{2}+9t^{3}+t^{4}\bigr),
\end{equation}
while the two single-variable specialisations give
\begin{equation}
\label{eq:FE2-sum-rules}
N(1,t) = 2\,(1-t)^{4}(1+t),
\qquad
N(t,1) = 12\,(1-t)^{4}.
\end{equation}
These translate into the row and column sum rules
\begin{equation}
\label{eq:FE2-row-col-sums}
\sum_{j}a_{j,k}
 = 2(-1)^{k}\biggl[\binom{4}{k}-\binom{4}{k-1}\biggr],
\qquad
\sum_{k}a_{j,k}
 = 12(-1)^{j}\binom{4}{j},
\end{equation}
with $\binom{4}{-1}=\binom{4}{5}=0$.

\subsubsection*{Generators and relations of the main branch}

The structure of the chiral ring on the main branch is encoded in
the plethystic logarithm of~\eqref{eq:unrhsfe4}:
\begin{equation}
\label{eq:FE2-PL-main}
\begin{aligned}
\mathrm{PL}\!\left[\mathcal{HS}^{\mathrm{main}}_{FE_2}\right](t_a,t_d)
= \;& \chi_{[1,0,0,0,0]_{so(10)}}\,t_a
    + \chi_{[0,0,0,0,1]_{so(10)}}\,t_d
    - t_a^{\,2}\\
  & - \chi_{[0,0,0,1,0]_{so(10)}}\,t_a t_d
    - \chi_{[1,0,0,0,0]_{so(10)}}\,t_d^{\,2}\\
  & + \mathcal{O}\!\bigl(t_a t_d\,(t_a+t_d)\bigr).
\end{aligned}
\end{equation}
At order $t_a$ one reads off $\{A_L,A_R\}$, jointly transforming in
the vector of $\mathfrak{so}(10)$; at order $t_d$ the bifundamental
$\Pi$, in the spinor.  These generators satisfy the quadratic
relations
\begin{equation}
\label{eq:FE2-main-relations}
\begin{aligned}
\Omega_{ab}\Omega_{cd}\,A_L^{\,ac}A_L^{\,bd}
 = \Omega_{AB}\Omega_{CD}\,A_R^{\,AC}A_R^{\,BD},\qquad &
  \bigl(1\ \text{relation at order } t_a^{\,2}\bigr),\\[2pt]
\Omega_{ac}\,A_L^{\,ab}\Pi^{\,cA}
 = \Omega_{BC}\,A_R^{\,AB}\Pi^{\,Cb},\qquad &
  \bigl(16\ \text{relations at order } t_a t_d\bigr),\\[2pt]
\Omega_{ab}\,\Pi^{\,aA}\Pi^{\,bB}
 = 0
 = \Omega_{AB}\,\Pi^{\,aA}\Pi^{\,bB},\qquad &
  \bigl(10\ \text{relations at order } t_d^{\,2}\bigr),
\end{aligned}
\end{equation}
with $a,b,\dots$ and $A,B,\dots$ denoting $USp(4)_L$ and $USp(4)_R$
indices respectively.  Since~\eqref{eq:FE2-PL-main} does not
terminate, the relations are not independent and the main branch is
not a complete intersection.

\subsubsection{Comparison with the 4d index}
\label{4d-ind-comp}

It is instructive to compare the $3d$ index of $FE_2$ with the
partition function of its $4d$ uplift on $\mathbf{S}^{3}\times \mathbf{S}^{1}$.  Since
the $4d$ theory reduces to the $3d$ one in the appropriate limit, we
expect a close agreement; in fact, as we show below, the two theories share the \emph{same} Hilbert series.  Discrepancies in the
full index -- coming from $\tfrac{1}{4}$-BPS states and other
non-chiral-ring contributions -- are expected, since the single-letter
indices of the two theories are distinct.

\paragraph{$4d$ superconformal R-charges.}
Denote by $U(1)_{\Delta}$ and $U(1)_{\tau}$ the global abelian
symmetries of the $4d$ theory, and parametrise the superconformal
R-charge as
\begin{equation}
\label{eq:FE2-4d-Rcharge}
R = R_{0} + \Delta\,q_{\Delta} + \tau\,q_{\tau},
\end{equation}
with $R_{0}$ the trial R-charge and $q_{\Delta},q_{\tau}$ the abelian
charges of the matter fields.  The mixing parameters $\Delta,\tau$
are fixed by $a$-maximisation~\cite{Intriligator_2003}; choosing the
trial R-charges as in the $3d$ quiver~\eqref{fe2}, we find
\begin{equation}
\label{eq:FE2-4d-Rcharges-num}
\Delta = 0.842786\ldots,\qquad \tau = 1.12026\ldots
\end{equation}
These values differ slightly from their $3d$ counterparts, as
expected: $a$-maximisation and $Z$-extremisation are genuinely
distinct procedures, and their near-agreement provides a
non-trivial consistency check.

\paragraph{The $4d$ unrefined index.}
To facilitate the comparison we fix the same R-charges
$\Delta=85/100$ and $\tau=12/10$ as in the $3d$ computation.  Up to
order $t^{4}$, the $4d$ unrefined index reads
\begin{equation}
\label{eq:FE2-4d-index}
\begin{aligned}
\mathcal{I}^{4d}_{FE_2}
= \;&1 + 10\,t^{4/5} + 16\,t^{17/20} + t^{11/10} + t^{3/2}
    + 54\,t^{8/5} + 144\,t^{33/20} + 126\,t^{17/10}\\
   &+ 10\,t^{9/5}(x+x^{-1})
    + 16\,t^{37/20}(x+x^{-1})
    + 10\,t^{19/10} - 47\,t^{2} - 16\,t^{41/20}\\
   &+ t^{21/10}(x+x^{-1}) + t^{11/5} + 209\,t^{12/5}
    + 720\,t^{49/20} + t^{5/2}\bigl(x+x^{-1}+1050\bigr)\\
   &+ 672\,t^{51/20} + t^{13/5}\bigl(99x+99x^{-1}+1\bigr)
    + 304\,t^{53/20}(x+x^{-1})\\
   &+ t^{27/10}\bigl(246x+246x^{-1}+54\bigr)
    + t^{14/5}\bigl(10x^{2}+10x^{-2}-450\bigr)\\
   &+ t^{57/20}\bigl(16x^{2}+16x^{-2}-720\bigr)
    + t^{29/10}\bigl(20x+20x^{-1}-210\bigr)
    + 16\,t^{59/20}(x+x^{-1})\\
   &+ t^{3}\bigl(-48x-48x^{-1}+11\bigr)
    - 16\,t^{61/20}(x+x^{-1})
    + t^{31/10}\bigl(x^{2}+x^{-2}-46\bigr)\\
   &- 16\,t^{63/20} + t^{16/5}\bigl(x+x^{-1}+660\bigr)
    + 2640\,t^{13/4}
    + t^{33/10}\bigl(10x+10x^{-1}+4951\bigr)\\
   &+ t^{67/20}\bigl(16x+16x^{-1}+5280\bigr)
    + t^{17/5}\bigl(529x+529x^{-1}+2772\bigr)\\
   &+ t^{69/20}\bigl(2144x+2144x^{-1}\bigr)
    + t^{7/2}\bigl(x^{2}+x^{-2}+3255x+3255x^{-1}+209\bigr)\\
   &+ 1872\,t^{71/20}(x+x^{-1})
    + t^{18/5}\bigl(154x^{2}+154x^{-2}+2x+2x^{-1}-2240\bigr)\\
   &+ t^{73/20}\bigl(464x^{2}+464x^{-2}-5280\bigr)\\
   &+ t^{37/10}\bigl(382x^{2}+382x^{-2}+153x+153x^{-1}-5173\bigr)\\
   &+ t^{15/4}\bigl(144x+144x^{-1}-1440\bigr)\\
   &+ t^{19/5}\bigl(10x^{3}+10x^{-3}-930x-930x^{-1}+54\bigr)\\
   &+ t^{77/20}\bigl(16x^{3}+16x^{-3}-1632x-1632x^{-1}\bigr)\\
   &+ t^{39/10}\bigl(30x^{2}+30x^{-2}-465x-465x^{-1}-301\bigr)\\
   &+ t^{79/20}\bigl(32x^{2}+32x^{-2}+48\bigr)\\
   &+ t^{4}\bigl(-48x^{2}-48x^{-2}+21x+21x^{-1}+2814\bigr)
    + \mathcal{O}\!\bigl(t^{81/20}\bigr),
\end{aligned}
\end{equation}
where we have set $tx=p$ and $t/x=q$, with $q,p$ the fugacities for
the $\mathbf{S}^{3}$ isometry group $Spin(4)\cong SU(2)_{q}\times SU(2)_{p}$.

\paragraph{Comparison with the $3d$ index.}
Three observations emerge when~\eqref{eq:FE2-4d-index} is set
alongside the $3d$ index~\eqref{eq:FE2-unref-index} (with the $3d$
flavour fugacities $\psi_{\tau},\chi_{\Delta}$ switched off).

First, the $x$-independent terms in~\eqref{eq:FE2-4d-index} coincide
exactly with the corresponding $t\to x$ terms of the $3d$ index: a
$4d$ coefficient $a_{r}\,t^{r}$ matches the $3d$ coefficient
$a_{r}\,x^{r}$ order by order.  These are precisely the contributions
from the lowest components of the gauge-invariant operators
parametrising the moduli space, i.e.\ the terms counted by the
Hilbert series.  Expanding our conjectured $3d$ Hilbert series for
small fugacities reproduces every such term in the $4d$ index with
identical coefficients, through every order we have checked -- strong
evidence that the two theories share the same Hilbert series.

Second, the $4d$ index contains infinitely many additional terms of
the form $t^{r}(x^{n}+x^{-n})$ with no $3d$ analogue.  Following
\cite{Gadde:2020yah}, these arise either from the action of the
spacetime derivatives $\partial_{+\dot{+}}$, $\partial_{+\dot{-}}$
on gauge-invariant operators, or from products of such operators
with the gauginos $\bar{\lambda}_{\dot{+}}$, $\bar{\lambda}_{\dot{-}}$.
Neither type of contribution parametrises the moduli space, and
neither is counted by the Hilbert series.

Third, a small number of terms appear in both indices but with
different coefficients.  These are accounted for by quadratic
combinations of the extra $4d$ letters just mentioned.  A
representative example: the $3d$ term $-2340\,x^{18/5}$ corresponds
in $4d$ to $-2240\,t^{18/5}$, differing by exactly $100$.  The $4d$
index also contains the earlier entry $10\,t^{9/5}(x+x^{-1})$ -- a
contribution from short multiplets that are not $\tfrac{1}{2}$-BPS,
hence absent from the Hilbert series.  Multiplying the $x$ term with the $x^{-1}$ one produces the term
$100\,t^{18/5}$ of schematic form
$\partial_{+\dot{+}}\partial_{+\dot{-}}(\text{op})$ or
$\bar{\lambda}_{\dot{+}}\bar{\lambda}_{\dot{-}}(\text{op})$, which
does not correspond to an independent chiral-ring element and is
therefore not counted by the HS.  Subtracting this contribution from the
$4d$ coefficient recovers the $3d$ one exactly, and we have verified
that the same mechanism accounts for every discrepancy up to the
order reported above.

In summary, the $3d$ and $4d$ $FE_2$ theories share the same Hilbert
series, providing strong evidence that their chiral rings are
isomorphic even though the two indices -- and hence the full
operator contents -- differ.

\subsubsection{An exceptional model: $\widetilde{FE}_2$}

Flipping the singlet $B_{2,1} \equiv v_1^2$ yields the theory that we denote by $\widetilde{FE}_2$.
This theory has an $SU(2)\times SU(6) \times U(1)$ UV global symmetry, with two fields in the $(\mathbf{2},\mathbf{6})$ and $(\mathbf{1},\mathbf{15})$ representations carrying the same $U(1)$ charge, together with the singlet $B_{1,2}$ in the $(\mathbf{1},\mathbf{1})$ representation.
This result can also be verified by $Z$-extremisation, which gives $\Delta = 0.834$ and $\tau = 2 - \Delta = 1.166$.
We claim that this theory flows to an IR fixed point with enhanced symmetry $E_6 \times U(1)$, as can be checked from the index.
The index for this flipped theory reads (with $\tau = 11/10$):
\be
 \mathcal{I}_{\widetilde{FE}_2} = 1+\mathbf{27} \chi  x^{9/10}+\frac{x^{13/10}}{\chi ^3}+\mathbf{\overline{351}'} \chi ^2 x^{9/5}-(\mathbf{78}+1) x^2+\frac{x^{13/5}}{\chi ^6}+\mathbf{3003} \chi ^3 x^{27/10}+O\left(x^{29/10}\right)
\ee
where we turned on a fugacity $\chi$ for the single $U(1)$ global symmetry.
The bold numbers denote characters of the $E_6$ global symmetry. At order $x^2$, for example, the adjoint representation $\mathbf{78}$ of $E_6$ is contributed by the conserved currents. We also find contributions from $1/2$-BPS operators $\chi_{[k,0,0,0,0,0]_{E_6}}x^{k \Delta}$ for $k\ge1$.
The moduli space of the theory consists of two branches with trivial intersection: a branch $\mathbb{C}$ freely generated by $B_{1,2}$ (appearing at order $x^{4-3\Delta}$ in the index), and a main branch generated by $27$ operators satisfying $27$ quadratic relations. Geometrically, the main branch is the cone over the rank-$2$ irreducible Hermitian symmetric space of compact type $E_6 /(SO(10) \times U(1))$.
We also write down the HS for the main branch, which can be extrapolated from the index. It reads:
\be 
    \begin{aligned}
        \mathcal{HS}_{\widetilde{FE}_2}(t \equiv x^{\Delta}) = & \;\sum_{k=0}^{\infty} \chi_{[k,0,0,0,0,0]_{E_6}} t^k \\
        \overset{unr}{=}& \; \frac{t^5+10 t^4+28 t^3+28 t^2+10 t+1}{(1-t)^{17}}
    \end{aligned}
\ee
This is exactly the HS of the cone over the complex Cayley space, studied for example in \cite{Razamat:2016gzx}.

Up to order $x^{10}$, the unrefined index (with respect to the global $E_6$ symmetry) reads:

\begin{equation}
\begin{split}
 \mathcal{I}_{\widetilde{FE}_2}(x; \chi) = \;&1
+27\chi x^{9/10}+\chi^{-3} x^{13/10}+351\chi^2 x^{9/5}-79 x^2
+\chi^{-6} x^{13/5}+3003\chi^3 x^{27/10} \\
&-1728\chi x^{29/10}+19305\chi^4 x^{18/5}-18954\chi^2 x^{19/5}+\chi^{-9} x^{39/10}
+2923 x^4 \\
&+378\chi^{-2} x^{21/5}+100386\chi^5 x^{9/2}-140608\chi^3 x^{47/10}+49842\chi x^{49/10}+\chi^{-12} x^{26/5}\\
&-729\chi^{-3} x^{53/10}+442442\chi^6 x^{27/5}-799227\chi^4 x^{28/5}+449658\chi^2 x^{29/5}-66834 x^6 \\
&-7317\chi^{-2} x^{31/5}+1706562\chi^7 x^{63/10}+351\chi^{-4} x^{32/5}
+\left(\chi^{-15} - 3729024\chi^5\right) x^{13/2} \\
&+2835508\chi^3 x^{67/10}-835434\chi x^{69/10}+42120\chi^{-1} x^{71/10}+5895396\chi^8 x^{36/5} \\
&+11490\chi^{-3} x^{73/10}-14916616\chi^6 x^{37/5}+729\chi^{-5} x^{15/2}+14009463\chi^4 \\
&+ x^{38/5} \left(\chi^{-18} - 5964867\chi^2\right) + x^{39/5}
+1046513 x^8+18559580\chi^9 x^{81/10} \\
&
-2754\chi^{-2} x^{41/5}-52700544\chi^7 x^{83/10}-7371\chi^{-4} x^{42/5}+57741255\chi^5 x^{17/2} \\
&
-1379\chi^{-6} x^{43/5}-30976062\chi^3 x^{87/10}
+8544987\chi x^{89/10}
+53965548\chi^{10} x^9 \\
&
+\left(\chi^{-21} - 1062126\chi^{-1}\right) x^{91/10}-168018786\chi^8 x^{46/5}-12873\chi^{-3} x^{93/10} \\
&+206545352\chi^6 x^{47/5}+756\chi^{-5} x^{19/2}-129114594\chi^4 x^{48/5}+702\chi^{-7} x^{97/10}\\
&+45121482\chi^2 x^{49/5}+146477916\chi^{11} x^{99/10}-9449098 x^{10}
+O\!\left(x^{101/10}\right)
\end{split}
\end{equation}

\subsubsection*{Index resummation for $\widetilde{FE}_2$}

Setting $t=\chi\,x^{9/10}$, the unrefined index may be written as
\begin{equation}
\mathcal{I}_{\widetilde{FE}_2}^{unr}(x,t) = \mathcal{I}^{(0)}(t) + x^2 \mathcal{I}^{(1)}(t) + x^4 \mathcal{I}^{(2)}(t) + \mathcal{O}(x^6)
\end{equation}
where the functions $I^{(k)}(t)$ can be resummed into closed rational forms:
\begin{align}
\mathcal{I}^{(0)}(t) &= \frac{t^5+10 t^4+28 t^3+28 t^2+10 t+1}{(1-t)^{17}} = \mathcal{HS}_{\widetilde{FE}_2}(t) \\[1ex]
\mathcal{I}^{(1)}(t) &= \frac{-79 t^6-306 t^5+63 t^4+644 t^3+63 t^2-306 t-79}{(1-t)^{18}}\\[1ex]
\mathcal{I}^{(2)}(t) &= \frac{-27 t^{10}-2751 t^9-1269 t^8+9477 t^7+886 t^6-8847 t^5}{(1-t)^{18} t^3}\\
&\phantom{{}={}}\frac{+288 t^4+2107 t^3+153 t^2-18 t+1}{(1-t)^{18} t^3}
\end{align}
In terms of the global $E_6$ symmetry, the functions $I^{(k)}(t)$ can be written as exact sums over irreducible representations, denoted by their Dynkin labels $\chi_{[a_1,a_2,a_3,a_4,a_5,a_6]}$. In this theory, the chiral ring $I^{(0)}(t)$ is generated by the representation $\mathbf{27}$ ($\chi_{[1,0,0,0,0,0]}$), while the current multiplet transforms in the adjoint $\mathbf{78}$ ($\chi_{[0,0,0,0,0,1]}$). The exact character combinations are

\begin{align}
\mathcal{I}^{(0)}(t) &= \sum_{k=0}^{\infty}\chi_{[k,0,0,0,0,0]}t^k \\[1ex]
\mathcal{I}^{(1)}(t) &= - \sum_{k=0}^{\infty} (\chi_{[k,0,0,0,0,1]}-\chi_{[k-2,1,0,0,0,0]})t^k \equiv - \sum_{k=0}^{\infty} C_k^{(1)}t^k   \\[1ex] 
\mathcal{I}^{(2)}(t) &= \frac{\chi_{[0,0,0,0,0,0]}}{t^3} + \sum_{k=0}^{\infty} C_k^{(2)}t^k 
\end{align}

The formula for $\mathcal{I}^{(1)}(t)$ incorporates the $U(1)_\tau$ singlet constraint through the Weyl character formula. When $k<2$, the subtracted representation $\chi_{[k-2,1,0,0,0,0]}$ has a negative or vanishing first Dynkin label, and its fate is determined by the Weyl alternation formula applied to $E_6$.

When $k=1$: the subtracted term is $\chi_{[-1,1,0,0,0,0]}$. The shifted weight is $[-1,1,0,0,0,0] + \rho = [0,2,1,1,1,1]$, whose first Dynkin label vanishes, placing it on the wall of the Weyl chamber associated with the simple root $\alpha_1$. The character vanishes identically, leaving
\begin{equation}
C_1^{(1)} = \chi_{[1,0,0,0,0,1]}
\end{equation}
of dimension $\mathbf{1728}$.

When $k=0$: the subtracted term is $\chi_{[-2,1,0,0,0,0]}$. The shifted weight is $[-2,1,0,0,0,0]+\rho = [-1,2,1,1,1,1]$, which has a negative first label. A single reflection $s_1$ across the first simple root acts by
\begin{equation}
[-1,2,1,1,1,1] \xrightarrow{s_1} [1,1,1,1,1,1] = \rho\,,
\end{equation}
the Weyl vector, which is the shifted weight of the trivial representation $\chi_{[0,0,0,0,0,0]}$. Since $s_1$ introduces a sign of $-1$, the Weyl formula gives $\chi_{[-2,1,0,0,0,0]} = -\chi_{[0,0,0,0,0,0]} = -\mathbf{1}$. Therefore:
\begin{equation}
C_0^{(1)} = \chi_{[0,0,0,0,0,1]} - \bigl(-\chi_{[0,0,0,0,0,0]}\bigr) = \mathbf{78} + \mathbf{1} = 79
\end{equation}
which reproduces the exact coefficient $79$ appearing in the numerator of $\mathcal{I}^{(1)}(t)$ at order $t^0$, consistent with the dimension of the conserved current multiplet of the $E_6$ theory together with a singlet.

In the unrefined limit, the dimensions of the combination $C_k^{(1)}$ follow from the rational generating function. Expanding $-\mathcal{I}^{(1)}(t) = \frac{79+306\,t-63\,t^2-644\,t^3-63\,t^4+306\,t^5+79\,t^6}{(1-t)^{18}}$, the first few values are
\begin{equation}
\dim C_k^{(1)} \Big|_{k=0,1,2,\ldots} = 79,\; 1728,\; 18954,\; \ldots
\end{equation}
This dimension is strictly positive for all $k \geq 0$, consistent with $C_k^{(1)}$ being a genuine character combination at every level.

Regarding $\mathcal{I}^{(2)}(t)$, the singular part contains only the singlet $\chi_{[0,0,0,0,0,0]}/t^3$, with the regular part $\sum_k C_k^{(2)}\,t^k$ beginning at $k=0$. Extracting the $t^0$ coefficient from the rational form yields
\begin{equation}
\dim C_0^{(2)} = 2923\,.
\end{equation}
This admits the decomposition
\begin{equation}
C_0^{(2)} = \chi_{[0,0,0,0,0,2]} + \chi_{[0,1,0,0,0,0]} - 2\,\chi_{[0,0,0,0,0,1]} - \chi_{[0,0,0,0,0,0]} = \mathbf{2430} + \mathbf{650} - 2 \times \mathbf{78} - \mathbf{1} = 2923\,.
\end{equation}
The dominant contribution $\chi_{[0,0,0,0,0,2]} = \mathbf{2430}$ is the highest-weight representation appearing in the symmetric square of the current multiplet: indeed $\mathrm{Sym}^2(\mathbf{78}) = \mathbf{2430} + \mathbf{650} + \mathbf{1}$, so $C_0^{(2)}$ is precisely this symmetric square with two copies of the adjoint $\mathbf{78}$ and two singlets subtracted, accounting for gauge-invariance and trace redundancies in the operator spectrum.

\subsubsection{Single branch moduli space: $FFE_{2}$}
As for the $FM_2$ theory, we would like to flip some or all of the singlets to obtain a theory whose moduli space has a single branch. As above, we flip both $B_{1,2}$ and $B_{2,1}$; we denote the doubly flipped theory by $FFE_{2}$, in parallel with the $FFM_{2}$ notation introduced in Section~\ref{sec:FFM2single}. Note that $FFE_{2}\neq\widetilde{FE}_{2}$: the latter flips only $B_{2,1}$ and has a two-branch moduli space, whereas the former flips both $B_{1,2}$ and $B_{2,1}$. The resulting unrefined index is given below, with $\Delta=97/100$ and $\tau=131/100$ chosen so that all operators lie above the unitarity bound; this choice should be checked through $Z$-extremisation:
\begin{equation*}
    \begin{split}
    \mathcal{I}^{flip}_{FE_2}=& 1 +x^{63/100} +10 x^{69/100} +16 x^{97/100} \\
    &+x^{5/4} +x^{63/50} +10 x^{33/25} +54x^{69/50} +16 x^{8/5} +144 x^{83/50} -16 x^{43/25} \\
    &+x^{47/25} +x^{189/100} +136 x^{97/50} +10 x^{39/20} -47 x^2 +54 x^{201/100} \\
    &-10 x^{103/50} +210 x^{207/100} +16x^{111/50} +16 x^{223/100} -16 x^{57/25} \\
    &+144 x^{229/100} -16 x^{117/50} +704 x^{47/20} -144 x^{241/100} +x^{5/2} +x^{251/100} \\
    &+x^{63/25} +136 x^{257/100} +10 x^{129/50} -x^{131/50} +1058 x^{263/100} +54 x^{66/25} \\
    &-586 x^{269/100} +210 x^{27/10} -54 x^{11/4} +660 x^{69/25} +16 x^{57/20} +16 x^{143/50} \\
    &+800 x^{291/100} +144 x^{73/25} -752 x^{297/100} +704 x^{149/50} +O\left(x^{3}\right)
    \end{split}
\end{equation*}
Assuming that the moduli space has a single branch, we can write the following HS for the full moduli space:
\begin{equation*}
    \begin{split}
        \mathcal{HS}(t_a,t_d,t_e,t_f)&= \sum_{n,k,l,m=0}^{\infty}[k,0,0,0,n]t_a^kt_d^nt_e^lt_f^m \\
        &\overset{unr}{=} 1 +x^{63/100} +10 x^{69/100} +16 x^{97/100} \\
        &+x^{5/4} +x^{63/50} +10 x^{33/25} +54x^{69/50} +16 x^{8/5} +144 x^{83/50} +x^{47/25} \\
        &+x^{189/100} +136 x^{97/50} +10 x^{39/20} +54 x^{201/100} +210 x^{207/100} \\
        &+16 x^{111/50} +16 x^{223/100} +144 x^{229/100} +720 x^{47/20} +x^{5/2} +x^{251/100} \\
        &+x^{63/25} +136 x^{257/100} +10x^{129/50} +1104 x^{263/100} +54 x^{66/25} +210 x^{27/10} \\
        &+660 x^{69/25} +16x^{57/20} +16 x^{143/50} +816 x^{291/100} +144 x^{73/25} \\
        &+720 x^{149/50} +O\left(x^{3}\right)
    \end{split}
\end{equation*}
where $t_a=x^{2-\tau},\;t_d={x^\Delta},\;t_e=x^{2\Delta-2+\tau},\;t_f=x^{2\Delta-\tau}$.
Again, we find nearly perfect agreement among the contributions from $1/2$-BPS operators, up to contributions from $1/4$-BPS operators that are not part of the chiral ring.

\section{Outlook and future directions}
The improved bifundamentals studied in this paper form a rich class of 3d $\mathcal{N}=2$ SCFTs whose moduli-space structure is far from fully understood. By analysing the superconformal index, we have inferred the operator spectrum, quantum relations, and Euler form of the Hilbert series, and have identified enhanced global symmetries for the $N=2$ cases. These results suggest several natural directions for further investigation.

\begin{itemize}
    \item It would be valuable to derive further constraints on the HS of the theories studied in this paper, potentially finding a closed-form expression for the coefficients in the numerator of the HS in its Euler form.
    
    \item Extending this analysis to more intricate objects, such as the asymmetric $\mathcal{S}$-wall \cite{Comi:2022aqo} and asymmetric improved bifundamentals, would be an interesting avenue for future research. In particular, it is natural to expect that a single branch moduli space arises also in this more general setting.
    
    \item Studying the HS for the two sides of the \emph{braid dualities} and \emph{basic moves} involving the various improved bifundamentals \cite{Benvenuti:2024mpn} might reveal interesting phenomena.
    
    \item We observed that most, though not all, of the moduli space branches studied exhibit Calabi-Yau properties, i.e., their HS in Euler form has a palindromic numerator. Understanding the connection between the geometry of the moduli space and the quantum relations satisfied by the chiral ring operators would be an intriguing problem to investigate further. Attempts in this direction have been performed in \cite{Eager:2018czl}.

    \item Since $FE_N$ admits a 4d $\mathcal{N}=1$ incarnation, a natural direction is to extend the present analysis to the 4d setting and determine whether the full Hilbert series and the pattern of quantum relations persist upon uplift. The partial comparison carried out in Section~\ref{4d-ind-comp} suggests this is the case, but a systematic study remains to be done.

    \item Finally, it would be interesting to study the 2d $\mathcal{N}=(2,2)$ theories obtained by compactifying the 3d improved bifundamentals on a circle. The resulting sigma models would have a cone over a coset space as their target, and computing their elliptic genus, as was done for cones over complex Grassmannians in \cite{Eager:2019zrc}, could shed light on the interplay between the chiral ring structure and the geometry of the target space.
\end{itemize}

We leave these directions for future study, hoping that they will lead to a more comprehensive understanding of the structure and properties of these theories.
\section*{Acknowledgement}
We are indebted to Anant Shri, Riccardo Comi, Sara Pasquetti, and Simone Rota for collaboration on related topics and useful discussions. We also thank Richard Eager for useful correspondence. SB is partially supported by MUR-PRIN grant No.~2022NY2MXY (Finanziato dall'Unione europea--Next Generation EU, Missione 4 Componente 1, CUP H53D23001080006).

\section*{AI use disclosure}
Claude (Anthropic) was used to proofread the manuscript and improve the grammar and syntax of the text. All content was reviewed and verified by the authors, who take full responsibility for the accuracy and integrity of the manuscript.

\pagebreak

\appendix 

\section{Hilbert series}
\label{appHS}
The moduli spaces we aim to study are complex cones $\mathcal{M} \subset \mathbb{C}^{n}$ over a projective base $\mathcal{B} \subset \mathbb{P}^{n-1}_{\mathbb{C}}$, defined as intersections of hypersurfaces. The Hilbert series (HS) encodes numerical properties of such a projective algebraic variety. It is not an intrinsic invariant: it depends on the specific projective embedding, equivalently on the choice of very ample line bundle \cite{Harris:1992}. The HS is built from the \textit{Hilbert function} of $\mathcal{B}$,
\[
h_{\mathcal{B}}: \mathbb{N} \to \mathbb{N},
\]
which records, in each degree $m$, the dimension of the space of degree‑$m$ polynomial functions on $\mathcal{B}$. Equivalently, $h_{\mathcal{B}}(m)$ is the codimension inside the space of all homogeneous polynomials of degree $m$ in $X_{1},\dots,X_{n}$ of the subspace of those vanishing on $\mathcal{B}$:
\[
h_{\mathcal{B}}(m) = \dim_{\mathbb{C}} S(\mathcal{B})_m,
\]
where $S(\mathcal{B}) = \mathbb{C}[X_{1}, \dots, X_{n}]/I$ is the homogeneous coordinate ring, $I$ is the homogeneous ideal generated by the polynomials defining $\mathcal{B}$, and the subscript $m$ denotes the $m$-th graded piece. The number of linearly independent hypersurfaces of degree $m$ containing $\mathcal{B}$ is complementary information, namely $\dim_{\mathbb{C}} I_{m} = \binom{n+m-1}{m} - h_{\mathcal{B}}(m)$.

The HS is the generating function for the dimensions of the graded pieces of $S(\mathcal{B})$:
\[
\mathcal{HS}(t; \mathcal{B}) = \sum_{m=0}^{\infty} \dim_{\mathbb{C}} S(\mathcal{B})_{m}\, t^{m} = \sum_{m=0}^{\infty} h_{\mathcal{B}}(m)\, t^{m}.
\]
By the Hilbert–Serre Theorem~\cite{Stanley:1978}, there exists a polynomial $P \in \mathbb{Z}[t]$ such that
\[
\mathcal{HS}(t; \mathcal{B}) = \frac{P(t)}{\prod_{i=1}^{n}\bigl(1 - t^{p_{i}}\bigr)},
\qquad p_{i} = \deg X_{i},
\]
sometimes called the \textit{Euler form}. Grouping the generators by their degree, with $q_{p}$ denoting the number of variables of degree $p$, this becomes
\[
\mathcal{HS}(t; \mathcal{B}) = \frac{P(t)}{\prod_{p}\bigl(1 - t^{p}\bigr)^{q_{p}}}.
\]
Since the order of the pole of $\mathcal{HS}$ at $t=1$ equals the Krull dimension of $S(\mathcal{B})$, which is $\dim_{\mathbb{C}}\mathcal{M} = \dim_{\mathbb{C}}\mathcal{B} + 1$, one may always reduce to the standard form
\[
\mathcal{HS}(t; \mathcal{B}) = \frac{h_{0} + h_{1} t + \dots + h_{d} t^{d}}{(1 - t)^{\,\dim_{\mathbb{C}}\mathcal{M}}},
\]
where $\dim_{\mathbb{C}}\mathcal{M}$ denotes the complex dimension of the affine cone $\mathcal{M} \subset \mathbb{C}^{n}$ and $\dim_{\mathbb{C}}\mathcal{B}$ that of the projective base $\mathcal{B} \subset \mathbb{P}^{n-1}_{\mathbb{C}}$. When $S(\mathcal{B})$ is Cohen--Macaulay, the coefficients $h_{0}, \dots, h_{d}$ are non‑negative integers with $h_{d} \neq 0$; their sequence $h(\mathcal{B}) = (h_{0}, \dots, h_{d})$ is called the \textit{h-sequence} (or \textit{h-vector}) of $\mathcal{B}$~\cite{Stanley:1978}.

We now introduce some terminology related to $h$-sequences.
\begin{itemize}
    \item \textbf{Unimodal}: An $h$-sequence is called \textit{unimodal} if $h_0 \leq h_1 \leq \dots \leq h_i \geq h_{i+1} \geq \dots \geq h_d$ for some $i < d$.
    \item \textbf{O-sequence}: An $h$-sequence is called an \textit{O-sequence} if $h_0 = 1$ and $h_{i+1} \leq h_i^{\langle i \rangle}$ for all $i \geq 1$. Here, for positive integers $i$ and $h$, $h^{\langle i \rangle}$ denotes the binomial expansion
    \[
    h^{\langle i \rangle} = \binom{m_i + 1}{i + 1} + \binom{m_{i-1} + 1}{i} + \dots + \binom{m_j + 1}{j + 1},
    \]
    where $m_i > m_{i-1} > \dots > m_j \geq j \geq 1$, and $0^{\langle i \rangle} := 0$. These integers are determined by the \textit{i}-binomial expansion of $h$:
    \[
    h = \binom{m_i}{i} + \binom{m_{i-1}}{i-1} + \dots + \binom{m_j}{j},
    \]
    which is unique and always exists.\footnote{E.g., for $h=156$ and $i =11$, we have the expansion: \[
\binom{13}{11} + \binom{12}{10} + \binom{10}{9} + \binom{8}{8} + \binom{7}{7} =156
\] 
and $156^{<11>}=182$.
}
    
    \item \textbf{Differentiable}: An $h$-sequence is \textit{differentiable} if its first difference $\Delta h = (h_0, h_1 - h_0, h_2 - h_1, \dots)$ is an O-sequence.
    \item \textbf{SI sequence}: An $h$-sequence is called a \textit{Stanley-Iarrobino sequence} (or SI sequence) if it is \textit{palindromic} (i.e., $h_i = h_{d-i}$) and its first half $(h_0, h_1, \dots, h_{\lfloor \frac{d}{2} \rfloor})$ is differentiable \cite{Stanley:1978, Harima:1995}. Every SI-sequence is unimodal.
\end{itemize}

Macaulay \cite{Macaulay:1926} proved the following theorem:

\begin{theorem}[Macaulay (1927)]
An $h$-sequence is an O-sequence \textit{if and only if} it is the $h$-vector of some standard graded Artinian algebra.
\end{theorem}

Let $A$ be a ring. An $A$-algebra $R$ is standard graded if $R_0=A$ and $R$ is generated by the elements of $R_1$ (i.e.\ by its degree-$1$ graded piece).
E.g., the polynomial ring $R=k[t_1,\dots, t_n]$ is a direct sum of $R_i$ consisting of homogeneous polynomials of degree $i$.
A ring is said to be Artinian if its (one sided) ideals satisfy the \textit{descending chain condition}, i.e. there is no infinite descending sequence of ideals. An Artinian algebra is an algebra over a commutative Artinian ring.

A theorem of Stanley \cite{Stanley:1978} states that a standard graded Cohen–Macaulay domain $R$ is Gorenstein if and only if the numerator $P(t)$ of its Hilbert series is palindromic, $t^{d}P(1/t)=P(t)$. Equivalently, $R$ is Gorenstein if and only if its Artinian reduction $R/(\theta_{1},\dots,\theta_{N+1})$ by a linear system of parameters is an Artinian Gorenstein algebra; the Hilbert function of such a reduction, which coincides with the $h$‑sequence of $\mathcal{B}$, is then called a \emph{Gorenstein sequence}. Every Gorenstein sequence is palindromic and an O‑sequence; it is, however, \emph{not} necessarily unimodal. Stanley's first example, $(1,13,12,13,1)$, has codimension $13$ and socle degree $4$~\cite{Stanley:1978,migliore:2015}. Gorenstein sequences are SI-sequences, and hence unimodal, in codimension $\leq 3$~\cite{Stanley:1978}; more generally, the weak or strong Lefschetz property implies unimodality~\cite{tHarima}. In higher codimension a Gorenstein sequence need not be an SI‑sequence: unimodality can fail~\cite{Stanley:1978,migliore:2015}, and even unimodal Gorenstein sequences can fail the differentiability condition on the first half~\cite{Migliore:2017}. Conversely, every SI‑sequence is realised as the $h$‑vector of an Artinian Gorenstein algebra~\cite{Migliore:2017,Cho:2011}.
A striking feature of most Hilbert series computed in this paper is the palindromicity of their numerators. Provided the corresponding coordinate rings are Cohen--Macaulay domains, Stanley's theorem then implies that they are Gorenstein~\cite{Stanley:1978}. In the affine-cone terminology commonly used here, this is the algebraic Calabi–Yau property.

As an example, consider the complex line $\mathcal{M} = \mathbb{C}$, parametrised by a single coordinate $x$. The $i$-th graded piece $S_i$ is generated by the monomial $x^i$. Thus, $\dim_{\mathbb{C}} S_i = 1$ for all $i \in \mathbb{N}$, and the Hilbert series becomes
\[
\mathcal{HS}(t; \mathbb{C}) = \frac{1}{1 - t}.
\]
Here, $\dim_{\mathbb{C}} \mathcal{M}=1=\dim_{\mathbb{C}} \mathcal{B} +1$, as the complex line is the affine cone over a point. This result generalises to $\mathcal{HS}(t; \mathbb{C}^n) = \frac{1}{(1 - t)^n}$.

Another algebraic quantity of interest is $\sum h_i$, which, for a standard graded Cohen--Macaulay coordinate ring, gives the \textit{degree} of the variety, i.e.\ the number of points in the intersection with a generic linear subspace of complementary dimension~\cite{Harris:1992}.

\subsection{Plethystic Exponential and Plethystic Logarithm}
\label{app:PE-PL}
An important mathematical tool employed in this paper is the \textit{Plethystic Exponential} (PE) \cite{Benvenuti:2006qr,Feng:2007ur}, which encodes symmetrisation in generating functions. Given a generating function $f(t)$, it is defined as
\[
\text{PE}[f(t)] = \exp\left(\sum_{k=1}^{\infty} \frac{1}{k} f(t^k) \right),
\]
where $\sum_{k=1}^{\infty} \frac{1}{k} f(t^k)$ accounts for all possible degrees of symmetrisation. This definition naturally extends to multiple variables: for a generating function $g(t_1, t_2, \dots, t_n)$, we replace $f(t^k)$ with $g(t_1^k, t_2^k, \dots, t_n^k)$. For example, if $f(t) = t$, the plethystic exponential generates the infinite symmetric power series
\[
\text{PE}[t] = 1 + t + t^2 + t^3 + \dots = \frac{1}{1 - t}.
\]

The inverse function, called the \textit{Plethystic Logarithm} (PL), is defined for a generating function $f(t)$ as
\[
\text{PL}[f(t)] = \text{PE}^{-1}[f(t)] = \sum_{k=1}^{\infty} \frac{\mu(k)}{k} \log(f(t^k)),
\]
where $\mu(k)$ is the Möbius function:
\[
\mu(k) := 
\begin{cases}
1 & \text{if } k = 1, \\
(-1)^n & \text{if } k \text{ is a product of } n \text{ distinct primes}, \\
0 & \text{otherwise}.
\end{cases}
\]

To illustrate its utility, consider the plethystic logarithm of the algebraic variety with generating function $f(t) = \frac{1 - t^2 - t^3 + t^4}{(1 - t)^2}$:
\[
\text{PL}[f(t)] = 2t - t^2 - t^3 + t^4 - t^5 + \dots.
\]
This expansion can be interpreted as follows: the variety is generated by two independent generators ($+2t$), subject to one quadratic ($-t^2$) and one cubic ($-t^3$) relation. These relations are not independent, but satisfy a higher-order syzygy ($+t^4$).

In general, the first terms in the expansion of the plethystic logarithm follow this hierarchy: independent generators appear with positive coefficients, followed by negative terms representing relations among these generators, and then higher-order syzygies. For varieties that are complete intersections, the higher-order syzygies do not appear, and the plethystic logarithm is a finite sum.

\section{Hilbert series for improved bifundamentals at \texorpdfstring{$N=3$}{n=3}}\label{AppN=3}

In the following sections, we study specific examples of moduli spaces of improved bifundamentals with rank $N=3$.

\subsection{\texorpdfstring{$FT_3$}{ftsu3} }

\be
\resizebox{0.75\hsize}{!}{
\begin{tikzpicture}[
    thick,
    gauge/.style={
        circle, draw=black, fill=white,
        minimum size=8mm, inner sep=1pt, font=\small
    },
    manifest/.style={
        rectangle, draw=blue!70!black, fill=blue!6,
        minimum size=8mm, inner sep=2pt,
        font=\small, text=blue!70!black
    },
    flavour/.style={
        rectangle, draw=red!70!black, fill=red!6,
        minimum size=7mm, inner sep=2pt,
        font=\small, text=red!70!black
    },
    arr/.style={-{Stealth[scale=0.8]}, shorten >=1pt, shorten <=1pt},
    barr/.style={draw=black!70}
]
\node[gauge]    (g1) at (0,0)   {$1$};
\node[gauge]    (g2) at (2.5,0) {$2$};
\node[manifest] (m1) at (5,0)   {$3$};
\draw[arr,barr,transform canvas={yshift= 3.5pt}] (g1) -- (g2);
\draw[arr,barr,transform canvas={yshift=-3.5pt}] (g2) -- (g1);
\node at (1.25, 0.45) {$b_1,\,\tilde{b}_1$};
\node at (1.25, 0) {\huge$\times$};
\draw[arr,barr,transform canvas={yshift= 3.5pt}] (g2) -- (m1);
\draw[arr,barr,transform canvas={yshift=-3.5pt}] (m1) -- (g2);
\node at (3.75, 0.45) {$b_2,\,\tilde{b}_2$};
\node at (3.75, 0) {\huge$\times$};
\draw (g2) to[out=65,in=115,looseness=3] (g2);
\node at (2.5, 0.95) {$a_2$};
\draw[blue!70!black] (m1) to[out=65,in=115,looseness=3] (m1);
\node[blue!70!black] at (5, 0.95) {$a_{3}$};
\node at (2.5,-1.5) {$\mathcal{W} =
    a_2 b_1\tilde{b}_1
    + b_2 (a_2 + a_{3})\tilde{b}_2
    + \sum_{i=1}^{2}\mathrm{Flip}[b_i\tilde{b}_i]$};
\node[right] at (6.5,-0.5) {%
    \renewcommand{\arraystretch}{1.4}
    \begin{tabular}{>{\small}c >{\small}c}
        \toprule
        $b_i,\,\tilde{b}_i$ & $\tau/2$ \\
        $a_i$                & $2-\tau$ \\
        \bottomrule
    \end{tabular}%
};
\end{tikzpicture}
}
\ee

We compute the Hilbert series and the superconformal index of $FT_3$, verify their consistency, and identify an IR symmetry enhancement.

\paragraph{Hilbert series.}
Assuming the moduli space is a complete intersection, the Hilbert series is generated by the two adjoints $A_L$ and $A_R$ of $SU(3)_L \times SU(3)_R$, subject to the quadratic and cubic relations $\mathrm{tr}_L(A_L^2) = \mathrm{tr}_R(A_R^2)$ and $\mathrm{tr}_L(A_L^3) = \mathrm{tr}_R(A_R^3)$:
\be\label{HSfTu3}
\CH\CS_{FT_3} = \mathrm{PE}\!\left[\bigl(\chi_{[1,1;0,0]_{su(3)_L\times su(3)_R}} + \chi_{[0,0;1,1]_{su(3)_L\times su(3)_R}}\bigr)\, x^{2-\tau} - x^{4-2\tau} - x^{6-3\tau}\right],
\ee
where $\tau$ parametrises the mixing of the trial R-symmetry with $U(1)_\tau$.

Setting all flavour characters to their dimensions and writing $t \equiv x^{2-\tau}$, the unrefined Hilbert series evaluates to
\be\label{eqftu3a}
\begin{aligned}
\CH\CS^{\mathrm{unr}}_{FT_3}
&= \frac{(1-t^2)(1-t^3)}{(1-t)^{16}} = \frac{1+2t+2t^2+t^3}{(1-t)^{14}} \\
&= 1 + 16t + 135t^2 + 799t^3 + 3724t^4 + 14553t^5 + 49588t^6 + 151300t^7 \\
&\quad + 421362t^8 + 1086572t^9 + 2623406t^{10} + O(t^{11}).
\end{aligned}
\ee

\paragraph{Superconformal index.}
$Z$-extremisation fixes $\tau = 1.318\dots$; for numerical evaluation of the index we use the rational approximation $\tau = 13/10$. The expansion then reads
\begin{equation}\label{eqftu3}
\begin{split}
\CI_{FT_3} = {}& 1 + 16\psi\, x^{7/10} + 135\psi^2 x^{7/5} - 17 x^2 + 799\psi^3 x^{21/10} - \psi^{-2} x^{13/5} \\
& - 319\psi\, x^{27/10} + 3724\psi^4 x^{14/5} - 79\psi^{-1} x^{33/10} - 2838\psi^2 x^{17/5} \\
& + 14553\psi^5 x^{7/2} - \psi^{-3} x^{39/10} - 287 x^4 - 16874\psi^3 x^{41/10} \\
& + 49588\psi^6 x^{21/5} + 210\psi^{-2} x^{23/5} + 2493\psi\, x^{47/10} - 77077\psi^4 x^{24/5} \\
& + 151300\psi^7 x^{49/10} + 1885\psi^{-1} x^{53/10} + 31505\psi^2 x^{27/5} - 291291\psi^5 x^{11/2} \\
& + 421362\psi^8 x^{28/5} - 210\psi^{-3} x^{59/10} + 2334\, x^6 + 188340\psi^3 x^{61/10} \\
& - 952952\psi^6 x^{31/5} + 1086572\psi^9 x^{63/10} + \psi^{-5} x^{13/2} - 2907\psi^{-2} x^{33/5} \\
& - 42769\psi\, x^{67/10} + 809381\psi^4 x^{34/5} - 2781064\psi^7 x^{69/10} + 2623406\psi^{10} x^7 \\
& + O\!\left(x^{71/10}\right).
\end{split}
\end{equation}

\paragraph{Consistency check.}
The Hilbert series is recovered from the index in the limit
\be
\begin{split}
\lim_{y \to 0} \CI_{FT_3}\!\left(x \to xy,\ \psi \to \psi\, y^{\tau - 2}\right)
&= \CH\CS^{\mathrm{unr}}_{FT_3}\!\left(x^{2-\tau}\psi\right) \\
&= 1 + 16\psi\, x^{2-\tau} + 135\psi^2 x^{4-2\tau} + 799\psi^3 x^{6-3\tau} + \cdots,
\end{split}
\ee
with agreement holding up to the computed order $x^7$; this confirms \eqref{eqftu3a} and \eqref{eqftu3} against each other.

\paragraph{IR symmetry enhancement.}
The coefficient $-17$ of $x^2$ in \eqref{eqftu3} signals an enhancement of the flavour symmetry in the IR,
\begin{equation*}
SU(3) \times U(1)^3 \;\longrightarrow\; SU(3) \times SU(3) \times U(1).
\end{equation*}
Indeed, conserved-current multiplets contribute $-\dim(G_{\mathrm{flavour}})\, x^2$ to the index, and $\dim\bigl(SU(3) \times SU(3) \times U(1)\bigr) = 17$ matches the enhanced rather than the manifest symmetry.
\subsection{\texorpdfstring{$FC_3$}{fcsu3} }

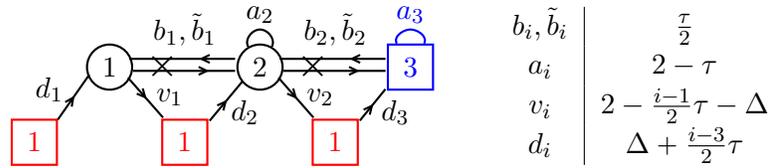
\be
\resizebox{\hsize}{!}{
\begin{tikzpicture}[
    thick,
    gauge/.style={
        circle, draw=black, fill=white,
        minimum size=8mm, inner sep=1pt, font=\small
    },
    manifest/.style={
        rectangle, draw=blue!70!black, fill=blue!6,
        minimum size=8mm, inner sep=2pt,
        font=\small, text=blue!70!black
    },
    flavour/.style={
        rectangle, draw=red!70!black, fill=red!6,
        minimum size=7mm, inner sep=2pt,
        font=\small, text=red!70!black
    },
    arr/.style={-{Stealth[scale=0.8]}, shorten >=1pt, shorten <=1pt},
    barr/.style={draw=black!70}
]
\node[gauge]    (g1) at (0,0)   {$1$};
\node[gauge]    (g2) at (2.5,0) {$2$};
\node[manifest] (m1) at (5,0)   {$3$};
\node[flavour] (f1) at (-0.8,-1.8) {$1$};
\node[flavour] (f2) at (1.7,-1.8)    {$1$};
\node[flavour] (f3) at (4.2,-1.8)  {$1$};
\draw[arr,barr,transform canvas={yshift= 3.5pt}] (g1) -- (g2);
\draw[arr,barr,transform canvas={yshift=-3.5pt}] (g2) -- (g1);
\node at (1.25, 0.45) {$b_1,\,\tilde{b}_1$};
\node at (1.25, 0) {\huge$\times$};
\draw[arr,barr,transform canvas={yshift= 3.5pt}] (g2) -- (m1);
\draw[arr,barr,transform canvas={yshift=-3.5pt}] (m1) -- (g2);
\node at (3.75, 0.45) {$b_2,\,\tilde{b}_2$};
\node at (3.75, 0) {\huge$\times$};
\draw (g2) to[out=65,in=115,looseness=3] (g2);
\node at (2.5, 0.95) {$a_2$};
\draw[blue!70!black] (m1) to[out=65,in=115,looseness=3] (m1);
\node[blue!70!black] at (5, 0.95) {$a_{3}$};
\draw[arr,barr] (f1) to[bend right=5] (g1);
\node at (-0.6,-0.7) {$d_1$};
\draw[arr,barr] (g1) to[bend right=5] (f2);
\node at (1.3,-0.7) {$v_1$};
\draw[arr,barr] (f2) to[bend right=5] (g2);
\node at (1.9,-0.7) {$d_2$};
\draw[arr,barr] (g2) to[bend right=5] (f3);
\node at (3.8,-0.7) {$v_2$};
\draw[arr,barr] (f3) to[bend right=5] (m1);
\node at (4.4,-0.7) {$d_3$};
\node at (2.5,-3.2) {$\mathcal{W} =
    \sum_{i=1}^{2}v_i b_i d_{i+1}
    + \mathfrak{M}^{+\,0} + \mathfrak{M}^{0\,+}
    + a_2 b_1\tilde{b}_1
    + b_2(a_2 + a_{3})\tilde{b}_2
    + \sum_{i=1}^{2}\mathrm{Flip}[b_i\tilde{b}_i]$};
\node[right] at (6.5,-0.5) {%
    \renewcommand{\arraystretch}{1.4}
    \begin{tabular}{>{\small}c >{\small}c}
        \toprule
        $b_i,\,\tilde{b}_i$ & $\tau/2$ \\
        $a_i$                & $2-\tau$ \\
        $v_i$                & $2-\tfrac{i-1}{2}\tau-\Delta$ \\
        $d_i$                & $\Delta+\tfrac{i-3}{2}\tau$ \\
        \bottomrule
    \end{tabular}%
};
\end{tikzpicture}
}
\ee

We compute the Hilbert series and superconformal index of $FC_3$, extrapolate a closed form for the unrefined Hilbert series, analyse its chiral ring, and verify the two computations against each other.

\paragraph{Chiral ring and Hilbert series.}
The chiral ring is generated by the bifundamental $\Pi$ and the adjoints $A_L$, $A_R$ of $SU(3)_L \times SU(3)_R$, subject to
\be
\mathrm{tr}_L(A_L^2) = \mathrm{tr}_R(A_R^2), \qquad
\mathrm{tr}_L(A_L^3) = \mathrm{tr}_R(A_R^3), \qquad
A_L\, \Pi = \Pi\, A_R.
\ee
These relations are not independent: the moduli space is not a complete intersection.

Refining in $SU(3)_L \times SU(3)_R$ characters (the $U(1)_\Delta \times U(1)_\tau$ charges are already encoded in the $x$-grading, so their fugacities would be redundant), the Hilbert series begins
\be\label{hsfc3}
\begin{aligned}
\CH\CS_{FC_3} ={}& 1 + \chi_{[1,0;0,1]}\, x^\Delta + \bigl(\chi_{[1,1;0,0]} + \chi_{[0,0;1,1]}\bigr)\, x^{2-\tau} + \bigl(\chi_{[2,0;0,2]} + \chi_{[0,1;1,0]}\bigr)\, x^{2\Delta} \\
&+ \bigl(\chi_{[2,1;0,1]} + \chi_{[0,2;0,1]} + \chi_{[1,0;0,1]} + \chi_{[1,0;1,2]} + \chi_{[1,0;2,0]}\bigr)\, x^{\Delta + 2 - \tau} \\
&+ \bigl(\chi_{[2,2;0,0]} + \chi_{[0,0;2,2]} + \chi_{[0,0;1,1]} + \chi_{[1,1;0,0]} + \chi_{[1,1;1,1]} + \chi_{[0,0;0,0]}\bigr)\, x^{4 - 2\tau} \\
&+ 165\, x^{3\Delta} + 639\, x^{2\Delta + (2-\tau)} + 1071\, x^{\Delta + (4-2\tau)} \\
&+ 799\, x^{6-3\tau} + 495\, x^{4\Delta} + 2236\, x^{3\Delta + (2-\tau)} + O\!\left(x^{201/50}\right),
\end{aligned}
\ee
where from the fourth line onward we display only the total dimension of the $SU(3)_L \times SU(3)_R$ representation content at each $x$-degree.

The chiral ring structure is confirmed by the plethystic logarithm
\be
\mathrm{PL}\!\left[\CH\CS_{FC_3}\right] = \chi_{[1,0;0,1]}\, t_d + \bigl(\chi_{[1,1;0,0]} + \chi_{[0,0;1,1]}\bigr)\, t_a - t_a^2 - t_a^3 - \chi_{[1,0;0,1]}\, t_a t_d + \cdots,
\ee
with $t_d \equiv x^\Delta$ and $t_a \equiv x^{2-\tau}$. The positive leading terms count the generators $\Pi$, $A_L$ and $A_R$; the negative terms $-t_a^2 - t_a^3$ and $-\chi_{[1,0;0,1]}\, t_a t_d$ encode the trace relations $\mathrm{tr}_L(A_L^k) = \mathrm{tr}_R(A_R^k)$ ($k=2,3$) and $A_L \Pi = \Pi A_R$, respectively. The infinitely many further terms reflect the fact that the moduli space is not a complete intersection.

\paragraph{Closed form for the unrefined Hilbert series.}
We extrapolate
\be
\CH\CS^{\mathrm{unr}}_{FC_3} \;=\; \frac{\displaystyle\sum_{j,k=0}^{6} a_{j,k}\, t_a^{\,j}\, t_d^{\,k}}{(1 - t_d)^9\, (1 - t_a)^{14}},
\ee
with coefficient matrix
\be
a_{j,k} \;\doteq\;
\begin{pmatrix}
 1 & 0 & 0 & 0 & 0 & 0 & 0 \\
 2 & -9 & 0 & 1 & 0 & 0 & 0 \\
 2 & -18 & 45 & -14 & 0 & 0 & 0 \\
 1 & -9 & 45 & -94 & 45 & -9 & 1 \\
 0 & 0 & 0 & -14 & 45 & -18 & 2 \\
 0 & 0 & 0 & 1 & 0 & -9 & 2 \\
 0 & 0 & 0 & 0 & 0 & 0 & 1 \\
\end{pmatrix}.
\ee
The matrix is palindromic, $a_{j,k} = a_{6-j,\,6-k}$, and its entries sum to zero, $\sum_{j,k} a_{j,k} = 0$.

\paragraph{Specialisations.}
Setting $t_a = t_d = t$ factorises the numerator as
\be
(t-1)^6 \bigl(t^6 + 8 t^5 + 26 t^4 + 39 t^3 + 26 t^2 + 8 t + 1\bigr),
\ee
while the specialisations $t_a = 1$ and $t_d = 1$ give $6(t-1)^6$ and $(t-1)^6$ respectively. The latter two fix the row and column sums of $a_{j,k}$:
\be
\sum_{j=0}^6 a_{j,k} = 6\,(-1)^k \binom{6}{k}, \qquad
\sum_{k=0}^6 a_{j,k} = (-1)^j \binom{6}{j}.
\ee

Two further truncations of the refined Hilbert series \eqref{hsfc3} are worth noting. Setting $t_d = 0$ reproduces, up to a rescaling of $x$, the Hilbert series of $FT_3$ --- the analogous statement does \emph{not} hold for the index. Setting $t_a = 0$ yields the freely-generated Hilbert series of $\mathbb{C}^9$.

\paragraph{Superconformal index.}
$Z$-extremisation fixes $\Delta = 0.90\dots$ and $\tau = 0.86\dots$; for numerical evaluation we adopt the rational approximations $\Delta = 9/10$ and $\tau = 86/100$. Refining only in the abelian fugacities $\psi_\tau$ and $\chi_\Delta$, the index reads
\begin{equation}
\begin{split}
\CI_{FC_3} ={}& 1 + 9 x^{9/10} \chi_\Delta + 16 x^{57/50} \psi_\tau - x^{43/25} \psi_\tau^{-2} \\
&- 9 x^{44/25} \chi_\Delta \psi_\tau^{-1} + 45 x^{9/5} \chi_\Delta^2 - 18 x^2 + 135 x^{51/25} \chi_\Delta \psi_\tau + 135 x^{57/25} \psi_\tau^2 \\
&- x^{129/50} \psi_\tau^{-3} - 9 x^{131/50} \chi_\Delta \psi_\tau^{-2} - 81 x^{133/50} \chi_\Delta^2 \psi_\tau^{-1} + 165 x^{27/10} \chi_\Delta^3 \\
&- 9 x^{141/50} \chi_\Delta^{-1} \psi_\tau^{-2} - 80 x^{143/50} \psi_\tau^{-1} - 324 x^{29/10} \chi_\Delta + 639 x^{147/50} \chi_\Delta^2 \psi_\tau \\
&- 45 x^{31/10} \chi_\Delta^{-1} - 336 x^{157/50} \psi_\tau + 1071 x^{159/50} \chi_\Delta \psi_\tau^2 - 9 x^{169/50} \chi_\Delta^{-1} \psi_\tau^2 \\
&+ 799 x^{171/50} \psi_\tau^3 + 9 x^{87/25} \chi_\Delta \psi_\tau^{-3} - 404 x^{89/25} \chi_\Delta^3 \psi_\tau^{-1} + 495 x^{18/5} \chi_\Delta^4 \\
&+ 9 x^{92/25} \chi_\Delta^{-1} \psi_\tau^{-3} + 146 x^{93/25} \psi_\tau^{-2} - 126 x^{94/25} \chi_\Delta \psi_\tau^{-1} - 2079 x^{19/5} \chi_\Delta^2 \\
&+ 2236 x^{96/25} \chi_\Delta^3 \psi_\tau + 90 x^{99/25} \chi_\Delta^{-1} \psi_\tau^{-1} - 576 x^4 - 3942 x^{101/25} \chi_\Delta \psi_\tau \\
&+ 4824 x^{102/25} \chi_\Delta^2 \psi_\tau^2 + 9 x^{21/5} \chi_\Delta^{-2} - 486 x^{106/25} \chi_\Delta^{-1} \psi_\tau - 3054 x^{107/25} \psi_\tau^2 \\
&+ x^{43/10} \psi_\tau^{-5} + 5985 x^{108/25} \chi_\Delta \psi_\tau^3 + 72 x^{219/50} \chi_\Delta^2 \psi_\tau^{-3} \\
&+ 224 x^{221/50} \chi_\Delta^3 \psi_\tau^{-2} - 1476 x^{223/50} \chi_\Delta^4 \psi_\tau^{-1} + 1287 x^{9/2} \chi_\Delta^5 \\
&- 135 x^{113/25} \chi_\Delta^{-1} \psi_\tau^3 + 9 x^{227/50} \chi_\Delta^{-1} \psi_\tau^{-4} + 3724 x^{114/25} \psi_\tau^4 \\
&- 64 x^{229/50} \psi_\tau^{-3} + 990 x^{231/50} \chi_\Delta \psi_\tau^{-2} + 1467 x^{233/50} \chi_\Delta^2 \psi_\tau^{-1} \\
&- 8660 x^{47/10} \chi_\Delta^3 + 6444 x^{237/50} \chi_\Delta^4 \psi_\tau + 9 x^{239/50} \chi_\Delta^{-2} \psi_\tau^{-3} \\
&+ 2838 x^{243/50} \psi_\tau^{-1} + 2664 x^{49/10} \chi_\Delta - 21357 x^{247/50} \chi_\Delta^2 \psi_\tau + 16200 x^{249/50} \chi_\Delta^3 \psi_\tau^2 \\
&+ 2115 x^{51/10} \chi_\Delta^{-1} - 50 x^{257/50} \psi_\tau - 28530 x^{259/50} \chi_\Delta \psi_\tau^2 + 25776 x^{261/50} \chi_\Delta^2 \psi_\tau^3 \\
&- 45 x^{131/25} \chi_\Delta^2 \psi_\tau^{-4} + 175 x^{132/25} \chi_\Delta^3 \psi_\tau^{-3} - x^{53/10} \chi_\Delta^{-3} \\
&+ 1386 x^{133/25} \chi_\Delta^4 \psi_\tau^{-2} + 144 x^{267/50} \chi_\Delta^{-2} \psi_\tau - 4410 x^{134/25} \chi_\Delta^5 \psi_\tau^{-1} \\
&- 2520 x^{269/50} \chi_\Delta^{-1} \psi_\tau^2 + x^{27/5} \bigl(3003\, \chi_\Delta^6 - 9\, \chi_\Delta^{-1} \psi_\tau^{-5}\bigr) \\
&- 18704 x^{271/50} \psi_\tau^3 + 26460 x^{273/50} \chi_\Delta \psi_\tau^4 - 1080 x^{137/25} \chi_\Delta \psi_\tau^{-3} \\
&+ 2187 x^{138/25} \chi_\Delta^2 \psi_\tau^{-2} + 10954 x^{139/25} \chi_\Delta^3 \psi_\tau^{-1} - 28008 x^{28/5} \chi_\Delta^4 \\
&+ 9 x^{281/50} \chi_\Delta^{-2} \psi_\tau^3 + 16182 x^{141/25} \chi_\Delta^5 \psi_\tau - 9 x^{141/25} \chi_\Delta^{-2} \psi_\tau^{-4} \\
&- 1071 x^{283/50} \chi_\Delta^{-1} \psi_\tau^4 - 90 x^{142/25} \chi_\Delta^{-1} \psi_\tau^{-3} + 14553 x^{57/10} \psi_\tau^5 \\
&- 3846 x^{143/25} \psi_\tau^{-2} + 8028 x^{144/25} \chi_\Delta \psi_\tau^{-1} + 31752 x^{29/5} \chi_\Delta^2 \\
&- 80070 x^{146/25} \chi_\Delta^3 \psi_\tau + 45090 x^{147/25} \chi_\Delta^4 \psi_\tau^2 - x^{147/25} \chi_\Delta^{-3} \psi_\tau^{-3} \\
&- 18 x^{148/25} \chi_\Delta^{-2} \psi_\tau^{-2} - 3456 x^{149/25} \chi_\Delta^{-1} \psi_\tau^{-1} + 17179\, x^6 + O\!\left(x^{301/50}\right).
\end{split}
\end{equation}

\paragraph{Consistency check.}
The full refined Hilbert series is recovered from the index in the limit
\be
\begin{aligned}
\lim_{y \to 0} \CI_{FC_3}&\!\left(x \to xy,\ \psi_\tau \to \psi_\tau\, y^{\tau - 2},\
  \chi_\Delta \to \chi_\Delta\, y^{-\Delta}\right) \\
&= \CH\CS^{\mathrm{unr}}_{FC_3}\!\left(x^{2-\tau}\psi_\tau,\ x^\Delta \chi_\Delta\right) \\
&= 1 + 9\, \chi_\Delta\, t_d + 16\, \psi_\tau\, t_a + 45\, \chi_\Delta^2\, t_d^2 \\
&\quad + 135\, \chi_\Delta \psi_\tau\, t_a t_d + 135\, \psi_\tau^2\, t_a^2 + 165\, \chi_\Delta^3\, t_d^3 + \cdots,
\end{aligned}
\ee
providing a nontrivial cross-check of the two computations.

\subsection{\texorpdfstring{$FH_3$}{fh3} }

Consider the mixed $USp(6)-SU(3)$ improved bifundamental.

\subsubsection*{Manifest $SU(3)$ global symmetry}
Consider the following UV frame, where the manifest global symmetry is $SU(3)$.

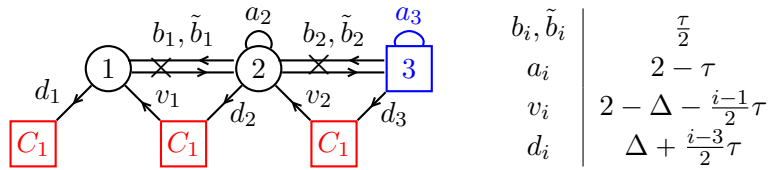
\be
\resizebox{\hsize}{!}{
\begin{tikzpicture}[
    thick,
    gauge/.style={
        circle, draw=black, fill=white,
        minimum size=8mm, inner sep=1pt, font=\small
    },
    manifest/.style={
        rectangle, draw=blue!70!black, fill=blue!6,
        minimum size=8mm, inner sep=2pt,
        font=\small, text=blue!70!black
    },
    flavour/.style={
        rectangle, draw=red!70!black, fill=red!6,
        minimum size=7mm, inner sep=2pt,
        font=\small, text=red!70!black
    },
    arr/.style={-{Stealth[scale=0.8]}, shorten >=1pt, shorten <=1pt},
    barr/.style={draw=black!70}
]
\node[gauge]    (g1) at (0,0)   {$1$};
\node[gauge]    (g2) at (2.5,0) {$2$};
\node[manifest] (m1) at (5,0)   {$3$};
\node[flavour] (f1) at (-0.8,-1.8) {$C_1$};
\node[flavour] (f2) at (1.7,-1.8)    {$C_1$};
\node[flavour] (f3) at (4.2,-1.8)  {$C_1$};
\draw[arr,barr,transform canvas={yshift= 3.5pt}] (g1) -- (g2);
\draw[arr,barr,transform canvas={yshift=-3.5pt}] (g2) -- (g1);
\node at (1.25, 0.45) {$b_1,\,\tilde{b}_1$};
\node at (1.25, 0) {\huge$\times$};
\draw[arr,barr,transform canvas={yshift= 3.5pt}] (g2) -- (m1);
\draw[arr,barr,transform canvas={yshift=-3.5pt}] (m1) -- (g2);
\node at (3.75, 0.45) {$b_2,\,\tilde{b}_2$};
\node at (3.75, 0) {\huge$\times$};
\draw (g2) to[out=65,in=115,looseness=3] (g2);
\node at (2.5, 0.95) {$a_2$};
\draw[blue!70!black] (m1) to[out=65,in=115,looseness=3] (m1);
\node[blue!70!black] at (5, 0.95) {$a_{3}$};
\draw[arr,barr] (g1) to[bend right=5] (f1);
\node at (-0.7,-0.7) {$d_1$};
\draw[arr,barr] (f2) to[bend right=5] (g1);
\node at (1.2,-0.7) {$v_1$};
\draw[arr,barr] (g2) to[bend right=5] (f2);
\node at (1.8,-0.7) {$d_2$};
\draw[arr,barr] (f3) to[bend right=5] (g2);
\node at (3.7,-0.7) {$v_2$};
\draw[arr,barr] (m1) to[bend right=5] (f3);
\node at (4.3,-0.7) {$d_3$};
\node at (2.5,-3.2) {$\mathcal{W} =
    \sum_{i=1}^{2}v_i b_i d_{i+1}
    + \mathfrak{M}^{\pm\,0} + \mathfrak{M}^{0\,\pm}
    + a_2 b_1\tilde{b}_1
    + b_2(a_2 + a_{3})\tilde{b}_2
    + \sum_{i=1}^{2}\mathrm{Flip}[b_i\tilde{b}_i]$};
\node[right] at (6.5,-0.5) {%
    \renewcommand{\arraystretch}{1.4}
    \begin{tabular}{>{\small}c >{\small}c}
        \toprule
        $b_i,\,\tilde{b}_i$ & $\tau/2$ \\
        $a_i$                & $2-\tau$ \\
        $v_i$                & $2-\Delta-\tfrac{i-1}{2}\tau$ \\
        $d_i$                & $\Delta+\tfrac{i-3}{2}\tau$ \\
        \bottomrule
    \end{tabular}%
};
\end{tikzpicture}
}
\ee

We compute the superconformal index of $FH_3$, extrapolate the unrefined Hilbert series from it, analyse the chiral ring, and verify consistency between the two quantities.

\paragraph{Superconformal index.}
The R-charges of the various fields are as indicated in the quiver diagram. Numerical $Z$-extremisation localises the minimum of the $\mathbf{S}^3$ partition function at $\Delta = \tau = 1.11\dots$; for evaluation of the index we adopt the rational approximation $\Delta = \tau = 11/10$. Refining only in the abelian fugacities $\psi_\tau$ and $\chi_\Delta$, the index reads
\begin{equation}
\begin{split}
\CI_{FH_3} ={}& 1 + 22 x^{9/10} \psi_\tau + 18 x^{11/10} \chi_\Delta + 252 x^{9/5} \psi_\tau^2 + x^2\bigl(378 \chi_\Delta \psi_\tau - 31\bigr) \\
& + x^{11/5}\bigl(-18 \chi_\Delta \psi_\tau^{-1} + 168 \chi_\Delta^2 - \psi_\tau^{-2}\bigr) \\
& + x^{27/10}\bigl(-18 \chi_\Delta^{-1} \psi_\tau^2 - 3 \chi_\Delta^{-2} \psi_\tau + 2001 \psi_\tau^3\bigr) \\
& + x^{29/10}\bigl(4140 \chi_\Delta \psi_\tau^2 - 3 \chi_\Delta^{-2} \psi_\tau^{-1} - 96 \chi_\Delta^{-1} - 773 \psi_\tau\bigr) \\
& + x^{31/10}\bigl(3378 \chi_\Delta^2 \psi_\tau - 18 \chi_\Delta^{-1} \psi_\tau^{-2} - 1002 \chi_\Delta - 135 \psi_\tau^{-1}\bigr) \\
& + x^{33/10}\bigl(-318 \chi_\Delta^2 \psi_\tau^{-1} - 18 \chi_\Delta \psi_\tau^{-2} + 1086 \chi_\Delta^3 - \psi_\tau^{-3}\bigr) \\
& + x^{18/5}\bigl(-378 \chi_\Delta^{-1} \psi_\tau^3 - 60 \chi_\Delta^{-2} \psi_\tau^2 + 12375 \psi_\tau^4\bigr) \\
& + x^{19/5}\bigl(31500 \chi_\Delta \psi_\tau^3 - 1614 \chi_\Delta^{-1} \psi_\tau - 9681 \psi_\tau^2\bigr) \\
& + x^{4}\bigl(35535 \chi_\Delta^2 \psi_\tau^2 - 17814 \chi_\Delta \psi_\tau + 102 \chi_\Delta^{-1} \psi_\tau^{-1} + 6 \chi_\Delta^{-2} \psi_\tau^{-2} - 2107\bigr) \\
& + x^{21/5}\bigl(20990 \chi_\Delta^3 \psi_\tau - 720 \chi_\Delta \psi_\tau^{-1} + 18 \chi_\Delta^{-1} \psi_\tau^{-3} - 12246 \chi_\Delta^2 + 142 \psi_\tau^{-2}\bigr) \\
& + x^{22/5}\bigl(-2902 \chi_\Delta^3 \psi_\tau^{-1} + 27 \chi_\Delta^2 \psi_\tau^{-2} + 18 \chi_\Delta \psi_\tau^{-3} + 5475 \chi_\Delta^4\bigr) \\
& + x^{9/2}\bigl(-4140 \chi_\Delta^{-1} \psi_\tau^4 - 582 \chi_\Delta^{-2} \psi_\tau^3 + 6 \chi_\Delta^{-3} \psi_\tau^2 + 63504 \psi_\tau^5\bigr) \\
& + x^{47/10}\bigl(187110 \chi_\Delta \psi_\tau^4 - 13278 \chi_\Delta^{-1} \psi_\tau^2 + 777 \chi_\Delta^{-2} \psi_\tau + 6 \chi_\Delta^{-3} - 82039 \psi_\tau^3\bigr) \\
& + x^{49/10}\bigl(260448 \chi_\Delta^2 \psi_\tau^3 - 185178 \chi_\Delta \psi_\tau^2 + 216 \chi_\Delta^{-2} \psi_\tau^{-1} + 6 \chi_\Delta^{-3} \psi_\tau^{-2} \\
& \phantom{{}+ x^{49/10}\bigl(} + 7332 \chi_\Delta^{-1} - 10905 \psi_\tau\bigr) \\
& + x^{51/10}\bigl(213000 \chi_\Delta^3 \psi_\tau^2 - 185424 \chi_\Delta^2 \psi_\tau + 450 \chi_\Delta^{-1} \psi_\tau^{-2} + 45 \chi_\Delta^{-2} \psi_\tau^{-3} \\
& \phantom{{}+ x^{51/10}\bigl(} + 5310 \chi_\Delta + 9797 \psi_\tau^{-1}\bigr) \\
& + x^{53/10}\bigl(102120 \chi_\Delta^4 \psi_\tau + 7428 \chi_\Delta^2 \psi_\tau^{-1} + 3780 \chi_\Delta \psi_\tau^{-2} + 18 \chi_\Delta^{-1} \psi_\tau^{-4} \\
& \phantom{{}+ x^{53/10}\bigl(} - 92526 \chi_\Delta^3 + 142 \psi_\tau^{-3}\bigr) \\
& + x^{27/5}\bigl(-31500 \chi_\Delta^{-1} \psi_\tau^5 - 3612 \chi_\Delta^{-2} \psi_\tau^4 + 118 \chi_\Delta^{-3} \psi_\tau^3 + 281358 \psi_\tau^6\bigr) \\
& + x^{11/2}\bigl(-18330 \chi_\Delta^4 \psi_\tau^{-1} + 2086 \chi_\Delta^3 \psi_\tau^{-2} + 276 \chi_\Delta^2 \psi_\tau^{-3} + 22968 \chi_\Delta^5 + \psi_\tau^{-5}\bigr) \\
& + x^{28/5}\bigl(924462 \chi_\Delta \psi_\tau^5 - 68370 \chi_\Delta^{-1} \psi_\tau^3 + 11226 \chi_\Delta^{-2} \psi_\tau^2 - 532379 \psi_\tau^4\bigr) \\
& + x^{29/5}\bigl(1494270 \chi_\Delta^2 \psi_\tau^4 - 1373922 \chi_\Delta \psi_\tau^3 + 93978 \chi_\Delta^{-1} \psi_\tau + 9 \chi_\Delta^{-2} + 30411 \psi_\tau^2\bigr) \\
& + x^{6}\bigl(1510568 \chi_\Delta^3 \psi_\tau^3 - 1731921 \chi_\Delta^2 \psi_\tau^2 + 248886 \chi_\Delta \psi_\tau - 8586 \chi_\Delta^{-1} \psi_\tau^{-1} \\
& \phantom{{}+ x^{6}\bigl(} - 192 \chi_\Delta^{-2} \psi_\tau^{-2} - 14 \chi_\Delta^{-3} \psi_\tau^{-3} + 139987\bigr) + O\!\left(x^{61/10}\right).
\end{split}
\end{equation}

\paragraph{Hilbert series.}
Extrapolating from the index up to order $x^9$, the unrefined Hilbert series reads
\begin{equation}\label{eqmb3}
\begin{split}
\CH\CS^{\mathrm{unr}}_{FH_3}(t_a,t_d) ={}&
1 + 22 t_a + 18 t_d \\
& + 252 t_a^2 + 378 t_a t_d + 168 t_d^2 \\
& + 2001 t_a^3 + 4140 t_a^2 t_d + 3378 t_a t_d^2 + 1086 t_d^3 \\
& + 12375 t_a^4 + 31500 t_a^3 t_d + 35535 t_a^2 t_d^2 + 20990 t_a t_d^3 + 5475 t_d^4 \\
& + 63504 t_a^5 + 187110 t_a^4 t_d + 260448 t_a^3 t_d^2 + 213000 t_a^2 t_d^3 + 102120 t_a t_d^4 + 22968 t_d^5 \\
& + 281358 t_a^6 + 924462 t_a^5 t_d + 1494270 t_a^4 t_d^2 + 1510568 t_a^3 t_d^3 \\
& \quad + 1003554 t_a^2 t_d^4 + 414936 t_a t_d^5 + 83504 t_d^6 \\
& + 1105863 t_a^7 + 3952872 t_a^6 t_d + 7148439 t_a^5 t_d^2 + 8408088 t_a^4 t_d^3 \\
& \quad + 6912477 t_a^3 t_d^4 + 3962388 t_a^2 t_d^5 + 1465944 t_a t_d^6 + 270504 t_d^7 \\
& + 3932379 t_a^8 + 15028200 t_a^7 t_d + 29661957 t_a^6 t_d^2 + 39114648 t_a^5 t_d^3 \\
& \quad + 37462866 t_a^4 t_d^4 + 26592852 t_a^3 t_d^5 + 13643535 t_a^2 t_d^6 + 4627944 t_a t_d^7 + 796518 t_d^8 \\
& + 12839750 t_a^9 + 51801750 t_a^8 t_d + 109659690 t_a^7 t_d^2 + 158154360 t_a^6 t_d^3 + \cdots,
\end{split}
\end{equation}
where $t_a \equiv x^{2-\tau}$ and $t_d \equiv x^\Delta$.

\paragraph{Closed form.}
We extrapolate the full closed form
\be
\CH\CS_{FH_3} \;=\; \frac{\displaystyle\sum_{j,k=0}^{9} a_{j,k}\, t_a^{\,j}\, t_d^{\,k}}{(1 - t_d)^{15}\,(1 - t_a)^{20}},
\ee
with coefficient matrix
\be
a_{j,k} \;\doteq\;
\begin{pmatrix}
 1 & 3 & 3 & 1 & 0 & 0 & 0 & 0 & 0 & 0 \\
 2 & -12 & -42 & -20 & 0 & 0 & 0 & 0 & 0 & 0 \\
 2 & -30 & 105 & 205 & 9 & -3 & 0 & 0 & 0 & 0 \\
 1 & -15 & 153 & -547 & -318 & 54 & 0 & 0 & 0 & 0 \\
 0 & 0 & 0 & -202 & 1191 & 75 & -59 & 3 & 0 & 0 \\
 0 & 0 & -3 & 59 & -75 & -1191 & 202 & 0 & 0 & 0 \\
 0 & 0 & 0 & 0 & -54 & 318 & 547 & -153 & 15 & -1 \\
 0 & 0 & 0 & 0 & 3 & -9 & -205 & -105 & 30 & -2 \\
 0 & 0 & 0 & 0 & 0 & 0 & 20 & 42 & 12 & -2 \\
 0 & 0 & 0 & 0 & 0 & 0 & -1 & -3 & -3 & -1 \\
\end{pmatrix}.
\ee
The matrix is antipalindromic, $a_{j,k} = -a_{9-j,\,9-k}$, so in particular $\sum_{j,k} a_{j,k} = 0$. This mirrors the analogous (palindromic) structure observed for the all-order Hilbert series of the $N=2$ theory and of $FC_3$.

The numerator $N(t_a, t_d) = \sum_{j,k=0}^{9} a_{j,k}\, t_a^j\, t_d^k$ was fixed by imposing the specialisations
\be
N(t,t) = (1-t)^9 \bigl(t^9 + 14 t^8 + 83 t^7 + 257 t^6 + 445 t^5 + 445 t^4 + 257 t^3 + 83 t^2 + 14 t + 1\bigr),
\ee
\be
N(t, 1) = 8(1-t)^9, \qquad N(1, t) = 6(1-t)^9.
\ee
The latter two fix the row and column sums of $a_{j,k}$:
\be
\sum_{j=0}^{9} a_{j,k} = 6\, (-1)^k \binom{9}{k}, \qquad
\sum_{k=0}^{9} a_{j,k} = 8\, (-1)^j \binom{9}{j}.
\ee
Setting $t_a = t_d = t$ then yields the completely unrefined Hilbert series,
\be
\CH\CS^{\mathrm{unr}}_{FH_3} \;=\; \frac{t^9 + 14 t^8 + 83 t^7 + 257 t^6 + 445 t^5 + 445 t^4 + 257 t^3 + 83 t^2 + 14 t + 1}{(1 - t)^{26}},
\ee
from which we read off that the moduli space is a complex variety of dimension $26$ and degree $1600$.

\paragraph{Chiral ring.}
The leading terms of the plethystic logarithm are
\be
\begin{aligned}
\mathrm{PL}\!\left[\CH\CS_{FH_3}\right]
={}& \chi_{(\Box,\,\mathbf{6})}\, t_d + \bigl(\chi_{(\mathrm{adj}_L,\,\mathbf{1})} + \chi_{(\mathbf{1},\,\mathrm{asym}_R)}\bigr)\, t_a \\
& - t_a^2 - t_a^3 - \chi_{(\mathrm{asym}_L,\,\mathbf{1})}\, t_d^2 - \chi_{(\Box,\,\mathbf{6})}\, t_a t_d + \cdots \\
\overset{\mathrm{unr}}{=}\;\,& 18\, t_d + 22\, t_a - t_a^2 - t_a^3 - 3\, t_d^2 - 18\, t_a t_d + \cdots .
\end{aligned}
\ee
The positive leading terms identify the chiral ring generators: the bifundamental $\Pi$ of $SU(3)_L \times USp(6)_R$ (contributing $\chi_{(\Box,\,\mathbf{6})}\, t_d$), the $SU(3)_L$ adjoint $A_L$ (contributing $\chi_{(\mathrm{adj}_L,\,\mathbf{1})}\, t_a$), and the $USp(6)$ traceless antisymmetric $A_R$ (contributing $\chi_{(\mathbf{1},\,\mathrm{asym}_R)}\, t_a$). These satisfy the relations (with their PL contributions shown to the right):
\be
\begin{aligned}
\Omega_{ab}\,\Pi^{a,A}\Pi^{b,B} &= 0,
&& -\chi_{(\mathrm{asym}_L,\,\mathbf{1})}\, t_d^2 \\
\mathrm{tr}_L A_L^k &= \mathrm{tr}_R A_R^k, \quad k = 2, 3,
&& -t_a^2 - t_a^3 \\
\Pi^{a,B}\,(A_L)^A{}_B &= \Pi^{b,A}\,A_R{}^{ba},
&& -\chi_{(\Box,\,\mathbf{6})}\, t_a t_d .
\end{aligned}
\ee
The infinitely many further terms in the PL show that these relations are not independent: the moduli space is not a complete intersection.

\paragraph{Consistency check.}
The full unrefined Hilbert series is recovered from the index in the limit
\be
\begin{split}
\lim_{y \to 0} \CI_{FH_3}&\!\left(x \to xy,\ \psi_\tau \to \psi_\tau\, y^{\tau - 2},\ \chi_\Delta \to \chi_\Delta\, y^{-\Delta}\right)
= \CH\CS^{\mathrm{unr}}_{FH_3}\!\left(x^{2-\tau}\psi_\tau,\, x^\Delta \chi_\Delta\right) \\
&= 1 + 22\, x^{9/10} \psi_\tau\, + 18\, x^{11/10} \chi_\Delta\, + 252\, x^{9/5}\psi_\tau^2\,   + 378\, x^{2} \chi_\Delta \psi_\tau\,  \\
&\quad + 168\, x^{11/5} \chi_\Delta^2\,+ 1086\, x^{33/10} \chi_\Delta^3\,  + \cdots ,
\end{split}
\ee

providing a cross-check of the two computations.

\subsection{\texorpdfstring{$FM_3$}{fmsu3} }
Consider the following theory

\be
\resizebox{1\hsize}{!}{
\begin{tikzpicture}[
    thick,
    gauge/.style={
        circle, draw=black, fill=white,
        minimum size=8mm, inner sep=1pt, font=\small
    },
    manifest/.style={
        rectangle, draw=blue!70!black, fill=blue!6,
        minimum size=8mm, inner sep=2pt,
        font=\small, text=blue!70!black
    },
    flavour/.style={
        rectangle, draw=red!70!black, fill=red!6,
        minimum size=7mm, inner sep=2pt,
        font=\small, text=red!70!black
    },
    arr/.style={-{Stealth[scale=0.8]}, shorten >=1pt, shorten <=1pt},
    barr/.style={draw=black!70}
]
\node[gauge]    (g1) at (0,0)   {$1$};
\node[gauge]    (g2) at (2.5,0) {$2$};
\node[manifest] (m1) at (5,0)   {$3$};
\node[flavour] (f1) at (-0.8,-1.8) {$1$};
\node[flavour] (f2) at (1.7,-1.8)  {$1$};
\node[flavour] (f3) at (4.2,-1.8)  {$1$};
\draw[arr,barr,transform canvas={yshift= 3.5pt}] (g1) -- (g2);
\draw[arr,barr,transform canvas={yshift=-3.5pt}] (g2) -- (g1);
\node at (1.25, 0.45) {$b_1,\,\tilde{b}_1$};
\node at (1.25, 0) {\huge$\times$};
\draw[arr,barr,transform canvas={yshift= 3.5pt}] (g2) -- (m1);
\draw[arr,barr,transform canvas={yshift=-3.5pt}] (m1) -- (g2);
\node at (3.75, 0.45) {$b_2,\,\tilde{b}_2$};
\node at (3.75, 0) {\huge$\times$};
\draw (g2) to[out=65,in=115,looseness=3] (g2);
\node at (2.5, 0.95) {$a_2$};
\draw[blue!70!black] (m1) to[out=65,in=115,looseness=3] (m1);
\node[blue!70!black] at (5, 0.95) {$a_{3}$};
\draw[arr,barr] (f1) to[bend left=4] (g1);
\draw[arr,barr] (g1) to[bend left=4] (f1);
\node[rotate=-30] at (-0.4,-0.9) {\huge$\times$};
\node at (-0.8,-0.55) {$d_1$};
\draw[arr,barr] (g1) to[bend left=4] (f2);
\draw[arr,barr] (f2) to[bend left=4] (g1);
\node at (1.05,-0.6) {$v_1$};
\draw[arr,barr] (f2) to[bend left=4] (g2);
\draw[arr,barr] (g2) to[bend left=4] (f2);
\node[rotate=-30] at (2.1,-0.9) {\huge$\times$};
\node at (1.7,-0.6) {$d_2$};
\draw[arr,barr] (g2) to[bend left=4] (f3);
\draw[arr,barr] (f3) to[bend left=4] (g2);
\node at (3.6,-0.6) {$v_2$};
\draw[arr,barr] (f3) to[bend left=4] (m1);
\draw[arr,barr] (m1) to[bend left=4] (f3);
\node at (5,-0.7) {$d_3$};
\node at (2.5,-3.8) {$\begin{aligned}\mathcal{W} =&\;
    \sum_{j=1}^{2}\left[v_j b_j d_{j+1}
    + \tilde{v}_j\tilde{b}_j\tilde{d}_{j+1}\right]
    + a_2 b_1\tilde{b}_1
    + b_2(a_2 + a_{3})\tilde{b}_2
    + \sum_{i=1}^{2}\mathrm{Flip}[b_i\tilde{b}_i]\\
    &+ \mathfrak{M}^{\pm\,0} + \mathfrak{M}^{0\,\pm}
    + \sum_{i=1}^{2}\mathrm{Flip}[d_i\tilde{d}_i]
    \end{aligned}$};
\node[right] at (6.5,-0.5) {%
    \renewcommand{\arraystretch}{1.4}
    \begin{tabular}{>{\small}c >{\small}c}
        \toprule
        $b_i,\,\tilde{b}_i$        & $\tau/2$ \\
        $a_i$                       & $2-\tau$ \\
        $v_i,\,\tilde{v}_i$        & $2-\Delta-\tfrac{i-1}{2}\tau$ \\
        $d_i,\,\tilde{d}_i$        & $\Delta+\tfrac{i-3}{2}\tau$ \\
        \bottomrule
    \end{tabular}%
};
\end{tikzpicture}
}
\ee

We compute the superconformal index of $FM_3$, extrapolate the Hilbert series of the main branch of the moduli space from it, analyse the chiral ring, and verify consistency between the two quantities. The main branch is generated by the bifundamentals $\Pi,\,\tilde\Pi$ and the adjoints $A_L,\,A_R$ of $SU(3)_L \times SU(3)_R$; all statements below pertain to this branch.

\paragraph{Superconformal index.}
$Z$-extremisation did not converge to adequate precision for $FM_3$; we therefore fix the R-charges by hand to $\tau = 5/11$ and $\Delta = 7/9$. Refining only in the abelian fugacities $\psi_\tau$ and $\chi_\Delta$, the index reads
\begin{equation}
\begin{split}
\CI_{FM_3}(x) ={}& 1 + 18 x^{7/9} \chi_\Delta + x^{89/99} \chi_\Delta^{-2} \psi_\tau^{-1} - x^{10/11} \psi_\tau^{-2} - 18 x^{122/99} \chi_\Delta \psi_\tau^{-1} \\
& + x^{134/99} \chi_\Delta^{-2} \psi_\tau^{-2} - x^{15/11} \psi_\tau^{-3} + 16 x^{17/11} \psi_\tau + 155 x^{14/9} \chi_\Delta^2 \\
& + 18 x^{166/99} \chi_\Delta^{-1} \psi_\tau^{-1} - 18 x^{167/99} \chi_\Delta \psi_\tau^{-2} + x^{178/99} \chi_\Delta^{-4} \psi_\tau^{-2} \\
& - x^{179/99} \chi_\Delta^{-2} \psi_\tau^{-3} + x^{197/99} \chi_\Delta^{-2} \psi_\tau - 19 x^2 - 306 x^{199/99} \chi_\Delta^2 \psi_\tau^{-1} \\
& - 18 x^{211/99} \chi_\Delta^{-1} \psi_\tau^{-2} + 18 x^{212/99} \chi_\Delta \psi_\tau^{-3} + x^{223/99} \chi_\Delta^{-4} \psi_\tau^{-3} \\
& - 2 x^{224/99} \chi_\Delta^{-2} \psi_\tau^{-4} + x^{25/11} \psi_\tau^{-5} + 270 x^{230/99} \chi_\Delta \psi_\tau + 870 x^{7/3} \chi_\Delta^3 \\
& + 17 x^{22/9} \chi_\Delta^{-2} + 56 x^{27/11} \psi_\tau^{-1} + 80 x^{244/99} \chi_\Delta^2 \psi_\tau^{-2} + 18 x^{85/33} \chi_\Delta^{-3} \psi_\tau^{-2} \\
& - 18 x^{256/99} \chi_\Delta^{-1} \psi_\tau^{-3} + x^{89/33} \chi_\Delta^{-6} \psi_\tau^{-3} - x^{269/99} \chi_\Delta^{-2} \psi_\tau^{-5} + 18 x^{274/99} \chi_\Delta^{-1} \psi_\tau \\
& - 668 x^{25/9} \chi_\Delta - 2410 x^{92/33} \chi_\Delta^3 \psi_\tau^{-1} + x^{26/9} \chi_\Delta^{-4} - 19 x^{287/99} \chi_\Delta^{-2} \psi_\tau^{-1} \\
& - 141 x^{32/11} \psi_\tau^{-2} + 159 x^{289/99} \chi_\Delta^2 \psi_\tau^{-3} + O\!\left(x^{298/99}\right).
\end{split}
\end{equation}

\paragraph{Main-branch Hilbert series.}
Extrapolating from the index, the first terms of the (unrefined) main-branch Hilbert series are
\be\label{hsfm3}
\begin{split}
\CH\CS_{FM_3}(t_a, t_d) ={}&
1 + 18 t_d + 16 t_a \\
& + 155 t_d^2 + 270 t_a t_d + 135 t_a^2 \\
& + 870 t_d^3 + 2174 t_a t_d^2 + 2142 t_a^2 t_d + 799 t_a^3 \\
& + 3676 t_d^4 + 11510 t_a t_d^3 + 16264 t_a^2 t_d^2 + 11970 t_a^3 t_d + 3724 t_a^4 \\
& + 12682 t_d^5 + 46308 t_a t_d^4 + 81900 t_a^2 t_d^3 + 86303 t_a^3 t_d^2 + 52920 t_a^4 t_d + 14553 t_a^5 \\
& + 37576 t_d^6 + 153344 t_a t_d^5 + 315963 t_a^2 t_d^4 + 416030 t_a^3 t_d^3 \\
&\quad + 364400 t_a^4 t_d^2 + 197064 t_a^5 t_d + 49588 t_a^6 \\
& + 98890 t_d^7 + 438920 t_a t_d^6 + 1010000 t_a^2 t_d^5 + 1547389 t_a^3 t_d^4 + 1690262 t_a^4 t_d^3  \\
& \quad + 1302259 t_a^5 t_d^2 + \cdots \\
& + 236665 t_d^8 + 1121648 t_a t_d^7 + 2805660 t_a^2 t_d^6 + 4796056 t_a^3 t_d^5 + \cdots \\
& + 523952 t_d^9 + 2617340 t_a t_d^8 + \cdots \\
& + 1086943 t_d^{10} + \cdots ,
\end{split}
\ee
where $t_a \equiv x^{2-\tau}$ and $t_d \equiv x^\Delta$.

Two specialisations of \eqref{hsfm3} are informative. Setting $t_d = 0$ kills the bifundamentals, leaving only the adjoint sector, and the resulting series matches that of $FT_3$:
\be
\CH\CS_{FM_3}(t_a \equiv t,\, t_d = 0) = 1 + 16 t + 135 t^2 + 799 t^3 + \cdots = \frac{(1+t)(1 + t + t^2)}{(1-t)^{14}}.
\ee
Setting $t_a = 0$ we have no a priori closed form. Fitting the ansatz
\be
\CH\CS_{FM_3}(t_a = 0,\, t_d \equiv t) = \frac{1 + a_1 t + a_2 t^2 + \cdots}{(1-t)^N}
\ee
against the computed terms of \eqref{hsfm3} fixes
\be
\CH\CS_{FM_3}(t_a = 0,\, t_d \equiv t) = \frac{1 + 8 t + 20 t^2 + 10 t^3 + t^4}{(1-t)^{10}}.
\ee
The numerator here is not palindromic, so this sector is not Calabi--Yau.

\paragraph{Closed form.}
Combining the computed terms of \eqref{hsfm3} with the structural properties listed in section \ref{fmsun_hs} (and reproduced below), we extrapolate
\be
\CH\CS_{FM_3}(t_a, t_d) = \frac{N(t_a, t_d)}{(1 - t_a)^{14}\,(1 - t_d)^{10}},
\qquad
N(t_a, t_d) = \sum_{j,k} a_{j,k}\, t_a^j\, t_d^k,
\ee
with coefficient matrix (the rows indexing $j = 0, \dots, 6$ and the columns $k = 0, \dots, 8$)
\be
a_{j,k} \;\doteq\;
\begin{pmatrix}
 1 &   8 &  20 &   10 &     1 &    0 &   0 &    0 &  0 \\
 2 &  -2 & -86 & -140 &   -16 &    2 &   0 &    0 &  0 \\
 2 & -20 &  23 &  360 &   272 &  -38 &   1 &    0 &  0 \\
 1 & -10 &  66 & -170 &  -574 & -170 &  66 &  -10 &  1 \\
 0 &   0 &   1 &  -38 &   272 &  360 &  23 &  -20 &  2 \\
 0 &   0 &   0 &    2 &   -16 & -140 & -86 &   -2 &  2 \\
 0 &   0 &   0 &    0 &     1 &   10 &  20 &    8 &  1
\end{pmatrix}.
\ee
The matrix is palindromic, $a_{j,k} = a_{6-j,\,8-k}$, and its entries sum to zero, $\sum_{j,k} a_{j,k} = 0$, mirroring the structure observed for the closed-form Hilbert series of $FC_3$ and $FH_3$.

The numerator satisfies the following specialisations:
\be
\begin{aligned}
N(t, 1) &= 40\,(1 - t)^6, \\
N(1, t) &= 6\,(1 - t)^6 (1 + t)^2, \\
N(t, t) &= (1 - t)^6 (1 + t)^2 \bigl(t^6 + 14 t^5 + 72 t^4 + 133 t^3 + 72 t^2 + 14 t + 1\bigr).
\end{aligned}
\ee
The first two in turn fix the row and column sums of $a_{j,k}$ (via the binomial expansion of $(1-t)^6$ and of $(1-t)^6(1+t)^2$, respectively).

\paragraph{Chiral ring.}
The leading terms of the plethystic logarithm are
\be
\begin{split}
\mathrm{PL}\!\left[\CH\CS_{FM_3}\right](t_a, t_d) ={}&
\bigl(\chi_{(\Box,\overline{\Box})} + \chi_{(\overline{\Box},\Box)}\bigr)\, t_d
+ \bigl(\chi_{(\mathrm{adj}_L,\mathbf{1})} + \chi_{(\mathbf{1},\mathrm{adj}_R)}\bigr)\, t_a \\
& - t_a^2 - t_a^3
- \bigl(\chi_{(\mathrm{adj}_L,\mathbf{1})} + \chi_{(\mathbf{1},\mathrm{adj}_R)}\bigr)\, t_d^2
- \bigl(\chi_{(\Box,\overline{\Box})} + \chi_{(\overline{\Box},\Box)}\bigr)\, t_a t_d + \cdots \\
\overset{\mathrm{unr}}{=}\;\,& 18\, t_d + 16\, t_a - t_a^2 - t_a^3 - 16\, t_d^2 - 18\, t_a t_d + \cdots,
\end{split}
\ee
where $\Box$ and $\overline{\Box}$ denote the (anti-)fundamental of $SU(3)$. The positive leading terms identify the generators: the bifundamentals $\Pi, \tilde\Pi$ of $SU(3)_L \times SU(3)_R$ contributing $\bigl(\chi_{(\Box,\overline{\Box})} + \chi_{(\overline{\Box},\Box)}\bigr)\,t_d$, and the adjoints $A_L, A_R$ contributing $\bigl(\chi_{(\mathrm{adj}_L,\mathbf{1})} + \chi_{(\mathbf{1},\mathrm{adj}_R)}\bigr)\,t_a$. They satisfy the relations (with their PL contributions shown on the right):
\be
\begin{aligned}
\bigl(\Pi \tilde\Pi\bigr)\big|_{\mathrm{traceless}} = 0 = \bigl(\tilde\Pi \Pi\bigr)\big|_{\mathrm{traceless}},
&& -\bigl(\chi_{(\mathrm{adj}_L,\mathbf{1})} + \chi_{(\mathbf{1},\mathrm{adj}_R)}\bigr)\, t_d^2 \\
\mathrm{tr}_L A_L^k = \mathrm{tr}_R A_R^k, \quad k = 2, 3,
&& -t_a^2 - t_a^3 \\
A_L\, \Pi = \Pi\, A_R,
&& -\chi_{(\Box,\overline{\Box})}\, t_a t_d \\
\tilde\Pi\, A_L = A_R\, \tilde\Pi,
&& -\chi_{(\overline{\Box},\Box)}\, t_a t_d .
\end{aligned}
\ee
The PL is not a finite sum --- higher syzygies appear --- so these relations are not independent and the main branch of the moduli space is not a complete intersection.

\paragraph{Consistency check.}
The unrefined main-branch Hilbert series is recovered from the index in the limit
\be
\begin{split}
\lim_{y \to 0} \CI_{FM_3}&\!\left(x \to xy,\ \psi_\tau \to \psi_\tau\, y^{\tau - 2},\ \chi_\Delta \to \chi_\Delta\, y^{-\Delta}\right)
= \CH\CS^{\mathrm{unr}}_{FM_3}\!\left(x^{2-\tau}\psi_\tau,\, x^\Delta \chi_\Delta\right) \\
&= 1 + 18\, x^{7/9} \chi_\Delta\, + 16\, x^{17/11} \psi_\tau\,  + 155\,x^{14/9} \chi_\Delta^2\, + 270\, x^{230/99} \chi_\Delta \psi_\tau\,  \\
&\quad + 135\, x^{34/11} \psi_\tau^2\,  + 870\,x ^{7/3} \chi_\Delta^3\,  + \cdots.
\end{split}
\ee

\subsection{\texorpdfstring{$FE_3$}{feusp6} }

The theory we consider has the following UV completion

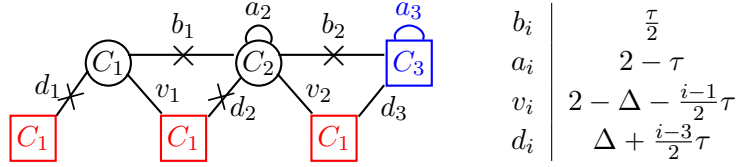
\be
\resizebox{\hsize}{!}{
\begin{tikzpicture}[
    thick,
    gauge/.style={
        circle, draw=black, fill=white,
        minimum size=8mm, inner sep=1pt, font=\small
    },
    manifest/.style={
        rectangle, draw=blue!70!black, fill=blue!6,
        minimum size=8mm, inner sep=2pt,
        font=\small, text=blue!70!black
    },
    flavour/.style={
        rectangle, draw=red!70!black, fill=red!6,
        minimum size=7mm, inner sep=2pt,
        font=\small, text=red!70!black
    },
    barr/.style={draw=black!70}
]
\node[gauge]    (g1) at (0,0)   {$C_1$};
\node[gauge]    (g2) at (2.5,0) {$C_2$};
\node[manifest] (m1) at (5,0)   {$C_3$};
\node[flavour] (f1) at (-0.8,-1.8) {$C_1$};
\node[flavour] (f2) at (1.7,-1.8)    {$C_1$};
\node[flavour] (f3) at (4.2,-1.8)  {$C_1$};
\draw[barr] (g1) -- (g2);
\node at (1.25, 0.45) {$b_1$};
\node at (1.25, 0) {\huge$\times$};
\draw[barr] (g2) -- (m1);
\node at (3.75, 0.45) {$b_2$};
\node at (3.58, 0) {\huge$\times$};
\draw (g2) to[out=65,in=115,looseness=3] (g2);
\node at (2.5, 0.95) {$a_2$};
\draw[blue!70!black] (m1) to[out=65,in=115,looseness=3] (m1);
\node[blue!70!black] at (5, 0.95) {$a_{3}$};
\draw[barr] (f1) -- (g1);
\node[rotate=-30] at (-0.42,-0.9) {\huge$\times$};
\node at (-0.75,-0.55) {$d_1$};
\draw[barr] (g1) -- (f2);
\node at (1.1,-0.6) {$v_1$};
\draw[barr] (f2) -- (g2);
\node[rotate=-30] at (2.05,-0.9) {\huge$\times$};
\node at (1.7,-0.6) {$d_2$};
\draw[barr] (g2) -- (f3);
\node at (3.55,-0.6) {$v_2$};
\draw[barr] (f3) -- (m1);
\node at (4.9,-0.9) {$d_3$};
\node at (2.5,-3.2) {$\mathcal{W} =
    b_1^2 a_2 + b_2^2(a_2 + a_3)
    + \sum_{i=1}^{2}\left[v_i b_i d_{i+1}
    + \mathrm{Flip}[b_i^2]\right]
    + \sum_{j=1}^{2}\mathrm{Flip}[d_j^2]
    + \mathfrak{M}^{1\,0} + \mathfrak{M}^{0\,1}$};
\node[right] at (6.5,-0.5) {%
    \renewcommand{\arraystretch}{1.4}
    \begin{tabular}{>{\small}c >{\small}c}
        \toprule
        $b_i$   & $\tau/2$ \\
        $a_i$   & $2-\tau$ \\
        $v_i$   & $2-\Delta-\tfrac{i-1}{2}\tau$ \\
        $d_i$   & $\Delta+\tfrac{i-3}{2}\tau$ \\
        \bottomrule
    \end{tabular}%
};
\end{tikzpicture}
}
\ee

We compute the superconformal index of $FE_3$, extrapolate the Hilbert series of the main branch of the moduli space from it, analyse the chiral ring, and verify consistency between the two quantities. The main branch is generated by the bifundamental $\Pi$ of $USp(6)_L \times USp(6)_R$ and by the two traceless-antisymmetric fields $A_L$, $A_R$ of $USp(6)_L$ and $USp(6)_R$ respectively; all statements below pertain to this branch.

\paragraph{Superconformal index.}
$Z$-extremisation did not converge to adequate precision for $FE_3$; we therefore fix the R-charges by hand to $\tau = 15/14$ and $\Delta = 13/12$.  This choice ensures that all chiral-ring operators sit above the unitarity bound and that the integer multiplicities of $\frac{1}{2}$-BPS terms (which alone enter the Hilbert series) are independent of small perturbations of $(\tau,\Delta)$, so the cross-check below is insensitive to the precise numerical values.  Refining only in the abelian fugacities $\psi_\tau$ and $\chi_\Delta$, the index up to order $x^3$ reads
\begin{equation}
\begin{split}
\CI_{FE_3}(x) ={}& 1 + x^{16/21} \chi_\Delta^{-2} \psi_\tau + x^{19/21} \chi_\Delta^{-2} \psi_\tau^{-1} + 28 x^{13/14} \psi_\tau + 36 x^{13/12} \chi_\Delta \\
& + x^{32/21} \chi_\Delta^{-4} \psi_\tau^2 + x^{5/3} \chi_\Delta^{-4} + 29 x^{71/42} \chi_\Delta^{-2} \psi_\tau^2 + x^{38/21} \chi_\Delta^{-4} \psi_\tau^{-2} \\
& + 29 x^{11/6} \chi_\Delta^{-2} + 36 x^{155/84} \chi_\Delta^{-1} \psi_\tau + 405 x^{13/7} \psi_\tau^2 + x^{83/42} \chi_\Delta^{-2} \psi_\tau^{-2} \\
& + 36 x^{167/84} \chi_\Delta^{-1} \psi_\tau^{-1} - 44 x^2 + 972 x^{169/84} \chi_\Delta \psi_\tau - x^{15/7} \psi_\tau^{-2} \\
& - 36 x^{181/84} \chi_\Delta \psi_\tau^{-1} + 638 x^{13/6} \chi_\Delta^2 + x^{16/7} \chi_\Delta^{-6} \psi_\tau^3 + x^{17/7} \chi_\Delta^{-6} \psi_\tau \\
& + 29 x^{103/42} \chi_\Delta^{-4} \psi_\tau^3 + x^{18/7} \chi_\Delta^{-6} \psi_\tau^{-1} + 30 x^{109/42} \chi_\Delta^{-4} \psi_\tau \\
& + 36 x^{73/28} \chi_\Delta^{-3} \psi_\tau^2 + 433 x^{55/21} \chi_\Delta^{-2} \psi_\tau^3 + x^{19/7} \chi_\Delta^{-6} \psi_\tau^{-3} \\
& + 30 x^{115/42} \chi_\Delta^{-4} \psi_\tau^{-1} + 36 x^{11/4} \chi_\Delta^{-3} + 362 x^{58/21} \chi_\Delta^{-2} \psi_\tau \\
& + 972 x^{233/84} \chi_\Delta^{-1} \psi_\tau^2 + 4031 x^{39/14} \psi_\tau^3 + x^{121/42} \chi_\Delta^{-4} \psi_\tau^{-3} \\
& + 36 x^{81/28} \chi_\Delta^{-3} \psi_\tau^{-2} - 44 x^{61/21} \chi_\Delta^{-2} \psi_\tau^{-1} + 768 x^{35/12} \chi_\Delta^{-1} \\
& - 764 x^{41/14} \psi_\tau + 13572 x^{247/84} \chi_\Delta \psi_\tau^2 + O\!\left(x^{253/84}\right) .
\end{split}
\end{equation}
The bifundamental contributes at order $x^\Delta$ and the two adjoints at order $x^{2-\tau}$, as expected. The term $-44\, x^2$ signals an IR enhancement of the flavour symmetry,
\begin{equation*}
SU(2)^3 \times USp(6) \times U(1)^2 \;\longrightarrow\; USp(6) \times USp(6) \times U(1)^2,
\end{equation*}
with $\dim\bigl(USp(6) \times USp(6) \times U(1)^2\bigr) = 21 + 21 + 1 + 1 = 44$ matching the enhanced (rather than the manifest) symmetry.

Beyond these, the index exhibits chiral singlets $B_{n,m}$ with R-charges
\be
R(B_{n,m}) = 2n - 2\Delta + (m - n)\tau, \qquad n, m \in \{1, 2, 3\}.
\ee
Aside from $B_{1,1}$ (removed from the spectrum as discussed in the main text), the singlets $B_{1,2}, B_{1,3}, B_{2,1}, B_{2,2}, B_{3,1}$ appear in the spectrum at orders $x^{k\,R(B_{n,m})}$ for $k = 1, 2, 3, \dots$.

\paragraph{Main-branch Hilbert series.}
Extrapolating from the index, the main-branch Hilbert series (unrefined, retaining only the two abelian fugacities) begins
\be
\begin{split}
\CH\CS_{FE_3}={}&
1 + 28\, t_a + 36\, t_d \\
& + 405\, t_a^2 + 972\, t_a t_d + 638\, t_d^2 \\
& + 4031\, t_a^3 + 13572\, t_a^2 t_d + 16611\, t_a t_d^2 + 7464\, t_d^3 \\
& + 31031\, t_a^4 + 130572\, t_a^3 t_d + 224133\, t_a^2 t_d^2 + 187732\, t_a t_d^3 + 65235\, t_d^4 \\
& + 196911\, t_a^5 + 972972\, t_a^4 t_d + 2087911\, t_a^3 t_d^2 + 2453100\, t_a^2 t_d^3 + 1588943\, t_a t_d^4 + 456704\, t_d^5 \\
& + 1072071\, t_a^6 + 5985252\, t_a^5 t_d + 15092951\, t_a^4 t_d^2 + 22179136\, t_a^3 t_d^3 \\
& \quad + 20157291\, t_a^2 t_d^4 + 10800776\, t_a t_d^5 + 2679281\, t_d^6 \\
& + 5147181\, t_a^7 + 31636332\, t_a^6 t_d + 90224617\, t_a^5 t_d^2 + 155914136\, t_a^4 t_d^3 \\
& \quad + 177318427\, t_a^3 t_d^4 + 133352820\, t_a^2 t_d^5 + 61676999\, t_a t_d^6 + 13593880\, t_d^7 \\
& + 22230936\, t_a^8 + 147676932\, t_a^7 t_d + 464196060\, t_a^6 t_d^2 + 907997124\, t_a^5 t_d^3 \\
& \quad + 1215116510\, t_a^4 t_d^4 + 1144031020\, t_a^3 t_d^5 + 742856373\, t_a^2 t_d^6 \\
& \quad + 305312868\, t_a t_d^7 + 61055281\, t_d^8 \\
& + 87687561\, t_a^9 + 621000432\, t_a^8 t_d + 2112294796\, t_a^7 t_d^2 + 4558313900\, t_a^6 t_d^3 \\
& \quad + 6909969942\, t_a^5 t_d^4 + 7659366316\, t_a^4 t_d^5 + 6228644954\, t_a^3 t_d^6 \\
& \quad + 3594881532\, t_a^2 t_d^7 + 1340726475\, t_a t_d^8 + 247112712\, t_d^9 \\
& + 319616076\, t_a^{10} + 2388074832\, t_a^9 t_d + 8670961286\, t_a^8 t_d^2 \\
& \quad + 20268965340\, t_a^7 t_d^3 + 33924376550\, t_a^6 t_d^4 + 42621254332\, t_a^5 t_d^5 \\
& \quad + 40823937118\, t_a^4 t_d^6 + 29517163820\, t_a^3 t_d^7 + 15462146865\, t_a^2 t_d^8 \\
& \quad + 5315606912\, t_a t_d^9 + 913996149\, t_d^{10} + \cdots,
\end{split}
\ee
where $t_a \equiv x^{2-\tau}$ and $t_d \equiv x^\Delta$.  We have computed this expansion up to and including total degree $14$; the terms of degree $11,\dots,14$ are not displayed, but they are reproduced by the closed form given below.

Two specialisations have closed form. Setting $t_d = 0$ retains only the adjoint sector, which --- as in the $FT_N$ theories --- is generated by $A_L$ and $A_R$ subject to $\mathrm{tr}_L A_L^k = \mathrm{tr}_R A_R^k$ for $k = 2, 3$. The resulting series is
\be
\CH\CS_{FE_3}(t_a \equiv t,\, t_d = 0) = 1 + 28 t + 405 t^2 + 4031 t^3 + \cdots = \frac{(1 - t^2)(1 - t^3)}{(1 - t)^{28}},
\ee
the power $28 = 2 \cdot (\dim \mathrm{Asym}_{USp(6)} - 1) = 2 \cdot 14$ counting the two copies of the traceless-antisymmetric. Setting $t_a = 0$, the computed terms fix
\begin{equation}
    \begin{aligned}
\CH\CS_{FE_3}(t_a = 0,\, t_d \equiv t) = & \quad \frac{1 + 14 t + 77 t^2 + 204 t^3 + 280 t^4 + 204 t^5 + 77 t^6 + 14 t^7 + t^8}{(1 - t)^{22}}  \\
 = & \quad \frac{(1 + t)^2 \bigl(1 + 12 t + 52 t^2 + 88 t^3 + 52 t^4 + 12 t^5 + t^6\bigr)}{(1 - t)^{22}} .
    \end{aligned}
\end{equation}
Both expressions have been checked against the explicit expansion through order $t^{14}$, obtained by taking the limit on the index computed order by order, as described in Section~\ref{sec:computation}.  The numerator of the second is palindromic, so this sector has the Calabi--Yau property.

\paragraph{Closed form.}
With the expansion available up to total degree $14$ the unrefined Hilbert series in a single fugacity is fixed outright, with no palindromicity assumption:
\begin{equation}
\begin{aligned}
\CH\CS_{FE_3}(t, t) = &\frac{(1 + t)^2 \bigl(1 + 26 t + 288 t^2 + 1716 t^3 + 5970 t^4 + 12545 t^5 + 16071 t^6 }{(1 - t)^{36}} \\
&\frac{+ 12545 t^7 + 5970 t^8 + 1716 t^9 + 288 t^{10} + 26 t^{11} + t^{12}\bigr)}{(1 - t)^{36}}
\end{aligned},
\end{equation}
\be
\CH\CS_{FE_3}(t,t) = 1 + 64\,t + 2015\,t^2 + 41678\,t^3 + 638703\,t^4 + 7756541\,t^5 + 77966758\,t^6 + \cdots .
\ee
That the numerator comes out palindromic is now a result rather than an input.

The refined numerator $N(t_a,t_d)$ in
\be
\CH\CS_{FE_3}(t_a, t_d) = \frac{\sum_{j,k} a_{j,k}\, t_a^j\, t_d^k}{(1 - t_a)^{26}(1 - t_d)^{22}}
\ee
has support $0\le j\le 12$, $0\le k\le 14$.  The degree-$14$ data determine every $a_{j,k}$ with $j+k\le 14$ directly; imposing palindromicity,
\be
a_{j,k} = a_{12-j,\,14-k},
\ee
fixes the remaining entries from coefficients of degree $\le 11$.  The two determinations overlap on the $39$ entries with $12 \le j+k \le 14$, giving $19$ non-trivial consistency checks, all of which are satisfied.  The numerator is therefore completely determined; writing $N(t_a,t_d) = \sum_{j=0}^{12} t_a^j\, n_j(t_d)$,
\be
\begin{aligned}
n_0 &= 1 + 14 t_d + 77 t_d^2 + 204 t_d^3 + 280 t_d^4 + 204 t_d^5 + 77 t_d^6 + 14 t_d^7 + t_d^8 ,\\
n_1 &= 2 - 8 t_d - 307 t_d^2 - 1602 t_d^3 - 3360 t_d^4 - 3294 t_d^5 - 1541 t_d^6 - 328 t_d^7 - 26 t_d^8 ,\\
n_2 &= 2 - 44 t_d + 59 t_d^2 + 3654 t_d^3 + 14589 t_d^4 + 21648 t_d^5 + 13705 t_d^6 + 3614 t_d^7 + 325 t_d^8 ,\\
n_3 &= 1 - 22 t_d + 459 t_d^2 - 1520 t_d^3 - 26473 t_d^4 - 70030 t_d^5 - 67060 t_d^6 \\
    &\quad - 24490 t_d^7 - 2720 t_d^8 + 14 t_d^9 + t_d^{10} ,\\
n_4 &= -10 t_d^2 - 2080 t_d^3 + 16723 t_d^4 + 112070 t_d^5 + 183050 t_d^6 + 104326 t_d^7 \\
    &\quad + 17747 t_d^8 - 160 t_d^9 - 26 t_d^{10} ,\\
n_5 &= -24 t_d^2 + 796 t_d^3 + 967 t_d^4 - 82390 t_d^5 - 267988 t_d^6 - 261506 t_d^7 \\
    &\quad - 78952 t_d^8 - 2062 t_d^9 + 556 t_d^{10} - 22 t_d^{11} + t_d^{12} ,\\
n_6 &= 3 t_d^2 - 30 t_d^3 - 2531 t_d^4 + 23868 t_d^5 + 202392 t_d^6 + 358324 t_d^7 \\
    &\quad + 202392 t_d^8 + 23868 t_d^9 - 2531 t_d^{10} - 30 t_d^{11} + 3 t_d^{12} ,
\end{aligned}
\ee
with the remaining rows given by $n_j(t_d) = t_d^{14}\, n_{12-j}(1/t_d)$ for $j = 7,\dots,12$.  The palindromicity of the full refined numerator implies that the main branch is Calabi--Yau, not merely its $t_a = 0$ locus.

The specialisations anticipated earlier are confirmed,
\be
N(t_a = 1, t_d \equiv t) = 6(1 - t)^{12}(1 + t)^2, \qquad
N(t_a \equiv t, t_d = 1) = 872(1 - t)^{12},
\ee
as are the two branch limits $N(t,0) = (1+t)(1+t+t^2)$ and $N(0,t) = (1+t)^2(1 + 12t + 52t^2 + 88t^3 + 52t^4 + 12t^5 + t^6)$, and the fully unrefined specialisation
\be
N(t,t) = (1-t)^{12}(1+t)^2\bigl(1 + 26 t + 288 t^2 + \cdots + 26 t^{11} + t^{12}\bigr).
\ee

\paragraph{Chiral ring.}
The leading terms of the plethystic logarithm are
\be
\begin{aligned}
\mathrm{PL}\!\left[\CH\CS_{FE_3}\right](t_a, t_d) ={}&
\bigl(\chi_{(\mathrm{asym}_L,\mathbf{1})} + \chi_{(\mathbf{1},\mathrm{asym}_R)}\bigr)\, t_a
+ \chi_{(\mathbf{6},\mathbf{6})}\, t_d \\
& - t_a^2 - t_a^3
- \bigl(\chi_{(\mathrm{asym}_L,\mathbf{1})} + \chi_{(\mathbf{1},\mathrm{asym}_R)}\bigr)\, t_d^2
- \chi_{(\mathbf{6},\mathbf{6})}\, t_a t_d + \cdots \\
\overset{\mathrm{unr}}{=}\;\,& 28\, t_a + 36\, t_d - t_a^2 - t_a^3 - 36\, t_a t_d - 28\, t_d^2 \\
& + 36\, t_d^3 + 43\, t_a t_d^2 + 36\, t_a^3 t_d + 197\, t_a^2 t_d^2 + 160\, t_a t_d^3 - 42\, t_d^4 + \cdots .
\end{aligned}
\ee
The positive leading terms identify the generators: the bifundamental $\Pi^{a,A}$ (contributing $\chi_{(\mathbf{6},\mathbf{6})}\, t_d$) and the two traceless-antisymmetric fields $A_L, A_R$, contributing $\chi_{(\mathrm{asym}_L,\mathbf{1})}\, t_a$ and $\chi_{(\mathbf{1},\mathrm{asym}_R)}\, t_a$. They satisfy the relations (with their PL contributions shown on the right):
\be
\begin{aligned}
\Omega_{ab}\, \Pi^{a,A}\Pi^{b,B} = 0 = \Omega_{AB}\, \Pi^{a,A}\Pi^{b,B},
&& -\bigl(\chi_{(\mathrm{asym}_L,\mathbf{1})} + \chi_{(\mathbf{1},\mathrm{asym}_R)}\bigr)\, t_d^2 \\
\mathrm{tr}_L A_L^k = \mathrm{tr}_R A_R^k, \quad k = 2, 3,
&& -t_a^2 - t_a^3 \\
\Omega_{ac}\, A_L^{\,ab}\, \Pi^{c,A} = \Omega_{BC}\, A_R^{\,AB}\, \Pi^{b,C},
&& -\chi_{(\mathbf{6},\mathbf{6})}\, t_a t_d .
\end{aligned}
\ee
The PL is not a finite sum --- the degree-$14$ data show non-vanishing terms at every order we can reach --- so the relations are not independent and the main branch of the moduli space is not a complete intersection.

\paragraph{Consistency check.}
The unrefined main-branch Hilbert series is recovered from the index in the limit
\be
\begin{split}
\lim_{y \to 0} \CI_{FE_3}&\!\left(x \to xy,\ \psi_\tau \to \psi_\tau\, y^{\tau - 2},\ \chi_\Delta \to \chi_\Delta\, y^{-\Delta}\right)
= \CH\CS^{\mathrm{unr}}_{FE_3}\!\left(x^{2-\tau}\psi_\tau,\, x^\Delta \chi_\Delta\right) \\
&= 1   + 28\, x^{13/14}\psi_\tau\, + 36\, x^{13/12}\chi_\Delta\, + 405\, x^{13/7} \psi_\tau^2\, \\
&\quad + 972\, x^{169/84}\chi_\Delta \psi_\tau\, + 638\, x^{13/6} \chi_\Delta^2\, + 7464\, x^{13/4} \chi_\Delta^3\, + \cdots,
\end{split}
\ee
and the coefficients match the Hilbert series order by order, up to total degree $14$.

\section{Hilbert series for improved bifundamentals at  \texorpdfstring{$N=4$}{n=4}}\label{AppN=4}

In the following section, we give the SCIs for the first few orders of the improved bifundamentals at $N=4$, and check (whenever possible) that our proposals for the HS agree with the single branch limits of the SCI.

\subsection{\texorpdfstring{$FT_4$}{ftsu4} }
The $FT_4$ theory is given by the following UV completion:

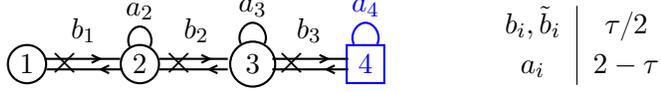
\be
\resizebox{\hsize}{!}{
\begin{tikzpicture}[
    thick,
    gauge/.style={
        circle, draw=black, fill=white,
        minimum size=8mm, inner sep=1pt, font=\small
    },
    manifest/.style={
        rectangle, draw=blue!70!black, fill=blue!6,
        minimum size=8mm, inner sep=2pt,
        font=\small, text=blue!70!black
    },
    flavour/.style={
        rectangle, draw=red!70!black, fill=red!6,
        minimum size=7mm, inner sep=2pt,
        font=\small, text=red!70!black
    },
    arr/.style={-{Stealth[scale=0.8]}, shorten >=1pt, shorten <=1pt},
    barr/.style={draw=black!70}
]
\node[gauge]    (g1) at (0,0)    {$1$};
\node[gauge]    (g2) at (2.5,0)  {$2$};
\node[gauge]    (g3) at (5,0)    {$3$};
\node[manifest] (m4) at (7.5,0)  {$4$};
\draw[arr,barr,transform canvas={yshift= 3.5pt}] (g1) -- (g2);
\draw[arr,barr,transform canvas={yshift=-3.5pt}] (g2) -- (g1);
\node at (1.25, 0.45) {$b_1$};
\node at (1.25, 0) {\huge$\times$};
\draw[arr,barr,transform canvas={yshift= 3.5pt}] (g2) -- (g3);
\draw[arr,barr,transform canvas={yshift=-3.5pt}] (g3) -- (g2);
\node at (3.75, 0.45) {$b_2$};
\node at (3.75, 0) {\huge$\times$};
\draw[arr,barr,transform canvas={yshift= 3.5pt}] (g3) -- (m4);
\draw[arr,barr,transform canvas={yshift=-3.5pt}] (m4) -- (g3);
\node at (6.25, 0.45) {$b_3$};
\node at (6.25, 0) {\huge$\times$};
\draw (g2) to[out=65,in=115,looseness=3] (g2);
\node at (2.5, 0.95) {$a_2$};
\draw (g3) to[out=65,in=115,looseness=3] (g3);
\node at (5, 0.95) {$a_{3}$};
\draw[blue!70!black] (m4) to[out=65,in=115,looseness=3] (m4);
\node[blue!70!black] at (7.5, 0.95) {$a_{4}$};
\node at (3.75,-1.5) {$\mathcal{W} =
    a_2 b_1\tilde{b}_1
    + \sum_{j=2}^{3} b_j (a_j + a_{j+1})\tilde{b}_j
    + \sum_{i=1}^{3}\mathrm{Flip}[b_i\tilde{b}_i]$};
\node[right] at (9,-0.5) {%
    \renewcommand{\arraystretch}{1.4}
    \begin{tabular}{>{\small}c >{\small}c}
        \toprule
        $b_i,\,\tilde{b}_i$ & $\tau/2$ \\
        $a_i$                & $2-\tau$ \\
        \bottomrule
    \end{tabular}%
};
\end{tikzpicture}
}
\ee

We compute the superconformal index and Hilbert series of $FT_4$, and verify their consistency.

\paragraph{Superconformal index.}
Fixing $\tau = 13/10$ and refining in the $U(1)_\tau$ fugacity $\psi$, the index reads
\begin{equation}\label{eqftu4}
\begin{split}
\CI_{FT_4}(x, \psi) ={}& 1 + 30\, \psi\, x^{7/10} + 464\, \psi^2 x^{7/5} - 31\, x^2 + 4929\, \psi^3 x^{21/10} - \psi^{-2} x^{13/5} \\
& - 899\, \psi\, x^{27/10} + 40424\, \psi^4 x^{14/5} + O\!\left(x^{31/10}\right).
\end{split}
\end{equation}

\paragraph{Hilbert series and chiral ring.}
The moduli space is a complete intersection: the chiral ring is generated by the adjoints $A_L$ and $A_R$ of $SU(4)_L$ and $SU(4)_R$, subject to the quadratic, cubic and quartic relations $\mathrm{tr}_L A_L^k = \mathrm{tr}_R A_R^k$ for $k = 2, 3, 4$. The refined Hilbert series is therefore
\be
\begin{aligned}
\CH\CS_{FT_4}(t) &= \mathrm{PE}\!\left[\chi_{[1,0,1]_{SU(4)_L}}\, t + \chi_{[1,0,1]_{SU(4)_R}}\, t - \sum_{j=2}^{4} t^j\right] \\
&\overset{\mathrm{unr}}{=} \frac{(1+t)(1 + t + t^2)(1 + t + t^2 + t^3)}{(1-t)^{27}} \\
&= 1 + 30\, t + 464\, t^2 + 4929\, t^3 + 40424\, t^4 + \cdots,
\end{aligned}
\ee
in agreement with \eqref{eqftu4}.

\paragraph{Consistency check.}
The unrefined Hilbert series is recovered from the index in the limit
\be
\lim_{y \to 0} \CI_{FT_4}\!\left(x \to xy,\, \psi \to \psi\, y^{\tau - 2}\right) = \CH\CS^{\mathrm{unr}}_{FT_4}\!\left(\psi\, x^{2-\tau}\right),
\ee
and the two expansions agree order by order.

\subsection{\texorpdfstring{$FC_4$}{FCSU4} }

The $FC_4$ theory has the following possible UV completion:

\be
\resizebox{\hsize}{!}{
\begin{tikzpicture}[
    thick,
    gauge/.style={
        circle, draw=black, fill=white,
        minimum size=8mm, inner sep=1pt, font=\small
    },
    manifest/.style={
        rectangle, draw=blue!70!black, fill=blue!6,
        minimum size=8mm, inner sep=2pt,
        font=\small, text=blue!70!black
    },
    flavour/.style={
        rectangle, draw=red!70!black, fill=red!6,
        minimum size=7mm, inner sep=2pt,
        font=\small, text=red!70!black
    },
    arr/.style={-{Stealth[scale=0.8]}, shorten >=1pt, shorten <=1pt},
    barr/.style={draw=black!70}
]
\node[gauge]    (g1) at (0,0)   {$1$};
\node[gauge]    (g2) at (2.5,0) {$2$};
\node[gauge]    (g3) at (5,0)   {$3$};
\node[manifest] (m1) at (7.5,0) {$4$};
\node[flavour] (f5) at (-0.8,-1.8) {$1$};
\node[flavour] (f2) at (1.7,-1.8)  {$1$};
\node[flavour] (f3) at (4.2,-1.8)  {$1$};
\node[flavour] (f4) at (6.7,-1.8)  {$1$};
\draw[arr,barr,transform canvas={yshift= 3.5pt}] (g1) -- (g2);
\draw[arr,barr,transform canvas={yshift=-3.5pt}] (g2) -- (g1);
\node at (1.25, 0.45) {$b_1,\,\tilde{b}_1$};
\node at (1.25, 0) {\huge$\times$};
\draw[arr,barr,transform canvas={yshift= 3.5pt}] (g2) -- (g3);
\draw[arr,barr,transform canvas={yshift=-3.5pt}] (g3) -- (g2);
\node at (3.75, 0.45) {$b_2,\,\tilde{b}_2$};
\node at (3.75, 0) {\huge$\times$};
\draw[arr,barr,transform canvas={yshift= 3.5pt}] (g3) -- (m1);
\draw[arr,barr,transform canvas={yshift=-3.5pt}] (m1) -- (g3);
\node at (6.25, 0.45) {$b_3,\,\tilde{b}_3$};
\node at (6.25, 0) {\huge$\times$};
\draw (g2) to[out=65,in=115,looseness=3] (g2);
\node at (2.5, 0.95) {$a_2$};
\draw (g3) to[out=65,in=115,looseness=3] (g3);
\node at (5, 0.95) {$a_3$};
\draw[blue!70!black] (m1) to[out=65,in=115,looseness=3] (m1);
\node[blue!70!black] at (7.5, 0.95) {$a_4$};
\draw[arr,barr] (f5) to[bend right=5] (g1);
\node at (-0.55,-0.7) {$d_1$};
\draw[arr,barr] (g1) to[bend right=5] (f2);
\node at (1.1,-0.7) {$v_1$};
\draw[arr,barr] (f2) to[bend right=5] (g2);
\node at (1.9,-0.7) {$d_2$};
\draw[arr,barr] (g2) to[bend right=5] (f3);
\node at (3.6,-0.7) {$v_2$};
\draw[arr,barr] (f3) to[bend right=5] (g3);
\node at (4.4,-0.7) {$d_3$};
\draw[arr,barr] (g3) to[bend right=5] (f4);
\node at (6.1,-0.7) {$v_3$};
\draw[arr,barr] (f4) to[bend right=5] (m1);
\node at (6.9,-0.7) {$d_4$};
\node at (5,-3.2) {$\mathcal{W} =
    \sum_{i=1}^{3}v_i b_i d_{i+1}
    + \left(\mathfrak{M}^{+00} + \mathfrak{M}^{0\,+0} + \mathfrak{M}^{00+}\right)
    + a_2 b_1\tilde{b}_1
    + \sum_{i=2}^{3} b_i(a_i + a_{i+1})\tilde{b}_i
    + \sum_{i=1}^{3}\mathrm{Flip}[b_i\tilde{b}_i]$};
\node[right] at (9,-0.5) {%
    \renewcommand{\arraystretch}{1.4}
    \begin{tabular}{>{\small}c >{\small}c}
        \toprule
        $b_i,\,\tilde{b}_i$ & $\tau/2$ \\
        $a_i$                & $2-\tau$ \\
        $v_i$                & $2-\tfrac{i-2}{2}\tau-\Delta$ \\
        $d_i$                & $\Delta+\tfrac{i-4}{2}\tau$ \\
        \bottomrule
    \end{tabular}%
};
\end{tikzpicture}
}
\ee

We compute the superconformal index and the Hilbert series of $FC_4$, analyse the chiral ring, and examine specialisations of the refined Hilbert series.

\paragraph{Superconformal index.}
Fixing $\tau = 3/4$ and $\Delta = 3/2$ and refining in the abelian fugacities $\psi_\tau$ and $\chi_\Delta$ for $U(1)_\tau$ and $U(1)_\Delta$, the index reads
\begin{equation}
\begin{split}
\CI_{FC_4} ={}& 1 + 30\, \psi_\tau\, x^{5/4} + x^{3/2}\bigl(16\, \chi_\Delta - \psi_\tau^{-2}\bigr) - 32\, x^2 + x^{9/4}\bigl(-16\, \chi_\Delta \psi_\tau^{-1} - \psi_\tau^{-3}\bigr) \\
& + 464\, \psi_\tau^2\, x^{5/2} + x^{11/4}\bigl(-16\, \chi_\Delta^{-1} \psi_\tau^{-3} + 464\, \chi_\Delta \psi_\tau - 30\, \psi_\tau^{-1}\bigr) \\
& + x^3\bigl(136\, \chi_\Delta^2 - 16\, \chi_\Delta \psi_\tau^{-2} - \psi_\tau^{-4}\bigr) + x^{13/4}\bigl(-176\, \chi_\Delta^{-1} \psi_\tau^{-1} - 930\, \psi_\tau\bigr) \\
& + x^{7/2}\bigl(16\, \chi_\Delta^{-1} \psi_\tau^{-4} - 960\, \chi_\Delta - 224\, \psi_\tau^{-2}\bigr) \\
& + x^{15/4}\bigl(-256\, \chi_\Delta^2 \psi_\tau^{-1} - 176\, \chi_\Delta^{-1} \psi_\tau + \psi_\tau^{-5} + 4929\, \psi_\tau^3\bigr) \\
& + x^4\bigl(6944\, \chi_\Delta \psi_\tau^2 + 192\, \chi_\Delta^{-1} \psi_\tau^{-2} - 826\bigr) \\
& + x^{17/4}\bigl(3824\, \chi_\Delta^2 \psi_\tau + 16\, \chi_\Delta^{-1} \psi_\tau^{-5} - 16\, \chi_\Delta^{-1} \psi_\tau^3 - 368\, \chi_\Delta \psi_\tau^{-1} + 482\, \psi_\tau^{-3}\bigr) \\
& + x^{9/2}\bigl(816\, \chi_\Delta^3 - 16\, \chi_\Delta^2 \psi_\tau^{-2} + 156\, \chi_\Delta^{-2} \psi_\tau^{-2} + 16\, \chi_\Delta \psi_\tau^{-4} - 3232\, \chi_\Delta^{-1} \\
& \phantom{{} + x^{9/2}\bigl(} + \psi_\tau^{-6} - 14174\, \psi_\tau^2\bigr) \\
& + x^{19/4}\bigl(36\, \chi_\Delta^{-2} \psi_\tau^{-5} + 736\, \chi_\Delta^{-1} \psi_\tau^{-3} - 21248\, \chi_\Delta \psi_\tau - 2139\, \psi_\tau^{-1}\bigr) \\
& + x^5\bigl(-16\, \chi_\Delta^{-3} \psi_\tau^{-2} - 11364\, \chi_\Delta^2 + 276\, \chi_\Delta^{-2} - 4608\, \chi_\Delta^{-1} \psi_\tau^2 + 160\, \chi_\Delta \psi_\tau^{-2} \\
& \phantom{{} + x^5\bigl(} + 40424\, \psi_\tau^4 + 31\, \psi_\tau^{-4}\bigr) \\
& + x^{21/4}\bigl(-2176\, \chi_\Delta^3 \psi_\tau^{-1} + 156\, \chi_\Delta^2 \psi_\tau^{-3} - 256\, \chi_\Delta^{-2} \psi_\tau^{-3} + 16\, \chi_\Delta \psi_\tau^{-5} \\
& \phantom{{} + x^{21/4}\bigl(} + 71440\, \chi_\Delta \psi_\tau^3 + 12864\, \chi_\Delta^{-1} \psi_\tau^{-1} + \psi_\tau^{-7} - 12544\, \psi_\tau\bigr) \\
& + x^{11/2}\bigl(\chi_\Delta^{-4} \psi_\tau^{-2} - 16\, \chi_\Delta^{-3} - 36\, \chi_\Delta^{-2} \psi_\tau^{-6} + 55544\, \chi_\Delta^2 \psi_\tau^2 + 156\, \chi_\Delta^{-2} \psi_\tau^2 \\
& \phantom{{} + x^{11/2}\bigl(} - 464\, \chi_\Delta^{-1} \psi_\tau^4 - 1168\, \chi_\Delta^{-1} \psi_\tau^{-4} - 1008\, \chi_\Delta + 16706\, \psi_\tau^{-2}\bigr) \\
& + x^{23/4}\bigl(22304\, \chi_\Delta^3 \psi_\tau + 3360\, \chi_\Delta^2 \psi_\tau^{-1} + 2632\, \chi_\Delta^{-2} \psi_\tau^{-1} - 16\, \chi_\Delta^{-1} \psi_\tau^{-7} \\
& \phantom{{} + x^{23/4}\bigl(} + 4608\, \chi_\Delta \psi_\tau^{-3} - 20256\, \chi_\Delta^{-1} \psi_\tau - 481\, \psi_\tau^{-5} - 150108\, \psi_\tau^3\bigr) \\
& + x^6\bigl(3876\, \chi_\Delta^4 + \chi_\Delta^{-4} - 16\, \chi_\Delta^{-3} \psi_\tau^{-6} + 1120\, \chi_\Delta^3 \psi_\tau^{-2} - 16\, \chi_\Delta^{-3} \psi_\tau^2 \\
& \phantom{{} + x^6\bigl(} + 100\, \chi_\Delta^2 \psi_\tau^{-4} - 92\, \chi_\Delta^{-2} \psi_\tau^{-4} - 16\, \chi_\Delta \psi_\tau^{-6} - 285120\, \chi_\Delta \psi_\tau^2 \\
& \phantom{{} + x^6\bigl(} - 6608\, \chi_\Delta^{-1} \psi_\tau^{-2} + 39860\bigr) + O\!\left(x^{25/4}\right).
\end{split}
\end{equation}

\paragraph{Hilbert series.}
Extrapolating from the index, the first terms of the Hilbert series (unrefined, retaining the two abelian fugacities) are
\be
\begin{split}
\CH\CS_{FC_4}(t_a, t_d) ={}&
1 + 30\, t_a + 16\, t_d \\
& + 464\, t_a^2 + 464\, t_a t_d + 136\, t_d^2 \\
& + 4929\, t_a^3 + 6944\, t_a^2 t_d + 3824\, t_a t_d^2 + 816\, t_d^3 \\
& + 40424\, t_a^4 + 71440\, t_a^3 t_d + 55544\, t_a^2 t_d^2 + 22304\, t_a t_d^3 + 3876\, t_d^4 \\
& + \cdots + 15504\, t_d^5 + \cdots,
\end{split}
\ee
where $t_a \equiv x^{2-\tau}$ and $t_d \equiv x^\Delta$.

Two specialisations have closed form. Setting $t_a = 0$ retains only the bifundamental sector and yields the freely-generated series
\be
\CH\CS_{FC_4}(t_a = 0,\, t_d \equiv t) = \frac{1}{(1 - t)^{16}} = 1 + 16\, t + 136\, t^2 + 816\, t^3 + \cdots .
\ee
Setting $t_d = 0$ retains only the adjoint sector, which --- as in $FT_4$ --- is subject to $\mathrm{tr}_L A_L^k = \mathrm{tr}_R A_R^k$ for $k = 2, 3, 4$, giving
\be
\CH\CS_{FC_4}(t_a \equiv t,\, t_d = 0) = \frac{(1 - t^2)(1 - t^3)(1 - t^4)}{(1 - t)^{30}} = 1 + 30\, t + 464\, t^2 + 4929\, t^3 + \cdots .
\ee

\paragraph{Chiral ring.}
The leading terms of the plethystic logarithm are
\be
\begin{split}
\mathrm{PL}\!\left[\CH\CS_{FC_4}\right](t_a, t_d) ={}&
\chi_{(\Box,\overline{\Box})}\, t_d + \bigl(\chi_{(\mathrm{adj}_L,\mathbf{1})} + \chi_{(\mathbf{1},\mathrm{adj}_R)}\bigr)\, t_a \\
& - t_a^2 - t_a^3 - t_a^4 - \chi_{(\Box,\overline{\Box})}\, t_a t_d + \cdots \\
\overset{\mathrm{unr}}{=}\;\,& 16\, t_d + 30\, t_a - t_a^2 - t_a^3 - t_a^4 - 16\, t_a t_d + \cdots .
\end{split}
\ee
The positive leading terms identify the generators: the bifundamental $\Pi$ of $SU(4)_L \times SU(4)_R$ (contributing $\chi_{(\Box,\overline{\Box})}\, t_d$) and the two adjoints $A_L$, $A_R$ (contributing $\bigl(\chi_{(\mathrm{adj}_L,\mathbf{1})} + \chi_{(\mathbf{1},\mathrm{adj}_R)}\bigr)\, t_a$). They satisfy the relations (with their PL contributions shown on the right):
\be
\begin{aligned}
A_L\, \Pi = \Pi\, A_R,
&& -\chi_{(\Box,\overline{\Box})}\, t_a t_d \\
\mathrm{tr}_L A_L^k = \mathrm{tr}_R A_R^k, \quad k = 2, 3, 4,
&& -t_a^2 - t_a^3 - t_a^4 .
\end{aligned}
\ee
The PL is not a finite sum, so these relations are not independent and the moduli space is not a complete intersection.

\paragraph{Consistency check.}
The unrefined Hilbert series is recovered from the index in the limit
\begin{equation}
\lim_{y \to 0} \CI_{FC_4}\!\left(x \to xy,\ \psi_\tau \to \psi_\tau\, y^{\tau - 2},\ \chi_\Delta \to \chi_\Delta\, y^{-\Delta}\right)
= \CH\CS^{\mathrm{unr}}_{FC_4}\!\left(x^{2-\tau}\psi_\tau,\, x^\Delta \chi_\Delta\right).
\end{equation}

\subsection{\texorpdfstring{$FH_4$}{fh4} }

The $FH_4$ theory has the following possible UV completion, where the $SU(4)$ symmetry is manifest while the $USp(8)$ symmetry is emergent:

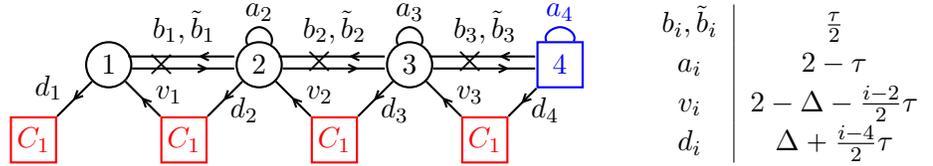
\be
\resizebox{\hsize}{!}{
\begin{tikzpicture}[
    thick,
    gauge/.style={
        circle, draw=black, fill=white,
        minimum size=8mm, inner sep=1pt, font=\small
    },
    manifest/.style={
        rectangle, draw=blue!70!black, fill=blue!6,
        minimum size=8mm, inner sep=2pt,
        font=\small, text=blue!70!black
    },
    flavour/.style={
        rectangle, draw=red!70!black, fill=red!6,
        minimum size=7mm, inner sep=2pt,
        font=\small, text=red!70!black
    },
    arr/.style={-{Stealth[scale=0.8]}, shorten >=1pt, shorten <=1pt},
    barr/.style={draw=black!70}
]
\node[gauge]    (g1) at (0,0)   {$1$};
\node[gauge]    (m1) at (2.5,0) {$2$};
\node[gauge]    (m2) at (5,0)   {$3$};
\node[manifest] (m3) at (7.5,0) {$4$};
\node[flavour] (f1) at (-0.8,-1.8) {$C_1$};
\node[flavour] (f2) at (1.7,-1.8)  {$C_1$};
\node[flavour] (f3) at (4.2,-1.8)  {$C_1$};
\node[flavour] (f4) at (6.7,-1.8)  {$C_1$};
\draw[arr,barr,transform canvas={yshift= 3.5pt}] (g1) -- (m1);
\draw[arr,barr,transform canvas={yshift=-3.5pt}] (m1) -- (g1);
\node at (1.25, 0.45) {$b_1,\,\tilde{b}_1$};
\node at (1.25, 0) {\huge$\times$};
\draw[arr,barr,transform canvas={yshift= 3.5pt}] (m1) -- (m2);
\draw[arr,barr,transform canvas={yshift=-3.5pt}] (m2) -- (m1);
\node at (3.75, 0.45) {$b_2,\,\tilde{b}_2$};
\node at (3.75, 0) {\huge$\times$};
\draw[arr,barr,transform canvas={yshift= 3.5pt}] (m2) -- (m3);
\draw[arr,barr,transform canvas={yshift=-3.5pt}] (m3) -- (m2);
\node at (6.25, 0.45) {$b_3,\,\tilde{b}_3$};
\node at (6.25, 0) {\huge$\times$};
\draw (m1) to[out=65,in=115,looseness=3] (m1);
\node at (2.5, 0.95) {$a_2$};
\draw (m2) to[out=65,in=115,looseness=3] (m2);
\node at (5, 0.95) {$a_3$};
\draw[blue!70!black] (m3) to[out=65,in=115,looseness=3] (m3);
\node[blue!70!black] at (7.5, 0.95) {$a_4$};
\draw[arr,barr] (g1) to[bend right=5] (f1);
\node at (-0.65,-0.7) {$d_1$};
\draw[arr,barr] (f2) to[bend right=5] (g1);
\node at (1.1,-0.7) {$v_1$};
\draw[arr,barr] (m1) to[bend right=5] (f2);
\node at (1.8,-0.7) {$d_2$};
\draw[arr,barr] (f3) to[bend right=5] (m1);
\node at (3.6,-0.7) {$v_2$};
\draw[arr,barr] (m2) to[bend right=5] (f3);
\node at (4.3,-0.7) {$d_3$};
\draw[arr,barr] (f4) to[bend right=5] (m2);
\node at (6.1,-0.7) {$v_3$};
\draw[arr,barr] (m3) to[bend right=5] (f4);
\node at (6.8,-0.7) {$d_4$};
\node at (5.4,-3.2) {$\mathcal{W} =
    \sum_{i=1}^{3}v_i b_i d_{i+1}
    + \mathfrak{M}^{\pm\,00} + \mathfrak{M}^{0\,\pm\,0} + \mathfrak{M}^{00\,\pm}
    + a_2 b_1\tilde{b}_1
    + \sum_{j=2}^{3} b_j(a_j + a_{j+1})\tilde{b}_j
    + \sum_{i=1}^{3}\mathrm{Flip}[b_i\tilde{b}_i]$};
\node[right] at (9,-0.5) {%
    \renewcommand{\arraystretch}{1.4}
    \begin{tabular}{>{\small}c >{\small}c}
        \toprule
        $b_i,\,\tilde{b}_i$ & $\tau/2$ \\
        $a_i$                & $2-\tau$ \\
        $v_i$                & $2-\Delta-\tfrac{i-2}{2}\tau$ \\
        $d_i$                & $\Delta+\tfrac{i-4}{2}\tau$ \\
        \bottomrule
    \end{tabular}%
};
\end{tikzpicture}
}
\ee

We compute the superconformal index of $FH_4$, extrapolate the Hilbert series of the main branch from it, and analyse the chiral ring. As in the $N=3$ case, the main branch is generated by the bifundamental $\Pi$ of $SU(4)_L \times USp(8)_R$ and by the fields $A_L$ in the adjoint of $SU(4)_L$ and $A_R$ in the traceless-antisymmetric of $USp(8)_R$.

A preliminary remark on monopoles: we retain only the zero-flux sector of the index. This is justified a posteriori by the fact that all the chiral-ring generators already appear in the zero-flux sector, so that contributions from nontrivial GNO fluxes can be dropped.

\paragraph{Superconformal index.}
Fixing $\tau = 3/4$ and $\Delta = 6/5$ and refining in the abelian fugacities $\psi_\tau$ and $\chi_\Delta$ for $U(1)_\tau$ and $U(1)_\Delta$, the (zero-flux) index reads
\begin{equation}
\begin{split}
\CI_{FH_4} ={}& 1 + 32\, \chi_\Delta\, x^{6/5} + 42\, \psi_\tau\, x^{5/4} - \psi_\tau^{-2} x^{3/2} - 32\, \chi_\Delta \psi_\tau^{-1} x^{39/20} - 59\, x^2 - \psi_\tau^{-3} x^{9/4} \\
& + 522\, \chi_\Delta^2 x^{12/5} + 1312\, \chi_\Delta \psi_\tau\, x^{49/20} + 902\, \psi_\tau^2 x^{5/2} - 32\, \chi_\Delta \psi_\tau^{-2} x^{27/10} - 41\, \psi_\tau^{-1} x^{11/4} \\
& - \psi_\tau^{-4} x^3 - 32\, \chi_\Delta^{-1} \psi_\tau^{-3} x^{61/20} - 6\, \chi_\Delta^{-2} \psi_\tau^{-2} x^{31/10} - 1014\, \chi_\Delta^2 \psi_\tau^{-1} x^{63/20} \\
& - 3152\, \chi_\Delta\, x^{16/5} - 2417\, \psi_\tau\, x^{13/4} - 388\, \psi_\tau^{-2} x^{7/2} - 384\, \chi_\Delta^{-1} \psi_\tau^{-1} x^{71/20} \\
& + \bigl(5792\, \chi_\Delta^3 - 16\, \chi_\Delta^{-2}\bigr) x^{18/5} + 20910\, \chi_\Delta^2 \psi_\tau\, x^{73/20} + 27520\, \chi_\Delta \psi_\tau^2 x^{37/10} \\
& + \bigl(\psi_\tau^{-5} + 13201\, \psi_\tau^3\bigr) x^{15/4} + 32\, \chi_\Delta^{-1} \psi_\tau^{-4} x^{19/5} + 10\, \chi_\Delta^{-2} \psi_\tau^{-3} x^{77/20} \\
& - 26\, \chi_\Delta^2 \psi_\tau^{-2} x^{39/10} - 384\, \chi_\Delta \psi_\tau^{-1} x^{79/20} - 796\, x^4 - 384\, \chi_\Delta^{-1} \psi_\tau\, x^{81/20} \\
& - 6\, \chi_\Delta^{-2} \psi_\tau^2 x^{41/10} + 32\, \chi_\Delta \psi_\tau^{-4} x^{21/5} + 554\, \psi_\tau^{-3} x^{17/4} + 208\, \chi_\Delta^{-1} \psi_\tau^{-2} x^{43/10} \\
& + x^{87/20}\bigl(-16384\, \chi_\Delta^3 \psi_\tau^{-1} - 130\, \chi_\Delta^{-2} \psi_\tau^{-1}\bigr) - 70242\, \chi_\Delta^2 x^{22/5} - 102016\, \chi_\Delta \psi_\tau\, x^{89/20} \\
& + \bigl(\psi_\tau^{-6} - 51135\, \psi_\tau^2\bigr) x^{9/2} + x^{91/20}\bigl(32\, \chi_\Delta^{-1} \psi_\tau^{-5} - 32\, \chi_\Delta^{-1} \psi_\tau^3\bigr) \\
& + 6\, \chi_\Delta^{-2} \psi_\tau^{-4} x^{23/5} + 654\, \chi_\Delta^2 \psi_\tau^{-3} x^{93/20} - 2208\, \chi_\Delta \psi_\tau^{-2} x^{47/10} - 11179\, \psi_\tau^{-1} x^{19/4} \\
& + \bigl(49207\, \chi_\Delta^4 - 11168\, \chi_\Delta^{-1}\bigr) x^{24/5} + x^{97/20}\bigl(226880\, \chi_\Delta^3 \psi_\tau - 570\, \chi_\Delta^{-2} \psi_\tau\bigr) \\
& + 428752\, \chi_\Delta^2 \psi_\tau^2 x^{49/10} + x^{99/20}\bigl(32\, \chi_\Delta \psi_\tau^{-5} + 393568\, \chi_\Delta \psi_\tau^3\bigr) \\
& + \bigl(148049\, \psi_\tau^4 + 692\, \psi_\tau^{-4}\bigr) x^5 + 2576\, \chi_\Delta^{-1} \psi_\tau^{-3} x^{101/20} \\
& + x^{51/10}\bigl(9952\, \chi_\Delta^3 \psi_\tau^{-2} + 1118\, \chi_\Delta^{-2} \psi_\tau^{-2}\bigr) + x^{103/20}\bigl(16\, \chi_\Delta^{-3} \psi_\tau^{-1} + 37500\, \chi_\Delta^2 \psi_\tau^{-1}\bigr) \\
& + 36560\, \chi_\Delta\, x^{26/5} + \bigl(\psi_\tau^{-7} - 9237\, \psi_\tau\bigr) x^{21/4} - 14688\, \chi_\Delta^{-1} \psi_\tau^2 x^{53/10} \\
& + x^{107/20}\bigl(158\, \chi_\Delta^{-2} \psi_\tau^{-5} - 242\, \chi_\Delta^{-2} \psi_\tau^3\bigr) + x^{27/5}\bigl(32\, \chi_\Delta^{-3} \psi_\tau^{-4} + 356\, \chi_\Delta^2 \psi_\tau^{-4}\bigr) \\
& + x^{109/20}\bigl(\chi_\Delta^{-4} \psi_\tau^{-3} + 19600\, \chi_\Delta \psi_\tau^{-3}\bigr) + 57639\, \psi_\tau^{-2} x^{11/2} \\
& + x^{111/20}\bigl(43424\, \chi_\Delta^{-1} \psi_\tau^{-1} - 180082\, \chi_\Delta^4 \psi_\tau^{-1}\bigr) + \bigl(946\, \chi_\Delta^{-2} - 966784\, \chi_\Delta^3\bigr) x^{28/5} \\
& + x^{113/20}\bigl(32\, \chi_\Delta^{-3} \psi_\tau - 2009458\, \chi_\Delta^2 \psi_\tau\bigr) + x^{57/10}\bigl(-32\, \chi_\Delta \psi_\tau^{-6} - 1944640\, \chi_\Delta \psi_\tau^2\bigr) \\
& + \bigl(-876\, \psi_\tau^{-5} - 743141\, \psi_\tau^3\bigr) x^{23/4} + x^{29/5}\bigl(-1312\, \chi_\Delta^{-1} \psi_\tau^4 - 2656\, \chi_\Delta^{-1} \psi_\tau^{-4}\bigr) \\
& + x^{117/20}\bigl(8832\, \chi_\Delta^3 \psi_\tau^{-3} - 708\, \chi_\Delta^{-2} \psi_\tau^{-3}\bigr) + x^{59/10}\bigl(112\, \chi_\Delta^{-3} \psi_\tau^{-2} + 47364\, \chi_\Delta^2 \psi_\tau^{-2}\bigr) \\
& + x^{119/20}\bigl(\chi_\Delta^{-4} \psi_\tau^{-1} + 71184\, \chi_\Delta \psi_\tau^{-1}\bigr) + \bigl(341568\, \chi_\Delta^5 - 41671\bigr) x^6 + O\!\left(x^{121/20}\right) .
\end{split}
\end{equation}

\paragraph{Main-branch Hilbert series.}
Extrapolating from the index, the first terms of the main-branch Hilbert series (unrefined, retaining the two abelian fugacities) are
\be
\begin{split}
\CH\CS_{FH_4}(t_a, t_d) ={}&
1 + 32\, t_d + 42\, t_a \\
& + 522\, t_d^2 + 1312\, t_a t_d + 902\, t_a^2 \\
& + 5792\, t_d^3 + 20910\, t_a t_d^2 + 27520\, t_a^2 t_d + 13201\, t_a^3 + \cdots ,
\end{split}
\ee
where $t_a \equiv x^{2-\tau}$ and $t_d \equiv x^\Delta$.

Two specialisations have closed form, and match the computed terms above. Setting $t_a = 0$ gives
\be
\CH\CS_{FH_4}(0, t_d) = \frac{(1 + t_d)^6}{(1 - t_d)^{26}} = 1 + 32\, t_d + 522\, t_d^2 + 5792\, t_d^3 + \cdots,
\ee
while setting $t_d = 0$ retains only the adjoint sector, subject to $\mathrm{tr}_L A_L^k = \mathrm{tr}_R A_R^k$ for $k = 2, 3, 4$:
\be
\CH\CS_{FH_4}(t_a, 0) = \frac{(1 + t_a)(1 + t_a + t_a^2)(1 + t_a + t_a^2 + t_a^3)}{(1 - t_a)^{39}} = 1 + 42\, t_a + 902\, t_a^2 + 13201\, t_a^3 + \cdots.
\ee

\paragraph{Chiral ring.}
The leading terms of the plethystic logarithm are
\be
\begin{split}
\mathrm{PL}\!\left[\CH\CS_{FH_4}\right](t_a, t_d) ={}&
\chi_{(\Box,\mathbf{8})}\, t_d + \bigl(\chi_{(\mathrm{adj}_L,\mathbf{1})} + \chi_{(\mathbf{1},\mathrm{asym}_R)}\bigr)\, t_a \\
& - t_a^2 - t_a^3 - t_a^4 - \chi_{(\mathrm{asym}_L,\mathbf{1})}\, t_d^2 - \chi_{(\Box,\mathbf{8})}\, t_a t_d + \cdots \\
\overset{\mathrm{unr}}{=}\;\,& 32\, t_d + 42\, t_a - t_a^2 - t_a^3 - t_a^4 - 6\, t_d^2 - 32\, t_a t_d + \cdots .
\end{split}
\ee
The positive leading terms identify the generators: the bifundamental $\Pi$ of $SU(4)_L \times USp(8)_R$ (contributing $\chi_{(\Box,\mathbf{8})}\, t_d$), the $SU(4)_L$ adjoint $A_L$ (contributing $\chi_{(\mathrm{adj}_L,\mathbf{1})}\, t_a$), and the $USp(8)$ traceless-antisymmetric $A_R$ (contributing $\chi_{(\mathbf{1},\mathrm{asym}_R)}\, t_a$). They satisfy the relations (with their PL contributions shown on the right):
\be
\begin{aligned}
\Omega_{ab}\, \Pi^{a,A}\Pi^{b,B} &= 0,
&& -\chi_{(\mathrm{asym}_L,\mathbf{1})}\, t_d^2 \\
\mathrm{tr}_L A_L^k &= \mathrm{tr}_R A_R^k, \quad k = 2, 3, 4,
&& -t_a^2 - t_a^3 - t_a^4 \\
\Pi^{a,B}\,(A_L)^A{}_B &= \Pi^{b,A}\,A_R{}^{ba},
&& -\chi_{(\Box,\mathbf{8})}\, t_a t_d .
\end{aligned}
\ee
The PL is not a finite sum, so these relations are not independent and the main branch of the moduli space is not a complete intersection.

\subsection{\texorpdfstring{$FM_4$}{fmsu4} }

The $FM_4$ theory has the following possible UV completion:

\be
\resizebox{1\hsize}{!}{
\begin{tikzpicture}[
    thick,
    gauge/.style={
        circle, draw=black, fill=white,
        minimum size=8mm, inner sep=1pt, font=\small
    },
    manifest/.style={
        rectangle, draw=blue!70!black, fill=blue!6,
        minimum size=8mm, inner sep=2pt,
        font=\small, text=blue!70!black
    },
    flavour/.style={
        rectangle, draw=red!70!black, fill=red!6,
        minimum size=7mm, inner sep=2pt,
        font=\small, text=red!70!black
    },
    arr/.style={-{Stealth[scale=0.8]}, shorten >=1pt, shorten <=1pt},
    barr/.style={draw=black!70}
]
\node[gauge]    (g1) at (0,0)   {$1$};
\node[gauge]    (g2) at (2.5,0) {$2$};
\node[gauge]    (g3) at (5,0)   {$3$};
\node[manifest] (m4) at (7.5,0) {$4$};
\node[flavour] (f1) at (-0.8,-1.8) {$1$};
\node[flavour] (f2) at (1.7,-1.8)  {$1$};
\node[flavour] (f3) at (4.2,-1.8)  {$1$};
\node[flavour] (f4) at (6.7,-1.8)  {$1$};
\draw[arr,barr,transform canvas={yshift= 3.5pt}] (g1) -- (g2);
\draw[arr,barr,transform canvas={yshift=-3.5pt}] (g2) -- (g1);
\node at (1.25, 0.45) {$b_1,\,\tilde{b}_1$};
\node at (1.25, 0) {\huge$\times$};
\draw[arr,barr,transform canvas={yshift= 3.5pt}] (g2) -- (g3);
\draw[arr,barr,transform canvas={yshift=-3.5pt}] (g3) -- (g2);
\node at (3.75, 0.45) {$b_2,\,\tilde{b}_2$};
\node at (3.75, 0) {\huge$\times$};
\draw[arr,barr,transform canvas={yshift= 3.5pt}] (g3) -- (m4);
\draw[arr,barr,transform canvas={yshift=-3.5pt}] (m4) -- (g3);
\node at (6.25, 0.45) {$b_3,\,\tilde{b}_3$};
\node at (6.25, 0) {\huge$\times$};
\draw (g2) to[out=65,in=115,looseness=3] (g2);
\node at (2.5, 0.95) {$a_2$};
\draw (g3) to[out=65,in=115,looseness=3] (g3);
\node at (5, 0.95) {$a_3$};
\draw[blue!70!black] (m4) to[out=65,in=115,looseness=3] (m4);
\node[blue!70!black] at (7.5, 0.95) {$a_4$};
\draw[arr,barr,bend left=8]  (f1) to (g1);
\draw[arr,barr,bend left=8] (g1) to (f1);
\node[rotate=-30] at (-0.4,-0.9) {\huge$\times$};
\node at (-0.9,-0.65) {$d_1$};
\draw[arr,barr,bend left=8]  (g1) to (f2);
\draw[arr,barr,bend left=8] (f2) to (g1);
\node at (1.1,-0.55) {$v_1$};
\draw[arr,barr,bend left=8]  (f2) to (g2);
\draw[arr,barr,bend left=8] (g2) to (f2);
\node[rotate=-30] at (2.1,-0.9) {\huge$\times$};
\node at (1.8,-0.55) {$d_2$};
\draw[arr,barr,bend left=8]  (g2) to (f3);
\draw[arr,barr,bend left=8] (f3) to (g2);
\node at (3.6,-0.65) {$v_2$};
\draw[arr,barr,bend left=8]  (f3) to (g3);
\draw[arr,barr,bend left=8] (g3) to (f3);
\node[rotate=-30] at (4.6,-0.9) {\huge$\times$};
\node at (4.3,-0.55) {$d_3$};
\draw[arr,barr,bend left=8]  (g3) to (f4);
\draw[arr,barr,bend left=8] (f4) to (g3);
\node at (6.1,-0.65) {$v_3$};
\draw[arr,barr,bend left=8]  (f4) to (m4);
\draw[arr,barr,bend left=8] (m4) to (f4);
\node at (7.55,-1.05) {$d_4$};
\node at (5.5,-3.6) {$\begin{aligned}\mathcal{W} =&\;
    \sum_{j=1}^{3}\!\left[v_j b_j d_{j+1}
    +\tilde{v}_j\tilde{b}_j\tilde{d}_{j+1}\right]
    + a_2 b_1\tilde{b}_1
    + \sum_{j=2}^{3} b_j(a_j+a_{j+1})\tilde{b}_j
    + \sum_{i=1}^{3}\mathrm{Flip}[b_i\tilde{b}_i]\\
    &+\mathfrak{M}^{\pm 00}+\mathfrak{M}^{0\pm 0}+\mathfrak{M}^{00\pm}
    + \sum_{i=1}^{3}\mathrm{Flip}[d_i\tilde{d}_i]
    \end{aligned}$};
\node[right] at (9,-0.5) {%
    \renewcommand{\arraystretch}{1.4}
    \begin{tabular}{>{\small}c >{\small}c}
        \toprule
        $b_i,\,\tilde{b}_i$  & $\tau/2$ \\
        $a_i$                & $2-\tau$ \\
        $v_i,\,\tilde{v}_i$  & $2-\Delta-\tfrac{i-2}{2}\tau$ \\
        $d_i,\,\tilde{d}_i$  & $\Delta+\tfrac{i-4}{2}\tau$ \\
        \bottomrule
    \end{tabular}%
};
\end{tikzpicture}
}
\ee

We compute the superconformal index of $FM_4$, extrapolate the Hilbert series of the main branch from it, and analyse the chiral ring. The main branch is generated by the bifundamentals $\Pi,\,\tilde\Pi$ and the adjoints $A_L,\,A_R$ of $SU(4)_L \times SU(4)_R$; all statements below pertain to this branch.

\paragraph{Superconformal index.}
Fixing $\tau = 4/9$ and $\Delta = 5/6$ and refining in the abelian fugacities $\psi_\tau$ and $\chi_\Delta$ for $U(1)_\tau$ and $U(1)_\Delta$, the index reads
\begin{equation}
\begin{split}
\CI_{FM_4} ={}& 1 + x^{7/9} \chi_\Delta^{-2} \psi_\tau^{-1} + 32\, x^{5/6} \chi_\Delta - x^{8/9} \psi_\tau^{-2} + x^{11/9} \chi_\Delta^{-2} \psi_\tau^{-2} - 32\, x^{23/18} \chi_\Delta \psi_\tau^{-1} \\
& - x^{4/3} \psi_\tau^{-3} + x^{14/9}\bigl(\chi_\Delta^{-4} \psi_\tau^{-2} + 30\, \psi_\tau\bigr) + 32\, x^{29/18} \chi_\Delta^{-1} \psi_\tau^{-1} + 498\, x^{5/3} \chi_\Delta^2 \\
& - 32\, x^{31/18} \chi_\Delta \psi_\tau^{-2} - x^{16/9} \psi_\tau^{-4} + x^{17/9} \chi_\Delta^{-2} \psi_\tau + x^2\bigl(\chi_\Delta^{-4} \psi_\tau^{-3} - 33\bigr) \\
& + x^{19/9}\bigl(-992\, \chi_\Delta^2 \psi_\tau^{-1} - 2\, \chi_\Delta^{-2} \psi_\tau^{-4}\bigr) + x^{20/9} \psi_\tau^{-5} + x^{7/3}\bigl(\chi_\Delta^{-6} \psi_\tau^{-3} + 31\, \chi_\Delta^{-2}\bigr) \\
& + x^{43/18}\bigl(928\, \chi_\Delta \psi_\tau + 32\, \chi_\Delta^{-3} \psi_\tau^{-2}\bigr) + x^{22/9}\bigl(\chi_\Delta^{-4} \psi_\tau^{-4} + 467\, \psi_\tau^{-1}\bigr) \\
& + x^{5/2}\bigl(5056\, \chi_\Delta^3 - 64\, \chi_\Delta^{-1} \psi_\tau^{-3}\bigr) + x^{23/9}\bigl(-2\, \chi_\Delta^2 \psi_\tau^{-2} - 3\, \chi_\Delta^{-2} \psi_\tau^{-5}\bigr) \\
& + 32\, x^{47/18} \chi_\Delta \psi_\tau^{-4} + x^{8/3}\bigl(\chi_\Delta^{-4} + \psi_\tau^{-6}\bigr) + 32\, x^{49/18} \chi_\Delta^{-1} \psi_\tau \\
& + x^{25/9}\bigl(\chi_\Delta^{-6} \psi_\tau^{-4} - 2\, \chi_\Delta^{-2} \psi_\tau^{-1}\bigr) - 1952\, x^{17/6} \chi_\Delta \\
& + x^{26/9}\bigl(-\chi_\Delta^{-4} \psi_\tau^{-5} - 790\, \psi_\tau^{-2}\bigr) + x^{53/18}\bigl(-14944\, \chi_\Delta^3 \psi_\tau^{-1} - 32\, \chi_\Delta^{-1} \psi_\tau^{-4}\bigr) \\
& + x^3\bigl(791\, \chi_\Delta^2 \psi_\tau^{-3} - \chi_\Delta^{-2} \psi_\tau^{-6}\bigr) + O\!\left(x^{55/18}\right).
\end{split}
\end{equation}

\paragraph{Main-branch Hilbert series.}
Extrapolating from the index, the first terms of the main-branch Hilbert series (unrefined, retaining the two abelian fugacities) are
\be
\begin{split}
\CH\CS_{FM_4}(t_a, t_d) ={}&
1 + 32\, t_d + 30\, t_a \\
& + 498\, t_d^2 + 928\, t_a t_d + 464\, t_a^2 \\
& + 5056\, t_d^3 + \cdots ,
\end{split}
\ee
where $t_a \equiv x^{2-\tau}$ and $t_d \equiv x^\Delta$.

Setting $t_d = 0$ retains only the adjoint sector, subject to $\mathrm{tr}_L A_L^k = \mathrm{tr}_R A_R^k$ for $k = 2, 3, 4$, and reproduces the $FT_4$ Hilbert series:
\be
\CH\CS_{FM_4}(t_a \equiv t,\, t_d = 0) = \frac{(1 + t)(1 + t + t^2)(1 + t + t^2 + t^3)}{(1 - t)^{27}} = 1 + 30\, t + 464\, t^2 + \cdots .
\ee

\paragraph{Chiral ring.}
The leading terms of the plethystic logarithm are
\be
\begin{split}
\mathrm{PL}\!\left[\CH\CS_{FM_4}\right](t_a, t_d) ={}&
\bigl(\chi_{(\Box,\overline{\Box})} + \chi_{(\overline{\Box},\Box)}\bigr)\, t_d
+ \bigl(\chi_{(\mathrm{adj}_L,\mathbf{1})} + \chi_{(\mathbf{1},\mathrm{adj}_R)}\bigr)\, t_a \\
& - t_a^2 - t_a^3 - t_a^4
- \bigl(\chi_{(\mathrm{adj}_L,\mathbf{1})} + \chi_{(\mathbf{1},\mathrm{adj}_R)}\bigr)\, t_d^2 \\
& - \bigl(\chi_{(\Box,\overline{\Box})} + \chi_{(\overline{\Box},\Box)}\bigr)\, t_a t_d + \cdots \\
\overset{\mathrm{unr}}{=}\;\,& 32\, t_d + 30\, t_a - t_a^2 - t_a^3 - t_a^4 - 30\, t_d^2 - 32\, t_a t_d + \cdots .
\end{split}
\ee
The positive leading terms identify the generators: the bifundamentals $\Pi, \tilde\Pi$ of $SU(4)_L \times SU(4)_R$ contributing $\bigl(\chi_{(\Box,\overline{\Box})} + \chi_{(\overline{\Box},\Box)}\bigr)\, t_d$, and the adjoints $A_L, A_R$ contributing $\bigl(\chi_{(\mathrm{adj}_L,\mathbf{1})} + \chi_{(\mathbf{1},\mathrm{adj}_R)}\bigr)\, t_a$. They satisfy the relations (with their PL contributions shown on the right):
\be
\begin{aligned}
\bigl(\Pi \tilde\Pi\bigr)\big|_{\mathrm{traceless}} = 0 = \bigl(\tilde\Pi \Pi\bigr)\big|_{\mathrm{traceless}},
&& -\bigl(\chi_{(\mathrm{adj}_L,\mathbf{1})} + \chi_{(\mathbf{1},\mathrm{adj}_R)}\bigr)\, t_d^2 \\
\mathrm{tr}_L A_L^k = \mathrm{tr}_R A_R^k, \quad k = 2, 3, 4,
&& -t_a^2 - t_a^3 - t_a^4 \\
A_L\, \Pi = \Pi\, A_R,
&& -\chi_{(\Box,\overline{\Box})}\, t_a t_d \\
\tilde\Pi\, A_L = A_R\, \tilde\Pi,
&& -\chi_{(\overline{\Box},\Box)}\, t_a t_d .
\end{aligned}
\ee
The PL is not a finite sum --- higher syzygies appear --- so these relations are not independent, and the main branch of the moduli space is not a complete intersection.

\subsection{\texorpdfstring{$FE_4$}{feusp8}}

The $FE_4$ theory has the following possible UV completion:

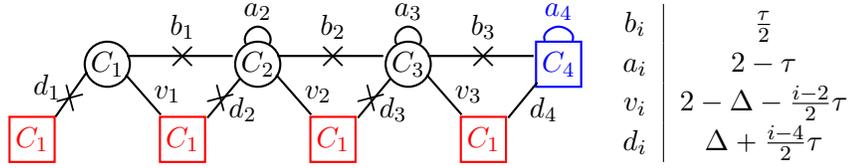
\be
\resizebox{\hsize}{!}{
\begin{tikzpicture}[
    thick,
    gauge/.style={
        circle, draw=black, fill=white,
        minimum size=8mm, inner sep=1pt, font=\small
    },
    manifest/.style={
        rectangle, draw=blue!70!black, fill=blue!6,
        minimum size=8mm, inner sep=2pt,
        font=\small, text=blue!70!black
    },
    flavour/.style={
        rectangle, draw=red!70!black, fill=red!6,
        minimum size=7mm, inner sep=2pt,
        font=\small, text=red!70!black
    },
    barr/.style={draw=black!70}
]
\node[gauge]    (g1) at (0,0)   {$C_1$};
\node[gauge]    (g2) at (2.5,0) {$C_2$};
\node[gauge]    (g3) at (5,0)   {$C_3$};
\node[manifest] (m1) at (7.5,0) {$C_4$};
\node[flavour] (f1) at (-0.8,-1.8) {$C_1$};
\node[flavour] (f2) at (1.7,-1.8)  {$C_1$};
\node[flavour] (f3) at (4.2,-1.8)  {$C_1$};
\node[flavour] (f4) at (6.7,-1.8)  {$C_1$};
\draw[barr] (g1) -- (g2);
\node at (1.25, 0.45) {$b_1$};
\node at (1.25, 0) {\huge$\times$};
\draw[barr] (g2) -- (g3);
\node at (3.75, 0.45) {$b_2$};
\node at (3.75, 0) {\huge$\times$};
\draw[barr] (g3) -- (m1);
\node at (6.25, 0.45) {$b_3$};
\node at (6.25, 0) {\huge$\times$};
\draw (g2) to[out=65,in=115,looseness=3] (g2);
\node at (2.5, 0.95) {$a_2$};
\draw (g3) to[out=65,in=115,looseness=3] (g3);
\node at (5, 0.95) {$a_3$};
\draw[blue!70!black] (m1) to[out=65,in=115,looseness=3] (m1);
\node[blue!70!black] at (7.5, 0.95) {$a_4$};
\draw[barr] (f1) -- (g1);
\node[rotate=-30] at (-0.42,-0.9) {\huge$\times$};
\node at (-0.75,-0.55) {$d_1$};
\draw[barr] (g1) -- (f2);
\node at (1,-0.6) {$v_1$};
\draw[barr] (f2) -- (g2);
\node[rotate=-30] at (2.05,-0.9) {\huge$\times$};
\node at (1.7,-0.6) {$d_2$};
\draw[barr] (g2) -- (f3);
\node at (3.45,-0.6) {$v_2$};
\draw[barr] (f3) -- (g3);
\node[rotate=-30] at (4.58,-0.9) {\huge$\times$};
\node at (4.22,-0.6) {$d_3$};
\draw[barr] (g3) -- (f4);
\node at (5.95,-0.6) {$v_3$};
\draw[barr] (f4) -- (m1);
\node at (7.4,-0.9) {$d_4$};
\node at (5.5,-3.0) {$\mathcal{W} =
    b_1^2 a_2
    + \displaystyle\sum_{i=2}^{3} b_i^2(a_i + a_{i+1})
    + \sum_{i=1}^{3}\!\left[v_i b_i d_{i+1}
    + \mathrm{Flip}[b_i^2] + \mathrm{Flip}[d_i^2]\right]$};
\node at (3.75,-4) {$+\,\bigl(\mathfrak{M}^{100}
    + \mathfrak{M}^{010} + \mathfrak{M}^{001}\bigr)$};
\node[right] at (9.0,-0.5) {%
    \renewcommand{\arraystretch}{1.4}
    \begin{tabular}{>{\small}c >{\small}c}
        \toprule
        $b_i$ & $\tau/2$ \\
        $a_i$ & $2-\tau$ \\
        $v_i$ & $2-\Delta-\tfrac{i-2}{2}\tau$ \\
        $d_i$ & $\Delta+\tfrac{i-4}{2}\tau$ \\
        \bottomrule
    \end{tabular}%
};
\end{tikzpicture}
}
\ee

We compute the superconformal index of $FE_4$, extrapolate the Hilbert series of the main branch from it, and analyse the chiral ring. The main branch is generated by the bifundamental $\Pi$ of $USp(8)_L \times USp(8)_R$ and by the two traceless-antisymmetric fields $A_L,\,A_R$ of $USp(8)_L$ and $USp(8)_R$ respectively; all statements below pertain to this branch.

\paragraph{Superconformal index.}
Fixing $\tau = 5/7$ and $\Delta = 9/8$ and refining in the abelian fugacities $\psi_\tau$ and $\chi_\Delta$ for $U(1)_\tau$ and $U(1)_\Delta$, the index reads
\begin{equation}
\begin{split}
\CI_{FE_4} ={}& 1 + x^{13/28} \chi_\Delta^{-2} \psi_\tau^{-1} + x^{13/14} \chi_\Delta^{-4} \psi_\tau^{-2} \\
&+ x^{29/28} \chi_\Delta^{-2} \psi_\tau + 64\, \chi_\Delta\, x^{9/8} + x^{33/28} \chi_\Delta^{-2} \psi_\tau^{-2} + 54\, \psi_\tau\, x^{9/7} \\
&+ x^{39/28} \chi_\Delta^{-6} \psi_\tau^{-3} - \psi_\tau^{-2} x^{10/7} + \chi_\Delta^{-4} x^{3/2} + 64\, \chi_\Delta^{-1} \psi_\tau^{-1} x^{89/56} \\
&+ x^{23/14} \chi_\Delta^{-4} \psi_\tau^{-3} + 55\, \chi_\Delta^{-2} x^{7/4} - 64\, \chi_\Delta \psi_\tau^{-1} x^{103/56} \\
&+ x^{13/7} \chi_\Delta^{-8} \psi_\tau^{-4} + x^{55/28} \chi_\Delta^{-6} \psi_\tau^{-1} - 74\, x^2 + 64\, \chi_\Delta^{-3} \psi_\tau^{-2} x^{115/56} \\
&+ x^{29/14} \chi_\Delta^{-4} \psi_\tau^2 + x^{59/28} \chi_\Delta^{-6} \psi_\tau^{-4} - \psi_\tau^{-3} x^{15/7} \\
&+ 64\, \chi_\Delta^{-1} \psi_\tau\, x^{121/56} + 56\, \chi_\Delta^{-4} \psi_\tau^{-1} x^{31/14} + 2026\, \chi_\Delta^2 x^{9/4} \\
&+ x^{65/28}\bigl(\chi_\Delta^{-10} \psi_\tau^{-5} + 55\, \chi_\Delta^{-2} \psi_\tau^2\bigr) + x^{33/14} \chi_\Delta^{-4} \psi_\tau^{-4} \\
&+ 3392\, \chi_\Delta \psi_\tau\, x^{135/56} + x^{17/7} \chi_\Delta^{-8} \psi_\tau^{-2} - 19\, \chi_\Delta^{-2} \psi_\tau^{-1} x^{69/28} \\
&+ 64\, \chi_\Delta^{-5} \psi_\tau^{-3} x^{141/56} + x^{71/28} \chi_\Delta^{-6} \psi_\tau - 64\, \chi_\Delta \psi_\tau^{-2} x^{143/56} \\
&+ x^{18/7}\bigl(\chi_\Delta^{-8} \psi_\tau^{-5} + 1484\, \psi_\tau^2\bigr) - 2\, \chi_\Delta^{-2} \psi_\tau^{-4} x^{73/28} \\
&+ 64\, \chi_\Delta^{-3} x^{21/8} - \chi_\Delta^{12} \psi_\tau^{18} x^{37/14} + 56\, \chi_\Delta^{-6} \psi_\tau^{-2} x^{75/28} \\
&+ 1967\, \psi_\tau^{-1} x^{19/7} + x^{39/14}\bigl(\chi_\Delta^{-12} \psi_\tau^{-6} + 56\, \chi_\Delta^{-4} \psi_\tau\bigr) \\
&+ x^{79/28}\bigl(\chi_\Delta^{-6} \psi_\tau^{-5} - \chi_\Delta^2 \psi_\tau^2\bigr) - \psi_\tau^{-4} x^{20/7} + 3392\, \chi_\Delta^{-1} x^{23/8} \\
&+ x^{81/28} \chi_\Delta^{-10} \psi_\tau^{-3} - 17\, \chi_\Delta^{-4} \psi_\tau^{-2} x^{41/14} - 4023\, \chi_\Delta^2 \psi_\tau^{-1} x^{83/28} \\
&+ 64\, \chi_\Delta^{-7} \psi_\tau^{-4} x^{167/56} + x^3 \chi_\Delta^{-8} - 128\, \chi_\Delta^{-1} \psi_\tau^{-3} x^{169/56} \\
&+ x^{85/28}\bigl(\chi_\Delta^{-10} \psi_\tau^{-6} + 1466\, \chi_\Delta^{-2} \psi_\tau\bigr) - \chi_\Delta^{-4} \psi_\tau^{-5} x^{43/14} \\
&+ 64\, \chi_\Delta^{-5} \psi_\tau\, x^{173/56} + x^{87/28} \chi_\Delta^{-6} \psi_\tau^3 - 8064\, \chi_\Delta\, x^{25/8} \\
&+ 56\, \chi_\Delta^{-8} \psi_\tau^{-3} x^{22/7} + 1895\, \chi_\Delta^{-2} \psi_\tau^{-2} x^{89/28} + 64\, \chi_\Delta^{-3} \psi_\tau^2 x^{179/56} \\
&+ x^{13/4}\bigl(\chi_\Delta^{-14} \psi_\tau^{-7} + 57\, \chi_\Delta^{-6}\bigr) \\
&+ x^{23/7}\bigl(\chi_\Delta^{-8} \psi_\tau^{-6} - 1918\, \psi_\tau\bigr) - 3\, \chi_\Delta^{-2} \psi_\tau^{-5} x^{93/28} \\
&+ 3456\, \chi_\Delta^{-3} \psi_\tau^{-1} x^{187/56} + x^{47/14}\bigl(\chi_\Delta^{-12} \psi_\tau^{-4} + 55\, \chi_\Delta^{-4} \psi_\tau^3\bigr) \\
&+ 42368\, \chi_\Delta^3 x^{27/8} + O\!\left(x^{95/28}\right).
\end{split}
\end{equation}

\paragraph{Main-branch Hilbert series.}
Extrapolating from the index, the first terms of the main-branch Hilbert series (unrefined, retaining the two abelian fugacities) are
\be
\begin{split}
\CH\CS_{FE_4}(t_a, t_d) ={}&
1 + 64\, t_d + 54\, t_a \\
& + 2026\, t_d^2 + 3392\, t_a t_d + 1484\, t_a^2 \\
& + 42368\, t_d^3 + 105378\, t_a t_d^2 + 91520\, t_a^2 t_d + 27665\, t_a^3 + \cdots ,
\end{split}
\ee
where $t_a \equiv x^{2-\tau}$ and $t_d \equiv x^\Delta$.

Two specialisations have closed form. Setting $t_d = 0$ retains only the adjoint sector, subject to $\mathrm{tr}_L A_L^k = \mathrm{tr}_R A_R^k$ for $k = 2, 3, 4$, and reproduces an $FT_4$-like structure:
\be
\CH\CS_{FE_4}(t_a \equiv t,\, t_d = 0) = \frac{(1 - t^2)(1 - t^3)(1 - t^4)}{(1 - t)^{54}} = 1 + 54\, t + 1484\, t^2 + 27665\, t^3 + \cdots,
\ee
where the power $54 = 2 \cdot (\dim \mathrm{Asym}_{USp(8)} - 1) = 2 \cdot 27$ counts the two copies of the traceless-antisymmetric. Setting $t_a = 0$ isolates the branch generated by the $64$ components of the bifundamental $\Pi$:
\be
\CH\CS_{FE_4}(t_a = 0,\, t_d \equiv t) = \frac{1 + 27\, t + 324\, t^2 + 2260\, t^3 + \cdots}{(1 - t)^{37}} = 1 + 64\, t + 2026\, t^2 + 42368\, t^3 + \cdots .
\ee

\paragraph{Chiral ring.}
The leading terms of the plethystic logarithm are
\be
\begin{aligned}
\mathrm{PL}\!\left[\CH\CS_{FE_4}\right](t_a, t_d) ={}&
\bigl(\chi_{(\mathrm{asym}_L,\mathbf{1})} + \chi_{(\mathbf{1},\mathrm{asym}_R)}\bigr)\, t_a
+ \chi_{(\mathbf{8},\mathbf{8})}\, t_d - t_a^2 - t_a^3 - t_a^4 \\
& - \bigl(\chi_{(\mathrm{asym}_L,\mathbf{1})} + \chi_{(\mathbf{1},\mathrm{asym}_R)}\bigr)\, t_d^2 - \chi_{(\mathbf{8},\mathbf{8})}\, t_a t_d + \cdots \\
\overset{\mathrm{unr}}{=}\;\,& 54\, t_a + 64\, t_d - t_a^2 - t_a^3 - t_a^4 - 54\, t_d^2 - 64\, t_a t_d + \cdots .
\end{aligned}
\ee
The positive leading terms identify the generators: the bifundamental $\Pi^{a,A}$ (contributing $\chi_{(\mathbf{8},\mathbf{8})}\, t_d$) and the two traceless-antisymmetric fields $A_L, A_R$, contributing $\chi_{(\mathrm{asym}_L,\mathbf{1})}\, t_a$ and $\chi_{(\mathbf{1},\mathrm{asym}_R)}\, t_a$. They satisfy the relations (with their PL contributions shown on the right):
\be
\begin{aligned}
\Omega_{ab}\, \Pi^{a,A}\Pi^{b,B} = 0 = \Omega_{AB}\, \Pi^{a,A}\Pi^{b,B},
&& -\bigl(\chi_{(\mathrm{asym}_L,\mathbf{1})} + \chi_{(\mathbf{1},\mathrm{asym}_R)}\bigr)\, t_d^2 \\
\mathrm{tr}_L A_L^k = \mathrm{tr}_R A_R^k, \quad k = 2, 3, 4,
&& -t_a^2 - t_a^3 - t_a^4 \\
\Omega_{ac}\, A_L^{\,ab}\, \Pi^{c,A} = \Omega_{BC}\, A_R^{\,AB}\, \Pi^{b,C},
&& -\chi_{(\mathbf{8},\mathbf{8})}\, t_a t_d .
\end{aligned}
\ee
The PL is not a finite sum, so these relations are not independent and the main branch of the moduli space is not a complete intersection.


\end{document}